\documentclass{IEEEoj}
\usepackage{cite}
\usepackage{multicol}
\usepackage{tablefootnote}

\usepackage{hyperref}
\hypersetup{
    bookmarks=true,         
    unicode=false,          
    pdftoolbar=true,        
    pdfmenubar=true,        
    pdffitwindow=false,     
    pdfstartview={FitH},    
    pdftitle={My title},    
    pdfauthor={Author},     
    pdfsubject={Subject},   
    pdfcreator={Creator},   
    pdfproducer={Producer}, 
    pdfkeywords={keyword1, key2, key3}, 
    pdfnewwindow=true,      
    colorlinks=true,       
    linkcolor=red,          
    citecolor=blue,
    filecolor=cyan,         
    urlcolor=magenta        
}
\usepackage[mathscr]{euscript}
\usepackage{amsmath,amssymb,amsfonts,mathtools}
\DeclareMathOperator{\sgn}{sgn} 

\usepackage{algorithmic}
\usepackage{color}
\usepackage{graphicx,psfrag,epsfig,epsf,latexsym,hhline}
\usepackage{pst-plot}
\usepackage{pstricks-add}
\usepackage{pifont}
\usepackage{amsthm}
\usepackage[linesnumbered,ruled,vlined]{algorithm2e}
\usepackage{array}
\usepackage{blindtext}
\usepackage{etoolbox}
\usepackage{comment}
\usepackage{epstopdf}
\usepackage{diagbox}
\usepackage{booktabs}

\usepackage{stmaryrd} 
\usepackage{algorithmic}
\usepackage{graphicx,color}
\usepackage{xcolor}
\usepackage{framed} 
\definecolor{shadecolor}{RGB}{220,220,220}

\usepackage{multirow}

\DeclareMathOperator*{\custommin}{min}
\DeclareMathOperator{\exteriorLambdahatboldfont}{{\boldsymbol{\hat{\Lambda}}}^{*}}
\DeclareMathOperator{\exteriorLambdaboldfont}{{\boldsymbol{\Lambda}}^{*}}
\DeclareMathOperator{\exteriorLambdaNoboldfont}{{\Lambda}^{*}}
\DeclareMathOperator*{\smax}{max^{\ast}}

\DeclareMathOperator{\bin}{bin}
\DeclareMathOperator{\supp}{supp}

\DeclareMathOperator{\w}{w}

\newcommand{\dm}{d_{\min}}
\newcommand{\wm}{w_{\min}}

\newcommand{\Ki}{\mathcal{K}_i}
\newcommand{\A}{\mathcal{A}}
\newcommand{\B}{\mathcal{B}}
\newcommand{\I}{\mathcal{I}}

\renewcommand{\H}{\mathcal{H}}
\renewcommand{\S}{\mathcal{S}}
\renewcommand{\P}{\mathcal{P}}

\newcommand{\J}{\mathcal{J}}

\newcommand{\M}{\mathcal{M}}

\newcommand{\C}{\mathcal{C}}
\newcommand{\Ci}{\mathcal{C}_i}

\newcommand{\E}{\mathcal{E}}

\newcommand{\bG}{\boldsymbol{G}}
\newcommand{\bB}{\boldsymbol{B}}
\newcommand{\bH}{\boldsymbol{H}}
\newcommand{\bP}{\boldsymbol{P}}

\newcommand{\by}{\boldsymbol{y}}

\newcommand{\bc}{\boldsymbol{c}}

\newcommand{\bzero}{\boldsymbol{0}}

\newcommand{\X}{\mathcal{X}}
\newcommand{\Y}{\mathcal{Y}}

\newcommand{\bzeros}{\boldsymbol{0}}

\newcommand{\bgi}{\boldsymbol{g}_i}
\newcommand{\bgj}{\boldsymbol{g}_j}
\newcommand{\bgm}{\boldsymbol{g}_m}

\newcommand{\bgh}{\boldsymbol{g}_h}
\newcommand{\bg}{\boldsymbol{g}}

\newcommand{\bGN}{\boldsymbol{G}_N}
\newcommand{\bu}{\boldsymbol{u}}
\newcommand{\bx}{\boldsymbol{x}}
\newcommand{\hu}{\hat{u}}

\newcolumntype{I}{!{\vrule width 3pt}}
\newlength\savedwidth
\newcommand\whline{\noalign{\global\savedwidth\arrayrulewidth
                           \global\arrayrulewidth 1.5pt}%
                  \hline
                  \noalign{\global\arrayrulewidth\savedwidth}}
\newlength\savewidth

\usepackage{textcomp}
\def\BibTeX{{\rm B\kern-.05em{\sc i\kern-.025em b}\kern-.08em
    T\kern-.1667em\lower.7ex\hbox{E}\kern-.125emX}}
\AtBeginDocument{\definecolor{ojcolor}{cmyk}{0.93,0.59,0.15,0.02}}
\def\OJlogo{\vspace{-14pt}\includegraphics[height=28pt]{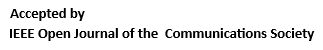}}
\begin{document}
\receiveddate{XX Month, XXXX}
\reviseddate{XX Month, XXXX}
\accepteddate{XX Month, XXXX}
\publisheddate{XX Month, XXXX}
\currentdate{XX Month, XXXX}

\title{Channel Coding Toward  6G:\\Technical Overview and Outlook}

\author{Mohammad Rowshan,~\IEEEmembership{Member,~IEEE}, Min Qiu,~\IEEEmembership{Member,~IEEE}, Yixuan Xie,~\IEEEmembership{Member,~IEEE}, Xinyi Gu,~\IEEEmembership{Student Member,~IEEE}, and Jinhong Yuan,~\IEEEmembership{Fellow,~IEEE}

\textit{(invited paper)}
}

\affil{School of Electrical Engineering and Telecommunications, University of New South Wales (UNSW), Sydney, Australia}
\corresp{CORRESPONDING AUTHOR: Mohammad Rowshan (e-mail: m.rowshan@unsw.edu.au).}
\authornote{The work was supported in part by the Australian Research Council (ARC) Discovery Project under Grant DP220103596, and in part by the ARC Linkage Project under Grant LP200301482.}
\markboth{Channel Coding for Beyond-5G \& 6G}{Rowshan \textit{et al.}}

\begin{abstract}
Channel coding plays a pivotal role in ensuring reliable communication over wireless channels. With the growing need for ultra-reliable communication in emerging wireless use cases, the significance of channel coding has amplified. Furthermore, minimizing decoding latency is crucial for critical-mission applications, while optimizing energy efficiency is paramount for mobile and the Internet of Things (IoT) communications. As the fifth generation (5G) of mobile communications is currently in operation and 5G-advanced is on the horizon, the objective of this paper is to assess prominent channel coding schemes in the context of recent advancements and the anticipated requirements for the sixth generation (6G).
In this paper, after considering the potential impact of channel coding on key performance indicators (KPIs) of wireless networks, we review the evolution of mobile communication standards and the organizations involved in the standardization, from the first generation (1G) to the current 5G, highlighting the technologies integral to achieving targeted KPIs such as reliability, data rate, latency, energy efficiency, spectral efficiency, connection density, and traffic capacity.
Following this, we delve into the anticipated requirements for potential use cases in 6G. The subsequent sections of the paper focus on a comprehensive review of three primary coding schemes utilized in past generations and their recent advancements: low-density parity-check (LDPC) codes, turbo codes (including convolutional codes), and polar codes (alongside Reed-Muller codes). Additionally, we examine alternative coding schemes like Fountain codes (also known as rate-less codes), sparse regression codes, among others. Our evaluation includes a comparative analysis of error correction performance and the performance of hardware implementation for these coding schemes, providing insights into their potential and suitability for the upcoming 6G era. Lastly, we will briefly explore considerations such as higher-order modulations and waveform design, examining their contributions to enhancing key performance indicators in conjunction with channel coding schemes.

\end{abstract}



\maketitle

\section{Introduction}\label{sec:intro}
Since the early 1980s, mobile communications have undergone a generational change almost every decade. With the emergence of new applications 
as well as the rapid technology advancements in hardware and computing power, 
the time difference between mobile communication generations is decreasing. Although 5G of mobile communications systems is already a commercial reality, there has been ongoing research in designing beyond 5G (B5G), and the sixth generation (6G) communication systems.

The emergence of new use cases drives the increasing demands on high data rates, high reliability, and low latency. To fulfill these visionary requirements, the next generation of wireless systems would require the allocation of new frequency bands and the development of new communication architectures.
The physical layer techniques will play an important role in the realization of the vision for 6G. Among them, channel coding, modulation, and signal waveforms are essential for the next generation of air interface design. In this paper, our focus is on the channel coding aspect of 6G networks.

Channel coding is essential in all communication systems to ensure reliable and efficient communications. It involves adding redundancy to the transmitted data in a controlled manner, allowing the receiver to detect and correct errors that occur during transmission and reception.
Wireless channels are noisy and unreliable. Therefore, data can be corrupted during transmission due to factors such as noise, interference, and channel fading. When errors in transmitted data cannot be corrected, retransmissions are required to obtain the correct data. Consequently, retransmissions can introduce a long delay, leading to poor end-user experience. Essentially, the role of channel coding is to ensure that the data received are the same as the data sent.  
Channel coding plays a critical role in improving the key performance indicators (KPIs) of a cellular network, including: 
\begin{itemize}
\item Reliability:
Channel coding is crucial to improve communication reliability by providing error detection and correction capabilities. It ensures that the data are reliably received in the presence of noise, interference, or fading conditions.
\item Throughput:
Channel coding affects throughput by influencing the error rate and reliability of data transmission. Efficient channel coding helps mitigate errors, reduce retransmissions, and improve overall throughput.
\item Latency:
Although channel coding adds processing delay due to encoding and decoding operations, efficient coding schemes contribute to minimizing the number of retransmissions. This reduction in retransmissions can help mitigate the overall latency.
\item Coverage:
Channel coding plays a critical role in extending coverage by enhancing the ability of the network to transmit data reliably over longer distances and in challenging radio environments.
\item Spectral Efficiency:
Efficient channel coding improves spectral efficiency by minimizing the impact of errors on the effective use of available frequency spectrum.
\item Energy Efficiency:
Channel coding can contribute to the reduction in power requirements at both the base station and user devices, as well as retransmissions,   consequently, resulting in an extension of battery life. 
\end{itemize}

As cellular networks have evolved from the second generation (2G) to the current 5G era, the path of channel coding has undergone significant advancements to address the challenges posed by varying channel conditions, increasing data rates, and the quest for ultra-reliable and low-latency communications. The coding schemes used in the 2G to 5G wireless communication standards are listed in TABLE \ref{tab:coding_1G_5G}.
\begin{table}[ht]
    \centering
    \caption{Coding Schmes from 1G to 5G.}
    \begin{tabular}{|l | l|} 
         \hline 
         Gen. &  Channel Coding \\ [0.5ex]  
         \hline\hline
         2G  & Cyclic Codes (FIRE/CRC), (Punctured) Convolutional Codes  \\
         \hline
         3G &  Convolutional Codes, Turbo Codes  \\
         \hline
         4G &  Tail Biting Convolutional Codes, Turbo Codes  \\
         \hline
         5G & Polar Codes, LDPC Codes  \\
         \hline
    \end{tabular}
    \label{tab:coding_1G_5G}
\end{table}

In this paper, we review the technologies involved in various generations of cellular networks to improve KPIs, with a particular focus on the channel coding schemes employed in each generation. We also review the envisioned requirements and the KPIs for 6G. Then, we turn our focus to channel coding schemes used in previous generations and cover new advances. We also consider other coding schemes developed recently. Then, we draw a conclusion. The list of contents is as follows.


{
\footnotesize
\setcounter{tocdepth}{1}
\tableofcontents
}

\begin{table*}[ht]
    {\centering
    \caption{Comparison of network specifications from 1G to 5G. 
    In this table, double comma ,, indicates repetition of the above.}
    \begin{tabular}{|p{0.4cm} p{0.9cm} p{1.0cm} p{0.95cm} p{1.7cm} p{1.7cm} p{0.8cm} p{1.8cm} p{2.2cm} p{2cm}|} 
         \hline 
         Gen. & 3GPP Release & Initial Rollout & System & Access Methodology & Peak Data Rate [bps] 
         & Latency [ms] & Modulation &  Freq. Band [MHz] & Carrier Spacing  [Hz]\\ [0.5ex]  %
         \hline\hline
         1G  & & & AMPS$^*$ & FDMA & &
         
         & FM (voice)\;\;\;\;\;\; FSK (control) & DL:824–849, UL:869–894 & 30k \\
         \hline
         2G  & 96 & 1991 & GSM & F/TDMA & 14.4k & & GMSK & DL:890–915, UL:935–960 & 200k \\
         2.5G & 97 & 2000 & GPRS & ,, & 171k
         &  & & & ,,\\
         2.75G & 98 & 2003 & EDGE & ,, & 236k & $<150$  & GMSK, 8-PSK & &  ,,\\
         \hline
         3G & 99,4 & 2003 & UMTS & DS-CDMA & 2M 
         & 150 & DL:QPSK, UL:BPSK & 890, 1900, 2100 & 5M \\
         3.5G & 5,6 & 2005 & HSPA & ,, & DL:14.4M, UL:5.76M & 100 & QPSK, 16QAM & ,, & ,,\\
         3.75G & 7 & 2008 & HSPA+ & ,, & DL:42M, UL:28M  & $<50$ & QPSK, 16QAM, 64QAM & ,, & ,,\\
         3.9G & 8,9 & 2009 & LTE & DL:OFDMA UL:SC-FDMA & DL:300M, UL:75M & $~10$ & ,, & 800, 1900, 2100 & 15k \\
         \hline
         4G & 10 & 2014 & LTE-A & DL:OFDMA UL:SC-FDMA & DL:3G, UL:1.5G & $<5$ & QPSK, 16QAM, 64QAM & 800, 1900, 2100, 2500, 2600 & 15k\\
         4.5G & 13 & 2016 & LTE-AP & ,, & DL:3G, UL:1.5G & $<5$ & ,, & 800, 1900, 2100, 2500, 2600, 3500 & 15k\\
         \hline
         5G & 15,16,17 & 2018 & NR & DL:CP-OFDMA, UL:DFT-S-OFDM  & DL:20G, UL:10G & $<1$ & QPSK, 16QAM, 64QAM, 256QAM, UL: $\pi/2$ BPSK & FR1:410-7125, FR2:24250-52600  &  FR1:15,30,60k, FR2:60,120,240k \\
         \hline
    \end{tabular}
    \label{tab:1G_5G}
    }\vspace{2pt}\\ 
    {$*$ There were various specifications implemented by different operators in different countries. See Section \ref{ssec:1G}.}\\
\end{table*}


\section{The 3rd Generation Partnership Project (3GPP)}
The 3rd generation partnership project (3GPP) is an umbrella term for a consortium of national (from Japan, USA, China, India and South Korea) or regional (from Europe) telecommunication standards organizations and other organizations to develop protocols for mobile telecommunications. As the name implies, this project was initially established to develop specifications for the 3rd generation (3G) of the cellular network based on 2G in December 1998 \cite{3gpp_about}. As the 3GPP standardization work is contribution-driven, companies can participate through their membership in organizational partners.
The 3GPP work is divided into three streams performed by three technical specification groups (TSGs), each of which consists of multiple working groups (WGs) \cite{3gpp_about}: Radio Access Networks (RAN), Services and Systems Aspects (SA), and Core Network and Terminals (CT). Among them, TSG RAN is responsible for the technical specifications of radio layer 1 (that is, physical layer, including multiplexing, channel coding, and error detection, the subject of this paper), layer 2, layer 3, etc.

\begin{figure*}
    \centering
    \includegraphics[width=1\textwidth]{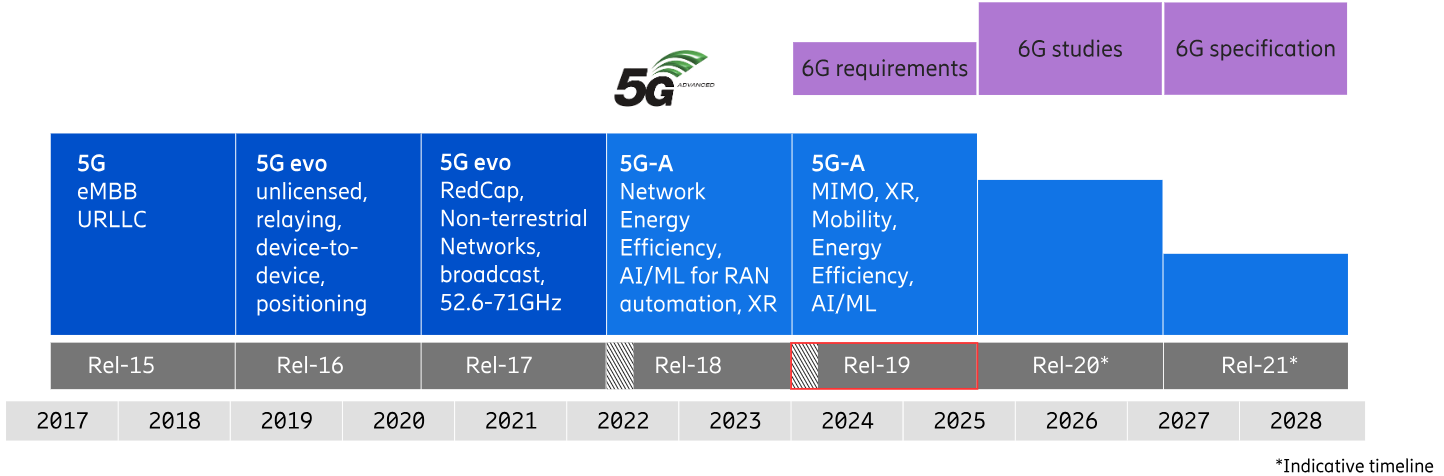}
    \caption{Roadmap for 5G standard releases towards 6G  \cite{ericsson_rel19}.}
    \label{fig:5G-A_6G_timeline}
\end{figure*}

3GPP standards are organized as \emph{Releases}. Each release consists of many individual technical specification and technical report documents, not only about new generations, but also including the revised documents of the older generations. For instance, Release 18 was concluded in late 2023 on 5G Advanced (5G-A). Fig. \ref{fig:5G-A_6G_timeline} illustrates the roadmap of 5G towards 6G based on 3GPP releases. 
These documents may go through many revisions over the years. The documents are available free of charge on 3GPP's portal \cite{3gpp_portal}.
Note that another collaboration between telecommunications associations called 3GPP2 existed until 2013. 3GPP2 was behind CDMA2000, the 3G upgrade to 2G cdmaOne networks used mainly in the United States. In this paper, our focus is only on the standards developed by 3GPP.

Furthermore, the requirements for future generations of international mobile telecommunications (IMT) are defined by the United Nations' International Telecommunication Union (ITU) radiocommunication sector (ITU-R).  For example, the requirements for 3G (called IMT-2000 \cite{series2003framework}), 4G (called IMT-Advanced \cite{series2009guidelines}), and 5G (called IMT-2020 \cite{series2015imt}) were issued in 1999, 2008, and 2015, respectively.
The technical specifications are then delegated to the 3GPP. The 3GPP prioritizes and divides specifications into releases according to the order in which new functionalities will be implemented and deployed in cellular networks.
The specifications in the 3GPP releases give operators and manufacturers confidence in their designs and investments.


\section{A Journey from 1G to 5G}\label{sec:1g_5g}
\begin{figure*}
    \centering
    \includegraphics[width=\textwidth]{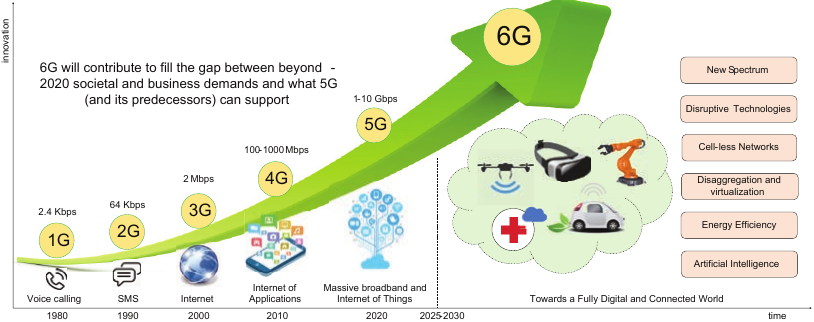}
    \caption{From 1G to 5G and beyond \cite{giordani2020toward}}
    \label{fig:1G-6G}
\end{figure*}

Since the first generation of mobile communication systems in the 1980s, different channel coding schemes and various technologies have been employed depending on the requirements and availability of the scheme(s) that meet the specifications. Table \ref{tab:1G_5G} summarizes the specifications of different generations of mobile networks given in the 3GPP releases.
As can be seen, every generation has a specific name in the column ``system" where the 5G is called ``New Radio" or NR. Fig. \ref{fig:1G-6G} illustrates the improvement in data rate, the use cases added over generations, and the innovations expected to achieve the 6G goals.

We will discuss the details and technologies that enable every generation to achieve these specifications as follows.

\subsection{First Generation (1G)}\label{ssec:1G} 
Although it was not called 1G at the time, the first generation of mobile cellular networks emerged in Japan in 1979 by Nippon Telegraph and Telephone (NTT) Corporation known as "Car Telephone Service" and later was rolled out by Ameritech in Chicago in 1983, known as the Advanced Mobile Phone System (AMPS) \cite{minoli2003telecommunications}. Soon after, other companies in different regions and countries launched their own networks, each following different specifications. AMPS was based on narrow-band analog frequency modulation (FM) with a usable audio frequency band of 300 Hz - 3 kHz with frequency division multiple access (FDMA) technology, where transmissions are separated in the frequency domain. Subscribers are assigned a pair of voice channels (forward and reverse) for the duration of their call. Analog cellular channels carried both voices using FM, operating between 824–849 MHz (from mobile stations to base stations) and 869-894 MHz (from base stations to the mobile stations). However, digital signaling based on binary frequency-shift keying (FSK)  was used to connect mobile customers to the base station as a control channel to place voice calls; these data were transmitted at 10 kbit/s. The 800 MHz band was split into a number of channels (395 voice, 21 control) with FDMA, where each channel was 30 kHz wide. The cells were able to cover a very large area (often a radius of $>$40 km depending on the terrain (land)). Although the large coverage had low infrastructure costs, it required the base station and the mobile terminal to transmit at high power to bridge large distances. 
Since voice signals were analog, they could not be efficiently compressed or protected by channel coding techniques like those used for digital data. This inherent constraint often resulted in compromised voice quality and security. 


\subsection{Second Generation (2G)}
The second generation of cellular networks, dubbed the global system for mobile communications (GSM), was developed by the European Telecommunications Standards Institute (ETSI) and commercially launched in Finland in 1991 \cite{minoli2003telecommunications}. 2G was an upgrade from 1G's analog radio signals to digital radio signals over a circuit-switched network optimized for full duplex voice telephony to provide a secure and reliable communication link. Furthermore, this generation adopted digital encrypted conversations using 64-bit A5/1 stream cipher and introduced data services such as short message services (SMS) and multimedia messaging services (MMS), Internet access, and also subscriber identity module (SIM) card to securely store the international mobile subscriber identity (IMSI) number and its related key, as well as to allow users to switch networks and handsets at will. The commonly used radio frequency band by the GSM is the 900 MHz and 1800 MHz bands.
GSM used a combination of time-division multiple access (TDMA) and frequency-division multiple access (FDMA). The bandwidth of 25 MHz (maximum) was divided into 124 carrier frequencies spaced 200 kHz apart using the FDMA scheme. However, carriers were divided into time slots, using the TDMA scheme, to be allocated to different users. Each time slot and the transmission made within it, called a GSM burst, lasted 0.577 ms. Every eight bursts were grouped and called a TDMA frame lasted approximately 4.615 ms, which formed a basic unit for the definition of logical channels, whereas one physical channel was one burst period in each TDAM frame. For synchronization purposes, the frames were organized into multiframes and the so-called superframes.

The modulation scheme in GSM is Gaussian minimum shift keying (GMSK).
The technologies involved in GSM expanded over time to include data communications, first by circuit-switched transport (as in the public switched telephone network, PSTN), then by packet-switched transport via the general packet radio service (GPRS), also known as 2.5G, which was commercially launched in 2000 and could be used for Internet connection.
In circuit switching, the bit delay is constant during a connection (as opposed to packet switching, where packet queues may cause varying and potentially indefinitely long packet transfer delays). The second extension of GSM called enhanced data rates for GSM Evolution (EDGE), also known as 2.75G, became operational in 2003 \cite{minoli2003telecommunications}.
In theory, the speed limit of GPRS is 115 kbps, but in most networks it is around 35 kbps.

EDGE as a single-carrier standard based on GSM can have a data rate up to 236 kbit/s (with an end-to-end latency of less than 150 ms) for 4 timeslots (for 8 timeslots, the theoretical maximum is 473.6 kbit/s) in packet-switching mode.
The theoretical maximum speed is 473 kbps for 8 timeslots, but it is typically limited to 135 kbps to conserve spectrum resources. This is four times as much traffic as GPRS. 
Therefore, EDGE could meet the 3G requirements defined by the ITU-R known as IMT-2000 \cite{series2003framework}.

Note that unlike 1G, in 2G the analogue voice signal was first compressed into a digital signal by source coding. As a result, the digitized signal can be protected by error correction codes. The channel codes used for GSM were block codes and convolutional codes with relatively simple structures and moderate performance. 

\begin{snugshade*}
EDGE, like GPRS, uses a rate adaptation algorithm that adapts the modulation and coding scheme (MCS) according to the quality of the radio channel, and thus adjusts the bit rate and robustness of data transmission. It also introduces {\em incremental redundancy} in which, instead of retransmitting disturbed packets, it sends more redundancy bits to be combined at the receiver. This increases the probability of correct decoding. EDGE uses 8 phase-shift keying (8PSK) in addition to GMSK, for the upper five of its nine MCSs. Since  every symbol carries 3 bits in 8PSK, EDGE can effectively triple the gross data rate offered by GSM. 

According to the ETSI's technical specification TS 45.003 \cite{3gppCoding2G}, channel coding in 2G is performed in two steps: In the first step, a block check sequence (BCS) is added for error detection. The BCS is either a 40-bit FIRE code or a 16-bit CRC with generator polynomial $g_{\mathrm{CRC16}}(D)=D^{16} + D^{12} + D^{5} + 1$. In the second step, four tail bits (TBs) are added and half-rate convolutional coding is performed for error correction. The convolutional codeword may be punctured to obtain the desired code rate. CS-1 to CS-4 are used to specify the length of the BCS and the puncturing rate (in CS-2 and CS-3 ) of the convolutional code, as shown in Table \ref{tbl:GSM_CSs}.  Note that in CS-4, where no convolutional coding is applied, we have the fastest but least reliable transmission, which is used for communications near a base transceiver station (BTS). Whereas the most robust coding scheme (CS-1) was used when the user equipment (UE) is further away from a BTS. Also, for most of the control channels, CS-1 is used. The bit rate per time slot increases from 8 kbps in CS-1 to 12, 14.4, and 20 kbps in CS-2, CS-3, and CS-4, respectively.
\end{snugshade*}
\setlength{\tabcolsep}{0.44em} 
\begin{table}[ht]
    \centering
    \caption{Convolutional coding parameters in GSM (2G)}
    \begin{tabular}{|c c l c c |} 
        \hline
        Scheme & Code rate & Generator Polynomial & BCS & Tail\\
        \hline
        CS-1 & 1/2 & $1+D^2+D^3$ & 40 & 4\\
        \hline
        CS-2 & $\sim$2/3 & $1+D^2+D^3+D^4+D^5$ & 16 & 4\\
        \hline
        CS-3 & $\sim$3/4 & $1+D^2+D^3+D^4+D^5+D^6$ & 16 & 4\\
        \hline
        CS-4 & 1 & & 16 & -\\
        \hline
    \end{tabular}
    \label{tbl:GSM_CSs}
\end{table}
\begin{snugshade*}
As GSM evolved and 8-PSK modulation was added to enhanced GPRS (EGPRS) in EDGE, the link quality control (LQC) was adapted to adjust the modulation and coding scheme (MCS) to the most suitable one as per the channel condition. There are nine MCSs (MCS-1 to MCS-9) for link adaptation, as shown in Table \ref{tbl:EGPRS_CSs}. The MCSs are divided into three families A, B, and C with the basic unit of payload 37 (and 34), 28, and 22 octets, respectively.
Different code rates within a family are achieved by transmitting a different number of payload units within one Radio Block. For families A and B, 1, 2, or 4 payload units are transmitted, for family C, only 1 or 2 payload units are transmitted.
Similar to GSM, a BCS is first added to every two payloads for error detection and then interleaved over two or four bursts, depending on the MCS. In the second step, six or twelve tail bits, depending on the MCS, are added, and then a rate-1/3 convolutional coding is performed for error correction, which is punctured to give the desired coding rate. The bit rate per time slot increases from 8.8 kbps in MCS-1 to 59.2 kbps in MCS-9.
\end{snugshade*}
\begin{table}[ht]
    \centering
    \caption{Convolutional coding parameters in EDGE (2.75G)}
    \begin{tabular}{|c c c c c c |} 
        \hline
        Scheme & Code rate & BCS & Tail Bits & Modulation & Family\\
        \hline
        MCS-1 & 53/100 & 12 & 6 & GMSK & C\\
        MCS-2 & 33/50 & 12 & 6 & & B\\
        MCS-3 & 17/20 & 12 & 6 & & A\\
        MCS-4 & 1 & 12 & 6 & & C\\
        \hline
        MCS-5 & 37/100 & 12 & 6 & 8-PSK & B\\
        MCS-6 & 49/100 & 12 & 6 &  & A\\
        MCS-7 & 19/25 & 2$\times$12 & 2$\times$6 &  & B\\
        MCS-8 & 23/25 & 2$\times$12 & 2$\times$6 &  & A\\
        MCS-9 & 1 & 2$\times$12 & 2$\times$6 &  & A\\
        \hline
    \end{tabular}
    \label{tbl:EGPRS_CSs}
\end{table}


\subsection{Third Generation (3G)}
The specifications of the third generation of the cellular network were defined in 3GPP Release 99 to meet the requirements of ITU-R IMT-2000. This generation was called the universal mobile telecommunications system (UMTS).
Unlike GSM and evolved GSM, it uses wideband code-division multiple access (W-CDMA) radio access technology to offer greater spectral efficiency and bandwidth to mobile network operators, hence UMTS is comparable with the CDMA2000 standard based on cdmaOne technology widely used in the United States as 2G. The modulation technique employed in UMTS is quadrature phase shift keying (QPSK).
3G added the 2100 MHz frequency band with 5 MHz bandwidth to the 2G frequency bands.

High-speed packet access (HSPA), also known as 3.5G, introduced with 3GPP Release 5, is a combination of uplink (known as HSUPA) and downlink (known as HSDPA) protocols that allow UMTS-based networks to have higher data rates and capacity. The key features of HSDPA are shared channel and multi-code transmission, higher-order modulation, QPSK, 16-quadrature amplitude modulation (16QAM), short transmission time interval (TTI), fast link adaptation and scheduling, and fast hybrid automatic repeat request (HARQ) with incremental redundancy, making retransmissions more effective. Another new feature is the high-speed medium access protocol (MAC-hs) in the base station.

Evolved high-speed packet access (HSPA+), also known as 3.75G, was introduced in 3GPP Release 7, further increased data rates (up to 42 Mbps in the downlink and 28 Mbps in the uplink with a single 5 MHz carrier) by adding 64QAM modulation, $2\times2$ multiple input multiple output (MIMO) technology.  With a dual carrier or dual cell (DC) where two base stations with different carrier frequencies are employed to communicate with user equipment, the data rate can be doubled. Note that higher-order modulation schemes such as 64QAM within a limited bandwidth (as bandwidth is a scarce and expensive resource) require higher receiver signal-to-noise or high signal-to-interference ratios, e.g., available in small cells or for user equipment (UE) close to a base station (BS, in the standard documents, it is called NodeB, eNB, or gNB, depending on the generation).


Originally, long term evolution (LTE), also known as 3.9G or pre-4G, was conceived as an IP-based wireless system used purely for carrying data traffic. Network carriers were supposed to provide voice communication through their concurrent 2G/3G networks or using VoIP. However, by popular request, Voice over LTE (VoLTE) was a standardized system for transferring voice traffic over LTE. Currently, the availability of Voice over LTE (VoLTE) depends on the carrier implementation. Theoretically, LTE networks should provide wireless data downlink speeds of up to 300 Mbps and uplink speeds of up to 75 Mbps \cite{holma2011lte}. Note that LTE was originally marketed as 4G, but it did not meet all 4G requirements, as defined by the ITU. Therefore, it was considered pre-4G, although it is still widely referred to as 4G.

The multiple access methodology used in LTE was orthogonal frequency-division multiple access (OFDMA) for the downlink and single-carrier frequency-division multiple access (SC-FDMA) for uplink. In SC-FDMA, each symbol is precoded by a discrete Fourier transform (DFT) before mapping to subcarriers; hence, unlike OFDMA, multiple subcarriers carry each data symbol. As a result, SC-FDMA offers (1) lower peak-to-average power ratio (PAPR) that benefits the mobile terminal in terms of transmit power efficiency and lower cost of the power amplifier, and (2)
spreading gain or frequency diversity gain in a frequency selective channel, hence called DFT-spread OFDM.

The capacity improvement of 3G over 2G is significant, which largely benefit from soft frequency reuse, fast power control, and turbo codes \cite{Xu22}. Turbo codes were proposed in 1993 \cite{397441}. They are the first coding schemes that have been practically demonstrated to approach the Shannon limit of a point-to-point channel with moderate decoding complexity. Because of this, turbo codes soon became the standard channel coding schemes for 3G. The invention of turbo codes was considered a major breakthrough in coding and communication theory.


\begin{snugshade*}
According to the 3GPP's technical specification TS 25.222 \cite{3gppCoding3G}, to be able to detect any errors that cannot be corrected by channel coding, cyclic redundancy check (CRC) bits of size 24, 16, 12, 8, or 0 (depending on the block size) are added to every transport block.
To form data blocks suitable for channel coding, the transport blocks in a transmission time interval (TTI) are serially concatenated. TTI is the time unit from set $\{$5 ms, 10 ms, 20 ms, 40 ms, and 80 ms$\}$ for the base station, known as eNodeB in 3G and 4G, to schedule UL and DL data transmissions.
After concatenation, if the length $X$ is larger than the maximum size of a data block, then the concatenated block is segmented and zero-padded to suit the appropriate channel coding scheme. The available channel coding schemes are convolutional coding, turbo coding, or no coding. The maximum size of a data block is $Z=504$ for convolutional coding, $Z=5114$ for turbo coding, and unlimited for no coding case. If the length of the concatenated block is not multiple of $Z$, filler bits, by default zeros, are added to the beginning of the first block. Hence, $K=\lceil X/Z \rceil$. Moreover, filler bits are added when the blocklength is less than 40 and turbo coding is employed. Hence, $K=40$ in this case.

Convolutional coding with constraint length 9 is performed by generator polynomials (in octal) $G_0=561$ and $G_1=753$ for code rate 1/2 and $G_0=557$, $G_1=663$, and $G_2=711$ for code rate 1/3. Before encoding, 8 zero-value tail bits are added. For turbo coding with code rate 1/3, a parallel concatenated convolutional coding with two constituent encoders and transfer function $G(D)=[1, g_1(D)/g_0(D)]$, where $g_0(D)=1+D^2+D^3$ and $g_1(D)=1+D+D^3$ are employed.

The coding parameters used in 3G are listed in Table \ref{tbl:UMTS_CSs}. Convolutional coding is used for short packets in control channels' signaling, while turbo coding is employed for longer packets. To realize a variable transmission rate and adjust the amount of data to fit the radio frames, rate-matching is performed as well.
\end{snugshade*}

\begin{table}[ht]
    \centering
    \caption{Coding parameters in UMTS (3G)}
    \begin{tabular}{|c c c|} 
        \hline
        Coding Scheme & Code Rate & Data Block Size, $K$\\
        \hline
        Convolutional Coding & 1/2 \& 1/3 & $\leq 504$\\
        \hline
        Turbo Coding & 1/3 & $[40,5114]$\\
        \hline
        No channel coding & 1 & unlimited\\
        \hline
    \end{tabular}
    \label{tbl:UMTS_CSs}
\end{table}

\subsection{Fourth Generation (4G)}
In 2008, the framework and objectives for the next generation of mobile communications were recommended in the International Mobile Telecommunications-Advanced (IMT-Advanced) by the ITU-R  \cite{series2009guidelines} marketed as 4G. Recall from the previous section that LTE is considered pre-4G as it did not meet all the 4G requirements defined in IMT-Advanced. 
The 4G specifications were determined in 3GPP Release 10.
This generation, known as LTE-Advanced (LTE-A), employed additional spectrum and frequency bands, namely around 600 MHz, 700 MHz, 1.7/2.1 GHz, 2.3 GHz, and 2.5 GHz. The LTE channel bandwidth can be 1.4, 3, 5, 10, 15, or 20 MHz. Taking into account about 10\% of the bandwidth as a guard band and 15 kHz for OFDM frequency spacing, the effective bandwidth for the 20 MHz bandwidth will be 18 MHz resulting in 18 MHz/15kHz = 1200 subcarriers. Now, since a physical resource block (PRB) consists of 12 consecutive subcarriers for one time slot (0.5 ms), then 1200/12 = 100 PRBs will be available for 20-MHz bandwidth.

Three technologies from LTE-Advanced - namely carrier aggregation, up to $4\times4$ MIMO in uplink and up to $8\times8$ MIMO in downlink, and 256QAM modulation in downlink - if used together and with sufficient aggregated bandwidth, can deliver maximum peak downlink speeds approaching 1 Gbps. Such networks are often described as ‘Gigabit LTE networks’ mirroring a term that is also used in the fixed broadband industry.
LTE-A carrier aggregation (CA) is a key technique used to increase the data rate and the capacity in both uplink and downlink by combining up to 5 individual carriers, called component carriers (CC), either in the same or different bands. Note that a similar technique was actually employed in HSPA+ named dual carrier (DC), as discussed earlier in the 3G section. The carrier aggregation (with 3 downlink carriers and 2 uplink carriers) is supported in both duplex schemes; frequency division duplex (FDD) and time division duplex (TDD). By considering the aggregation of up to 5 carriers, the maximum bandwidth of 20 MHz in LTE increases to 100 MHz in LTE-A.
Note that mobile worldwide interoperability for microwave access (Mobile WiMAX) based on the IEEE 802.16m standard (a.k.a. WirelessMAN-Advanced) intended to compete with LTE-A as a candidate for 4G cellular networks by fulfilling the ITU-R IMT-Advanced requirements. However, it was not well established at the time.

LTE-Advanced Pro (LTE-AP), also known as 4.5G, was 3GPP Release 13 as an evolution of the LTE-A cellular standard that supports data rates in excess of 3 Gbps using 32-carrier aggregation. It also introduces the concept of License-Assisted Access, which allows for the sharing of licensed and unlicensed spectrum. Similarly to LTE-A, the aggregation of up to 32 carriers in LTE-AP increases the maximum bandwidth of 20 MHz in LTE up to 640 MHz.
Furthermore, it incorporates several new technologies that were later used in 5G, such as 256-QAM, Massive MIMO, LTE-Unlicensed, and LTE IoT. Hence, 4.5G facilitated the early migration of existing networks to enhancements promised with the full 5G standard. Recall that 2.75G and 3.9G played a similar role in migrations from 2G to 3G and from 3G to 4G, respectively.
To reduce energy consumption, 4G networks were designed with improved network optimization techniques, including strategies such as expansion of the cell range, optimization of the sleep mode, and switching of base stations to minimize power consumption while maintaining coverage and quality of the network.

Channel coding of LTE/LTE-A reuses the turbo codes from 3G traffic channels and convolutional codes from 3G physical control. In contrast to the 3G channel coding, turbo codes were enhanced with improved performance and lower decoding complexity. In addition, tail-biting convolutional codes were introduced to reduce the overhead as previously zero tail bits were added for termination.

\begin{snugshade*}The channel coding scheme used for the transport blocks in 4G, which is specified in Evolved Universal Terrestrial Radio Access (E-UTRA), is turbo coding with code rate of $R=1/3$. This scheme incorporates two 8-state constituent encoders (identical with 3G) and a contention-free quadratic permutation polynomial (QPP) interleaver. Furthermore, trellis termination is used for turbo coding. Prior to turbo coding, transport blocks are segmented into byte-aligned segments with a maximum information block size of 6144 bits. 

According to the 3GPP's technical specification TS 36.212 \cite{3gppCodingLTE}, the transport blocks (TBs) are protected by appended CRC bits with a length of 24 (calculated by type-A generator polynomial, CRC24A). The detected error by CRCs is reported to higher layers. The transport block is segmented into code blocks (CBs) of no larger than 6144 bits, where each CB is protected by type-B 24-bit CRC (CRC24B). If the code block is smaller than 40 bits, filling bits are appended. The CRC bits for control channels are shorter as broadcast channel (BCH) and physical downlink control channel (PDCCH) use 16-bit CRC, while physical uplink control channel (PUCCH) employs 8-bit CRC. The channel coding scheme for the aforementioned control channels is tail-biting convolutional coding (TBCC) whereas the transport channels (TrCH) use turbo coding.
Tail-biting convolutional coding with constraint length 7 is performed by generator polynomials (in octal) $G_0=133$, $G_1=171$, and $G_2=165$ for the code rate 1/3. The details of the coding schemes used in 4G are summarized in Table \ref{tbl:E-UTRA_CSs}. 
\end{snugshade*}

\begin{table}[ht]
    \centering
    \caption{Coding parameters in E-UTRA (4G)}
    \begin{tabular}{|c c c|} 
        \hline
        Coding Scheme & Code Rate & Data Block Size, $K$\\
        \hline
        Tail-biting Convolutional Coding & 1/3 & \\ 
        \hline
        Turbo Coding & 1/3 & $[40,6144]$\\
        \hline
        Block Code & 1/16, variable & \\
        \hline
        Repetition Code & 1/3 & \\
        \hline
    \end{tabular}
    \label{tbl:E-UTRA_CSs}
\end{table}

\subsection{Fifth Generation (5G)}
\begin{figure}[ht]
    \centering
    \includegraphics[width=\columnwidth]{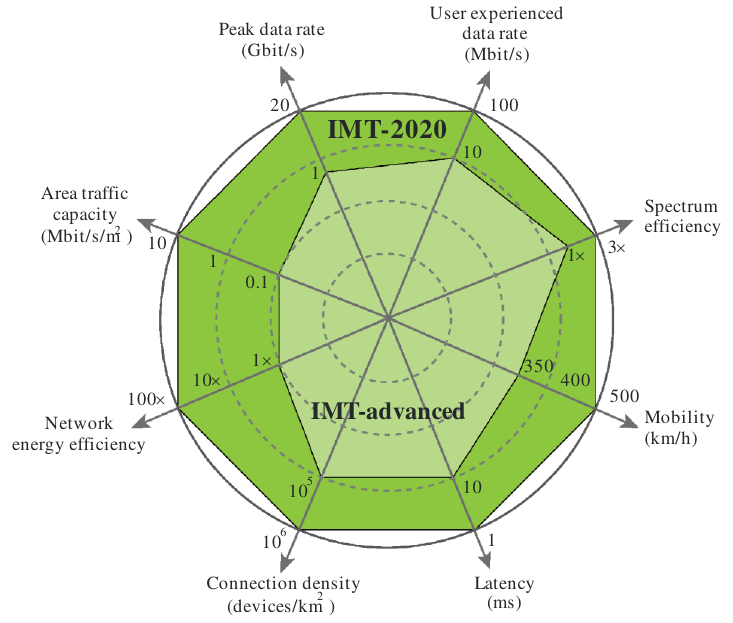}
    \caption{Key capabilities from IMT-Advanced (for 4G) to IMT-2020 (for 5G)\cite{series2015imt}.}
    \label{fig:5G_reqs}
\end{figure}

The ITU framework and objectives for 5G cellular networks, devices, and services, were recommended in the IMT-2020 vision by ITU-R in 2015 \cite{series2015imt}, to address the requirements of emerging applications. Fig. \ref{fig:5G_reqs} summarizes the requirements in terms of 8 KPIs. As can be seen, throughput (the peak data rate) of up to 20 Gbps in downlink and 10 Gbps in uplink aims to answer the growing demand for high data rates. This is a significant increase compared to IMT-Advanced (requirements for 4G/LTE-A in Release 14), which offers 1 Gbps in downlink and 50 Mbps in uplink. The latency of 1 ms (compared to 30 - 50 ms in 4G) will allow near real-time responses, and the connection density of 1000 devices per square kilometer (100 times more than 4G) will meet the growing demand for IoT devices and sensors.

5G specifications have been integrated into 3GPP releases 15, 16, and 17. 
Full deployment of the 5G capabilities defined in IMT-2020 requires the implementation of totally new networks, significant investments by operators, and considerable elapsed time to enable a full rollout. To ease the migration path, 3GPP defined 5G NR non-stand alone (NSA) in release 15 leveraging existing LTE infrastructure. The throughput of existing macro cells can be increased by adding additional MIMO layers, and the spectrum can be dynamically shared between 4G LTE and 5G NR. Operators can use existing spectrum in the so-called “MIMO sweet spot”, around 3.5GHz.


The 5G frequency spectrum is divided into frequency range 1 (FR1) spanning from 450 MHz to 7.125 GHz (previously up to 6 GHz, hence this range is still known as sub-6 GHz), and millimeter-wave (mm-Wave) range (FR2) spanning from 24.25 GHz to 52.6 GHz, as well as unlicensed spectrum. 
The maximum bandwidths in sub-6 GHz and the mm-Wave range are 100 MHz and 400 MHz, respectively. There are 7 subcarrier spacings as $\Delta f=2^\mu \cdot 15$ kHz, $\mu=0,\dots,6$. The subcarrier spacing of $\Delta f=2^0\cdot15$ (the same as LTE) and $2^1\cdot15$ kHz is used only in sub-6 GHz, and the subcarrier spacing of $2^3\cdot15$ kHz is used only in the mm-Wave range, while 60 kHz can be used in both ranges.
Thus, 5G includes the previous cellular spectrum and further expands it. The additional spectrum addresses the physical limitations associated with throughput and bandwidth.
4G band plans accounted for 5 MHz to 20 MHz of bandwidth per channel, whereas the 5G FR1 allows for 5 to 100 MHz of bandwidth per channel. As bandwidth is directly proportional to maximum throughput, the 5 times increase in bandwidth relates to roughly a 5 times increase in throughput. Furthermore, 3GPP Release 15 established new waveforms and the addition of $\pi/2$ binary phase shift keying (BPSK) as a modulation method in the uplink with the aim of further reducing the peak-to-average power ratio (PAPR) and increasing the power efficiency of the RF amplifier at lower data rates. Additional waveforms are discrete Fourier transform spread orthogonal frequency division multiplexing (DFT-S-OFDM) for FR1 and the cyclic prefix OFDM (CP-OFDM) for FR2.

\begin{figure}[ht]
    \centering
    \includegraphics[width=\columnwidth]{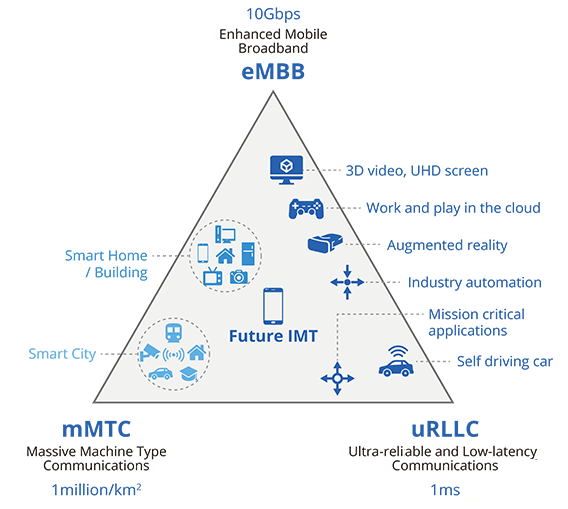}
    \caption{5G use cases \cite{series2015imt}.}
    \label{fig:5G_usgaes}
\end{figure}

The key technologies reviewed above were introduced to meet the requirements of new use cases. 
The 5G services and applications can be classified into three main use scenarios: enhanced mobile broadband (eMBB), ultra-reliable low-latency communications (URLLC) and massive machine-type communications (mMTC), as illustrated in Fig. \ref{fig:5G_usgaes}.
\begin{itemize}
\item eMBB focuses on delivering significantly higher data rates, increased network capacity, and improved user experiences compared to previous generations of mobile networks. It enables applications that require high-speed and high-bandwidth connectivity, such as ultra-high-definition video streaming, virtual reality (VR), augmented reality (AR), online gaming, and immersive multimedia experiences. eMBB provides users with ultra-fast downloads, seamless video streaming, and enhanced browsing capabilities. However, note that the latency requirements of eMBB depend on the specific service types. For example, stream-type services require a high data rate and can tolerate a delay of $50$ to $100$ ms. Interactive services, on the other hand, have a more stringent requirement for latency, e.g., $5$ to $10$ ms.

\item URLLC in 5G is a set of technologies and features designed to provide extremely reliable and low-latency communication for mission-critical applications. URLLC ensures that time-sensitive and critical data are transmitted with ultra-low latency and high reliability. Some of the key technologies involved in URLLC are network slicing, edge computing, beamforming, and massive MIMO, quality-of-service (QoS) mechanisms, redundancy, and error correction, time-sensitive networking (TSN), and network synchronization.

\item MTC in 5G involves various technologies and features specifically designed to provide efficient massive connectivity for IoT devices. It is needed to support a large number of devices that transmit data sporadically and without coordination with other devices or the network, that is, asynchronously. This can lead to interference between devices and the network, leading to a reduction in the data rate and reliability of communications. 
Furthermore, since MTC devices are often powered by batteries or energy harvesting devices, they need to have low power consumption, which requires a low computational complexity transmitter/receiver, so that they can operate for long periods of time without having to be recharged. Applications and services of MTC include both low-rate and high-rate data collection, and some delay-insensitive control-type services. Here are some key technologies involved in MTC: 
\begin{itemize}
\item NB-IoT: 
NB-IoT is a low-power, wide-area (LPWA) technology optimized for IoT applications. It enables long-range communication, extended battery life, and deep indoor coverage. NB-IoT operates in licensed spectrum, offering improved security and quality of service for IoT devices. 
\item LTE-M (Long-Term Evolution for Machines): 
LTE-M is another LPWA technology in 5G designed for IoT applications. It provides higher data rates compared to NB-IoT, making it suitable for applications that require more bandwidth. LTE-M offers improved mobility support, voice communication capabilities, and reduced power consumption. 
\item Device-to-device (D2D) communication: 
D2D communication in 5G enables direct communication between nearby IoT devices without routing through the network infrastructure. It improves communication efficiency, reduces latency, and conserves network resources. D2D communication is particularly beneficial for applications that require local coordination and peer-to-peer interaction. 
\end{itemize}
\end{itemize}



\begin{figure}[ht]
    \centering
    \includegraphics[width=\columnwidth]{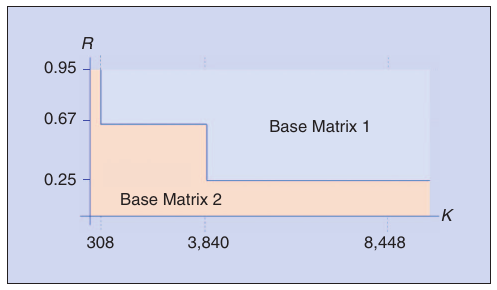}
    \caption{Key capabilities from IMT-Advanced to IMT-2020 \cite{hui2018channel}.}
    \label{fig:ldpc_K_R}
\end{figure}

\begin{snugshade*}
5G supports new channel coding schemes. According to 3GPP's technical specification TS 38.212 \cite{3gppCoding5G},
5G incorporates Low-Density Parity-Check (LDPC) codes for data channels including physical downlink shared channel (PDSCH) and physical uplink shared channel (PUSCH), replacing turbo codes in 4G. Likewise, for control channels (to protect the downlink control information (DCI), the uplink control information (UCI), and the system information in the physical broadcast channel (PBCH)), Polar codes are introduced in lieu of the TBCC used in 4G. The key advantages of LDPC codes (compared to turbo codes) are improved performance with very low error floors, reduced decoding complexity and latency, better power and area efficiency, and support of multi-Gbps data rates. Polar coding yields better performance at moderate payload sizes (in the order of $K \leq 250$ bits); however, it comes at the cost of higher complexity compared to TBCC. Note that the minimum supported payload size $K$ in 5G polar codes is 12 bits. For the control information sequence $K<12$ in the uplink, the short block codes employed in 4G, namely, repetition codes, simplex codes and Reed-Muller codes, are used.

Targeting good performance and decoding latency for the full range of code rates and information block sizes, 5G supports two LDPC base matrices (which are, in turn, constructed using a photograph, see Section \ref{sec:ldpc} for more information). Base matrix 1 was optimized for large information block sizes $K$ and high code rates $R\geq 1/3$. On the other hand, base matrix 2 is suitable for small information block sizes and lower code rates than base matrix 1, down to $1/5$, which is lower than the code rate of 4G turbo codes (where for lower than $1/3$, repetition is used).
Fig. \ref{fig:ldpc_K_R} illustrates the switching points between the two base matrices based on the pair of $(K,R)$. For $K$ larger than the maximum information block size of the base matrices (shown in Table \ref{tbl:5G_coding_params}), code block segmentation is used.

5G polar codes employ 6, 11, and 24 CRC bits for error detection with the corresponding generator polynomials $g_{\mathrm{CRC6}}(D)=D^6+D^5+1$, $g_{\mathrm{CRC11}}(D)=D^{11}+D^{10}+D^9+D^5+1$, and $g_{\mathrm{CRC24C}}(D)= D^{24}+D^{23}+D^{21}+D^{20}+D^{17}+D^{15}+D^{13}+D^{12}+D^8+D^4+D^2+D+1$. However, the 5G LDPC codes use 16 and 24 CRC bits with generator polynomials $g_{\mathrm{CRC16}}(D)=D^{16}+D^{12}+D^5+1$, $g_{\mathrm{CRC24A}}(D)=D^{24}+D^{23}+D^{18}+D^{17}+D^{14}+D^{11}+D^{10}+D^7+D^6+D^5+D^4+D^3+D+1$, and $g_{\mathrm{CRC24B}}(D)=D^{24}+D^{23}+D^6+D^5+D+1$.

Table \ref{tbl:5G_coding_params} lists the parameters of the coding schemes in the 5G standard.
\end{snugshade*}

\begin{table}[ht]
    \centering
    \caption{Channel coding schemes in 5G}
    \begin{tabular}{| c | c | p{24pt} | p{32pt} | p{30pt}| p{49pt}|} 
        \hline
         & Channel & Coding Scheme & Info. block, $A$ & CRC+PC & Encoded block, $G$\\
        \hline
        \multirow{3}{*}{\rotatebox{90}{Downlink}} & PDSCH & LDPC & [80,8448] & 16,24B & $[384,6480]$\\ 
         & PDCCH & Polar & $[12, 140]$ & 24C & $[A\!+\!24,8192]$\\ 
         & PBCH & Polar & 32 & 24C & 864\\ \hline

        \multirow{4}{*}{\rotatebox{90}{Uplink}} & PUSCH & LDPC   & $[48,3840]$ & 24A,24B & $[192,15360]$ \\ 
         & \multirow{2}{*}{PUCCH} & \multirow{2}{*}{Polar} & $[12,19]$ & 6+3 & $[A\!+\!9,8192]$\\   &  &  & $[20,1706]$ & 11 & $[A\!+\!11,16385]$\\ \hline
    \end{tabular}
    \label{tbl:5G_coding_params}
\end{table}

In a recent milestone, 3GPP finalized Release 18, signaling the advent of 5G Advanced. The features integrated into 5G Advanced are poised to elevate the performance of 5G networks, providing augmented support for services such as extended reality (XR), indoor positioning, and non-terrestrial networks. 
Building upon the foundation laid by 5G Advanced, Release 19 will focus on enhancing performance and addressing vital requirements in commercial 5G deployments. The evolution of 5G Advanced is expected to progress throughout this decade. Concurrently, standardization efforts for 6G are projected to intensify with Releases 20 and 21 commencing in 2025, aligning with the timeline depicted in Fig. \ref{fig:5G-A_6G_timeline}.

Table \ref{tab:KPI_tech} summarizes the role of technologies introduced in every generation, which were discussed in this section, in improving the KPIs of the mobile communication network. Note that the improvement is considered relative to the previous generation. As can be seen, the major contribution of channel coding is improving channel reliability in a wireless link, although a low-complexity decoder can also improve energy efficiency and latency. Furthermore, higher order modulation, adaptive modulation, and coding schemes, and HARQ, in collaboration with channel coding, can influence other KPIs except for connection density.

\setlength{\tabcolsep}{0.44em} 
\begin{table}[h]
\small
\caption{The major contribution of the technologies used in 3G to 5G standards to the KPIs. }
\label{tab:KPI_tech}
\begin{tabular}{|l|l|l|l||l|l|l|l|l|l|l|}
\hline
                       &  \rotatebox{90}{3G} & \rotatebox{90}{4G} & \rotatebox{90}{5G} & \rotatebox{90}{Reliability} & \rotatebox{90}{Data Rate} & \rotatebox{90}{Connection Density} & \rotatebox{90}{Spectral Efficiency} & \rotatebox{90}{Energy Efficiency} & \rotatebox{90}{Latency} \\ \hline
Channel Coding    & X  & X  & X  & X         &             &                    &                    &                     &                   \\ \hline
Adaptive Modul. \& Coding & X  & X  & X  & X         & X           &                    & X                   &                     &                \\ \hline
Higher Order Modulation  &  X & X  & X  &   & X           &                    & X                   &                     &               \\ \hline
HARQ                    & X  & X  & X  & X         &             &                    &                     &                     & X               \\ \hline\hline
Packet Switching        & X  &    &    &           &             &                    &                     & X                   & X                \\ \hline
IP-Based Architecture   &    & X  & X  &           &             &                    &                     &                     & X               \\ \hline
Enhanced Packet Core    &    & X  & X  &           &             &                    &                     &                     & X               \\ \hline\hline
Discontinuous Tx/Rx     &    & X  & X  &           &             &                    &                     & X                   &                 \\ \hline
Power-Saving Techniques &    &    & X  &           &             & X                  &                     & X                   &                \\ \hline\hline
New Network Architecture    & X  &   X &  X  &           &             &     X               &                     & X                   &                 \\ \hline
QoS Management          &    & X  & X  &           &             &                    &                     &                     & X               \\ \hline
Dynamic Spectrum Sharing&    &    & X  &           &             &                    &                     & X                   &                 \\ \hline
Network Optimization    &    &    & X  &           &             &                    &                     & X                   &                 \\ \hline
Edge Computing          &    &    & X  &           &             &                    &                     &                     & X                 \\ \hline
Network Slicing         &    &    & X  & X         &             & X                  &                     &                     &                   \\ \hline\hline
Small Cells             &    & X  & X  &    X     &    X        & X                  &                     &                     & X               \\ \hline
Network Densification   &    &    & X  &           &             &                    &                     & X                   &                   \\ \hline\hline
Wideband CDMA           & X  &    &    & X         & X           & X                  & X                   &                     &                   \\ \hline
OFDMA                   &    & X  & X  &   X        &  X           & X                  &   X                 &                     &                   \\ \hline
Carrier Aggregation     &    & X  & X  &           & X           &                    &                     &                     &                   \\ \hline\hline
MIMO                    &    & X  & X  & X         & X           & X                  & X                   &                     &                   \\ \hline
Beamforming             &    &  X  & X  & X         &             &                    &                     &                     &                   \\ \hline
mm-Wave     &    &    & X  &           & X           &                    &                     &                     &                   \\ \hline
Massive MIMO            &    &    & X  &     X      &            X & X                  &             X       &                     &                   \\ \hline\hline
Narrowband IoT          &    &    & X  &           &             & X                  &                     & X                   &                \\ \hline
LTE-M                   &    &    & X  &           & X           & X                  &                     & X                   &                \\ \hline
\end{tabular}
\end{table}

\section{6G Use Cases, Requirements and Challenges}\label{sec:6G_reqs}
\begin{figure}
    \centering
    \includegraphics[width=\columnwidth]{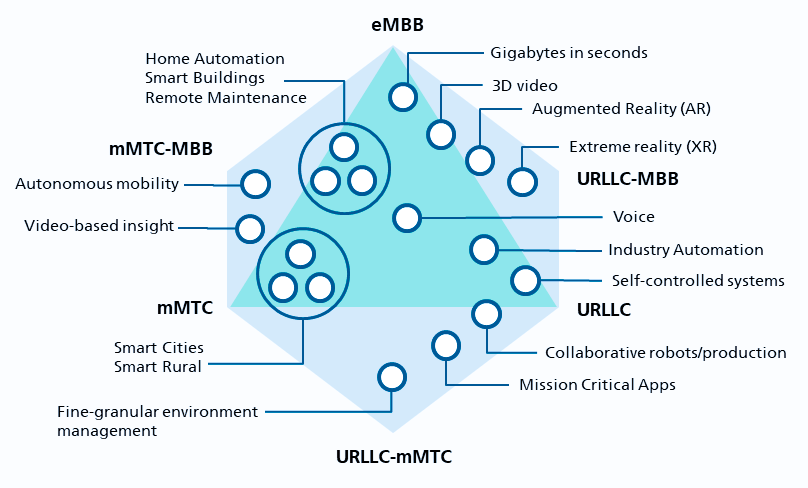}
    \caption{6G use cases \cite{fraunhofer_6g}.}
    \label{fig:6G_usgaes}
\end{figure}

Since the early discussions on 6G, various organizations and projects have been identifying the potential 6G use cases. Among them, the 6G Flagship research program led the way, publishing a white paper in 2019 \cite{latva2019key} that identified the first set of use cases for the various types of devices expected at the time of 6G commercialization. The major communications companies followed suit, releasing their 6G white papers since 2020. Collaborative European projects also began in 2021, including Hexa-X, which aims to develop a 6G vision and an intelligent fabric of technology enablers. One of Hexa-X's first deliverables was a comprehensive set of 6G use cases \cite{d2021expanded}. Other organizations, such as the Next Generation Mobile Network Alliance (NGMN) \cite{ngmn_6g}, have also published white papers on 6G use cases.
The use cases suggested for 6G are shown in Fig. \ref{fig:6G_usgaes}.
While the potential use cases for 6G technology are still in the early stages of development, the current deployment of 5G networks can help us to identify a set of highly promising directions that current commercial networks cannot meet.

The classification of 6G use cases illustrated in Fig. \ref{fig:6G_usgaes} aims to change the way we think about network transformation. This does not exclude the further evolution of 5G use cases, but rather complements them with a longer-term perspective that addresses the communication needs of 2030s.
Starting with 4G MTC and 5G URLLC, we have seen a strong diversification of service classes beyond simply increasing network capacity. This has led to the need for specific characteristics of the network that cannot be met with the same infrastructure.
While we still expect to see more devices, higher network capacity, and more reliability or lower latency, the next generation of applications will require a combination of these capabilities. For example, new eMBB services such as Augmented Reality (AR) and eXtended Reality (XR) will require high reliability, low latency, and network capacity. Similarly, mMTC will need to extend itself to high reliability and low latency to handle fine-grained automatic environment management and increase network capacity for autonomous mobility and video-based insight generation.

Based on potential requirements, the key driver for 6G would be the extended network capacity, which may require new Terahertz (THz) spectrum-based technologies. This is necessary to support capacities of 4 Tbit/s for AR/XR, the under 100 $\mu$s delay for industrial or holographic presence, a 7-nines (i.e. 99.99999\%) reliability, or localization with a precision of less than 1 cm.
Even with increasing computing capacity in terminals, the network will need to be significantly enhanced to support these use cases.
Note that mobile data traffic has increased by 50\%-100\% every year over the past decade.  This growth is expected to continue further as connected devices, in particular sensors, cars, and home devices, become more and more popular, and the demand for emerging applications will grow exponentially. This alone implies that 6G should address the volume of mobile traffic of from 100 times to 1000 times more than 5G.
The increase in the volume of data traffic would require increased energy consumption at the base stations and network nodes. Therefore, we need to significantly improve the energy efficiency of connected devices by reducing the energy use per transported/processed data to keep the power consumption per device comparable to 5G. As some new use cases such as AR/VR, and holographic telepresence or the need for 8K resolution in communications are expected to come into play, the increase in data volume is followed by the need for a remarkably higher end-user experienced data rate of up to 100 Gbps.
However, the increase in the number of connected devices would imply the need for an increase in device density per square meter, and consequently an increase in the required capacity to address the number of devices exchanging high data volume. As a result, the cells should be smaller to meet these demands. An example could be a stadium full of spectators using AR glasses.


\begin{table}
    \caption{Comparision between KPIs of 5G  and 6G}
    \begin{tabular}{|l|c|c|}
        \hline \multicolumn{1}{|c|}{ Key Performance Indicator } & $5 \mathrm{G}$ & $6 \mathrm{G}$ \\
        \hline Peak Data Rate & $20 \mathrm{~Gb} / \mathrm{s}$ & $1 \mathrm{~Tb} / \mathrm{s}$ \\
        \hline Experienced Data Rate & $0.1 \mathrm{~Gb} / \mathrm{s}$ & $1 \mathrm{~Gb} / \mathrm{s}$ \\
        \hline Peak Spectral Efficiency & $30 \mathrm{~b} / \mathrm{s} / \mathrm{Hz}$ & $60 \mathrm{~b} / \mathrm{s} / \mathrm{Hz}$ \\
        \hline Exp. Spectral Efficiency & $0.3 \mathrm{~b} / \mathrm{s} / \mathrm{Hz}$ & $3 \mathrm{~b} / \mathrm{s} / \mathrm{Hz}$ \\
        \hline Maximum Bandwidth & $1 \mathrm{GHz}$ & $100 \mathrm{GHz}$ \\
        \hline Area Traffic Capacity & $10 \mathrm{Mb} / \mathrm{s} / \mathrm{m}^{2}$ & $1 \mathrm{~Gb} / \mathrm{s} / \mathrm{m}^{2}$ \\
        \hline Connection Density & $10^{6} \mathrm{devices} / \mathrm{km}^2$ & $10^7 \mathrm{devices} / \mathrm{km}^{2}$ \\
        \hline Energy Efficiency & $\mathrm{N} / \mathrm{A}$ & $1 \mathrm{~Tb} / \mathrm{J}$ \\
        \hline Latency & $1 \mathrm{~ms}$ & $100 \mathrm{us}$ \\
        \hline Jitter & $\mathrm{N} / \mathrm{A}$ & $1 \mathrm{us}$ \\
        \hline Reliability & $1 \times 10^{-5}$ & $1 \times 10^{-7}$ \\
        \hline Mobility & $500 \mathrm{Km} / \mathrm{h}$ & $1000 \mathrm{Km} / \mathrm{h}$ \\
        \hline
    \end{tabular}
\end{table}

\begin{figure}
    \centering
    \includegraphics[width=\columnwidth]{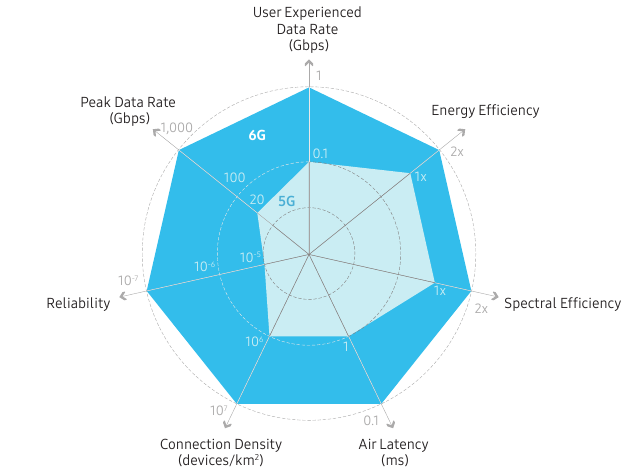}
    \caption{The key performance requirements from 5G to 6G \cite{samsung_6g}.}
    \label{fig:6G_KPIs}
\end{figure}

Toward defining the official specifications for 6G, ITU-R has already formed a group on IMT toward 2030 and beyond, with the aim of completing the study on the 6G vision by the end of 2023. The development process is expected to begin with technical studies in 2026 and the first specifications are aimed to be released by 2028. The 6G networks are expected to be commercially deployed from 2030 \cite{orange_6g}.


\subsection{Challenges in Channel Coding for 6G}
The upcoming generation of cellular networks is expected to impose even more rigorous demands in terms of higher data rates, increased reliability, and lower latency when compared to the current 5G networks. Hence, it is imperative to design new channel coding schemes or improve the ones available to meet those future requirements and KPIs. In the subsequent discussion, we discuss some of the formidable challenges associated with this endeavor. 


First, the complexity of codes that perform well under practical constraints such as limited processing delay and high spectral efficiency is still a major hurdle for low-power implementations in integrated circuits. There is a serious need for new methods that simplify code design, construction, storage, and decoder implementation. In particular, the new channel coding schemes will be required to encode and decode data at very high speeds to support the high data rate, for example 1 Tb/s, in 6G. Meanwhile, both code structures and low-complexity decoders should be designed to achieve performance closer to the Shannon limit than the coding schemes in 5G. In addition, channel coding in next-generation cellular networks will be required to handle excessive interference in the presence of a massive number of users and devices, along with signal detection at the receiver end.

The other major research challenge is to design robust channel coding schemes with short blocklength for delay-sensitive services. From an information-theoretic point of view, short blocklength codes are less reliable, such that error-free transmission is no longer guaranteed. Hence, an increase in the error probability can increase the need for retransmissions, which is also not desirable for time-sensitive applications requiring ultra-low latencies. On the other hand, codes with longer block lengths have a larger time/computational complexity, which implies an increase in transmission and processing latency. In addition, the optimal decoders for short blocklength codes generally have high computational complexity, leading to large processing latency. To this end, new short blocklength codes with low-complexity decoders need to be designed to meet the target error probability while satisfying the stringent latency constraints.
As integrated/joint/coexisting sensing and communication is envisioned for 6G, channel coding faces a number of challenges when employed in communication-centric sensing and sensing-centric communications. These challenges include: 1) balancing communication and sensing performance: channel coding schemes need to balance between communication performance (e.g., data rate, reliability) and sensing performance (e.g., accuracy, resolution). For example, a channel coding scheme optimized for communication performance may not be well suited for sensing-centric communications, where sensing accuracy is paramount. 2) dealing with noise and interference: Joint sensing and communications applications are often subject to high levels of noise and interference. This can make it difficult for channel coding schemes to reliably decode the transmitted signal, which can affect both communication and sensing performance. 3) supporting dynamic channel conditions: The channel conditions in joint sensing and communication applications can be very dynamic, due to factors such as the movement of sensor nodes and the presence of obstacles in the environment. This can make it difficult for channel coding schemes to maintain reliable communication and sensing. 4) limited resources: Sensor nodes often have limited resources, such as battery life and processing power. This can make it difficult to implement complex channel coding schemes.

\subsection{Potential Coding Schemes for 6G}
Improving the channel coding scheme results in the improvement of the reliability (directly) and the spectral efficiency (directly) by allowing more data to be transmitted per unit of (frequency) bandwidth. Moreover, the adoption of higher-order modulation further improves the data rate and spectral efficiency.
Hence, the coding and modulation schemes have impacts on three key performance indicators; reliability, data rate, and spectral efficiency. These three performance indicators must be dramatically improved in 6G according to Section \ref{sec:6G_reqs} to make the new applications and use cases possible. The responsibility for improving these indicators is not only on channel coding and modulation schemes, as other components in the physical layer also play a role.
Historically, developing and adapting a new coding scheme into a standard takes more than a decade. The reason lies in the difficulty of advancing this field as we approach the theoretical performance bound. It is not a long time since the 3GPP adopted new coding schemes in 2016.
We will consider the coding schemes already in the 5G standard in Sections \ref{sec:ldpc} and \ref{sec:polar}; namely, LDPC codes and polar codes. We will investigate these coding schemes in detail, review recent advances, and compare them from different angles. We also consider turbo codes and convolutional codes used in 3G and 4G in Section \ref{sec:turbo}. The coding schemes in the upcoming sections will be presented in chronological order, reflecting their development and incorporation into standards over time.
Furthermore, other coding schemes such as lattice codes, rateless codes, and sparse regression codes are reviewed in Section \ref{sec:other}.
\subsection{Finite Block Length Coding Performance}\label{BLER_limit}

The Shannon limit is the theoretical maximum rate, which is called \emph{channel capacity}, that can be achieved with arbitrarily small errors by using a code with very long (infinite) blocklength. However, practical systems only allow finite blocklength coding such that the Shannon limit may not be achievable. That is, as we decrease the blocklength, the coding gain is reduced, and consequently the gap to the Shannon limit increases. In this section, we review an approximation called \emph{normal approximation (NA)} for the performance of finite blocklength codes \cite{polyanskiy2010channel}.

Consider the real AWGN channel with noise variance 1. For this channel, consider using a length-$n$ code $\mathcal{X}$ with each codeword $\boldsymbol{x}\in\mathcal{X}$ satisfying the maximal power constraint $\|\boldsymbol{x}\|^2 \leq n\rho$, where $\rho$ denotes the power or the Signal-to-Noise Ratio (SNR). Under the constraint that the average decoding error probability or the block error rate (BLER) does not exceed $\epsilon$, the following rate is achievable \cite{polyanskiy2010channel}
\begin{equation}\label{eq:ppv_rate}
R\approx C(\rho)-\sqrt{\frac{V(\rho)}{n}} Q^{-1}\left(\epsilon\right)+\frac{\log _2(n)}{2 n},
\end{equation}
where $Q(x)=\int^{\infty}_x\frac{1}{\sqrt{2\pi}}e^{-\frac{t^2}{2}}dt$ is the standard $Q$-function, $C(\rho)$ and $V(\rho)$ are the Gaussian capacity and dispersion functions, respectively, and
\begin{equation}
V(\rho)=(\log_2 e)^2 \cdot\frac{\rho(\rho+2)}{2(\rho+1)^2} \; \text{bits}^2\;\text{per channel use}.
\end{equation}
Note that \eqref{eq:ppv_rate} is known as the NA. 



The NA has been shown to be a valid asymptotic approximation for the achievability bound (i.e., random coding union bound \cite[Th. 16]{polyanskiy2010channel}) and converse bound (i.e., metaconverse \cite[Th. 26]{polyanskiy2010channel}). The achievability bound is intended as a performance that can be achieved by a suitable encoding/decoding couple, while a converse bound is intended as a performance that outperforms any choice of the encoding/decoding couple. However, the computation of the achievability and converse bounds becomes very difficult when $n$ is not small. Hence, the NA is often used as the performance benchmark due to its simpler computation complexity.

Now, assume that each transmitted symbol $x$ of the length-$n$ sequence $\boldsymbol{x}$ is i.i.d. over the BPSK modulation and $y$ is the received noisy symbol. This channel model is also known as the binary-input AWGN (BI-AWGN) channel. The BLER upper bound $\epsilon$ can be evaluated by rearranging \eqref{eq:ppv_rate} as \cite{erseghe2016coding,COSKUN201966}
\begin{equation}
\epsilon \approx Q\left(\sqrt{\frac{n(C_\text{b}(\rho)-R)+\frac{\log_2n}{2}}{n V_\text{b}(\rho)}}\right),
\end{equation}
where $R=\frac{k}{n}$ is the code rate with $k$ being the source blocklength, $C_\text{b}(\rho)$ and $V_\text{b}(\rho)$ are the BIAWGN capacity and dispersion functions, respectively. Let $P_{Y|X}(y|x)$ and $P_Y(y)$ denote the channel transition probability and the probability density function of the channel output $Y$, respectively. Specifically, $C_\text{b}(\rho)$ and $V_\text{b}(\rho)$ can be computed by using the information density $i(X;Y) = \log_2 \frac{P_{Y|X}(y|x)}{P_Y(y)}$ such that by \cite{COSKUN201966} we have
\begin{align}
C_\text{b}(\rho) =& \mathbb{E}[i(X;Y)] \\
 =& \frac{1}{\sqrt{2\pi}}\int e^{-\frac{z^2}{2}}\left( 1-\log_2 (1+e^{-2\rho+2z\sqrt{\rho}})\right)dz,
\end{align}
\begin{align}
&V_\text{b}(\rho) = \text{Var}[i(X;Y)] \\
 =& \frac{1}{\sqrt{2\pi}}\int e^{-\frac{z^2}{2}}\left( 1-\log_2 (1+e^{-2\rho+2z\sqrt{\rho}})-C_\text{b}(\rho)\right)^2dz.
\end{align}
The extension of NA to high-order modulations has been investigated in \cite{7127525,JSACQiu22}.

Given the code parameters and the channel SNR, the design of coding schemes should target the BLER predicted by the NA. However, as we shall observe in the following sections, some well-known codes can perform very close to the NA when the blocklength is short.

\section{Turbo Codes}\label{sec:turbo}

Turbo codes or parallel concatenated convolutional codes (PCCCs) were introduced in 1993 by Berrou, Glavieux, and Thitimajshima \cite{397441}. The invention of turbo codes marked a major breakthrough in coding theory \cite{1270546}. It was the first class of codes that was practically demonstrated to achieve the near-Shannon-limit performance with modest decoding complexity. Owing to the close-to-capacity performance and moderate decoding complexity, turbo codes have been the standard channel coding in standard 3G \cite{CDMAFDD25.212,CDMATDD25.222} and 4G mobile communication standards \cite{LTE136212}. In addition, turbo codes have also been adopted in the IEEE 802.16 WiMAX (worldwide interoperability for microwave access) \cite{WIMAX_turbo} and DVB-RCS2 (2nd generation digital video broadcasting - return channel via satellite) \cite{DVB_RCS2_turbo} standards. For the list of standards that have adopted turbo codes, we refer the reader to Table 4 in \cite[Ch. 5.3.3]{BENEDETTO201453}. During the study phase of 5G NR, an enhanced version of turbo codes with better waterfall and error floor performance than LTE turbo codes, was one of the candidate channel coding schemes \cite{R1-167413}. Besides error correction coding, the concept of the turbo principle has been applied to the decoding of product codes \cite{705396,7505623} and iterative detection \cite{4460060506,774855}.

In this section, we first review the properties of turbo codes and their component codes, i.e., convolutional codes. Then, we provide a comprehensive survey covering state-of-the-art designs on interleavers, puncturing patterns, and decoding algorithms. Several variants of turbo codes and future research directions are presented in the end.


\subsection{Convolutional Codes}\label{ssec:conv}
Convolutional codes \cite{1057464} are the building blocks of turbo codes. Besides, convolutional codes were adopted as the coding schemes in 3G UMTS \cite{CDMAFDD25.212,CDMATDD25.222} and 4G LTE standards \cite{LTE136212} for control channels. Unlike block codes, the information and codeword sequences for convolutional codes may or may not be terminated \cite{9780470276839}.

\subsubsection{Encoding of Convolutional Codes}
A rate-$k/n$ convolutional code is specified by $k \times n$ generator polynomials, which form the generator matrix
\begin{align}
\boldsymbol{G}(D)=\begin{bmatrix}
\boldsymbol{g}^{(1)}_1(D) & \boldsymbol{g}^{(2)}_1(D) & \ldots & \boldsymbol{g}^{(n)}_1(D)\\
\boldsymbol{g}^{(1)}_2(D) & \boldsymbol{g}^{(2)}_2(D) & \ldots & \boldsymbol{g}^{(n)}_2(D)\\
 \vdots&  \vdots & \ldots & \vdots\\
\boldsymbol{g}^{(1)}_k(D) & \boldsymbol{g}^{(2)}_k(D) & \ldots & \boldsymbol{g}^{(n)}_k(D)\\
\end{bmatrix},
\end{align}
where $D$ is known as a unit delay operator. Consider $k$ input sequences $\boldsymbol{u}_1,\ldots,\boldsymbol{u}_k$. For $i\in\{1,\ldots,k\}$, the $i$-th input sequence can be represented by $\boldsymbol{u}^{(i)}(D)=\sum^{\infty}_{l=0}u^{(i)}_lD^l$. A collection of these sequences can be arranged into $\boldsymbol{u}(D)=[\boldsymbol{u}^{(1)}(D)\;\ldots\;\boldsymbol{u}^{(k)}(D)]$. The encoder output consists of $n$ polynomials $\boldsymbol{c}(D)=[\boldsymbol{c}^{(1)}(D)\;\ldots\;\boldsymbol{c}^{(n)}(D)]=\sum^n_{j=1}D^{j-1}\boldsymbol{c}^{(j)}(D^n)$, which is generated by
\begin{align}
\boldsymbol{c}(D)=\boldsymbol{u}(D)\boldsymbol{g}(D).
\end{align}
For $j\in\{1,\ldots,n\}$, the $j$-th code sequence is generated by
\begin{align}
\boldsymbol{c}^{(j)}(D)=\sum^k_{i=1}\boldsymbol{u}^{(i)}(D)\boldsymbol{g}^{(j)}_i(D).
\end{align}
The memory of the encoder is defined as
\begin{align}
m\triangleq \max_{i\in\{1,\ldots,k\},j\in\{1,\ldots,n\}} \left\{\text{deg}\left(\boldsymbol{g}^{(j)}_i(D)\right)  \right\}.
\end{align}

For turbo codes, \emph{recursive systematic convolutional (RSC)} codes are mostly employed as the component codes. A recursive convolutional code means that at least one entry in $\boldsymbol{G}(D)$ is a rational function or equivalently the encoder realization has feedback. In addition, it is common to represent the generator matrix in \emph{octal} notation. For example, the generator polynomial for the LTE turbo code $G(D)=[1,\frac{1+D+D^3}{1+D^2+D^3}]$ can be represented by $[1,15/13]_8$. In addition to turbo codes, convolutional codes have been used as the precoding stage for polar codes, resulting in polarization-adjusted convolutional (PAC) codes as shown in Section \ref{sec:polar}-\ref{ssec:pac}.

\subsubsection{Trellis Diagram and Termination}
A convolutional encoder can be modeled as a finite state machine for which the input-output relation and state transition can be described by a state-transition table or a state diagram. The time evolution of the state diagram of the state machine can be described by a trellis diagram \cite{9780470276839}. For a rate-$k/n$ convolutional code with memory $m$ and $K$ information symbols, the trellis diagram contains $2^m$ states and $\frac{K}{k}+1$ time instants. The connection between two states is called a branch. Essentially, it visualizes how the output bits of the current state $s$ are computed based on the input bits and the previous state $s$.

In the case of packet transmission, termination of convolutional codes is required. There are several approaches for termination. The first method is \emph{direct truncation}, which stops the encoding process when all the information bits have been applied to the encoder input. Although it does not have rate loss, the final state is unknown to the decoder, leading to some performance degradation on the convolutional code. That said, this method is employed for the convolutional precoding of PAC codes, see Section \ref{sec:polar}-\ref{ssec:pac}. The second method is called \emph{zero termination} \cite{525138}. It adds $m$ additional bits to the original message to force the encoder trellis to the all-zero state. The turbo codes in 3G UMTS \cite{CDMAFDD25.212,CDMATDD25.222} and 4G LTE standards \cite{LTE136212}. However, this termination method leads to a small rate loss. In addition, the termination bits are not turbo coded \cite[Ch. 3.2]{BENEDETTO201453}. The third approach is the \emph{tail-biting termination} \cite{1096498}, which does not incur any rate loss. Specifically, the initial state of the trellis diagram is obtained based on the input sequence such that the final state is the same as the initial state. Tail-biting termination has been adopted in the turbo codes in WIMAX \cite{WIMAX_turbo}, DVB-RCS2 \cite{DVB_RCS2_turbo}, and the enhanced turbo codes originally proposed for 5G NR \cite{R1-167413}. The drawback is that it requires an additional step to determine the initial state compared to the first and second methods.

		\begin{figure}[ht]
				\centering
				\includegraphics[width=\linewidth]{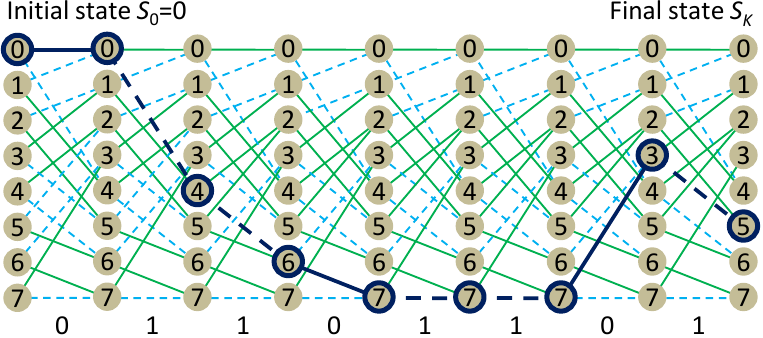}
				\caption{Tail-biting encoding step 1: initial state $S_0=0$ and the final state $S_K=5$ \cite{R1-167413}.} \label{fig:circular_enc_1}
			\end{figure}

			\begin{figure}[ht]
				\centering
				\includegraphics[width=\linewidth]{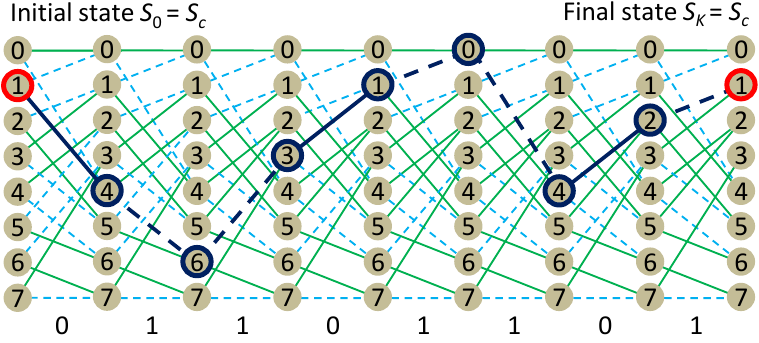}
				\caption{Tail-biting encoding step 2: initial state $S_0=1$ and the final state $S_K=1$ \cite{R1-167413}.} \label{fig:circular_enc_2}
			\end{figure}

As an example, we show the two-step tail-biting encoding for a $K$ bit message $\boldsymbol{u}$ with component encoder with generator polynomial $[1,15/13]_8$ in Figs. \ref{fig:circular_enc_1}-\ref{fig:circular_enc_2}. Assume that $K=8$ and $\boldsymbol{u}=[0,1,1,0,1,1,0,1]$. In the first step, the message bits are encoded from the initial state $S_0=0$ to the final state $S_K=5$. Then, by using Table \ref{tab:circular_state_enc}, we find the value of the circular state $S_C$. In the second step, the encoder starts from the initial state $S_0=S_C = 1$ and finishes in the final state $S_K=S_C = 1$. The contents of the circulation state table depend on the code memory and the recursion polynomial. The computation principle is described in \cite{738095,904537}, and \cite[Ch. 5.5.1]{978-2-8178-0039-4}. Note also that for the example in Figs. \ref{fig:circular_enc_1}-\ref{fig:circular_enc_2}, the circulation exists if and only if $K$ is not a multiple of 7.

\begin{table}[ht]
 \centering
 \caption{Table for circular states $S_C$ of the RSC code $[1,15/13]_8$ as a function of $K\bmod 7$ and the final state obtained in the first encoding step \cite{R1-167413}.}\label{tab:circular_state_enc}
\begin{tabular}{|c|c|c|c|c|c|c|}
\hline
\diagbox{$S_K$}{$K \bmod 7$}& 1 & 2 & 3 & 4 & 5 & 6 \\ \hline
0 & 0 & 0& 0& 0& 0& 0 \\ \hline
1 & 6 &4 & 3& 2& 5& 7 \\ \hline
2 & 3 & 5& 4& 6&7 & 1 \\ \hline
3 & 5 & 1& 7& 4& 2& 6 \\ \hline
4 & 7 & 2& 1& 5&6 & 3 \\ \hline
5 &1  & 6& 2& 7& 3& 4 \\ \hline
6 & 4 & 7& 5& 3& 1& 2 \\ \hline
7 & 2 & 3& 6& 1& 4& 5 \\ \hline
\end{tabular}
\end{table}

\subsubsection{Decoding of Convolutional Codes}\label{sec:conv_dec}
The decoding of convolutional codes can be realized by two popular algorithms: the \emph{Viterbi algorithm (VA)} \cite{1054010} and the \emph{Bahl–Cocke–Jelinek–Raviv (BCJR) algorithm} \cite{1055186}. The VA was recognized as an ML decoder in \cite{1450960}. To enable soft-output, \cite{64230} introduced soft-output VA (SOVA). We note that tail-biting convolutional codes (TBCC) with wrap-around VA (WAVA) \cite{1237439} have been adopted in 4G LTE for control channel \cite{LTE136212}. Notably, for $(n,k)=(128,64)$, the TBCC with $m=14$ under WAVA has the best performance among all other short codes \cite{COSKUN201966}. When using CRC-aided list Viterbi decoding \cite{577040}, TBCCs also remain competitive \cite{9709319}.

The BCJR decoder is the bit-wise maximum a posteriori (MAP) decoder \cite{1055186}, which has been widely used as the decoding algorithm for the convolutional component codes of turbo codes in most communication standards. It was shown in \cite{524253} that turbo codes with iterative BCJR decoding have a gain of 0.7 dB of those with iterative SOVA. For the interest of turbo codes, we only describe the BCJR algorithm here. Particularly, we consider the \emph{log domain} implementation of the BCJR algorithm (Log-MAP) \cite{524253} as it allows efficient implementation.

\textbf{Log-MAP decoding:} Recall that $\boldsymbol{u}_t$ and $\boldsymbol{c}_t$ are the information and codeword sequences at time $t=1,\ldots,K$. Let $\boldsymbol{x}_t\in\{-1,1\}^n$ be the transmitted BPSK modulated codeword at time $t$ and $\boldsymbol{y}_t$ be the corresponding received vector.
The branch metric of the trellis edge departing from state $s'$ at time $t-1$ to state $s$ at time $t$ is computed as
\begin{align}
\gamma_t(s',s) =u_t(s',s)L_{\text{A}}(u_t)+\sum^n_{j=1}c^{(j)}_t(s',s)L(y^{(j)}_t|x^{(j)}_t),
\end{align}
where $(s',s)$ indicates the association to the state transition from $s'$ to $s$, $L_{\text{A}}(u_t)$ is the \emph{a priori} LLR of $u_t$, and $L(y^{(j)}_t|x^{(j)}_t)$ is the $j$-th channel LLR at time $t$. In turbo decoding, $L_{\text{A}}(u_t)$ is obtained from the extrinsic information from another component decoder as shown in Section \ref{sec:turbo}-\ref{ssec:turbo_props}\ref{sec:turbo_dec_intro}. For stand-alone convolutional decoding without \emph{a priori} information, $L_{\text{A}}(u_t)=0$. The forward and backward metrics of the log-MAP decoder, denoted by $\alpha$ and $\beta$, are computed in \eqref{eq:metric_dec} and \eqref{eq:metric_dec2}, respectively, where
\begin{align}
\alpha_t(s)=&\smax_{s'}(\alpha_{t-1}(s')+\gamma_t(s',s)),\label{eq:metric_dec}\\
\beta_{t-1}(s')=&\smax_{s}(\beta_{t}(s)+\gamma_t(s',s)), \label{eq:metric_dec2}
\end{align}
where the max star function is defined as
\begin{align}\label{eq:max_star}
\smax(x,y) \triangleq \ln(e^x+e^y)=\max(x,y)+\ln(1+e^{-|x-y|}).
\end{align}
The soft-decision output for the LLR of $u_t$ is
\begin{align}\label{eq:turbo_dec_ut}
L(u_t) = &\smax_{(s',s):u_t=1}(\alpha_{t-1}(s')+\gamma_t(s',s)+\beta_{t}(s))  \nonumber\\
&-\smax_{(s',s):u_t=0}(\alpha_{t-1}(s')+\gamma_t(s',s)+\beta_{t}(s)),
\end{align}

Assume that the encoder is initialized and terminated to the zero state. We have the following initial conditions for the forward and backward metrics
\begin{align}
\alpha_0(s)=\beta_k(s)=\left\{ \begin{array}{l}
0, \;s=0 \\
-\infty, \;s\neq 0\\
\end{array}\right..
\end{align}



%

\subsection{Properties of Turbo Codes}\label{ssec:turbo_props}
In this section, we first introduce the encoding and decoding of turbo codes. In addition, we will also show that turbo codes can be represented by a tanner graph. Finally, we will discuss the distance properties of turbo codes.


\subsubsection{Encoding of Turbo Codes}
\begin{figure}[ht]
				\centering
				\includegraphics[width=0.6\linewidth]{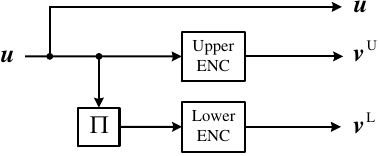}
				\caption{Turbo code encoder.} \label{fig:PCC_enc}
			\end{figure}

As shown in Fig. \ref{fig:PCC_enc}, the encoder consists of two RSC encoders and one interleaver.

A length $K$ information sequence $\boldsymbol{u}$ is encoded by the upper convolutional encoder. The upper encoder outputs a length $N_1$ codeword $\boldsymbol{c}_1=[\boldsymbol{u},\boldsymbol{v}^{\text{U}}]$, where $\boldsymbol{v}^{\text{U}}$ denotes the parity bits generated by the upper encoder. Meanwhile, the information sequence $\boldsymbol{u}$ is interleaved and become $\Pi(\boldsymbol{u})$, where $\Pi(.)$ represents the interleaving function. The interleaved information sequence is encoded by the lower convolutional encoder and becomes a length $N_2$ codeword $\boldsymbol{c}_2=[\Pi(\boldsymbol{u}),\boldsymbol{v}^{\text{L}}]$, where $\boldsymbol{v}^{\text{L}}$ represents the parity bits generated by the lower encoder. The final turbo codeword to be transmitted is $\boldsymbol{c}=[\boldsymbol{u},\boldsymbol{v}^{\text{U}},\boldsymbol{v}^{\text{L}}]$. The rate of turbo code is
\begin{align}
R=\frac{K}{N_1+N_2-K}=\frac{1}{1/R_1+1/R_2-1},
\end{align}
where $R_1=\frac{K}{N_1}$ and $R_2=\frac{K}{N_2}$ are the code rates of constituent convolutional codes \cite{1516279}.

To increase the code rate, a puncturer is required to reduce the codeword bits. Denote by $\rho\in[\frac{1}{(1/R_1+1/R_2-1)},1]$ the portion of surviving bits after puncturing. The code rate of a punctured turbo code is
\begin{align}
R=\frac{1}{(1/R_1+1/R_2-1)\rho}.
\end{align}
Both interleaving and puncturing patterns affect the waterfall and error floor performance of a turbo code. The design of interleaving and puncturing patterns will be introduced in Sections \ref{sec:int_design}-\ref{sec:punc_design}.

\subsubsection{Decoding of Turbo Codes}\label{sec:turbo_dec_intro}
\begin{figure}[ht]
				\centering
				\includegraphics[width=0.9\linewidth]{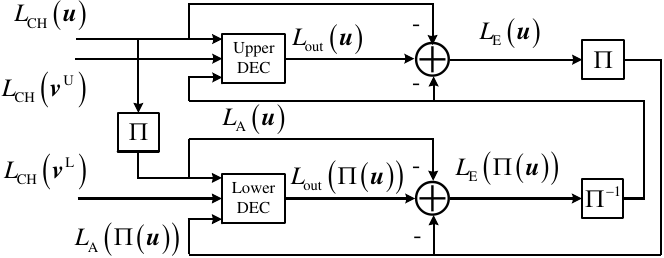}
				\caption{Turbo code decoder.} \label{fig:PCC_dec}
			\end{figure}

The diagram of a turbo decoder is depicted in Fig. \ref{fig:PCC_dec}. The turbo decoder consists of two constituent soft-in soft-out (SISO) decoders, an interleaver $\Pi$ and a deinterleaver $\Pi^{-1}$. The decoding of turbo codes is performed iteratively between the two SISO decoders.


Let $L_{\text{CH}}(.)$, $L_{\text{A}}(.)$, and $L_{\text{E}}(.)$ represent the channel, a-priori, and extrinsic LLRs, respectively. Moreover, let $L_{\text{in}}(.)$ and $L_{\text{out}}(.)$ represent the input and output LLRs of the BCJR decoder, respectively. The SISO decoding function is denoted by $D_{\text{SISO}}(.)$. At the $\ell$-th iteration, $\ell\in\{1,\ldots,\ell_{\max}\}$, the inputs to the upper SISO decoder are
\begin{align}
L^{(\ell)}_{\text{in}}(\boldsymbol{u})=& L^{(\ell)}_{\text{A}}(\boldsymbol{u})+L_{\text{CH}}(\boldsymbol{u}) \\
=&\Pi^{-1}\left(L^{(\ell-1)}_{\text{E}}(\Pi(\boldsymbol{u}))\right)+L_{\text{CH}}(\boldsymbol{u}),\\
L^{(\ell)}_{\text{in}}(\boldsymbol{v}^{\text{U}})=&L_{\text{CH}}(\boldsymbol{v}^{\text{U}}) \label{eq:turb_ext2}.
\end{align}
The key idea is that the extrinsic information from the lower decoder from the previous iteration is used as the \emph{a priori} information at the upper decoder. The outputs of the upper SISO decoder are
\begin{align}
\left[L^{(\ell)}_{\text{out}}(\boldsymbol{u}),L^{(\ell)}_{\text{out}}(\boldsymbol{v}^{\text{U}})\right]=& D^{\text{U}}_{\text{SISO}}\left(L^{(\ell)}_{\text{in}}(\boldsymbol{u}),L^{(\ell)}_{\text{in}}(\boldsymbol{v}^{\text{U}})\right), \\
L^{(\ell)}_{\text{E}}(\boldsymbol{u})=&L^{(\ell)}_{\text{out}}(\boldsymbol{u})-L^{(\ell)}_{\text{in}}(\boldsymbol{u}) \label{eq:turbo_ext1}.
\end{align}
Similarly, the extrinsic information from the upper SISO decoder is used as the \emph{a priori} information at the lower SISO decoder as follows
\begin{align}
L^{(\ell)}_{\text{in}}(\Pi(\boldsymbol{u}))=& L^{(\ell)}_{\text{A}}(\Pi(\boldsymbol{u}))+L_{\text{CH}}(\Pi(\boldsymbol{u})) \\
=& \Pi(L^{(\ell)}_{\text{E}}(\boldsymbol{u}))+L_{\text{CH}}(\Pi(\boldsymbol{u})),\\
L^{(\ell)}_{\text{in}}(\boldsymbol{v}^{\text{L}})=&L_{\text{CH}}(\boldsymbol{v}^{\text{L}}).
\end{align}
The outputs of the lower SISO decoders are
\begin{align}
\left[L^{(\ell)}_{\text{out}}(\Pi(\boldsymbol{u})),L^{(\ell)}_{\text{out}}(\boldsymbol{v}^{\text{L}})\right]=& D^{\text{L}}_{\text{SISO}}\left(L^{(\ell)}_{\text{in}}(\Pi(\boldsymbol{u})),L^{(\ell)}_{\text{in}}(\boldsymbol{v}^{\text{L}})\right), \\
L^{(\ell)}_{\text{E}}(\Pi(\boldsymbol{u}))=&L^{(\ell)}_{\text{out}}(\Pi(\boldsymbol{u}))-L^{(\ell)}_{\text{in}}(\Pi(\boldsymbol{u})). \label{eq:turbo_ext2}
\end{align}
Finally, the hard-decision estimation is performed based on the \emph{a posteriori} LLR of $\boldsymbol{u}$, which is
\begin{align}
L(\boldsymbol{u})=L_{\text{CH}}(\boldsymbol{u})+L_{\text{E}}(\boldsymbol{u})+\Pi^{-1}(L_{\text{E}}(\Pi(\boldsymbol{u}))).
\end{align}

It was proved in \cite{4658707} that the output L-values of a turbo decoder cannot grow to infinity. Hence, the optimal stopping rule is to stop iterations when the output probabilities do not change anymore \cite{4658707}.

\subsubsection{Graph Representation}
\begin{figure}[ht]
				\centering
				\includegraphics[width=0.8\linewidth]{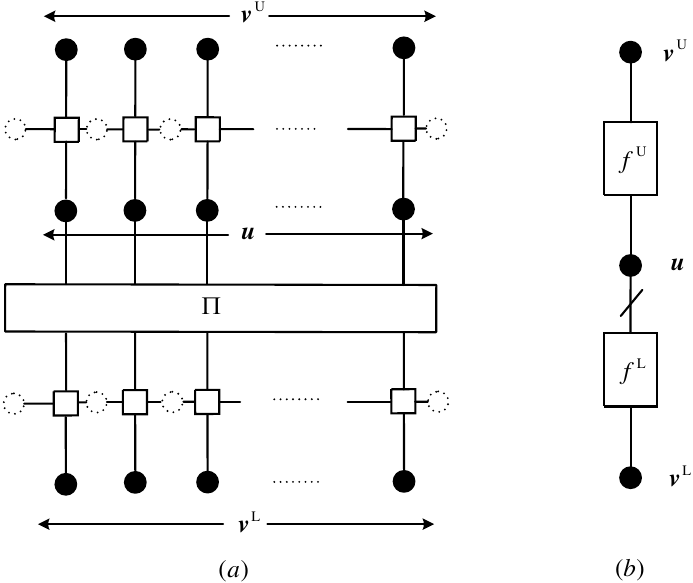}
				\caption{(a) Factor graph representation of a turbo code. (b) Compact graph representation of a turbo code.} \label{fig:turbo_FG}
			\end{figure}

Turbo codes are a class of codes on graphs. The graph representations of turbo codes can help to simplify their analysis. The factor graph \cite{910572} of a turbo code is shown in Fig. \ref{fig:turbo_FG}(a). The information and parity nodes can be regarded as variable nodes (VNs), which are represented by black circles. The trellis factor nodes are presented by white boxes. In addition, the states of the convolutional encoder are regarded as hidden VNs, represented by white circles with dash lines since they do not correspond to code bits.

To simplify the factor graph representation, \cite{8002601} introduced a compact graph representation, which is shown in Fig. \ref{fig:turbo_FG}(b). The main idea is that each sequence of information and parity bits are represented by a single variable node while the trellis factor nodes are represented by a single factor node, e.g., $f^{\text{U}}$ and $f^{\text{L}}$. The interleaving function is represented by a line that crosses the edge connecting node $\boldsymbol{u}$ and lower factor node $f^{\text{L}}$.

\subsubsection{Distance Bound}
It was analytically shown in \cite{TLZ0101} that there exists a length-$N$ turbo code $\mathcal{C}$ whose minimum distance satisfies
\begin{align}
d_{\min}(\mathcal{C}) \geq \alpha \log N,
\end{align}
where $\alpha>0$ is a constant depending on the types of constituent encoders. Meanwhile, \cite{1317115} proved that the minimum distance of a length-$N$ turbo code $\mathcal{C}$ is upper bounded by the inequality
\begin{align}
d_{\min}(\mathcal{C})  \leq O(\log N).
\end{align}
For a length-$N$ turbo code $\mathcal{C}$ with two state-2 constituent convolutional codes, its distance is upper bounded by \cite[Ch. 9.2]{9780470276839}
\begin{align}
d_{\min}(\mathcal{C}) \leq 6\log N.
\end{align}
The above results indicate that designing turbo codes with a large minimum distance may be achieved by using constituent codes with small states.

\subsection{Tools for Design and Analysis}\label{ssec:turbo_tool}
In this section, we introduce several tools to analyze both waterfall and error floor performance of turbo codes. These tools have been used for designing the component codes, interleavers, and puncturers for turbo codes. As we will see in the later sections, using a combination of different tools may be required.

\subsubsection{Distance Evaluation}\label{turbo_dis}
We introduce three main approaches that are commonly used to evaluate the distance of turbo codes in the literature.

\emph{1a) Distance Spectrum Search:} Conventionally, the distance of turbo codes is based on searching the free distance of the constituent convolutional codes \cite{45271,1302309} or turbo codes \cite{924864}. Particularly, \cite{924864} introduced an algorithm to determine the total or partial enumeration of a turbo codeword with input weight smaller than or equal to a given minimum distance. This method is based on the use of constrained subcodes, i.e., a subset of a code defined via constraints on the edges of its trellis. An improved method of \cite{924864} with lower computational complexity was introduced in \cite{1391183}. However, these methods are more suitable for small blocklengths and small minimum distances.

\emph{1b) Error Impulse Method}: The second approach is based on the error correction capability of the decoder, which is known as the error impulse method \cite[Ch. 7.6.2]{978-2-8178-0039-4}. It first considers transmitting an all-zero codeword, which becomes a vector $\boldsymbol{x}$ with all of its elements being $-1$ after BPSK mapping. Then, it introduces an error impulse to the $i$-th symbol of the systematic part of an all-zero codeword, i.e., $x_i=-1+A_i$ for some amplitude $A_i$, where $i\in\{1,\ldots,K\}$. Let $A^*_i$ be the maximum amplitude such that the decoded codeword is an all-zero sequence. Then, we have $d_{\min}(\mathcal{C})=\min_{i\in\{1,\ldots,K\}}\{ A^*_i\}$ if the decoder is an ML decoder. Since the turbo decoder is not an ML decoder, this method produces only an approximation of the true minimum distance. An improved method was proposed in \cite{1312511}, where a high amplitude error impulse is placed in a specific information bit position of $\boldsymbol{x}$ before decoding, i.e., set $P(c_i=1)=1$ or $\ln\frac{P(c_i=1)}{P(c_i=0)}=\infty$. In addition, AWGN noise is added to the all-zero codeword to help the decoder converge towards a low-weight codeword concurrent to the all-zero. An upper bound on the minimum distance for all codewords having a 1 in the specific data index being tested can be found. Several methods based on multiple error impulses that follow the same strategy were reported and compared in \cite{5755853}, which increase the accuracy of the turbo code minimum distance estimation at the cost of increased complexity.

\emph{1c) Input-Parity Weight Enumerator Function}: The third approach is based on analyzing the weight enumerator functions (WEFs) of turbo codes \cite[Ch. 4.3]{Vucetic2000}, \cite[Ch 6.9]{Richardson:2008:MCT:1795974}. The first step is to identify the transfer matrix associated with the state transitions of each constituent convolutional code $\boldsymbol{M}$, where the element in the $i$-th row and $j$-th column $\boldsymbol{M}[i,j]$ corresponds to the trellis branch from the $i$-th state to the $j$-th state, $i,j\in\{1,\ldots,s\}$, and $s$ denotes the number of states. Then, we derive the average input-parity WEF (IP-WEF) of turbo codes by using the transfer matrix. Consider a rate-$1/3$ turbo code $\mathcal{C}(N,K)$ with interleaver length $K$ and two identical rate-$1/2$ constituent convolutional codes as an example. Assume that the trellis is initialized and terminated to the all-zero state. The average IP-WEF for the upper and lower constituent codes is defined as the following polynomial
\begin{equation}\label{eq:IP_WEF}
A(I,P)=\boldsymbol{M}^{K}[1,1]=\sum_{i}\sum_{P}A_{i,p}I^iP^p,
\end{equation}
which shows that there are $A_{i,p}$ codewords with weight-$i$ input bits and weight-$p$ parity bits. The coefficients of the average IP-WEF of turbo code $\mathcal{C}(N,K)$ is obtained by averaging over all possible permutation \cite{485713}
\begin{equation}
A_{i,p}(\mathcal{C})=\frac{\sum_{p'}A_{i,p'}\cdot A_{i,p-p'}}{\binom{K}{i}},
\end{equation}
where $A_{i,p'}$ and $A_{i,p-p'}$ are obtained from the IP-WEF of convolutional component codes in \eqref{eq:IP_WEF}.

After obtaining the distance profile of a turbo code $\mathcal{C}$, the BER and FER can be evaluated as
\begin{align}
\text{BER}(\mathcal{C}) \leq& \sum^K_{i=1}\sum^{N-K}_{p=1} \frac{i}{N}A_{i,p}(\mathcal{C})\cdot Q\left( \sqrt{2(i+p)R\frac{E_b}{N_0}}\right),\\
\text{FER}(\mathcal{C}) \leq& \sum^K_{i=1}\sum^{N-K}_{p=1} A_{i,p}(\mathcal{C})\cdot Q\left( \sqrt{2(i+p)R\frac{E_b}{N_0}}\right).
\end{align}

\subsubsection{Density Evolution}\label{sec:turbo_DE}
The asymptotic decoding threshold of turbo codes with iterative BCJR decoding can be analyzed by using density evolution (DE). DE naturally assumes infinite blocklength and ideal random interleaving. Although DE has been proposed to analyze the performance of LDPC codes a decade ago \cite{910577}, the application of DE to the design and analysis of turbo codes and other turbo-like codes only started to gain some attention recently \cite{8002601}.

\emph{2a) Density Evolution on the Binary Erasure Channel}: For the binary erasure channel (BEC), the exact DE equations for turbo codes can be derived to track the evolution of the erasure probability with the number of decoding iterations. First, the exact transfer functions between input and output erasure probabilities on both information and parity bits for a rate-$k/n$ convolutional code under BCJR decoding are derived by following \cite{1258535}
\begin{align}
p^{\text{ext}}_l=f_l(p_1,\ldots,p_n),
\end{align}
where $p_l$, $l\in\{1,\ldots,n\}$, is the input erasure probability of the $l$-th code bit, $p^{\text{ext}}_l$ denotes the output extrinsic erasure probability of the $l$-th code bit, and $f_l(.)$ is the transfer function of the BCJR decoder for the $l$-th code bit. The DE equations of a turbo code can then be easily obtained from the derived transfer functions of the underlying convolutional codes \cite{5577801,8002601}. Consider a rate-$1/3$ turbo code as an example. Let $\epsilon$ be the erasure probability of the BEC. The DE equations for the information bit and parity bit at the $\ell$-th iteration are
\begin{align}\label{eq:turbo_DE}
\left\{ \begin{array}{{l}}
p^{(\ell+1)}_{\text{U,s}}=f^{\text{L}}_{\text{s}}\left( \epsilon \cdot f^{\text{U}}_{\text{s}}\left( \epsilon \cdot p^{(\ell)}_{\text{U,s}},p^{(\ell)}_{\text{U,q}}\right),p^{(\ell)}_{\text{L,q}}\right), \\
p^{(\ell)}_{\text{U,q}} = p^{(\ell)}_{\text{L,q}} = \epsilon
\end{array} \right.
\end{align}
where $f^{\text{U}}_{\text{s}}$ and $f^{\text{L}}_{\text{s}}$ are the transfer functions for the information bit at the upper and lower BCJR decoders, respectively, $p^{(\ell)}_{\text{U,s}}$ is the input erasure probability for the information bit at the upper BCJR decoder, $p^{(\ell)}_{\text{U,q}}$ and $p^{(\ell)}_{\text{L,q}}$ are the input erasure probabilities for the parity bit at the upper and lower BCJR decoders, respectively. The belief propagation (BP) decoding threshold\footnote{The decoding of turbo-like codes comprises BCJR decoding for convolutional component codes while the message exchange between BCJR component decoders follows the extrinsic message passing rule. Hence, we refer to the threshold under iterative message passing decoding with BCJR component decoding as BP threshold.} of the turbo code on the BEC is defined as
\begin{equation}\label{eq:turbo_BP}
\epsilon^{*} \triangleq \sup\{\epsilon>0 |\lim_{\ell \rightarrow \infty}p^{(\ell)}_{\text{U,s}}= 0  \}.
\end{equation}
Having obtained the BEC decoding threshold, the finite blocklength performance on the BEC can be predicted by using the scaling law in \cite{5174521}.


\emph{2b) Density Evolution on the AWGN Channel}: The DE analysis of turbo codes on the AWGN channel is similar to that on the BEC, except that the densities instead of probabilities are tracked. Let $a$ represent the channel density experienced by the information bits, which is the same as for the parity bits. The DE equation for the information bits at the $\ell$-th iteration is \cite[Lemma 6.33]{Richardson:2008:MCT:1795974}
\begin{align}
p^{(\ell+1)}_{\text{U,s}}=f^{\text{L}}_{\text{s}}\left(a \star f^{\text{U}}_{\text{s}}\left(a \star p^{(\ell)}_{\text{U,s}},a\right),a\right),
\end{align}
where $\star$ denotes the convolution operation. However, the transfer function of the BCJR decoder $f(.)$ on the AWGN channel is difficult to obtain \cite{1523540}. Hence, the DE analysis of turbo codes on the AWGN channel can only be performed by Monte Carlo simulation \cite{e23020240}. Alternatively, given a BP threshold on the BEC $\epsilon^*$, the computation of BP threshold on the AWGN channel $\sigma^*$ can be performed by the following approximation \cite{e23020240}
\begin{align}
\sigma^* \approx C_{\text{G}}^{-1}(1-\epsilon^*),
\end{align}
where $C_{\text{G}}(\sigma)$ is the capacity of the binary-input AWGN channel with noise following $\mathcal{N}(0,\sigma^2)$.

\subsubsection{Extrinsic Information Transfer (EXIT) Chart}\label{sec:exit_turbo}
EXIT chart is a well-known technique to visualize the exchange of extrinsic information between a pair of SISO constituent decoders \cite{957394,6517051}. The characteristics of exchange information transfer are based on mutual information. EXIT chart can be used for designing the component convolutional codes for turbo codes based on the waterfall performance on the AWGN channel.

Assume BPSK signaling. Let $I_{\text{A}} \triangleq I(X;L_{\text{A}})$ denote the average \emph{a priori} information input to the decoder, which is also the mutual information between the transmitted symbol $X$ and the log-likelihood ratio (LLR) of the \emph{a priori} decoder input associated to that symbol. Moreover, $L_{\text{A}}$ can be modeled as $L_{\text{A}}=\mu_{\text{A}}x+z_{\text{A}}$, where $z_{\text{A}}\sim \mathcal{N}(0,\sigma^2_{\text{A}})$, and $\mu_{\text{A}}=\frac{\sigma^2_{\text{A}}}{2}$ \cite{957394}, which can be evaluated as
\begin{align}
I_{\text{A}}=1-\int^{\infty}_{\infty}\frac{1}{\sqrt{2\pi\sigma_{\text{A}}}}e^{-\frac{\left(l-\frac{\sigma^2_{\text{A}}}{2}\right)^2}{2\sigma^2_{\text{A}}}}\log(1+e^l)dl.
\end{align}
The closed-form expression for $I_{\text{A}}$ can be found on \cite[Appendix]{1291808}. Let $I_{\text{E}} \triangleq I(X;L_{\text{E}})$ denote the average extrinsic information output from the decoder, which is also the mutual information between the transmitted symbol $X$ and the LLR of the extrinsic decoder output associated to that symbol. It can be evaluated as \cite{Hagenauer04}
\begin{align}
I_{\text{E}} =1-\mathbb{E}[\log(1+e^{-L_{\text{E}}})].
\end{align}

For a given SNR, we express $I_{\text{E}}$ as a function of $I_{\text{A}}$ as follows
\begin{align}
I^{\text{U}}_{\text{E}}(\ell) =& T^{\text{U}}(I^{\text{U}}_{\text{A}}(\ell)), \\
I^{\text{L}}_{\text{E}}(\ell) =& T^{\text{L}}(I^{\text{L}}_{\text{A}}(\ell)),
\end{align}
where the superscripts U and L indicate the upper and lower decoders, respectively, $T(.)$ is the constituent convolutional decoder transfer function which is determined by Monte Carlo simulation. During decoding, the extrinsic output of one decoder is forwarded to the other decoder and becomes its \emph{a priori} input, such that
\begin{align}
I^{\text{L}}_{\text{A}}(\ell) =& I^{\text{U}}_{\text{E}}(\ell), \\
I^{\text{U}}_{\text{A}}(\ell+1)=&I^{\text{L}}_{\text{E}}(\ell).
\end{align}
To analyze the iterative decoding process, we can plot two EXIT curves, i.e., $I^{\text{U}}_{\text{E}}$ versus $I^{\text{U}}_{\text{A}}$ and $I^{\text{L}}_{\text{A}}$ versus $I^{\text{L}}_{\text{E}}$, on the same plot. The decoding is successful if $(I^{\text{U}}_{\text{E}},I^{\text{L}}_{\text{E}})$ reach $(1,1)$ and both curves do not intersect. The minimum required SNR for this to happen is the BP threshold on the AWGN channel. An example of the EXIT chart will be provided in Sec. \ref{sec:turbo}-\ref{sec:punc_design}\ref{sec:turbo_punc_exit}.

\subsection{Constituent Codes Design}\label{ssec:turbo_c}
The 3G \cite{CDMAFDD25.212,CDMATDD25.222} and 4G \cite{LTE136212} turbo codes, as well as the enhanced turbo codes \cite{R1-167413} all have two 8-state convolutional codes with generator polynomial $[1,15/13]_8$. In addition, the enhanced turbo codes adopt the convolutional generator polynomial $[1,15/13,17/13]_8$ for rates between $1/5$ and $1/3$ \cite{R1-167413}. The design of constituent convolutional codes affects a turbo code's threshold and error floor performance. To this end, the design criteria can be divided into distance-based and threshold-based.

For the distance-based design approach \cite{485713,494303}, one can adopt the methods in Section \ref{sec:turbo}-\ref{ssec:turbo_tool}\ref{turbo_dis} to find the constituent convolutional codes that lead to a larger minimum distance for the turbo code. To focus on the convolutional encoder design, it is commonly assumed the use of uniform random interleaving \cite{494303}. Essentially, the distance-based design criteria place more emphasis on the error floor region of turbo codes.

For the threshold-based approach, the choice of convolutional component codes can be determined by using either DE or EXIT charts \cite{957394} to find the turbo code ensemble that has the best decoding threshold. Both DE and EXIT charts assume infinite blocklength and ideal interleaving. At rate $1/3$, the convolutional component codes of several turbo ensembles that have larger BEC thresholds than the LTE turbo ensemble were reported in \cite[p117]{Measson2006thesis}. In addition, for rates $1/2$ and $1/3$, turbo ensembles that have AWGN thresholds within 0.3 dB from the capacity were reported in \cite{957394}.

%

\subsection{Interleaver Design}\label{sec:int_design}
The first purpose of interleaving is the time-spreading of errors that could be produced in bursts over the transmission channel. Secondly, the interleaver design has a great impact on the error correction performance of
the turbo code and especially on its minimum Hamming distance.
Denote the bits input to the turbo code interleaver by $\boldsymbol{u}=[u_0,\ldots,u_{K-1}]$. The output bits of the interleaver is then denoted by $\Pi(\boldsymbol{u})=[u_{\Pi(0)},\ldots,u_{\Pi(K-1)}]$.

Interleavers can be divided into random interleavers and deterministic interleavers \cite{4282133}. Random interleavers are usually used for analysis or simulation purposes \cite{525138,1315677}. From the implementation point of view, deterministic interleavers are more favorable compared to random interleavers. In what follows, we first present several criteria for interleaver design. Then, we introduce three popular deterministic interleaver classes. For these interleavers, only a few parameters rather than the whole interleaver indices need to be stored.

\subsubsection{Minimum Spread}
When designing an interleaver, an important parameter we need to consider is the minimum spread. It is defined as \cite{Dolinar95,Crozier2000}
\begin{align}
S_{\min} \triangleq \min_{i_1 \neq i_2 \in \{1,\ldots,K\}}|i_1-i_2|_K+|\Pi(i_1)-\Pi(i_2)|_K,
\end{align}
where $|a-b|_K\triangleq \min\{|a-b|,K-|a-b| \}$. The minimum spread measures the minimum cumulated spatial distance between two bits before and after interleaving. It was shown in \cite{1495840} that $S_{\min} \leq \lfloor\sqrt{2K}\rfloor$. A larger $S_{\min}$ can yield a larger minimum Hamming distance for turbo codes.

\subsubsection{Puncture-Constrained Design}\label{sec:turbo_pcd_int}
In LTE, puncturing and interleaver are separately designed such that the interleaver does not change with code rates \cite{Xu22}. However, when puncturing information bits, the effective puncturing pattern for the interleaved information sequence may be catastrophic or nearly catastrophic \cite{4658735}. An effective way to combat this issue is to constrain the interleaver such that the detrimental arrangements of information bits in the lower code are not possible. There are two types of puncturing constraints on interleavers: fully-constrained and partially-constrained \cite{4658735}.

\emph{2a) Fully Puncture-Constrained Interleaver}: When the interleaver is fully constrained under the information bits puncturing pattern, the effective puncturing pattern for the interleaved information sequence is the same as the actual puncturing pattern used for the uniterleaved information sequence. In other words, for an information puncturing mask of length $M$, all bits at offset $k \in\{1,\ldots,M\}$ in the mask, go to positions that are also at offset $k$. In effect, the information bits are divided into $M$ sets and are forced to be interleaved only within their own sets. This can avoid bad information puncturing patterns in the lower component codeword. However, this can increase the difficulty in searching for the interleaver that achieves a given spread.

\emph{2a) Partially Puncture-Constrained Interleaver}: Partially puncture-constrained interleavers include fully puncture-constrained interleavers as special cases. The punctured (unpunctured) indices at the upper encoder are interleaved only to the punctured (unpunctured) indices at the lower encoder. In effect, the information bits are divided into two sets (rather than $M$) and interleaved within these sets. Partially puncture-constrained interleavers typically have less difficulty in achieving a given spread and distance than fully puncture-constrained interleavers. However, for rate-compatible designs, fully puncture-constrained interleavers are preferred \cite{6133842}.

\subsubsection{Correlation Girth}

In the turbo decoding process, there are two types of information exchanges contributing to the error correcting process. The first one is the extrinsic information exchange between the two component decoders via the interleaver. The other one is the local information exchange between neighboring symbols within each component decoder due to the convolutional nature of the component codes. These two types of exchanges create some dependency loops between
sets of symbols, which can be represented by cycles in a correlation graph \cite{8214245,8625316}, in the same way as in the Tanner graph for LDPC codes \cite{1056404}.

\begin{figure}[ht]
				\centering
				\includegraphics[width=\linewidth]{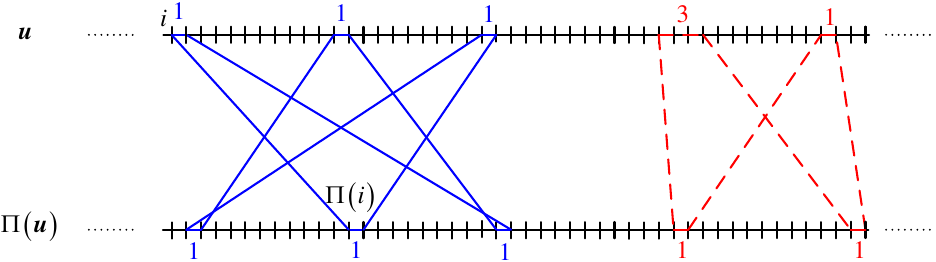}
				\caption{A correlation graph with two correlation cycles with length-6.} \label{fig:turbo_CG}
			\end{figure}

To see this, notice that the output information bit at position $i$ from the upper decoder depends on the received bit at the same position. Moreover, it is also affected by the received bits at positions in the vicinity of $i$. In addition, it also depends on the \emph{a priori} information provided by the lower decoder from position $\Pi(i)$. Similar observation can also be made for the output information bit at position $\Pi(i)$ from the lower decoder. As a result, a correlation graph can be established to design turbo code interleavers. An example of a correlation graph with two length-6 cycles is shown in Fig. \ref{fig:turbo_CG}.

Based on the correlation graph, it is natural to think about designing the interleaver by improving the minimum girth in the graph and reducing the number of minimum girths \cite{8214245}. Note also that in Fig. \ref{fig:turbo_CG}, for the cycle in the dashed red line, each symbol benefits from other symbols coming from three other trellis sections. In contrast, for the cycle in the blue solid line, each symbol benefits from symbols coming from five single distant trellis sections. As pointed out by \cite{8625316}, the first case has a higher code diversity such that a larger number of different and distant trellis sections participate in the cycle. Hence, \cite{8625316} introduced a criterion of maximizing the number of non-contiguous trellis sections participating in short correlation cycles. In addition, the number of symbols that participate in different short correlation cycles should be minimized \cite{8625316}.

\subsubsection{Protograph-Based Design}

	\begin{figure}[ht]
				\centering
				\includegraphics[width=0.8\linewidth]{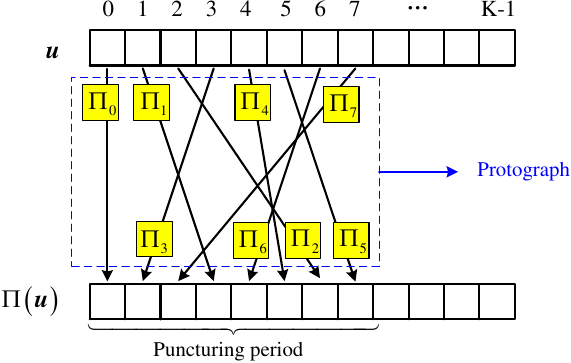}
				\caption{A protograph for puncturing period $M=8$.} \label{fig:PB_interleaver}
			\end{figure}

The concept of protograph \cite{PLDPC03} was applied to interleaver design under periodic information puncturing patterns \cite{8214245}. The protograph defines a set of inter-period permutations. Specifically, given an information puncturing pattern with period $M$, the protograph is represented by $M$ different sub-interleavers $\Pi_0,\ldots,\Pi_{M-1}$, where the $i$-th sub-interleaver $\Pi_i$, $i\in\{0,\ldots,M-1\}$, ensures that the symbols at a position within a puncturing period in $\boldsymbol{u}$ are interleaved (or connected) to a specified position within a puncturing period in $\Pi(\boldsymbol{u})$. This, combined with other design constraints, can also significantly limit the search space for the interleaver design. Fig. \ref{fig:PB_interleaver} shows an example of a protograph for puncturing period $M=4$. The connection design will be discussed in Section \ref{sec:turbo}-\ref{sec:int_design}\ref{sec:arp}.

%

\subsubsection{Quadratic Permutation Polynomial (QPP) Interleaver}
One of the popular classes of deterministic interleavers is based on QPP over integer rings \cite{1377495}. The interleaving function satisfies the following quadratic form
\begin{align}
\Pi(i)=(f_1\cdot i+f_2\cdot i^2)\bmod K,
\end{align}
where $f_1$ and $f_2$ are the interleaver coefficients. These polynomial coefficients are selected based on the maximization of the minimum Hamming distance of a subset of low-weight input sequences with weights of the form $2n$, for small $n>0$.

QPP interleavers have been shown to have large minimum Hamming distances \cite{4036316} and spread \cite{4215151}. In addition, \cite{4215151} established many useful properties regarding the spread to reduce the search space in searching for QPP interleavers with the largest available spread. Later, \cite{Trifina19} provides new searching methods for finding QPP interleavers for a target spread with reduced complexity. More importantly, QPP interleavers can be designed to have contention-free property, allowing a high degree of freedom for parallel processing \cite{1603790}. Due to these advantages, QPP interleaver has been adopted in the LTE standard turbo coding whose interleaver coefficients $f_1$ and $f_2$ can be found in Tables 5.1.3-3 in \cite{LTE136212}.

\subsubsection{Dithered Relative Prime (DRP) Interleaver}
DRP interleavers \cite{957178} were among the best-known interleavers for turbo codes according to \cite{4215151}. A DRP interleaver consists of the following interleaving stages \cite{957178}
\begin{align}
\Pi_1(i)=&F\lfloor i/F\rfloor+f_{(i\bmod F)}, \\
\Pi_2(i)=&(s+P \cdot i) \bmod K, \\
\Pi_3(i)=&G\lfloor i/G\rfloor+g_{(i\bmod G)}, \\
\Pi(i) =&\Pi_1(\Pi_2(\Pi_3(i))),
\end{align}
where $\boldsymbol{f}$ and $\boldsymbol{g}$ are the read and write dither vectors with lengths $F$ and $G$, respectively, $P$ is the regular permutation period, and $s$ is a constant shift. Note that $K$ must be a multiple of both $F$ and $G$. Let $E$ be the least common
multiple of $F$ and $G$. Then, the DRP interleaving function can be simplified \cite{957178}
\begin{align}
U(i) =& (\Pi(i)-\Pi(i-1))\bmod K , \\
\Pi(i)=&(\Pi(i-1)+U(i \bmod E))\bmod K,
\end{align}
where $\boldsymbol{U}=[U_{(0)},\ldots,U_{(E-1)}]$ is the index increments vector that needs to be stored and $\Pi(0)$ is arbitrary and can be set to 0. DRP interleavers also allow a parallelism of degree $C$ equal to any multiple of the dither vector length, provided that $C$ remains a factor of $K$ \cite{Crozier11}.

The design criteria for DRP interleaver parameters are similar to that for a QPP interleaver. First, a regular interleaver with high scattering properties is identified. Then, the dither vectors of the DRP interleaver are selected in order to maximize the minimum Hamming distance of a subset of low-weight input sequences \cite{Crozier05}.
To ensure that the interleaver also performs well under puncturing, \cite{6133842} and \cite{6364028} designed partially and fully puncture-constrained DRP interleavers , respectively, based on distance search and error impulse method as introduced in Section \ref{sec:turbo}-\ref{ssec:turbo_tool}\ref{turbo_dis}. Notably, turbo codes with 4-state convolutional constituent codes and rate larger than $2/3$ can have decoding thresholds within 0.1 dB to the BI-AWGN capacity under fully puncture-constrained DRP interleavers \cite{6364028}.


%

\subsubsection{Almost Regular Permutation (ARP) Interleaver}\label{sec:arp}
ARP interleavers were originally proposed in \cite{1312507}. It is based on a regular permutation of period $P$ and a vector of shifts $\boldsymbol{S} = [S_{(0)},\ldots,S_{(Q-1)}]$, where $Q$ represents the number of shifts. The interleaving function is given by
\begin{align}
\Pi(i)=(P\cdot i +S_{(i \bmod Q)}) \bmod K.
\end{align}
Note that $K$ must be a multiple of $Q$. Moreover, $P$ and $K$ must be mutually prime numbers. A major advantage of ARP interleaving is that it naturally offers a parallelism degree of $Q$ in the decoding process \cite[Ch. 3.3.2]{BENEDETTO201453}. It is worth noting that ARP interleavers were adopted in the IEEE 802.16 WiMAX \cite{WIMAX_turbo} and DVB-RCS2 \cite{DVB_RCS2_turbo} standards.

It is worth pointing out that the ARP interleavers adopted in enhanced turbo codes \cite{R1-167413} are protograph based \cite{8214245}, where the values of $Q$, $P$, and $[S_{(0)},\ldots,S_{(Q-1)}]$ for various $K$ and $R$ can be found in Tables 6.1.2, 7.1.2, 7.1.3, 7.2.3, and 7.2.4 in \cite{R1-167413}. Significant gains in terms of better waterfall and error floor performance over LTE interleaving and puncturing were reported in \cite{R1-167414}. To design the protograph-based ARP interleavers, the first step is to classify the positions from the least reliable information bit to the most reliable bit based on their distance spectrum of the punctured RSC code \cite{8214245}. The protograph involves connecting the least reliable bit position of one RSC code to the most reliable bit position of the other one, the second least reliable bit position of one RSC code to the second most reliable bit position of the other one, and so on \cite{8214245}. This spreads the error correction capability of the turbo code over the whole information block.

To allow the error floor of turbo codes to approach the union bound based on their Hamming distance spectrum, \cite{8625316} incorporated additional criteria in terms of minimizing the multiplicity of small correlation cycles, the number of symbols that participate in different short correlation cycles, and maximizing the number of non-contiguous trellis sections participating in short correlation cycles.

\subsubsection{Equivalence between ARP, QPP, and DRP Interleavers}
Interestingly, it was analytically shown in \cite{6948262} that DRP and QPP interleavers can be expressed in terms of ARP interleavers. Later, the equivalence between cubic permutation polynomial and APR interleavers has been shown \cite{7744593}. These results imply that ARP interleavers are capable of achieving at least the same interleaving properties and the same distance spectra as QPP or DRP interleavers. Consequently, a unified design on the interleaverd for turbo codes becomes possible.



\subsection{Puncturer Design}\label{sec:punc_design}
Puncturing can be classified into random puncturing and deterministic puncturing. Random puncturing is usually used for decoding threshold analysis. In this section, we focus on deterministic puncturing. Moreover, we consider a periodic puncturing pattern. Note that the performance difference between the periodic puncturing and the non-periodic puncturing becomes negligible when the puncturing period is large \cite{5371462}.

Consider a rate-$1/3$ turbo code. The puncturing pattern or puncturing mask can be represented by
\begin{align}
\boldsymbol{p}_{i} = [p_{i,0},p_{i,1},\ldots,p_{i,M_i-1}],
\end{align}
where $i\in\{s,U_p,L_p\}$ indicates that the puncturing pattern is for information, upper, and lower parity sequences, respectively, and $M_i$ is the puncturing period such that the puncturing is performed for every $M_i$ bit. Note that $M_i$ should be a divisor of the information length $K$. Moreover, each element in $\boldsymbol{p}_{i}$ only takes either 0 or 1, meaning that the corresponding code bit is not transmitted or transmitted, respectively. After puncturing, the code rate becomes
\begin{align}
R = \frac{1}{\frac{\sum^{M_s-1}_{j=0}p_{s,j}}{M_s}+\frac{\sum^{M_{U_p}-1}_{j=0}p_{U_p,j}}{M_{U_p}}+\frac{\sum^{M_{L_p}-1}_{j=0}p_{L_p,j}}{M_{L_p}}}.
\end{align}

Conventionally, the puncturing patterns were designed based on the distance properties of turbo codes, see Section \ref{sec:turbo}-\ref{ssec:turbo_c}\ref{turbo_dis}, \cite[Sec. 3]{6133842}, and the references therein. Apart from the distance-based designs, we also present other puncturing design criteria in the following.

\subsubsection{Puncturer Design Based on Density Evolution}
The design of puncturing patterns can be carried out by using DE on the BEC \cite{5577801}. The puncturing pattern is determined by optimizing the BEC decoding threshold. The derivation of DE equations mostly follows Section \ref{sec:turbo}-\ref{ssec:turbo_tool}\ref{sec:turbo_DE}, except that the fixed puncturing pattern is incorporated into the transfer functions of the constituent convolutional codes under BCJR decoding. To illustrate the key idea, consider a rate-$1/2$ convolutional code with parity puncturing pattern $\boldsymbol{p}=[p_1,\ldots,p_M]$ with period $M$. Then, the transfer function of the BCJR decoder for the $l$-th code bit, $l\in\{1,2\}$ is given by \cite{5577801}
\begin{align}
f_{\boldsymbol{p},l}(x_1,x_2) = \frac{1}{M-1}\sum^{M-1}_{j=0}f_{p_j,l}(x_1,x_2),
\end{align}
where $f_{p_j,l}(x_1,x_2)$ is the transfer function for the $l$-th code bit under the condition of whether the corresponding parity bit is punctured ($p_j=0$) or not ($p_j=1$). It is worth noting that the DE based approach can detect a catastrophic puncturing case where infinite error events occur \cite{5577801}.


\subsubsection{Puncturer Design Based on EXIT Charts}\label{sec:turbo_exit_punc}
Puncturing patterns can also be designed by the EXIT chart analysis \cite{4658735,8214245}. When only parity bits are punctured, the convergence behavior of the resultant turbo codes can be predicted by the EXIT charts \cite{957394}. However, when information bits are punctured, the distribution of the extrinsic information related to punctured information positions is different from that related to the unpunctured counterpart \cite{4658735}. Rather than relying on the Gaussian approximation of the \emph{a priori} information $I_{\text{A}}$ in the conventional EXIT chart analysis, the \emph{a priori} mutual information within the actual turbo decoding is measured by Monte Carlo simulations \cite{8214245}. The best puncturing pattern in terms of convergence performance is the one providing the closest crossing point $(I^{\text{U}}_{\text{E}},I^{\text{L}}_{\text{E}})$ to $(1,1)$.

\subsubsection{Puncturer Design Based on Distance and Threshold criteria}\label{sec:turbo_punc_exit}
For some configurations of block sizes and code rates, the current LTE puncturing does not lead to a good error floor and decoding threshold due to the undesirable interactions between the puncturer and the interleaver \cite{4656994}. To address this issue, a joint design of puncturing and interleaving is necessary \cite{8214245}. More importantly, the design of puncturing patterns needs to take into account both distance and threshold criteria.

	\begin{figure}[ht]
				\centering
				\includegraphics[width=0.8\linewidth]{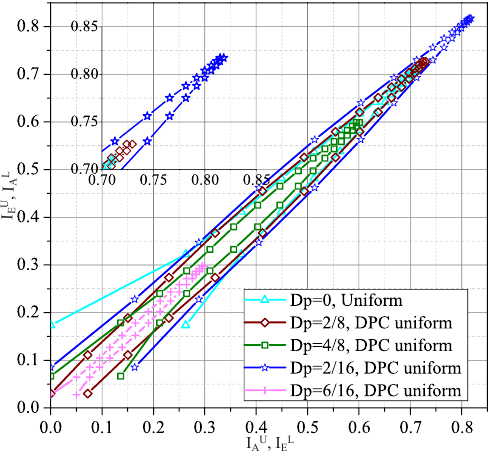}
				\caption{Extrinsic information exchange between two constituent codes of the LTE turbo code at $E_b/N_0=1.6$ dB. Five puncturing patterns with different data puncturing ratios are applied to reach code rate $2/3$ \cite{8214245}.} \label{fig:Turbo_EXIT_dpc}
			\end{figure}

To ensure achieving a good minimum distance and decoding threshold at the same time, a set of candidate puncturing patterns are selected according to the distance criteria as in Section \ref{sec:turbo}-\ref{ssec:turbo_tool}\ref{turbo_dis}. In \cite{8214245}, the FAST algorithm \cite{45271} was employed to evaluate the truncated distance spectrum of each constituent convolutional code punctured by a periodic pattern of period $M$. Moreover, the FAST algorithm needs to run $M$ times, where each time
starts from position $j\in\{0,\ldots,M-1\}$ in the puncturing pattern. The resulting $M$ distance spectra are accumulated to obtain the Hamming distance spectrum of the punctured convolutional code. The set of puncturing pattern candidates is selected by finding the largest distance values in the first spectrum terms and the minimal number of codewords at these distances. Then, the EXIT analysis following Section \ref{sec:turbo}-\ref{ssec:turbo_tool}\ref{sec:turbo_exit_punc} is carried out for these puncturing patterns. For example, Fig. \ref{fig:Turbo_EXIT_dpc} shows the EXIT curves for the LTE turbo code with uniform interleaving under data puncturing constraint (DPC uniform) \cite{8214245}, where puncturing patterns with different data puncturing ratios $D_p$ are applied. From Fig. \ref{fig:Turbo_EXIT_dpc}, we see that the puncturing pattern with $D_p=2/16$ has the best convergence performance. Finally, the puncturing pattern that has the best trade-off between performance in the waterfall and error floor regions is selected. Under the constraints introduced by the selected puncturing pattern (see Section \ref{sec:turbo}-\ref{sec:int_design}\ref{sec:turbo_pcd_int}), the interleaver that achieves the best turbo code Hamming distance spectrum is selected. The puncturing patterns for the enhanced turbo codes proposed for 5G can be found in Tables 6.1.1, 7.1.1, 7.2.1, and 7.2.2 in \cite{R1-167413}.

\subsection{Turbo Code Decoding Algorithms}
In Section \ref{sec:turbo}-\ref{ssec:turbo_props}\ref{sec:turbo_dec_intro}, we know that the turbo decoder is an iterative BCJR decoder. However, even though the BCJR decoder is recursive, it poses implementation challenges because of the necessity of non-linear functions, and a large number of addition and multiplications \cite{524253}. In this section, we introduce several variants of turbo decoding algorithms. Note that we focus on the AWGN channel in this section. Recently, machine learning based turbo decoders have also been proposed in \cite{8815400,9145754,9834589} to improve the robustness and adaptivity of non-AWGN channels.

\subsubsection{The Log-MAP Decoder}
As introduced in Section \ref{sec:turbo}-\ref{ssec:conv}\ref{sec:conv_dec}, the Log-MAP decoder is the log domain implementation of the original BCJR decoder. By converting into the logarithmic domain, the Log-MAP algorithm replaces the multiplications and additions from the original BCJR decoder by additions and the max star functions defined in \eqref{eq:max_star}.

	\begin{figure}[ht]
				\centering
				\includegraphics[width=\linewidth]{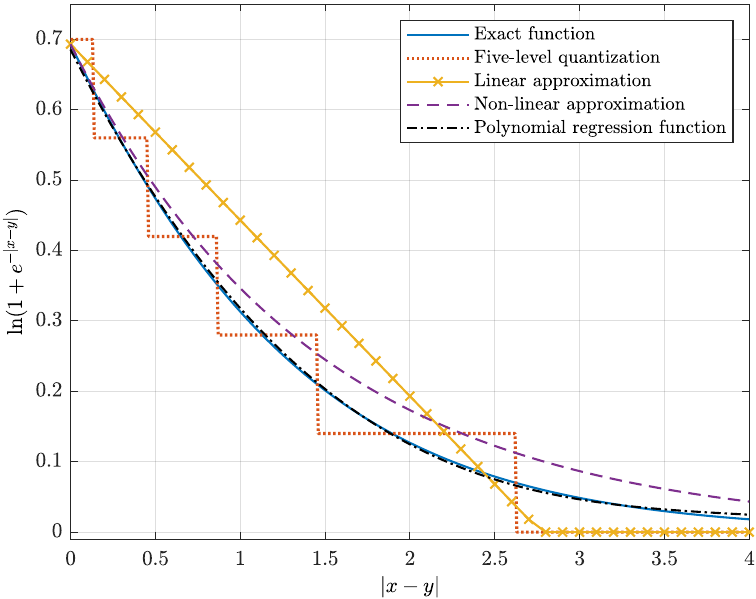}
				\caption{Plot of $\ln(1+e^{-|x-y|})$, showing the exact values and four approximated values.} \label{fig:max_star_quantize}
			\end{figure}

Notice that the term in \eqref{eq:max_star}, i.e., $\ln(1+e^{-|x-y|})$, is called the correction term. Since direct computation is costly to implement in hardware, it is often desirable to approximate the correction term. Since the value of $\ln(1+e^{-|x-y|})$ is always in the range $[0,0.69]$ \cite{7131434}, it can be easily pre-calculated and stored in a lookup table for efficient implementation \cite{ett4460080202}. Specifically, a table size of eight is usually sufficient to keep the same performance as the original MAP decoder \cite{524253}. In \cite{851673}, a linear approximation for the correction term was suggested as follows
\begin{equation}
\ln(1+e^{-|x-y|}) \approx \max\left(\log2-\frac{|x-y|}{4},0\right).
\end{equation}
Inspired by the linear approximation, \cite{1603379} introduced the following non-linear approximation with higher accuracy
\begin{equation}
\ln(1+e^{-|x-y|}) \approx \frac{\log2}{2^{|x-y|}}.
\end{equation}
In \cite{7247502}, a more accurate approximation based on polynomial regression functions was proposed as follows
\begin{align}
\ln(1+e^{-|x-y|}) \approx \left\{\begin{array}{l}
-0.0098(x-y)^3 + 0.1164(x-y)^2 \\
-0.474(x-y) + 0.6855, \; x_y\leq 5 \\
0, \qquad\qquad \text{otherwise}
\end{array}\right..
\end{align}

Fig. \ref{fig:max_star_quantize} shows the comparison between the exact correction term and the approximated correction terms computed by the above four methods.

\subsubsection{The Max-Log-MAP Decoder}
The Max-Log-MAP decoder simplifies the Log-MAP decoder by approximating the max star functions in \eqref{eq:metric_dec}-\eqref{eq:turbo_dec_ut} as \cite{524253}
\begin{align}
\smax(x,y)=\ln(e^x+e^y) \approx \max(x,y).
\end{align}
At the cost of some performance degradation, the Max-Log-MAP algorithm does not require the computation of the correction terms in the max star function. Moreover, it is the most currently used in hardware implementations of turbo decoders \cite{BENEDETTO201453}. An interesting advantage of the Max-Log-MAP decoder compared to the original MAP decoding is that the estimation of signal-to-noise ratio is not required \cite{848410}. This makes the Max-Log-MAP decoder more robust to the channel estimation error than the Log-MAP decoder.

To reduce the performance loss, \cite{Vogt00} suggested multiplying the extrinsic information, i.e., \eqref{eq:turbo_ext1} and \eqref{eq:turbo_ext2}, by a scaling factor $s^{(\ell)}\in[0,1]$ before being used by a SISO decoder. For easy hardware implementation, the scaling factor is fixed to 0.7 \cite{Vogt00} or 0.75 \cite{4656994} for all iterations. For the decoding of the enhanced turbo codes \cite{R1-167414}, the following scaling factor is used: $s^{(1)}=0.6$, $s^{(\ell_{\max})}=1$, $s^{(\ell)}=0.7,\forall \ell\in\{2,\ldots,\ell_{\max}-1\}$. One may also design the scaling factors adaptive to SNR as in \cite{Claussen05}.

\subsubsection{The Local SOVA Decoder}
To improve the decoding throughput, high-radix decoding, i.e., several successive symbols are decoded at once, is often considered \cite{7137638}. However, for the Max-Log-MAP algorithm, the computational complexity increases rapidly with the radix orders. Inspired by the early discovery on the equivalence between the Max-Log-MAP decoding and SOVA decoding in \cite{673659}, the local SOVA decoder was introduced in \cite{8960428} as a low complexity alternative to the Max-Log-MAP decoder. It relies on a new low-complexity soft-output calculation unit that applies a path-based decoding variant of the Max-Log-MAP algorithm. It was revealed that the soft output generated by the Max-Log-MAP algorithm is a special case of the local SOVA method \cite{8960428}. Hence, the local SOVA decoder can achieve the same performance as that of the Max-Log-MAP decoder but with lower complexity.

The local SOVA decoder uses the same recursive metrics calculations in \eqref{eq:metric_dec} and \eqref{eq:metric_dec2}. However, the LLRs are computed using path-based local update rules known as the Hagenauer rule \cite{64230} and the Battail rule \cite{Battail87,595033}. Consider a radix-$2^q$ trellis where there are $2^q$ branches coming in and out of a state $s$ at a time index. For each trellis section, a path $P_s$ for state $s$ is defined as
\begin{equation}
P_s \triangleq \{M_s, u_s, L_s\}\in \mathbb{R} \times \{0,1\}^q\times \{\mathbb{R}^+\}^q,
\end{equation}
where $M_s$ is the path metric, $u_s \triangleq \{u_{s,0},\ldots,u_{j,q-1}\}$ is the $q$ hard decisions labeling this path, and $L_s \triangleq \{L_{s,0},\ldots,L_{s,q-1}\}$ are the reliability values associated with the hard decisions.  Two paths $P_a$ and $P_b$ can be merged into path $P_c$ according to
\begin{align}
M_c =& f_0(M_a,M_b) \triangleq \max(M_a,M_b), \label{eq:turbo_path _merge}\\
u_c(l) = & f_1(u_a(l),u_b(l)), \forall l \in\{0,\ldots,q-1\}, \label{eq:LSOVA_f1}
\end{align}
where $f_1$ selects the hard decision of the winning path resulting from \eqref{eq:turbo_path _merge}. Define $p \triangleq \arg\max( M_a,M_b)$, $p' \triangleq \arg\min( M_a,M_b)$, and $\Delta_{p,p'} \triangleq M_p-M_{p'}$. The reliability of $u_c(l)$ is updated as
\begin{align}
L_c(l) =&f_2(L_a(l),L_b(l)), \forall l\in\{0,\ldots,q-1\}, \label{eq:LSOVA_f2} \\
=&\min(L_p(l),\Delta_{p,p'}+\boldsymbol{1}(u_a(l) = u_b(l))\cdot L_{p'}(l)).\label{eq:LSOVA_update_rule}
\end{align}
where $\boldsymbol{1}(.)$ is the indicator function, and \eqref{eq:LSOVA_update_rule} is the result of combining both the Hagenauer rule \cite{64230} and the Battail rule \cite{Battail87,595033} for the cases $u_a(l) \neq u_b(l)$ and $u_a(l) = u_b(l)$, respectively.

Motivated by the feature of convolutional codes that all trellis paths merge to the maximum likelihood path after some trellis steps, \cite{9594265} further proposed a low-complexity local SOVA decoder. Specifically,  the reliability values $L$ for all previous trellis steps during the computation of the recursion metrics for each new radix-$2^q$ trellis segment are updated instead of being recursively computed. To see this, let $P^{\text{f}}_1(i),\ldots,P^{\text{f}}_{2q}(i)$ be the paths at trellis position $i$ to be merged to compute $P^{\text{f}}_{\text{out}}(i+q)$ at trellis position $i+q$. The merge operation computes
\begin{align}
M^{\text{f}}_{\text{out}} =& f_0(M_1,\ldots,M_{2q}) = \max(M^{\text{f}}_1,\ldots,M^{\text{f}}_{2q}), \\
u^{\text{f}}_{\text{out}}(l) =& f_1(u^{\text{f}}_1(l),\ldots,u^{\text{f}}_{2q}(l)), \forall l \in\{1,\ldots,i+q\},
\end{align}
where $f_1$ is defined in \eqref{eq:LSOVA_f1} below. The reliability values are updated via
\begin{equation}
L^{\text{f}}_{\text{out}}(l) = f_2(L^{\text{f}}_1(l),\ldots,L^{\text{f}}_{2q}(l)), \forall l \in\{1,\ldots,i+q\},
\end{equation}
where $f_2$ is defined in \eqref{eq:LSOVA_f2}. The merge operations of backward recursions are performed similarly. Compared to the original local SOVA algorithm, a complexity reduction of the add-compare-select units in the order of $50\%$ can be achieved at the price of 0.2 dB performance degradation \cite{9594265}.

\subsubsection{CRC-Aided Turbo Decoding}
In the LTE standard \cite{LTE136212}, a CRC code is used as the outer error detection on top of a turbo code. Hence, it is natural to leverage CRC to improve the performance of turbo decoding \cite{5755908,4273779,7475891,9594119}.

\emph{4a) Flip and Check Decoding}: The most intuitive method is the Flip and Check (FC) algorithm proposed in \cite{7475891}. The turbo decoder first iterates $\ell_{\min}$ times without using CRC verification. At iteration $\ell_{\min}+1$ and when the CRC check fails, the reliability of the $k$-th information bit, $k\in \{1,\ldots,K\}$, is characterized according to the following extrinsic-information-based metric
\begin{equation}
\Delta^{(\ell)}_{\text{E}}(u_k) \triangleq |L^{(\ell)}_{\text{E}}(u_k)+\Pi^{-1}(L^{(\ell)}_{\text{E}}(u_{\Pi(k)}))|,
\end{equation}
where we adopt the same notations of the extrinsic information as in Section \ref{sec:turbo}-\ref{ssec:turbo_props}\ref{sec:turbo_dec_intro}. A set of test patterns $\boldsymbol{\tau}_j$, $j\in\{1,\ldots,2^{q_{\text{FC}}}-1\}$, are generated by identifying the least $q_{\text{FC}}$ reliable bits based on the smallest values of $\Delta^{(\ell)}_{\text{E}}(u_k)$. We denote by $\boldsymbol{\hat{u}}^{(\ell)}$ the least $q_{\text{FC}}$ reliable bits after $\ell$ turbo decoding iterations. Then, the candidate CRC codeword is generated by flipping the decoded values of those bits
\begin{equation}
\boldsymbol{\tilde{u}}_j = \boldsymbol{\hat{u}}^{(\ell)}\oplus \boldsymbol{\tau}_j,
\end{equation}
where $\oplus$ denotes the modulo 2 sum. The CRC check is performed on all candidates $\boldsymbol{\tilde{u}}_j$ for $j\in\{1,\ldots,2^{q_{\text{FC}}}-1\}$. The one that passes the CRC check is kept. Finally, the turbo decoding process and the FC principle are repeated, until the CRC is verified or $\ell = \ell_{\max}$. To reduce the undetected error rate due to wrongly codewords satisfying the CRC, \cite{9594119} proposed a new reliability metric based on choosing the smallest Euclidean distance between the CRC codeword after flipping and the received information.

\emph{4b) Blind Candidate Decoding}: Recently, a new method called Blind Candidate Decoding (BCD) is proposed in \cite{9594119}, which can combine with the FC principle above. The key idea is to generate a set of candidates within an Euclidean sphere around the original received information LLR $L^{(0)}(\boldsymbol{y})$. Let $G_{\gamma}(k)$ be a length $K$ vector filled with equally spaced values centred at 0
\begin{equation}
G_{\gamma}(k) = \gamma\frac{2k}{K-1}-1, k\in \{1,\ldots,K\},
\end{equation}
where $\gamma$ is a parameter that controls the norm $d_{\gamma}=\gamma^2E_K$ of vector $G_{\gamma}$, and $E_K$ is a constant given by \cite{9594119}
\begin{equation}
E_K = \frac{K}{3}\left(\frac{2}{K-1}+1\right).
\end{equation}
Then, generate $n_c$ number of candidates as
\begin{align}
L^{(\ell)}(\tilde{y}_{k,i}) =& L^{(0)}(y_{k,i})+G_{\gamma}(\Pi_{i}(k)), \;k=1,\ldots,K, \nonumber \\
 &i = 1,\ldots,n_c,
\end{align}
where $\Pi_{i}(.)$ is a permutation function for indices set $\{1,\ldots,K\}$ randomly generated for each candidate $i\in \{1,\ldots,n_c\}$, and $n_c$ denotes the number of candidates. Each candidate LLR vector $L^{(\ell)}(\boldsymbol{\tilde{y}_{i}})$ is input to the turbo decoder with CRC check. The optimal distance $d_{\gamma}$ for a given code configuration and SNR is determined by Monte-Carlo simulation \cite{9594119}.

\subsection{High-Throughput Turbo Decoder Hardware Architectures}\label{sec:turbo_hardware}
The BCJR algorithm is serial in nature due to the recursive calculation of the state metrics. As a result, it requires a relatively large amount of memory for storing the state metrics, and its throughput can be very limited. To this end, several works have proposed new hardware architectures for high-throughput turbo decoders beyond 100 Gb/s.

\subsubsection{Parallel MAP (PMAP) Architecture}
\begin{figure}[ht]
				\centering
				\includegraphics[width=0.5\linewidth]{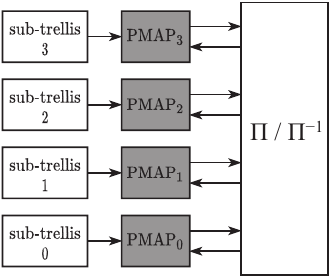}
				\caption{The PMAP architecture with $P=4$ \cite{Vinh2021thesis}.} \label{fig:PMAP}
			\end{figure}

Turbo decoders with a PMAP architecture divide a length-$K$ trellis into $P$ sub-trellis of length $K_P$. Each sub-trellis is decoded independently by a sub-decoder core in parallel \cite{Thul05}. Additionally, each sub-decoder core splits the sub-blocks further into smaller blocks with size $W$ called windows \cite{Benedetto96}, to enable a parallel processing of the forward and backward recursions. The PMAP architecture with $P=4$ is illustrated in Fig. \ref{fig:PMAP}.

Let $n_{\text{HI}}$ be the number of half iterations performed by the decoding process. By neglecting the I/O latency and the latency due to metric initialization, the latency of the turbo decoder with PMAP architecture can be estimated in terms of clock cycles as $(K_P+W)n_{\text{HI}}$ and the throughput is $\frac{Kf}{(K_P+W)n_{\text{HI}}}$ \cite{Vinh2021thesis}, where $f$ is the maximum operating frequency. With a fixed frame size $K$, the throughput increases with the number of parallel sub-trellises $P$. The asymptotic throughput of the PMAP architecture is $\lim_{k \rightarrow \infty}\frac{1}{(K_P/K+W/K)n_{\text{HI}}}=\frac{Pf}{n_{\text{HI}}}$. Implementations with current silicon technologies achieve a throughput in the order of single-digit Gb/s \cite{6325191,6847747}. Note that the maximum degree of parallelism is limited since the decoding of small sub-blocks leads to an error correction performance loss \cite{9120779}. In addition, the amount of hardware resources increases by a factor $P$ since each sub-trellis is processed by an independent SISO decoder \cite{Vinh2021thesis}.

\subsubsection{Pipelined MAP (XMAP) Architecture}
Given $P$ sub-trellises of length $K_P$, the XMAP architecture processes a single sub-trellis at a time \cite{886725,7593077}. Sub-trellises are time-multiplexed into a pipeline consisting of a chain of computation units (branch metric unit (BMU), add-compare-select unit (ACSU), and soft-output unit (SOU)) connected through pipeline registers. As a result, for each clock cycle, the decoder can produce $K_P$ soft-output values. Note that the XMAP architecture only differs from the PMAP architecture in the way each individual sub-trellis is processed.

By neglecting the I/O latency and the latency due to metric initialization, the latency of the turbo decoder with XMAP architecture in clock cycles is $(K_P+P-1)n_{\text{HI}}$ and the throughput is $\frac{Kf}{(K_P+P-1)n_{\text{HI}}}$ \cite{Vinh2021thesis}. A throughput of over 1 Gb/s has been demonstrated in \cite{6779698,7593077}. Since the XMAP core consists of a chain of computation units set up in a pipeline fashion, its complexity increases linearly with the sub-trellis length $K_P$. Similarly to PMAP, the maximum degree of parallelism of XMAP is also limited due to the decoding of small sub-blocks leading to an error correction performance loss \cite{9120779}.

\subsubsection{Fully Parallel MAP (FPMAP) Architecture}
\begin{figure}[ht]
				\centering
				\includegraphics[width=\linewidth]{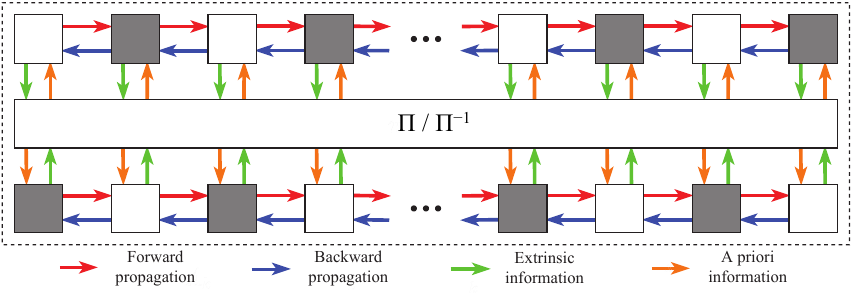}
				\caption{The FPMAP architecture \cite{Vinh2021thesis}.} \label{fig:FPMAP}
			\end{figure}

The FPMAP architecture \cite{7137638} can be seen as an extreme case of the PMAP architecture with $P=K$ such that the size of the sub-trellises is reduced to $K_P=1$. It uses a shuffled decoding scheme \cite{1402639}. It employs $2K$ processing elements (PE). Each PE computes the branch metrics, the forward and backward state metrics, and the extrinsic information for one trellis step in one clock cycle. The calculated state metrics at the border are then exchanged with neighboring PEs. Meanwhile, the calculated extrinsic information is fed to the interleaved/deinterleaved PEs. Consequently, a complete turbo code iteration is processed in parallel in each clock cycle. For the LTE turbo codes, the use of the odd-even QPP interleaver helps split the PEs into two groups, where each group consists of $K/2$ PEs of the component SISO decoder and can be processed independently. The FPMAP architecture is illustrated in Fig. \ref{fig:FPMAP}.

By neglecting the I/O latency, the decoder latency in clock cycles of the FPMAP is $2n_{\text{I}}$, where $n_{\text{I}}$ is the number of full iterations. The throughput of the FPMAP decoder can be calculated as $\frac{Kf}{2n_{\text{I}}}$ \cite{Vinh2021thesis}. Although the FPMAP can achieve a high throughput, the price to pay is a decrease in area efficiency compared to the PMAP architectures \cite{Vinh2021thesis}. In addition, the combination of sub-trellis size of 1 and shuffle decoding degrades the error correction performance of the decoder \cite{8625377}.

\subsubsection{Iteration Unrolled XMAP (UXMAP) Architecture}
\begin{figure}[ht]
				\centering
				\includegraphics[width=\linewidth]{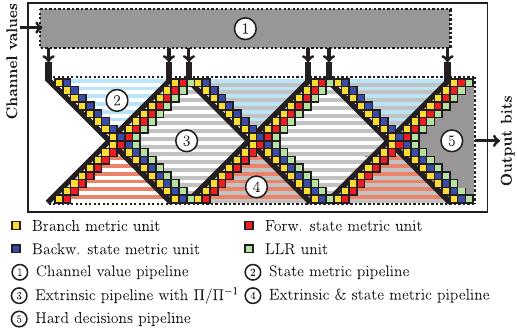}
				\caption{Iteration unrolled pipelined decoder architecture \cite{8625377}.} \label{fig:UXMAP}
			\end{figure}

Further pipelining of the decoding by unrolling the individual half iteration of the turbo decoding leads to the fully pipelined UXMAP decoder architecture \cite{8625377}. This allows for the output of a complete decoded frame per clock cycle resulting in a very high throughput, which is only limited by the achievable clock frequency and frame size. The architecture is illustrated in Fig. \ref{fig:UXMAP}. By neglecting the I/O latency and the latency due to the metric initialization, the throughput of the UXMAP decoder architecture is $Kf$. The latency of the decoder can be derived in clock cycles as $n_{\text{HI}}(T_{\text{BMU}}+K_P+T_{\text{SOU}})$, where $T_{\text{BMU}}$ and $T_{\text{SOU}}$ are the number of clock cycles required for processing the first BMU and the last SOU, respectively. Moreover, the ACSU takes one clock cycle to finish a trellis section, thus, it takes KP clock cycles to finish a sub-trellis.

Employing high radix schemes in the fully pipelined UXMAP architecture has a particular impact. Increasing the number of trellis sections processed in a clock cycle leads to a reduction in the number of pipeline stages for all the pipeline registers (state metrics, channel, and extrinsic information). Hence, using radix-4 instead of radix-2 yields an area
saving of about $50\%$ and halves the overall pipeline latency. However, as we increase the radix order, the area overhead in the computation units also increases. In addition, \cite{9120779} investigated the rate and frame flexibility aspect of the UXMAP. A throughput of 409.6 Gb/s was reported in \cite{9053453}.

\begin{figure}[ht]
				\centering
				\includegraphics[width=0.8\linewidth]{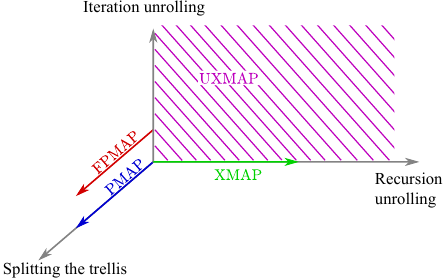}
				\caption{Parallelism in turbo decoder hardware architectures \cite{9053453}.} \label{fig:turbo_parallel}
			\end{figure}

Finally, different methods of parallel processing are illustrated in Fig. \ref{fig:turbo_parallel}.

\subsection{Turbo Coded Modulations}\label{sec:turbo_CM}
Unlike LDPC codes, turbo coded modulations receive less attention. Among those works, most of them focus on bit-interleaved turbo coded modulation (BITCM) schemes \cite{1703814,5407616}. Alternatively, one can use non-binary turbo codes \cite{Klaimi2019thesis} constructed from convolutional codes over rings \cite{Massey89NBCC} to map codeword symbols to modulation symbols directly.

BITCM follows the principles of BICM, where the turbo codeword is interleaved and mapped to higher-order modulations. \cite{1703814} introduced a greedy algorithm for designing bit interleavers to lower the error floor of BITCM on the AWGN channel. It requires an accurate list of low-weight turbo codewords as the input to the algorithm. \cite{5407616} exploited unequal error protection caused by the binary labeling of the equally spaced constellations in the interleaver design. By using Gaussian approximation for the L-values in QAM and the generalized transfer function of a code, the union bound of the coded bit error rate was derived and used as the metric for bit interleaver design \cite{5407616}. It can be seen that these methods can also be applied to the bit interleaver design for other codes.

\subsection{Other Variants of Turbo Codes}
In the following two subsections, we introduce several variants of turbo codes that may be the candidate channel coding schemes for future communication systems. In this subsection, we introduce irregular turbo codes and serially concatenated convolutional codes. In the next subsection, we introduce different classes of spatially coupled turbo codes.

\subsubsection{Irregular Turbo Codes}
Irregular turbo codes were first introduced in \cite{866413} as a generalization of the regular turbo codes \cite{397441}. It was numerically demonstrated in \cite{1023328} that a rate-$1/3$ irregular turbo code with $[1,21/37]_8$ convolutional component codes can achieve a threshold within 0.03 dB from the BI-AWGN capacity. \cite{5577801} adopted density evolution to design capacity-approaching irregular turbo codes and periodic puncturing patterns on the BEC.

The conventional regular turbo encoder with two identical constituent convolutional encoders can be seen that the information bit is repeated twice, interleaved, and fed to the constituent encoder. In this regard, an irregular turbo encoder consists of a non-uniform repetition, an interleaver, and an RSC component encoder. First, the information bits can be divided into $d$ classes with $d=2,\ldots,d_{\max}$, where $d_{\max}$ is the maximum bit-node degree. The number of bits in class $d$ is a fraction $f_d$ of the total number of information bits at turbo encoder input. Moreover, each information bit in class $d$ is repeated $d$ times. After irregular repetition, the length-$K$ information sequence becomes a length-$N$ sequence. As a result, the following equalities hold
\begin{equation}
\sum^{d_{\max}}_{d=2}f_d = 1, \sum^{d_{\max}}_{d=2}d\cdot f_d = \bar{d},N=K\sum^{d_{\max}}_{d=2}d\cdot f_d = K\bar{d},
\end{equation}
where $\bar{d}$ denotes the average bit-node degree. The length-$N$ sequence is interleaved and fed into a constituent convolutional encoder with rate $R_C$. The code rate of the irregular turbo code is
\begin{equation}
R = \frac{K}{K+\frac{N}{R_C}-N}=\frac{1}{1+\left(\frac{1}{R_C}-1\right)\bar{d}}.
\end{equation}

\subsubsection{Serially Concatenated Convolutional Codes (SCCCs)}
Serially concatenated convolutional codes (SCCCs) were introduced in \cite{669119}. Its encoder consists of an outer and inner RSC component encoder and an interleaver.


Consider a rate-$1/4$ SCCC with two rate-$1/2$ constituent encoders. A length $K$ information sequence $\boldsymbol{u}$ is encoded by the outer encoder to produce the parity sequence $\boldsymbol{v}^{\text{O}}$. Then, the sequences $\boldsymbol{u}$ and $\boldsymbol{v}^{\text{O}}$ are multiplexed and interleaved become $\Pi([\boldsymbol{u},\boldsymbol{v}^{\text{O}}])$. Then, the inner encoder takes this interleaved sequence as the input and generates the parity sequence $\boldsymbol{v}^{\text{I}}$. The transmitted codeword is $\boldsymbol{c}=[\boldsymbol{u},\boldsymbol{v}^{\text{O}},\boldsymbol{v}^{\text{I}}]$. In general, PCCCs have better waterfall performance than SCCCs. However, in some cases and with a careful puncturing design, SCCCs can achieve a lower error floor and comparable waterfall performance compared to PCCCs \cite{5201021}.

\subsection{Spatially Coupled Turbo Codes}\label{ssec:sc_turbo}
Recently, spatially coupled turbo codes have gained some interest since the work of \cite{8002601}. It has been demonstrated both theoretically and numerically that spatial coupled turbo codes outperform their uncoupled counterpart in terms of better waterfall and error floor performance \cite{8002601,8631116,8368318,9328182,9851473}. This makes them appealing to future communication systems where both close-to-capacity and lower error floor performance will be required. In this section, we first introduce several important properties of spatially coupled turbo codes. Then, we introduce several classes of spatially coupled turbo codes, including the one that has been proved to be capacity-achieving.

\subsubsection{Properties of Spatially Coupled Turbo Codes}\label{sssec:sc_turbo_prop}
Spatial coupling has been mainly applied to PCCCs \cite{397441}, SCCCs \cite{669119}, and braided convolutional codes (BCCs) \cite{5361461}. These spatially coupled turbo codes are constructed by applying spatial coupling on the systematic component encoders, which share some structural similarity with spatially coupled product-like codes \cite{6074908,8425763,9812617,9843869}. One of the appealing features of such a construction is that the encoding of spatially coupled turbo codes can be performed in a \emph{streaming} fashion. Note that this is different from spatially coupled LDPC codes \cite{5571910,5695130}, where the spatial coupling is defined based on their parity-check matrices. The main idea is to construct powerful long turbo codes by using short turbo component codes. For the rest of this subsection, we let $L$ and $m$ represent the coupled chain length and coupling memory, respectively.

\begin{figure}[ht]
				\centering
				\includegraphics[width=\linewidth]{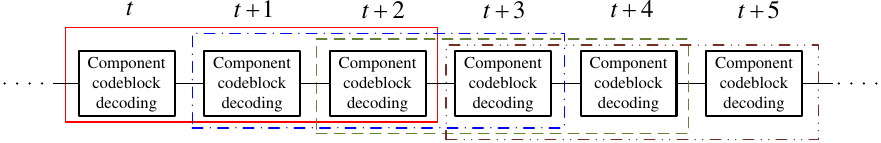}
				\caption{Sliding window decoding with window size $W=3$.} \label{fig:SC_TC_WDec}
			\end{figure}

\emph{1a) Sliding Window Decoding}: It is worth noting that the decoding of spatially coupled turbo codes can leverage sliding window decoding \cite{6374679}, Within the decoding window size $W<L$, the component codewords at time $t,\ldots,t+W-1$ are decoded by using the constituent decoder, e.g, turbo decoder. The windowed decoder outputs the decoded codeword at time $t$ and moves to the next decoding window from time $t+1$ to $t+W$. The windowed decoding process continues up to the point when the coupled code chain is terminated. As a result, the decoding delay of the coupled codes is \cite{7296605}
\begin{equation}
\mathcal{L}= W \cdot K,
\end{equation}
where $K$ denotes the component code information length. One can see that the sliding window decoder enables a continuous streaming fashion compared to decoding the conventional block codes. An example of the sliding window decoding with window size $W=3$ is illustrated in Fig. \ref{fig:SC_TC_WDec}.

\emph{1b) Threshold Saturation}:
The DE recursion of most spatially coupled turbo ensembles on the BEC with erasure $\epsilon$ can be described by the following \cite{8002601}
\begin{equation}\label{eq:SC_TC_DE}
x_t^{(\ell)} =\frac{1}{1+m}\sum_{j=0}^mf\left( \frac{1}{1+m}\sum_{k=0}^m g\left(x_{t+j-k}^{(\ell-1)}\right) ; \epsilon\right),
\end{equation}
where $t\in\{1,\ldots,L\}$ denotes the time instant or spatial position and $(f,g)$ forms a scalar admissible system \cite[Def. 1]{6325197} defined by the recursion
\begin{equation}\label{eq:turbo_sas}
x^{(\ell)} = f(g(x^{(\ell-1)});\epsilon).
\end{equation}
From Section \ref{sec:turbo}-\ref{ssec:turbo_tool}\ref{sec:turbo_DE}, we know that \eqref{eq:turbo_sas} is the DE recursion of the underlying uncoupled turbo code ensemble such that the DE recursion in \eqref{eq:turbo_DE} is a special case of \eqref{eq:turbo_sas}. By \cite[Th. 1]{6325197}, the spatially coupled turbo codes defined by the recursion in \eqref{eq:SC_TC_DE} have the threshold saturation property \cite{5695130}: the suboptimal BP decoding threshold converges to the optimal MAP decoding threshold as $L \rightarrow \infty$, $m \rightarrow \infty$ and $L \gg m$. This implies that one can design good spatially coupled turbo codes by simply increasing $L$ and $m$ without the need for meticulous optimization as in irregular turbo codes. Since the MAP thresholds of the coupled turbo ensembles and the corresponding uncoupled ensembles are identical \cite{6325197,6887298}, we can design spatially coupled turbo codes by optimizing the MAP threshold of the underlying uncoupled turbo codes.


\emph{1c) MAP Decoding Threshold}:
On the BEC, the MAP threshold of the uncoupled turbo ensemble $\epsilon_{\text{MAP}}$ can be computed by using the area theorem \cite[Lemma 4.4]{Measson2006thesis}
\begin{align}\label{eq:MAP_find}
R  = \int_{\epsilon_{\text{MAP}}}^1 R \bar{p}(\epsilon)+(1-R)\bar{q}(\epsilon) d \epsilon,
\end{align}
where $\bar{p}(\epsilon)$ and $\bar{q}(\epsilon)$ denote the average extrinsic erasure probability for information bits and parity bits, respectively. They are obtained from the fixed point solutions of the DE equations for information and parity bits, respectively, e.g., the solution to equation $x = f(g(x;\epsilon))$ or equivalently the value of $x^{(\ell = \infty)}$ from \eqref{eq:turbo_sas}. Strictly speaking, the MAP threshold given by the area theorem is an upper bound. However, we opt to drop the term ``upper bound'' for simplicity as various works \cite{8002601,8631116,8368318,9328182,9851473} show that the BP thresholds of the coupled turbo ensembles converge to the upper bound of their MAP thresholds.

\begin{figure}[ht]
				\centering
				\includegraphics[width=\linewidth]{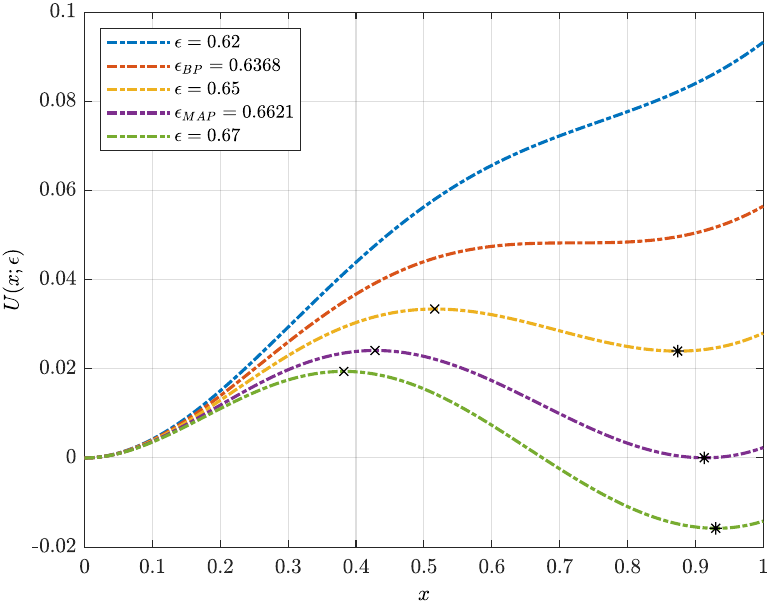}
				\caption{The potential function of a LTE turbo ensemble. The minimum unstable fixed point $u(\epsilon)$ and the energy gap $\min_{x\in[u(\epsilon),1]}U(x;\epsilon)$ for $\epsilon>\epsilon_{\text{BP}}$ are represented by $\times$ and $*$, respectively.} \label{fig:potential_func_LTE}
			\end{figure}

Alternatively, we can obtain the so-called potential threshold \cite[Def. 6]{6325197} since it coincides with the MAP threshold \cite{8002601}. To do so, we first obtain the potential function \cite[Def. 2]{6325197} of the scalar admissible system in \eqref{eq:turbo_sas} as
\begin{equation}\label{eq:turbo_pot2}
U(x;\epsilon) = xg(x)-G(x)-F(g(x);\epsilon),
\end{equation}
where $F(x;\epsilon) = \int_0^x f(z;\epsilon) dz$ and $G(x) = \int_0^x g(z) dz$. Then the potential threshold $\epsilon_\text{c}$ is obtained as \cite{6325197,6887298}
\begin{align}
\epsilon_\text{c} = \sup\left\{\epsilon \in [0,1]:\min_{x\in[u(\epsilon),1]}U(x;\epsilon)\geq 0,u(\epsilon)>0 \right\},
\end{align}
where
\begin{align}
u(\epsilon) = \sup\left\{\tilde{x}\in[0,1]: f( g(x);\epsilon )<x,x\in (0,\tilde{x})\right\},
\end{align}
is the minimum unstable fixed point for $\epsilon>\epsilon_\text{s}$, and $\epsilon_\text{s}$ is single system threshold \cite[Def. 4]{6325197}
\begin{align}
\epsilon_\text{s} = \sup\left\{\epsilon \in [0,1]:U'(x;\epsilon)>0,\forall x \in (0,1] \right\},
\end{align}
which is also the BP threshold defined in \eqref{eq:turbo_BP}. In general, the computation of the potential threshold from the potential function is simpler than that of the MAP threshold from the area theorem for turbo codes. This can be seen by noting that the integration in \eqref{eq:MAP_find} is over the fixed point solutions of the DE equations of turbo codes. Thus, the closed-form expressions are very difficult to derive. As an example, the potential function of the uncoupled LTE turbo ensemble is shown in Fig. \ref{fig:potential_func_LTE} for several values of $\epsilon$.

Estimating the MAP threshold and proving threshold saturation on the AWGN channel for turbo codes are difficult. The only progress so far is due to the recent work in \cite{9756446}, which relies on Monte Carlo simulation to compute the MAP threshold from the generalized area theorem \cite[Th. 1]{5290273} and observes threshold saturation for SC-SCCCs on the AWGN channel numerically. In fact, various works \cite{8002601,8631116,8368318,9328182,9851473} suggest that the good performance of spatially coupled turbo codes from the BEC can be carried over to the AWGN channel.


\emph{1d) Minimum Distance}: Consider an uncoupled turbo-like code $\mathcal{C}'$ which can belong to PCCCs, SCCCs, and BCCs. Let $\mathcal{C}$ represent the spatially coupled turbo-like code with $\mathcal{C}'$ as the component code. Assume that the permutations of $\mathcal{C}$ are time-invariant and satisfy certain conditions. Moreover, both $\mathcal{C}$ and $\mathcal{C}'$ have the same length. The minimum distances of the coupled and uncoupled codes satisfy \cite{8631116}
\begin{equation}\label{eq:sc_tc_dmin}
d_{\min}(\mathcal{C}) \geq d_{\min}(\mathcal{C}').
\end{equation}
Thus, spatial coupling either preserves or improves the minimum distance of turbo codes. That said, the exact analysis of the error floor performance of spatially coupled turbo codes is difficult. By \eqref{eq:sc_tc_dmin}, one can use the minimum distance of the uncoupled turbo codes as the lower bound to study the minimum distance behavior of the coupled codes \cite{8631116}.



\subsubsection{Spatially Coupled PCCCs (SC-PCCCs)}
SC-PCCCs \cite{8002601} are obtained by performing spatial coupling on the PCCC encoders. The design rate of SC-PCCCs is the same as that of the underlying uncoupled PCCCs. 

The encoding procedures for SC-PCCCs are as follows. An information sequence $\boldsymbol{u}_t$ at time $t$ is divided into $m+1$ subsequences with equal length, i.e., $\boldsymbol{u}_t=[\boldsymbol{u}^{\text{U}}_{t,0},\ldots,\boldsymbol{u}^{\text{U}}_{t,t+m}]$, where $t\in\{1,\ldots,L\}$. Meanwhile, at the lower encoder, $\boldsymbol{u}_t$ is interleaved becoming $\Pi_t(\boldsymbol{u}_t)$ and also divided into $m+1$ subsequences with equal length, i.e., $\Pi_t(\boldsymbol{u}_t)=[\boldsymbol{u}^{\text{L}}_{t,t},\ldots,\boldsymbol{u}^{\text{L}}_{t,t+m}]$, where $\Pi_t(.)$ denotes the interleaving function at time $t$ before coupling. The coupling is performed such that inputs to the upper and lower convolutional encoders at time $t$ are $\Pi^{\text{U}}_t([\boldsymbol{u}^{\text{U}}_{t-m,t},\ldots,\boldsymbol{u}^{\text{U}}_{t,t}])$ and $\Pi^{\text{L}}_t([\boldsymbol{u}^{\text{L}}_{t-m,t},\ldots,\boldsymbol{u}^{\text{L}}_{t,t}])$, respectively, where $\Pi^{\text{U}}_t(.)$ and $\Pi^{\text{L}}_t(.)$ are the permutation functions of the upper and lower encoders, respectively, at time $t$. The codeword obtained at time $t$ is $\boldsymbol{c}_t=[\boldsymbol{u}_t,\boldsymbol{v}^{\text{U}}_t,\boldsymbol{v}^{\text{L}}_t]$, where $\boldsymbol{v}^{\text{U}}_t$ and $\boldsymbol{v}^{\text{L}}_t$ are the parity sequences as the result of upper and lower convolutional component encoding at time $t$. It is worth mentioning that the encoding of SC-PCCCs can be performed in parallel, i.e., encoding $L$ information sequences in parallel.

Density evolution analysis shows that SC-PCCCs have a strictly larger BP threshold than uncoupled PCCCs on the BEC \cite{8002601}. Moreover, the decoding threshold of SC-PCCCs improves when employing convolutional component codes with larger states. Most importantly, by using the potential function argument \cite{6325197}, it was also analytically shown that SC-PCCCs have threshold saturation property \cite{8002601}. In addition, the required coupling memory for observing threshold saturation numerically is small, i.e., $m\leq 2$. In other words, a small coupling memory is sufficient for achieving the optimal MAP decoding performance. However, the BP threshold of SC-PCCCs still has a noticeable gap to the BEC capacity when punctured.

	\begin{figure}[ht]
				\centering
				\includegraphics[width=\linewidth]{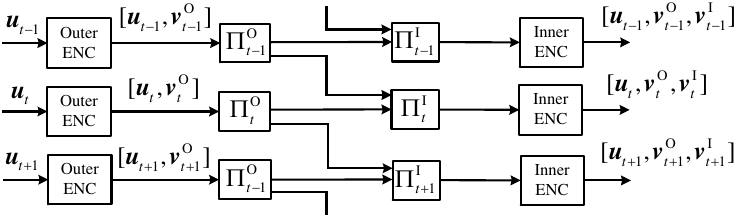}
				\caption{SC-SCCC encoding with coupling memory $m=1$.} \label{fig:SCSCC_enc}
			\end{figure}

\subsubsection{Spatially Coupled SCCCs (SC-SCCCs)}
The encoder of SC-SCCCs \cite{8002601} with $m=1$ is illustrated in Fig. \ref{fig:SCSCC_enc}. The design rate of SC-SCCCs is the same as that of the underlying uncoupled SCCCs.

The encoding procedures for SC-PCCCs are as follows. Similar to SC-PCCCs, at time $t$, information sequence $\boldsymbol{u}_t$ is encoded by the outer convolutional encoder, and parity sequence $\boldsymbol{v}^{\text{O}}_t$ is obtained. Both sequences $[\boldsymbol{u}_t,\boldsymbol{v}^{\text{O}}_t]$ are interleaved and divided into $m+1$ equal length subsequences, i.e., $\Pi^{\text{O}}_t([\boldsymbol{u}_t,\boldsymbol{v}^{\text{O}}_t]) = \boldsymbol{\tilde{v}}^{\text{O}}_{t,t},\ldots,\boldsymbol{\tilde{v}}^{\text{O}}_{t,t+m}$, where $\Pi^{\text{O}}_t(.)$ is the permutation function of the outer component code at time $t$. The coupling is performed such that the input to the inner convolutional encoder at time $t$ is $\Pi^{\text{I}}_t([\boldsymbol{\tilde{v}}^{\text{O}}_{t-m,t},\ldots,\boldsymbol{\tilde{v}}^{\text{O}}_{t,t}])$, where $\Pi^{\text{I}}_t(.)$ is the permutation function of the inner component code at time $t$. The parity sequence generated by the inner decoder at time $t$ is $\boldsymbol{v}^{\text{I}}_t$. Finally, the codeword at time $t$ is $\boldsymbol{c}_t = [\boldsymbol{u}_t,\boldsymbol{v}^{\text{O}}_t,\boldsymbol{v}^{\text{I}}_t]$. Different from SC-PCCCs, the encoding of SC-SCCCs can only be sequential.

Interestingly, although SCCCs have a worse BP threshold than PCCCs, after coupling SC-SCCCs have a much better BP threshold than SC-PCCCs when $m$ is large \cite{5695130}. Density evolution results show that the BP threshold SC-SCCCs is within 0.001 to the BEC capacity for a wide range of code rates under random parity puncturing. The superior performance of SC-SCCCs over SC-PCCCs is due to the threshold saturation property as well as the fact that uncoupled SCCCs have a larger MAP threshold than PCCCs \cite{5695130}. Simulation results in \cite{8631116} show that SC-SCCCs have a lower error floor than SC-PCCCs. The performance of SC-SCCCs can be further improved by coupling a fraction of inner parity sequences from the previous time instant \cite{9775672}. In addition, hardware architectures of SC-SCCC decoders are recently investigated in \cite{9721673}

\subsubsection{Spatially Coupled BCCs (SC-BCCs)}

		\begin{figure*}[ht]
				\centering
				\includegraphics[width=\linewidth]{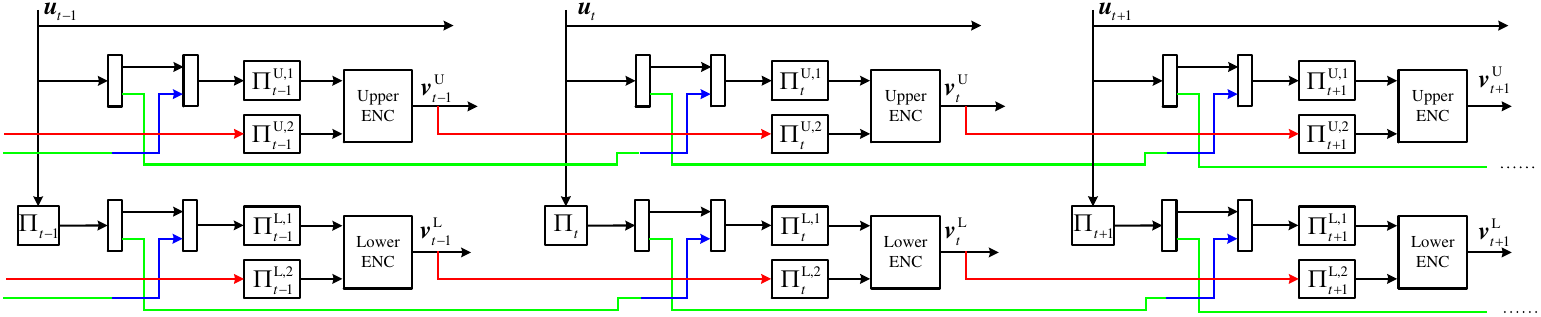}
				\caption{SC-BCC encoding with coupling memory $m=1$.} \label{fig:SCBCC_enc}
			\end{figure*}

BCCs \cite{5361461} are inherently spatially coupled codes with coupling memory $m=1$ \cite{8002601}. Spatially coupled BCCs refer to the generalization of BCCs to a large coupling memory \cite{8002601}. In most cases, rate-$2/3$ convolutional codes are selected as the component codes for SC-BCCs, leading to rate $1/3$ SC-BCCs. In addition, there are two types of SC-BCCs according to \cite{8002601}. Here, we only consider the type II SC-BCCs as the type I SC-BCCs can be seen as a special case of type II SC-BCCs without spatial coupling on information sequences. The encoder of type II SC-BCCs with $m=1$ is depicted in Fig. \ref{fig:SCSCC_enc}.

The encoding of SC-BCCs is performed as follows. At time $t$, the upper parity sequences $\boldsymbol{v}^{\text{U}}_t$, and the lower parity sequence $\boldsymbol{v}^{\text{L}}_t$, are divided into $m$ equal length subsequences, respectively, such that $\boldsymbol{v}^{\text{U}}_t=[\boldsymbol{v}^{\text{U}}_{t,t},\ldots,\boldsymbol{v}^{\text{U}}_{t,t+m}]$ and $\boldsymbol{v}^{\text{L}}_t=[\boldsymbol{v}^{\text{L}}_{t,t},\ldots,\boldsymbol{v}^{\text{U}}_{t,t+m}]$. SC-BCCs consist of information and parity coupling, where the coupled information and parity sequences are the first and second inputs of the convolutional component encoders, respectively. Specifically, the information coupling part is the same as in SC-PCCCs such that the upper and lower coupled information becomes the first inputs of the upper and lower encoders, respectively. In addition, the parity coupling is performed such that the parity sequences $[\boldsymbol{v}^{\text{L}}_{t-m,t-1},\ldots,\boldsymbol{v}^{\text{L}}_{t-1,t-1}]$ and $[\boldsymbol{v}^{\text{U}}_{t-m,t-1},\ldots,\boldsymbol{v}^{\text{U}}_{t-1,t-1}]$ are interleaved and become the second inputs of the upper and lower encoders, respectively. At time $t$, the component codeword is $\boldsymbol{c}_t = [\boldsymbol{u}_t,\boldsymbol{v}^{\text{U}}_t,\boldsymbol{v}^{\text{L}}_t]$. The encoding of SC-BCCs is sequential in nature.

It was analytically shown that SC-BCCs have the threshold saturation property. Moreover, the MAP threshold of SC-BCCs is within 0.001 to the BEC capacity for rates between $1/2$ to $9/10$ \cite{8002601}. When $m$ is small, e.g., $m=1$, SC-BCCs have larger BP thresholds than SC-PCCCs and SC-SCCCs \cite{8002601}. In addition, \cite{8631116} shows that SC-BCCs exhibit a linear minimum distance growth rate, which is faster than that of SC-PCCCs and SC-SCCCs \cite{8631116}. Recently, the research on mitigating error propagation in sliding window decoding of SC-BCCs was carried out in \cite{9165839}.

\subsubsection{Partially Information-Coupled Turbo Codes}
Partially information-coupled turbo codes (PIC-TCs) were introduced in \cite{8368318,9328182} to improve the error performance of transport block (TB)-based HARQ in LTE. Instead of using a very long code to encode the entire information of a TB into a codeword, in PIC-TCs the information sequence of a TB is divided into several small sub-sequences. Each sub-sequence as well as a part of the information bits from consecutive sub-sequences are encoded into a component codeword. In other words, a faction of information bits are shared between consecutive component codewords. This introduced coupling between component codewords improves the reliability of the transmitted TBs while the spatial coupling nature of PIC-TCs allows low latency decoding via sliding window decoding. It is worth noting that the coupling of PIC-TCs is on the turbo code level while the coupling of SC-PCCCs is on the convolutional code level. In other words, existing turbo encoders and decoders can be directly employed in PIC-TCs without changing the architectures, whereas SC-PCCCs require some modifications.

	\begin{figure}[ht]
				\centering
				\includegraphics[width=0.8\linewidth]{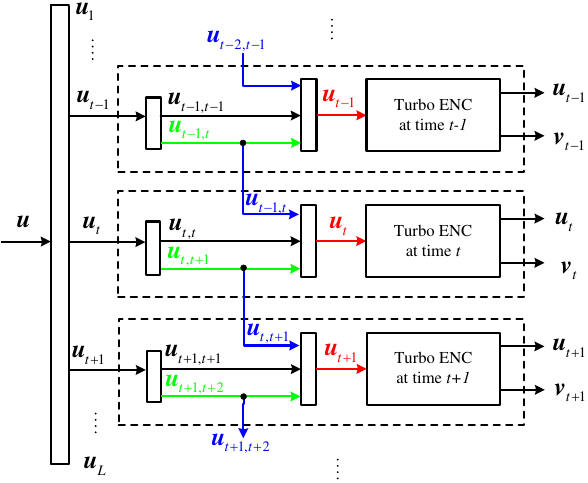}
				\caption{PIC-TC encoding with $m=1$.} \label{fig:PIC_TC_enc}
			\end{figure}	

The encoder diagram of PIC-TCs with $m=1$ is shown in Fig. \ref{fig:PIC_TC_enc}. At time $t$, the information sequence $\boldsymbol{u}_t$ is decomposed into $m$ equal length subsequences $\boldsymbol{u}_{t,t},\ldots,\boldsymbol{u}_{t,t+m}$. The coupling is performed such that the input sequence of the turbo encoder at time $t$ is $ [\boldsymbol{u}_{t-m,t},\ldots,\boldsymbol{u}_{t-1,t},\boldsymbol{u}_{t}]$. We denote by $\lambda$ the ratio of the length of the coupled information sequence over the total information length. The \emph{coupling ratio} $\lambda$ is an important parameter that affects the code rate, waterfall, and error floor performance of PIC-TCs. After the turbo encoding, we obtain the component codeword $\boldsymbol{c}_t = [\boldsymbol{u}_t,\boldsymbol{v}_t]$, where $\boldsymbol{v}_t$ is the parity sequence generated by the turbo encoder at time $t$. The design rate of PIC-TCs is
\begin{equation}
R = \frac{R_0(1-\lambda)}{1-\lambda R_0},
\end{equation}
where $R_0$ is the code rate of the mother turbo code.

It was demonstrated in \cite{9328182} that PIC-TCs can achieve a larger BEC decoding threshold than SC-PCCCs with or without puncturing. One can also generalize PIC-TCs by coupling parity sequences to attain a decoding threshold within 0.0002 of the BEC capacity for code rates ranging from $1/3$ and $9/10$ \cite{9328182}. Extension of PIC-TCs by employing duo-binary convolutional component codes \cite{1516279} has been investigated in \cite{9174156}. However, it requires a large coupling memory for these codes to achieve a BP threshold close to the MAP threshold.

\subsubsection{Generalized SC-PCCCs (GSC-PCCCs) and Capacity-Achieving}
To improve the decoding threshold, \cite{9517979,9851473} generalized SC-PCCCs by allowing a fraction of information bits to be repeated $q$ times before performing coupling and component code encoding. The resultant codes are called generalized SC-PCCCs (GSC-PCCCs). In fact, GSC-PCCCs inherit many useful properties from PIC-TCs and SC-PCCCs, such as that the repeated and coupled information bits are protected by component turbo codewords at multiple time instants and the threshold saturation property.

			\begin{figure}[ht]
				\centering
				\includegraphics[width=0.8\linewidth]{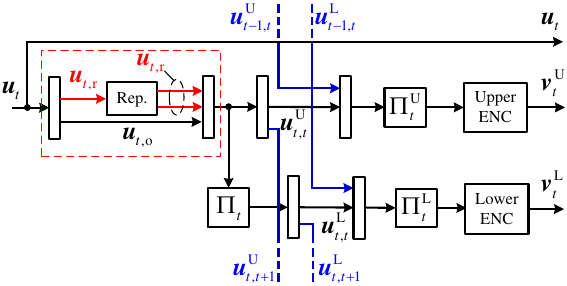}
				\caption{GSC-PCCCs encoding with coupling memory $m=1$. Note that the encoder of SC-PCCCs is without the structures inside the dash line box.} \label{fig:GSC_PCC_enc}
			\end{figure}

The encoding of GSC-PCCCs with $m=1$ is illustrated in Fig. \ref{fig:GSC_PCC_enc}. At time $t$, $\boldsymbol{u}_t$ is decomposed into $\boldsymbol{u}_{t,\text{r}}$ and $\boldsymbol{u}_{t,\text{o}}$. Sequence $\boldsymbol{u}_{t,\text{r}}$ is repeated $q$ times and combined with $\boldsymbol{u}_{t,\text{o}}$ to form sequence $[\boldsymbol{u}_{t,\text{r}},\ldots,\boldsymbol{u}_{t,\text{r}},\boldsymbol{u}_{t,\text{o}}]$. We define $\lambda$ the \emph{repetition ratio} as the length of $\boldsymbol{u}_{t,\text{r}}$ over the length of $[\boldsymbol{u}_{t,\text{r}},\ldots,\boldsymbol{u}_{t,\text{r}},\boldsymbol{u}_{t,\text{o}}]$. The resultant sequence is then decomposed into $m+1$ sequences of equal length, denoted by $\boldsymbol{u}^\text{U}_{t,t+j}$, $j=0,\ldots,m$. The information sequence $\boldsymbol{u}^\text{U}_{t,t+j}$ is used as a part of the input of the upper convolutional encoder at time $t+j$. Then, the coupling is performed in the same way as for SC-PCCCs. The design rate of GSC-PCCCs is
\begin{equation}
R=\frac{1-(q-1)\lambda}{\frac{1}{R_0}-(q-1)\lambda},
\end{equation}
where $R_0$ is the code rate of the mother turbo code. In addition, the encoding of GSC-PCCCs can also be performed in parallel.

Note that SC-PCCCs can be regarded as a special case of GSC-PCCCs with $q=1$. Most importantly, it was rigorously proved that the rate-$R$ GSC-PCCC ensemble with $R \in[ 1/(q(1/R_0-1)+1),1)$ and 2-state convolutional component codes under suboptimal BP decoding achieves at least a fraction $1-\frac{R}{R+q}$ of the BEC capacity for repetition factor $q\geq2$, where the multiplicative gap vanishes as $q$ tends to infinity \cite{9851473}. This indicates that GSC-PCCCs can achieve a threshold all the way to the BEC capacity by increasing $q$. To the best of our knowledge, this is the first class of turbo codes that are proved to be capacity-achieving.

\subsection{Comparisons: Performance and Complexity}
In this section, we first compare the performance between interleavers and punctures designed for the enhanced turbo codes and those with LTE turbo codes. Then, we illustrate the impacts of different convolutional component codes as well as decoders on the error performance. In addition, a comparison between convolutional codes and turbo codes at short blocklength is provided. The performance of different turbo decoders and the comparison between different hardware implementations of high throughput turbo decoders are discussed. Finally, we compare the BER between different classes of spatially coupled turbo codes.

\subsubsection{Interleavers and Puncturing Patterns Comparison}
\begin{figure}[ht]
				\centering
				\includegraphics[width=\linewidth]{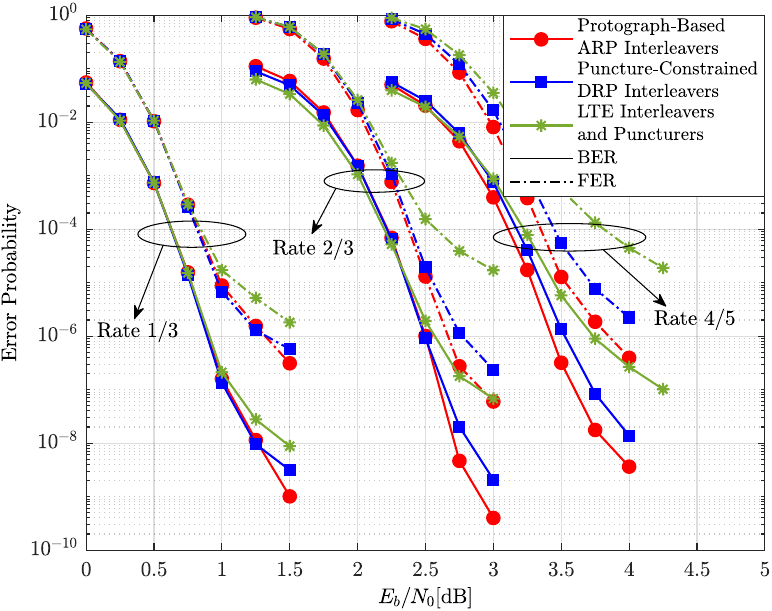}
				\caption{BER and FER comparison between turbo codes with information $K=1504$ and different interleavers and puncturing patterns on the AWGN channel.} \label{fig:Turbo_Int_punc_compare_1}
			\end{figure}

\begin{table}[h]
    \centering
     \caption{Simulation parameters for evaluation the performance of various interleavers and puncturing patterns.}
    \label{tab:sim1_int_punc}
    \begin{tabular}{|c|c|c|c|c|c|}
        \hline Channel & \multicolumn{3}{|c|}{ AWGN } \\
        \hline Modulation & \multicolumn{3}{|c|}{ BPSK } \\
        \hline Generator Polynomial & \multicolumn{3}{|c|}{$[1,15/13]_8$} \\
        \hline Interleavers
         & ARP & DRP & LTE  \\
        \hline Code rate & \multicolumn{3}{|c|}{$1/3,2/3,4/5$} \\
        \hline Information Length
         & \multicolumn{3}{|c|}{
        $1504$ } \\
        \hline Decoding Algorithm & \multicolumn{3}{|c|}{ Log-MAP }\\
         \hline Maximum Decoding Iterations & \multicolumn{3}{|c|}{16}\\
        \hline
    \end{tabular}
\end{table}
We consider a turbo code with generator polynomial $[1,15/13]_8$. The turbo decoder is the iterative Log-MAP decoder with 16 maximum iterations. The BER and FER of turbo codes with rates $1/3$, $2/3$, and $4/5$ with the protograph-based ARP interleavers and puncturers in \cite{8214245}, the puncture-constrained DRP interleavers and punctures in \cite{6133842}, and LTE interleavers and punctures \cite{LTE136212}, are shown in Fig. \ref{fig:Turbo_Int_punc_compare_1}. Note that the interleaver lengths for the ARP and LTE interleavers are 1504 while for the DRP interleaver is 1512. The simulation parameters are summarized in Table \ref{tab:sim1_int_punc}.

Thanks to the joint interleaving and puncturing design, both DRP and ARP interleavers lead to better waterfall and error floor performance than the LTE interleavers. Moreover, the protograph-based ARP interleavers achieve the best overall performance among all three interleavers. In particular, at rate $4/5$, the protograph-based ARP interleaver provides a gain about 0.1 dB over the DRP interleaver and 0.4 dB over the LTE interleaver at a FER of $10^{-4}$. Therefore, the new ARP interleavers designed for the enhanced turbo codes can provide substantial performance gains in both low and high rates.

\subsubsection{Component code Comparison}
\begin{figure}[ht]
				\centering
				\includegraphics[width=\linewidth]{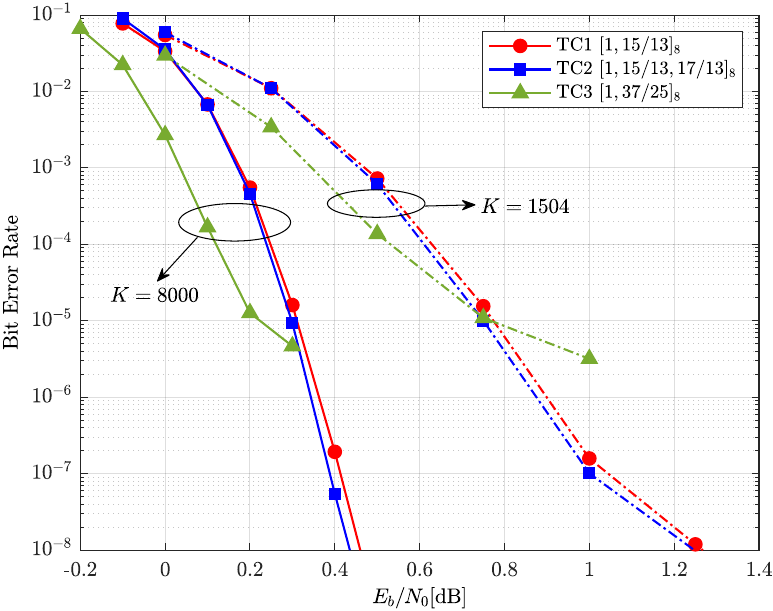}
				\caption{BER comparison between turbo codes with rate-$1/3$, information length $K\in\{1504,8000\}$, and different convolutional component codes.} \label{fig:Turbo_conv_compare_1}
			\end{figure}

\begin{table}[h]
    \centering
     \caption{Simulation parameters for evaluation the impacts on the choice of convolutional component codes.}
    \label{tab:sim2_comp_code}
    \begin{tabular}{|c|c|c|c|c|c|}
        \hline Channel & \multicolumn{3}{|c|}{ AWGN } \\
        \hline Modulation & \multicolumn{3}{|c|}{ BPSK } \\
        \hline  & \multicolumn{3}{|c|}{$[1,15/13]_8$} \\
         Generator Polynomial & \multicolumn{3}{|c|}{$[1,15/13,17/13]_8$} \\
          & \multicolumn{3}{|c|}{$[1,37/25]_8$} \\
        \hline Interleavers
         &\multicolumn{3}{|c|}{ ARP}  \\
        \hline Code rate & \multicolumn{3}{|c|}{$1/3$} \\
        \hline Information Length
         & \multicolumn{3}{|c|}{
        $1504,8000$ } \\
        \hline Decoding Algorithm & \multicolumn{3}{|c|}{ Log-MAP }\\
         \hline Maximum Decoding Iterations & \multicolumn{3}{|c|}{16}\\
        \hline
    \end{tabular}
\end{table}
We investigate the impacts of convolutional component codes on the performance of turbo codes. We consider two information lengths $K=1504, 8000$. Same as in the previous section, we set the maximum iteration of the Log-MAP turbo decoder to be 16. The BER of turbo codes with generator polynomials $[1,15/13]_8$, $[1,15/13,17/13]_8$, and $[1,37/25]_8$, which we referred to as TC1, TC2, and TC3, respectively, are shown in Fig. \ref{fig:Turbo_conv_compare_1}. All turbo codes with the same information length adopt the same ARP interleavers from \cite{R1-167413}. Since TC2 is with rate $1/5$, we use the puncturing patterns from the third row of Table 2.8 in \cite{Bohorquez2015thesis} to achieve rate $1/3$. The simulation parameters are summarized in Table \ref{tab:sim2_comp_code}.

Fig. \ref{fig:Turbo_conv_compare_1} shows that TC3 has the best waterfall performance. However, it suffers from a higher error floor. A tailored interleaver design for TC3 is necessary to lower its error floor. It is also interesting to note that TC2 achieves slightly better waterfall and error floor performance than TC1. This implies that a low-rate turbo code with carefully designed puncturing patterns can outperform a high-rate turbo code without puncturing.

\subsubsection{Short blocklength Convolutional Codes and Turbo Codes Comparison}
\begin{figure}[ht]
				\centering
				\includegraphics[width=\linewidth]{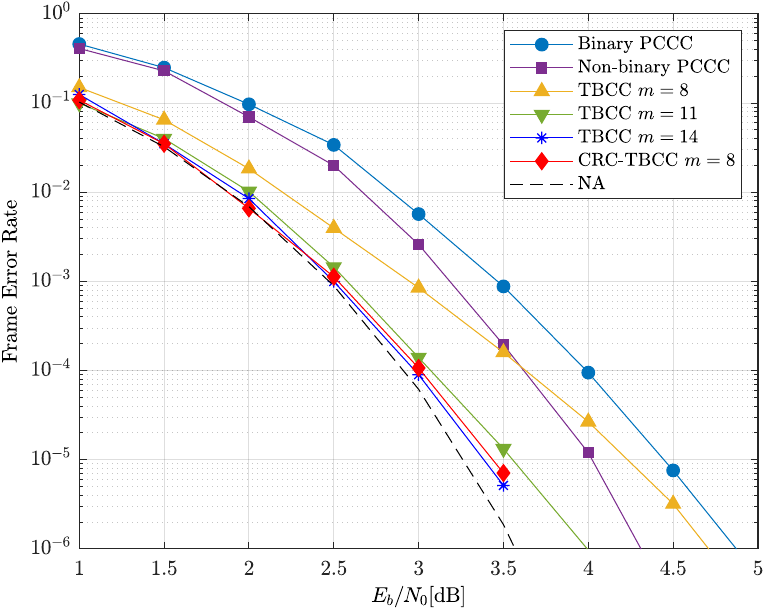}
				\caption{FER Comparison between convolutional and turbo codes with rate $1/2$ and information length $K=64$.} \label{fig:short_turbo_conv_compare}
			\end{figure}

\begin{table}[h]
    \centering
     \caption{Simulation parameters for evaluation the performance of short convolutional and turbo codes.}
    \label{tab:sim3_conv_turbo}
    \begin{tabular}{|c|c|c|c|c|c|}
        \hline Channel & \multicolumn{3}{|c|}{ AWGN } \\
        \hline Modulation & \multicolumn{3}{|c|}{ BPSK } \\
        \hline Coding schemes  & \multicolumn{3}{|c|}{Binary TC, non-binary TC, TBCC, CRC-TBCC} \\
        \hline Code rate & \multicolumn{3}{|c|}{$1/2$} \\
        \hline Information Length
         & \multicolumn{3}{|c|}{
        $64$ } \\
        \hline Decoding Algorithm & \multicolumn{3}{|c|}{ Log-MAP, WAVA, CA-LVA }\\
        \hline
    \end{tabular}
\end{table}
We compare the error performance between turbo codes and convolutional codes with information length $K=64$ and rate $1/2$. We consider a binary turbo with $m=4$ tail-biting RSC component codes from \cite{6692087} and a non-binary turbo code with $m=1$ tail-biting RSC component codes over $\mathbb{F}_{256}$ from \cite{6502170}, where all component codes are under BCJR decoding. We further consider three TBCCs with memories $m=8,11,14$ under WAVA \cite{COSKUN201966}. In addition, we include a TBCC with $m=8$ under CRC-aided list Viterbi algorithm (CRC-LVA) with a 10-bit CRC optimized based on distance spectrum \cite{9709319}. The simulation parameters are summarized in Table \ref{tab:sim3_conv_turbo}. Fig. \ref{fig:short_turbo_conv_compare} shows the FER versus $E_b/N_0$ in dB for the aforementioned candidate codes. Normal approximation (NA) (see Section \ref{BLER_limit}) with the third order term \cite{5452208,7056434} based on BPSK signaling is used as the benchmark.

As can be seen in this figure, the TBCC with $m=14$ closely approaches the NA and is within 0.1 dB at a FER of $10^{-5}$. The TBCC under CRC-LVA provides comparable performance to the TBCC with $m=14$ but with $98\%$ reduction in decoding complexity \cite[p20]{9709319}. At short blocklength, TBCC can significantly outperform turbo codes provided that $m$ is very large. However, at a FER of $10^{-6}$, the non-binary turbo code is about 0.4 dB better than the TBCC with $m=8$. Finally, the binary turbo code has the worst FER performance but the lowest decoding complexity. The above promising results confirm that TBCCs are very powerful short blocklength codes.

\subsubsection{Turbo Decoders Comparison}
\begin{figure}[ht]
				\centering
				\includegraphics[width=\linewidth]{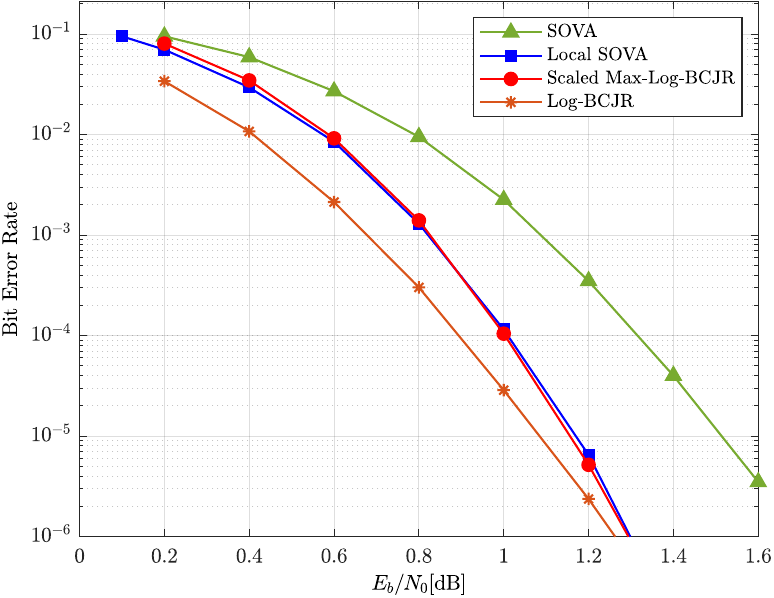}
				\caption{BER comparison between different turbo decoders for decoding rate-$1/3$ LTE turbo codes with information length $K=1056$.} \label{fig:TC_dec}
			\end{figure}

We compare the error correction performance of the Log-MAP decoder \cite{524253}, scaled Max-Log-MAP decoder \cite{524253} with a fixed scaling factor of 0.75, the local SOVA decoder with the third configuration in \cite[Sec. IV-C]{8960428} and the original SOVA decoder \cite{64230}. The simulations were carried out for rate-$1/3$ LTE turbo codes with $K=1056$. The maximum number of iterations for each decoder is set to 5.5, where one convolutional component decoding is regarded as a half iteration. The BER curves for all decoders are shown in Fig. \ref{fig:TC_dec}.

We see that the performance of the local SOVA decoder and the scaled Max-Log-MAP decoder is almost identical. However, the local SOVA has a lower computational complexity than the scaled Max-Log-MAP decoder when high-radix decoding is employed \cite{8960428}. It was shown that for radix-4 and radix-8, using the local-SOVA
reduces the complexity by $27\%$ compared to the Max-Log-MAP decoder \cite{Vinh2021thesis}. Moreover, the local SOVA decoder outperforms the original SOVA decoder and has about 0.1 dB loss compared to the Log-MAP decoder. Hence, the local SOVA decoder provides a good trade-off between error correction performance and computational complexity.

\subsubsection{Comparison Between Different Decoder Hardware Architectures}
We compare the turbo decoder architectures UXMAP, PMAP, FPMAP, and XMAP with a throughput of more than 1 Gb$/$s reported in the literature. The decoding is the scaled Max-Log-MAP decoder with a scaling factor of 0.75. Table \ref{tab:turbo_dec_compare} shows different implementation results of the above four types of architectures, where the term ``parallelism'' refers to the number of bits decoded in parallel. Since results from the literature are
reported for different technology nodes, we provide a scaling to 28 nm technology in the caption of Table \ref{tab:turbo_dec_compare}. To this end, we cap the frequency scaling to a reasonable 1000 MHz, which allows to preserve single cycle accesses to static random access memory. In addition, except for the UXMAP architectures \cite{9053453,9120779} that consider ARP interleavers \cite{8214245}, the rest of the implementations focus on the LTE interleavers.

\begin{table*}[ht]
 \centering
 \caption{Comparison of implementation results for different turbo decoder architectures. Frequency scaling to 28 nm (capped at 1000 MHz): $^{\dag}$ 2.52; $^{\ddag}$ 1.95; $^{\S}$ 1.46. Area scaling to 28 nm: $^{\flat}$ 0.40; $^{\natural}$ 0.51; $^{\sharp}$ 0.69.}\label{tab:turbo_dec_compare}
\begin{tabular}{|c|c|c|c|c|c|c|c|}
\hline
Implementation & \cite{9053453} & \cite{9120779} & \cite{6847747} & \cite{6695789} & \cite{7378273} & \cite{6779698} & \cite{7593077} \\ \hline
Architecture & \multicolumn{2}{c|}{Radix-4 UXMAP}  & Radix-2 PMAP & Radix-4 PMAP& Radix-2 FPMAP &    \multicolumn{2}{c|}{Radix-4 XMAP} \\ \hline
Information length & 512 & 128/64/32 &6144 & 6144 & 6144& 6144& 6144 \\ \hline
Rate & $1/3$ & $1/3$ &$1/3$  & 0.95 & $1/3$& $1/3 $& 0.94 \\ \hline
Parallelism & - & 128 & 64& 32& 6144&64 & 32 \\ \hline
Channel values quantization & 6 bits& 6 bits & 7 bits& - & 4 bits &5 bits & 6 bits\\ \hline
Extrinsic values quantization& - & 7 bits & 7 bits & - & 6 bits &6 bits & 7 bits\\ \hline
State metrics quantization& - & - & 9 bits & 10 bits & 6 bits & 10 bits & 11 bits\\ \hline
Branch metrics quantization& - & - & 8 bits & - & 6 bits & 9 bits & -\\ \hline
Max iterations & 2.5 & 4 & 5.5 & 5.5& 39& 5.5 & 7 \\ \hline
Technology & 28 nm & 28 nm & 90 nm $^{\dag\flat}$ & 65 nm $^{\ddag\natural}$ & 65 nm $^{\ddag\natural}$ &45 nm $^{\S\sharp}$ & 28 nm \\ \hline
Frequency [MHz] & 800 & 800  & 625 (1000)& 410 (1000)& 100 (252)& 600 (1000)& 625 \\ \hline
Throughput [Gb/s] &409.6 & 102.4 & 3.3 (5.29)& 1.01 (2.47)& 15.8 (39.86)& 1.67 (3.2)& 1.13 \\ \hline
Area [mm$^2$] & 30 & 16.54 & 19.75 (2.44)& 2.49 (0.55)& 109 (24.09)& 2.43 (1.03)& 0.49 \\ \hline
Area efficiency [Gb/s/mm$^2$] & 13.65 & 6.19 & 0.17 (2.17) & 0.41 (4.49)& 0.14 (1.65)& 0.69 (2.68)& 2.32 \\ \hline
\end{tabular}
\end{table*}

Although with scaling to 28 nm technology, neither the PMAP decoders nor the XMAP decoders have a throughput close to 15 Gb/s. The FPMAP implementation from \cite{7378273} can achieve a throughput close to 40 Gb/s when scaled to 28 nm technology. In contrast, the UXMAP can achieve a throughput of more than 100 Gb/s and also the highest area efficiency.

\subsubsection{Comparison Between Spatially Coupled Turbo Codes and Uncoupled Turbo Codes}
\begin{figure}[ht]
				\centering
				\includegraphics[width=\linewidth]{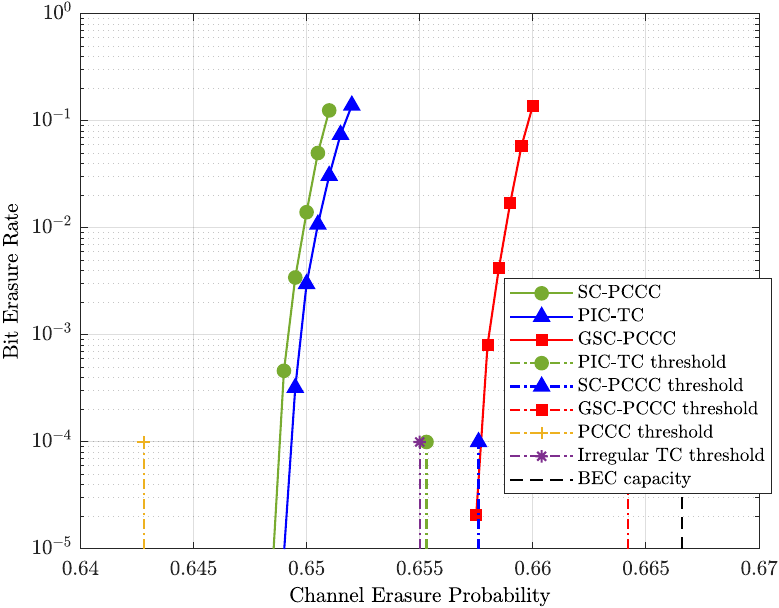}
				\caption{BER Comparison between rate-$1/3$ spatially coupled turbo codes and uncoupled turbo codes on the BEC.} \label{fig:SC_TC_bec_1}
			\end{figure}

We compare spatially coupled turbo codes and uncoupled turbo codes on the BEC. We fix all code rates to be $1/3$. For the uncoupled codes, we consider a regular turbo code with generator polynomial $[1,15/13]_8$ and the irregular turbo codes from \cite[Fig. 4]{5577801} that have the largest BEC threshold. For the spatially coupled turbo codes, we consider SC-PCCCs \cite{8002601}, PIC-TCs with coupling ratio $\lambda=0.4$ \cite[Table II]{9328182}, and GSC-PCCC with repetition ratio $\lambda=0.165$ and repetition factor $q=4$ \cite[Table I]{9851473}. All coupled codes have convolutional component codes with generator polynomial $[1,5/7]_8$, component code information length $K=10000$, coupling memory $m=1$, and coupling length $L=100$. Moreover, all coupled codes are under full decoding of the entire spatial code chain whereas the maximum component code decoding iterations are 20. In addition, we adopt random interleavers and periodic parity puncturing patterns for all coupled codes. The bit erasure rates are shown in Fig. \ref{fig:SC_TC_bec_1}. Note that for uncoupled turbo codes, we plot their BEC decoding thresholds, representing their asymptotic performance as the blocklengths go to infinity. The decoding thresholds for the coupled codes and the BEC capacity are also included in the same figure.

It can be seen that all coupled codes under finite blocklength outperform the regular turbo code with infinite blocklength at a BER of $10^{-5}$. Moreover, all coupled codes have decoding thresholds close to the BEC capacity. Most notably, the GSC-PCCC under finite blocklength also has a noticeable performance gain over the irregular turbo code with infinite blocklength. It is worth noting that the performance of all coupled codes can be further improved by increasing $K$ and $m$. Hence, spatial coupling provides new degrees of freedom in designing good codes based on existing component codes.

\subsubsection{Spatially Coupled Turbo Codes Comparison}
\begin{figure}[ht]
				\centering
				\includegraphics[width=\linewidth]{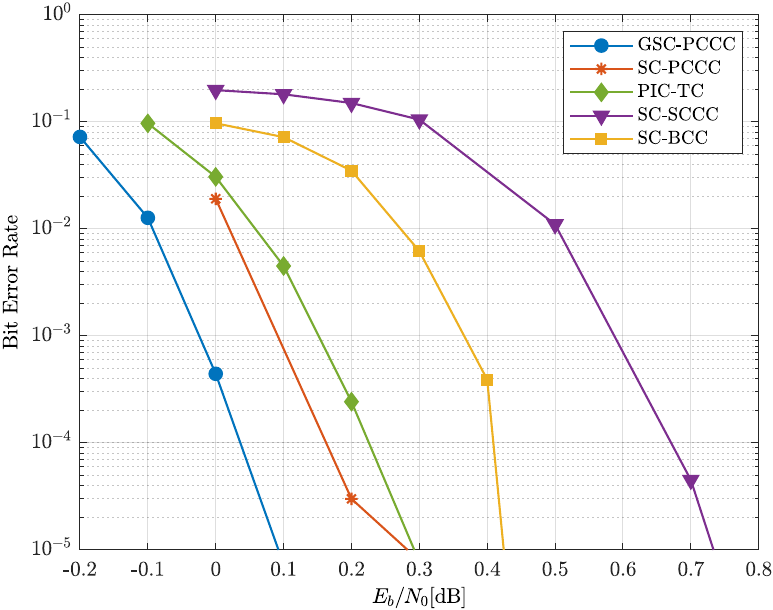}
				\caption{BER Comparison between different spatially coupled turbo codes with $K=1000$, $m=1$, and $L=100$ under sliding window decoding with window size 10.} \label{fig:C_TC_WD_compare_1}
			\end{figure}

We compare the BER performance between various coupled codes, including SC-PCCCs \cite{8002601}, SC-SCCCs \cite{8002601}, SC-BCCs \cite{8002601}, PIC-TCs with $\lambda=0.24$ \cite{9328182}, and GSC-PCCCs with $\lambda=0.11, q=4$ \cite{9851473}, on the AWGN channel. All coupled codes have $K=1000$, $m=1$, $L=100$, and are under sliding window decoding with window size 10. The convolutional component codes for SC-PCCCs, PIC-TCs, and GSC-PCCCs are with generator polynomial $[1,15/13]_8$, for SC-SCCCs are $[1,5/7]_8$, and for SC-BCCs are $\begin{bmatrix} 1 & 0 & 1/7\\ 0&1 & 5/7 \end{bmatrix}_8$. For PIC-TCs and GSC-PCCCs, we adopt the LTE interleavers and periodic parity puncturing patterns. The BER is shown in Fig. \ref{fig:C_TC_WD_compare_1}, where the BER plots for SC-PCCCs, SC-SCCCs, and SC-BCCs are taken from \cite{8228241,8900893}.

The coupled codes whose component codes are PCCCs or turbo codes have better performance than those with other component codes. Among them, the GSC-PCCC has the best waterfall performance. However, the SC-SCCC and SC-BCC have better error floor performance as their BER slopes at $10^{-5}$ are steeper compared to the other three coupled codes. Note that the finite length performance of spatially coupled turbo codes depends on various factors such as $K$, $m$, and the decoding window size, interleavers, the choice of coupling bit indices, etc. \cite{9448689}. Thus, the design of spatially coupled turbo codes to attain the best finite length performance for a given decoding latency \cite{7296605} is still an ongoing research topic.

\begin{figure}[ht]
				\centering
				\includegraphics[width=\linewidth]{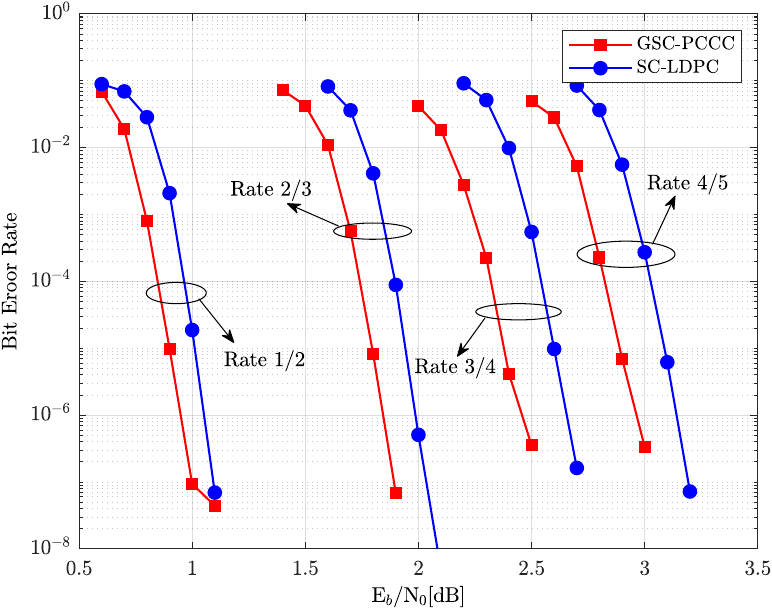}
				\caption{BER Comparison between GSC-PCCCs with $m=1$ and SC-LDPC codes with $m=2$ at different rates. For both codes, the component code information length and coupling length are $K=1000$ and $L=50$, respectively.} \label{fig:C_TC_LDPC_compare_1}
			\end{figure}

Finally, we present the comparison between spatially coupled turbo codes and SC-LDPC codes. Since GSC-PCCCs achieve the best BER as shown in Fig. \ref{fig:C_TC_WD_compare_1}, we pick them as the candidate codes and set $q=2$, $\lambda=0.335$, $k=1000$, $m=1$, and $L=50$. The benchmark regular $(3,6)$ protograph SC-LDPC codes are constructed by following \cite{7152893}, with $m=2$, $L=50$, and a lifting factor of 1000. In addition, we consider four code rates, i.e., $1/2$, $2/3$, $3/4$, and $4/5$. The BER results of rate-$1/2$ codes are available in \cite{9851473}. For other code rates, the GSC-PCCCs are interleaved and punctured by using the ARP interleavers and puncturers in \cite{8214245}, respectively. For SC-LDPC codes, we note that randomly punctured SC-LDPC codes are capacity-approaching provided that the mother codes are capacity-approaching \cite{7353121}. However, we use fixed and period puncturing patterns following the design in \cite[Section VII-A]{7932507}, which have slightly better error performance than random puncturing patterns. The maximum intra-block and inter-block decoding iterations for the GSC-PCCCs are set to 20 while the maximum BP decoding iterations for the SC-LDPC codes are set to 1000. The BER performance of these codes on the AWGN channel is shown in Fig. \ref{fig:C_TC_LDPC_compare_1}.

It can be seen that GSC-PCCCs have better waterfall performance than SC-LDPC codes for all the considered code rates. Interestingly, the performance gains of GSC-PCCCs over SC-LDPC codes at high rates are larger than those at low rates. Hence, the interleavers and puncturers designed for uncoupled turbo codes are also effective for coupled turbo codes. However, SC-LDPC codes have better error floor than GSC-PCCCs. Since the candidate GSC-PCCCs have a fraction of information bits repeated twice before component turbo encoding, additional design criteria are required to obtain the optimal interleavers and puncturers in this case.

\subsection{New Research Directions}
In light of all the above sections, the following new directions can be considered to further improve the classes of turbo codes suitable for future communication systems.
\begin{itemize}
\item As shown in Fig. \ref{fig:Turbo_conv_compare_1}, the choice of convolutional component codes affects the waterfall and error floor of the resultant turbo codes. It is worth extending the puncture-constrained ARP interleavers design methods from \cite{8214245} to turbo codes with other convolutional component codes that have better decoding thresholds.
\item Various works have demonstrated the necessity of joint interleaving and puncturing pattern designs to ensure that the resultant turbo codes have a good waterfall and error floor performance. However, to fulfill the threshold requirements, most designs use the EXIT charts which rely on Monte Carlo simulation. Alternatively, one can explore new design methods based on the density evolution for which the exact transfer function can be derived.
\item It has already been demonstrated that the local SOVA decoder can achieve the same performance as the Max-Log-MAP decoder with lower complexity. It is worth to investigate efficient hardware implementations of high radix schemes for the local SOVA decoder with ultra-high throughputs, e.g., over 1 Tb/s.
\item Spatial coupling can boost the performance of turbo codes without the need for meticulous optimization as in irregular codes. Several classes of spatially coupled turbo codes exhibit the threshold saturation property such that the threshold under suboptimal decoding can approach to that under the optimal MAP decoding. In this regard, it would be interesting to investigate whether the use of suboptimal Max-Log-MAP or the local SOVA algorithm as the component code decoders can still have threshold saturation. Finally, the designs of interleavers, puncturers, and the choice of coupling bits for spatially coupled turbo codes require further investigation.
\end{itemize}

\section{Low-Density Parity-Check (LDPC) Codes}\label{sec:ldpc}

LDPC codes, invented by Gallager in the early 1960s \cite{gallager_low-density_1962} and rediscovered by MacKay and Neal in the mid-1990s \cite{mackay_good_1995,mackay_near_1996,mackay_good_1999}, are a class of linear block codes capable of performing extremely close to the channel capacity. 
More importantly, this performance is achieved under the iterative \emph{belief propagation} (BP) decoding \cite{j_pearl_probabilistic_1988} (also known as message-passing (MP) decoding), which has complexity linear in the blocklength of the codes.
Quasi-cyclic (QC) LDPC codes are a type of LDPC codes with notable importance in the efficient encoding and practical implementation of the corresponding decoder due to their compact representation of the parity-check matrix and high level of parallelism in the decoder architecture.
In the new era of mobile communications, where high data rates, low latency, and extremely reliable transmission become more demanding,
QC-LDPC codes have been selected as one of the channel coding schemes in the 5th generation NR standards by 3GPP for the usage scenarios of eMBB \cite{noauthor_etsi_nodate}. 

In this section, we will start with the fundamentals of various LDPC code ensembles and the design techniques. 
This will be followed by a comprehensive examination of different designs of QC-LDPC codes and variations of BP decoding. 
Furthermore, we will discuss the currently standardized LDPC codes and provide an overview of the state-of-the-art implementations of LDPC codes for achieving ultra-high throughput in future applications. 
The section concludes by presenting several research directions for future 6G.

\subsection{LDPC Block Codes}
An $\left[N, K\right]$ LDPC code is defined by an $M \times N$ binary parity-check matrix $\boldsymbol{H}$, where $M = N - K$ represents the number of parity bits, $N$ is the code length, and $K$ is the length of the information bits.
The rate of the code is $R = K/N$.
The code is the set of all length-$N$ binary vectors $\boldsymbol{c}$, called \emph{codewords}, satisfying $\boldsymbol{cH}^T = \boldsymbol{0}$.
Such a code is said to be \emph{linear} in that a linear combination of codewords yields another codeword. 
The \emph{Tanner graph} \cite{tanner_recursive_1981} representation of an LDPC code is a bipartite graph consisting of a set $\left\{v_0, v_1, v_2, \ldots, v_{N-1}\right\}$ of $N$ variable nodes (VNs) and a set $\left\{c_0, c_1, c_2, \ldots, c_{M-1}\right\}$ of $M$ check nodes (CNs). 
An edge connecting VN $v_j$ to CN $c_i$ if and only if $\boldsymbol{H}_{i,j} = 1$.  
The number of $1$s of each row of $\boldsymbol{H}$ is the degree of the corresponding CN. Similarly, the number of $1$s of each column of $\boldsymbol{H}$ is the degree of the corresponding VN.

In LDPC code design, an \emph{ensemble} of codes is a family of codes that is characterized by the same set of parameters.
The parity-check matrix $\boldsymbol{H}$ of the $\left[N, K\right]$ LDPC code is (randomly) chosen from the ensemble of $M\times N$ binary matrices.
Any code from the ensemble will have a code rate equal to or greater than $R = 1- M/N$, and hence $R$ is commonly known as the \emph{design rate}.
Let $\lambda(x)=\sum^{d_v}_{d=1} \lambda_d x^{d-1}$ and $\rho(x)=\sum^{d_c}_{d=1} \rho_d x^{d-1}$ be the \emph{degree distribution} of an LDPC code, where $d_v$ and $d_c$ denote the maximum VN and CN degrees, respectively. The constant $\lambda_d$ represents the fraction of edges of the Tanner graph which are connected to VNs of degree $d$, and $\rho_d$ represents the fraction of edges of the Tanner graph which are connected to CNs of degree $d$.

\begin{figure}
    \centering
    \includegraphics[width=1.0\linewidth]{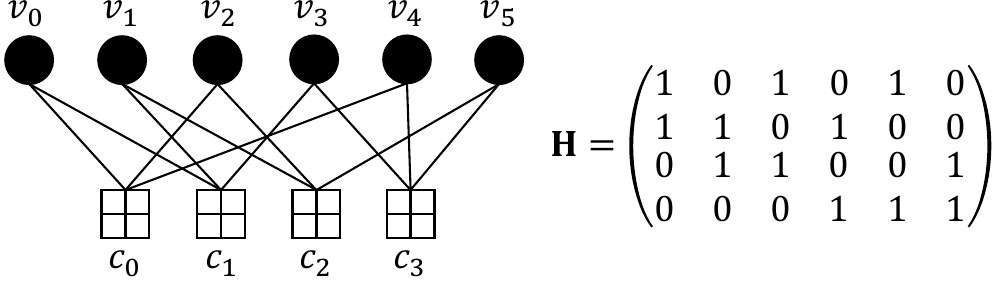}
    \caption{An example of an LDPC code in Tanner graph and matrix representations.}
    \label{LDPC_Tanner_Graph}
\end{figure}
Fig. \ref{LDPC_Tanner_Graph} shows an example of Tanner graph and matrix representation of a $[N,K] = [6,3]$ LDPC code.
The code has a constant VN and CN degrees of 2 and 3, respectively, which corresponds to a constant number of $1$s in each column and each row of the parity-check matrix $\boldsymbol{H}$.
Since the rank of $\boldsymbol{H}$ is 3, the rate of this code is $1/2$ which is larger than the design rate of this code $R=1 - 4/6 = 1/3$.

\subsubsection{Regular Ensemble}
An LDPC code is said to be regular if all the $M\times N$ parity-check matrix $\boldsymbol{H}$ has the same number of 1s in each row and each column. 
The degree distribution of a regular ensemble is defined as $\lambda(x)=x^{d_v-1}$ and $\rho(x)=x^{d_c-1}$ with $\lambda_{d_v} = \rho_{d_c}=1$, or $(d_v,d_c)$ for short. 
The design rate of a regular LDPC ensemble is $R=1-d_v/d_c$.

\subsubsection{Irregular Ensemble}
An irregular LDPC code has a range of VN and CN degrees, that is,  $\boldsymbol{\lambda} = (\lambda_1,\lambda_2,\ldots,\lambda_{d_v})$ and $\boldsymbol{\rho} = (\rho_1,\rho_2,\ldots,\rho_{d_c})$ 
%
with $\sum_d\lambda_d = 1$ and $\sum_d\rho_d = 1$.
The design rate of the ensemble is then
$
R = 1-\int^1_0\rho(x)/\int^1_0\lambda(x).
$
It is well-known from the literature that irregular LDPC codes can approach the channel capacity if properly designed \cite{richardson_capacity_2001,richardson_design_2001,luby_improved_2001,chung_design_2001}.

\subsubsection{Protograph Ensemble}
\label{sec:protograph_ensemble}
An ensemble of protograph LDPC codes \cite{thorpe_low-density_2003} is defined by a Tanner graph with a relatively small number of VNs and CNs, namely \emph{protograph}. 
A protograph $\mathscr{P}=\left(\mathscr{V}, \mathscr{C}, \mathscr{E}\right)$ consists of a set of variable nodes $\mathscr{V} = \mathscr{V}_{pun}\cup\mathscr{V}_{tran}$, a set of check nodes $\mathscr{C}$, and a set of edges $\mathscr{E}$, each VN and CN nodes is of its own type. 
Note that $\mathscr{V}_{pun}$ and $\mathscr{V}_{tran}$ denote the set of punctured VNs and the set of transmitted VNs, respectively.
Each edge $e\in \mathscr{E}$ connects a variable node of type $v \in \mathscr{V}$ to a check node of type $c \in \mathscr{C}$. 
The protograph is equivalently described by an $m_p\times n_p$ non-binary integer matrix $\boldsymbol{B}_{p}$, namely \emph{protomatrix}, with $m_p = \vert\mathscr{C}\vert$ and $n_p = \hat{n}_{pu}+\hat{n}_t = \vert \mathscr{V}\vert$, where $\hat{n}_{pu}$ and $\hat{n}_{t}$ denote the number of punctured and transmitted columns in $\boldsymbol{B}_{p}$, respectively. 
Each entry $\boldsymbol{B}_{p}(i,j)$, $i=0,1,\cdots,m_p-1$, $j=0,1,\cdots,n_p-1$, represents the number of edges connecting variable node type $v_j$ to check node type $c_i$.
The design rate of the protograph ensemble is $R=(n_p-m_p)/\hat{n}_t$.

\textit{3a) Graph Representation:} The Tanner graph of a protograph-based LDPC code with length $N = Zn_p$ is obtained by \emph{lifting} the protograph $\mathscr{P}$.
The lifting process is described as follows: each edge in the protograph becomes a bundle of $Z$ edges, connecting $Z$ copies of a VN to $Z$ copies of a CN. 
The connections within each bundle are then permuted between the variable and check node pairings. 
This process is equivalent to replacing every element in $\boldsymbol{B}_{p}$ by an $Z\times Z$ square matrix, while the connections between different node types remain unchanged. 
The square matrix can be considered as a permutation matrix of each edge type that connects a VN type to a CN type.
The lifted Tanner graph is also known as \emph{derived graph}.
Since $\boldsymbol{B}_{p}$ can contain integers greater than $1$, it is usually required that the lifting factor $Z$ is greater than or equal to the largest value in $\boldsymbol{B}_{p}$ so that no overlapping edges in the resulting derived graph after the permutation. 
A protograph code ensemble is defined by randomizing over all possible permutations during the lifting process.

\begin{figure}
    \centering
    \includegraphics[width=1.0\linewidth]{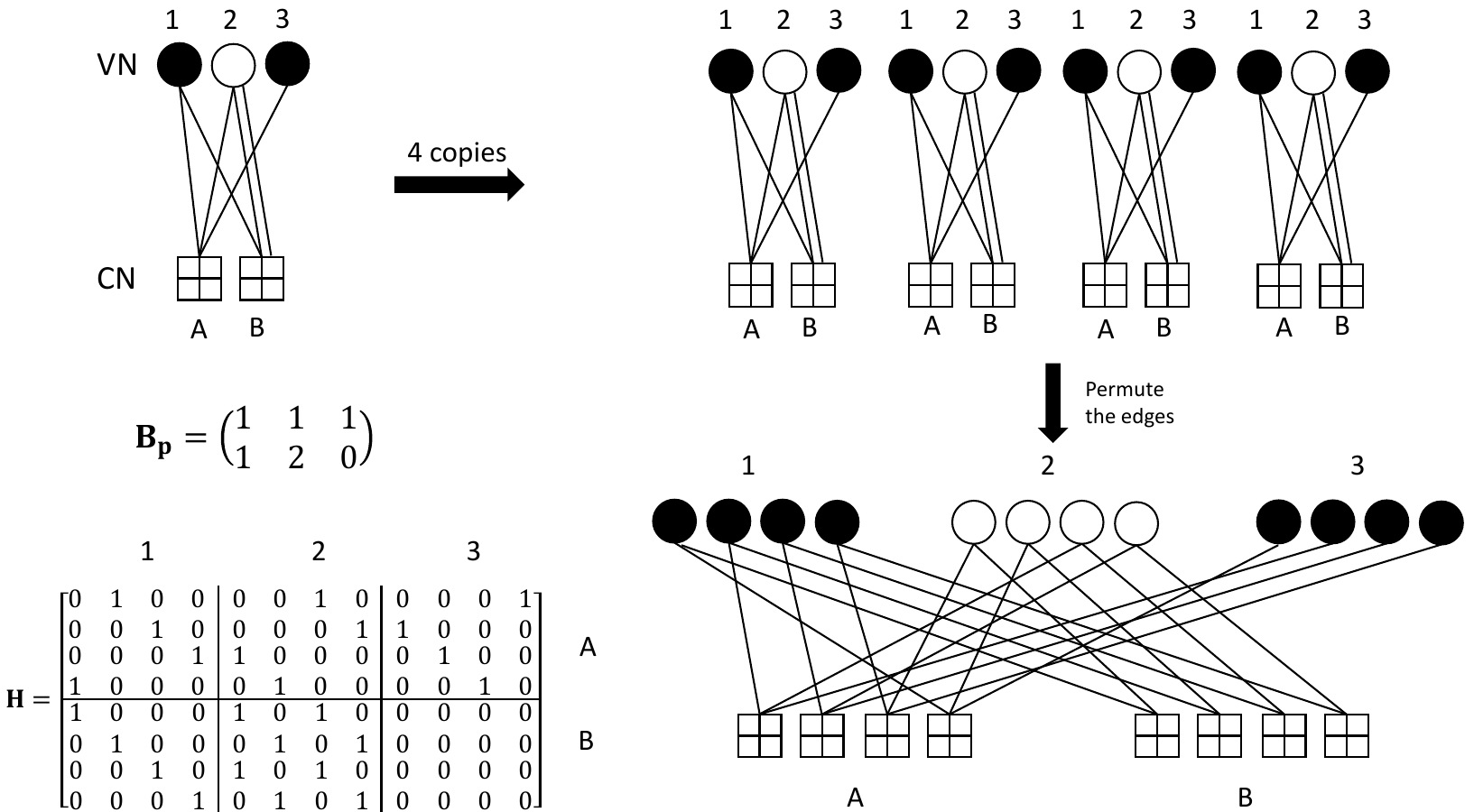}
    \caption{An example construction of a protograph LDPC code.}
    \label{protoLDPC_Tanner_Graph}
\end{figure}
Fig. \ref{protoLDPC_Tanner_Graph} illustrates the process of obtaining a derived graph through the lifting of a protomatrix $\boldsymbol{B}_{p}$ with $m_p = 2$ and $n_p = 3$.
The design rate of the code is $R = (3-2)/(3-1) = 1/2$ since the VN 2 is punctured. 
The parity-check matrix $\boldsymbol{H}$ is obtained by lifting (replacing) each entry in $\boldsymbol{B}_{p}$ with a $4\times 4$ square matrix. 
The identity matrix and its shifted version are used in this example to perform the lifting of non-zero entries in $\boldsymbol{B}_{p}$ and the zero matrix is used for 0s.

\textit{3b) Minimum Distance:} To determine whether or not the minimum distance of typical LDPC codes in a protograph ensemble increases linearly with code length $N$, the normalized logarithmic asymptotic weight distribution $r(\delta)$ for a given protograph $\mathscr{P}$ can be expressed as \cite{divsalar_protograph_2005}
\begin{align}
\hat{n}_{t}r(\delta) = \max_{\boldsymbol{\delta}_t:\vert\boldsymbol{\delta}_t\vert = \hat{n}_t\delta}\left\{\sum_{i=1}^{m_p}\phi^{c_i}(\boldsymbol{\delta}_i) - \sum_{j = 1}^{n_p}\left(d_{v_j}-1\right)\mathscr{H}(\delta_j)\right\},
\end{align}
where $\delta$ denotes the normalized weight of the protograph, $\hat{n}_t$ denotes the transmitted VNs of the protograph and $d_{v_j}$ denotes the degree of the $j$-th VN.
The subvector $\boldsymbol{\delta}_t = (\delta_1,\delta_2,\ldots, \delta_{\hat{n}_t})$ denote the normalized partial weights of the transmitted VNs, and $\mathscr{H}(x) = -x \ln x - (1 - x) \ln(1 - x)$ is the binary entropy function.
Moreover, the function \[\phi^{c_i}\left(\boldsymbol{\delta}_i\right) = \limsup_{Z\rightarrow \infty}\frac{\ln{A^{c_i}_{Z\boldsymbol{\delta}_i}}}{Z}\] denotes the normalized logarithmic asymptotic weight distribution for CN $c_i$ with normalized partial weight vector $\boldsymbol{\delta}_i$, where $A^{c_i}_{Z\boldsymbol{\delta}_i}$ is the weight enumerator of the partial weights of the $d_{c_i}$ VNs connected to check node $c_i$.
The first zero-crossing of the function $r(\delta)$ at $\delta = \delta_{min}>0$ is called the \emph{typical minimum distance ratio}, which shows a high probability that the minimum distance of most LDPC codes in
the ensemble increases linearly with $N$ with proportionality constant $\delta_{min}$.

\begin{figure}
    \centering
    \includegraphics[width=1.0\linewidth]
{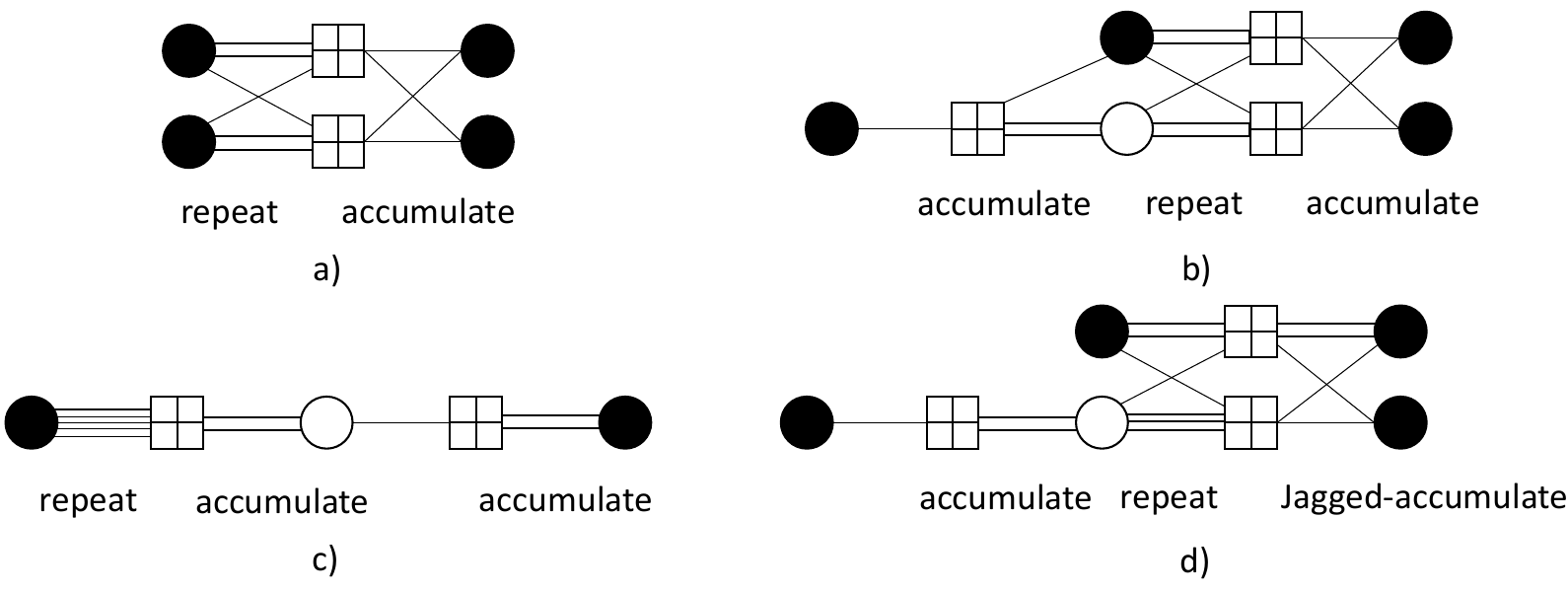}
    \caption{Protograph for popular code constructions: a) repeat-accumulate (RA) structure; b) accumulate-repeat-accumulate (ARA) structure; c) repeat-accumulate-accumulate (RAA) structure; d) accumulate-repeat -jagged-accumulate (AR4JA) structure. Puncturing VNs (denoted by a white circle) ensures that all code ensembles have a design rate of $1/2$.}
    \label{fig:proto_ra_family}
\end{figure}

\textit{3c) Types of Protograph Codes:} Protograph code properties and design methods were studied in \cite{divsalar_protograph_2005,divsalar_construction_2006, liva_quasi-cyclic_2008,abu-surra_enumerators_2011}, among many other works.
Various examples of popular code designs were developed based on the protograph framework.
Fig. \ref{fig:proto_ra_family} shows the protograph of several popular code constructions.
The protograph for a rate $1/2$ systematic repeat accumulate (RA) code \cite{d_divsalar_h_jin_r_mceliece_coding_1998} is shown in Fig. \ref{fig:proto_ra_family}-a).
If an accumulator is cascaded at the input of the systematic RA code, a performance improvement in the waterfall can be achieved, at the expense of a modest increase in the decoding complexity.
This type of code is called an accumulate-repeat-accumulate (ARA) code \cite{abbasfar_accumulate-repeat-accumulate_2007} and is shown in Fig. \ref{fig:proto_ra_family}-b).
Alternatively, the use of another accumulator at the output of an RA code can lead to better performance in the error floor region.
This is called a repeat-accumulate-accumulate (RAA) structure and is shown in Fig. \ref{fig:proto_ra_family}-c).
Fig \ref{fig:proto_ra_family}-d) shows another code structure known as the accumulate-repeat-jagged-accumulate (ARJA) which has good performance in the waterfall and error floor regime.

Furthermore, the protograph LDPC ensemble has also been widely adopted in the design of quasi-cyclic (QC) LDPC codes in various communication standards such as WiFi, e.g., \cite{echard_spl_2001}, WiMAX and 5G New Radio (NR), e.g., \cite{li_algebra-assisted_2018}.
The standardized AR4JA codes \cite{noauthor_low_2007}, which belong to the family of ARJA codes, are also used for next-generation deep-space communications.

The protograph-based raptor-like (PBRL) LDPC codes \cite{chen_protograph-based_2015, ranganathan_quasi-cyclic_2019} are a class of LDPC codes that have the basic structure of Raptor codes \cite{shokrollahi_raptor_2006} and LT codes \cite{luby_lt_2002}, where the raptor-like code structure enables the design of rate-compatible protograph LDPC code families with efficient encoding and decoding.  
The protograph of a PBRL LDPC code ensemble can be described as
\begin{align}
\boldsymbol{B_{\text{PBRL}}} = 
\left[
\begin{matrix}
\boldsymbol{B}_{\text{HRC}} & \boldsymbol{0} \\
\boldsymbol{B}_{\text{IRC}} & \boldsymbol{I}
\end{matrix}
\right],
\end{align}
 where:
 \begin{itemize}
\item $\boldsymbol{B}_{\text{HRC}}$: protograph of a highest-rate code (HRC),
\item $\boldsymbol{B}_{\text{IRC}}$: protograph of an incremental redundancy code (IRC),
\item $\boldsymbol{0}$: all-zeros matrix,
\item $\boldsymbol{I}$: identity matrix.  
 \end{itemize}
 
The overall protograph is lifted to produce the derived code. 
After lifting, the HRC portion of the code structure is identical to the precode in a Raptor code. 
Similar to Raptor codes, where a precoded $[N,K]$ block code is coded by an additional LT code with a specific degree distribution, the degree one VN in $\boldsymbol{I}$ can be efficiently encoded as modulo-2 sums of the precode symbols by $\boldsymbol{B}_{\text{HRC}}$ in the manner similar to the LT code in Raptor codes. 
The encoding of PBRL codes can be divided into two steps, 1) encoding of the precode, in this case is the HRC code $\boldsymbol{B}_{\text{HRC}}$, and 2) the encoding of IRC code, which involves XOR operations only.

\subsection{Design and Analysis Tools for LDPC Codes}\label{sec:ldpc_tools}
\subsubsection{Density Evolution}
As the code length $N\rightarrow \infty$, the \emph{asymptotic decoding threshold} of an LDPC ensemble represents the \emph{capacity} of the ensemble, which distinguishes between \emph{reliable} and \emph{unreliable} communication in the limit of infinite blocklength codes. 
Similar to the design of turbo codes described in Section \ref{sec:turbo}, the asymptotic decoding threshold of an LDPC code can be calculated using \emph{density evolution} (DE) \cite{richardson_capacity_2001}, and hence, to obtain the optimal degree distributions $\lambda(x)$ and $\rho(x)$ of an LDPC code ensemble. 
For a given degree distribution pair, DE calculates the decoding threshold by tracking the evolution of the messages passed between VNs and CNs during the iterative BP decoding process.
For the binary erasure channel (BEC), the erasure probability $\epsilon$ is used as the metric to measure the reliability of the decoded messages. For the binary-input additive white Gaussian noise (BI-AWGN) channel, the probability density of the messages is used. 
The optimal degree distribution of an LDPC ensemble can be determined such that the probability of error converges to zero for the largest value of the channel parameter, e.g., the erasure probability $\epsilon$ for the BEC channel and the noise standard deviation $\sigma$ for the BI-AWGN channel.

\textit{1a) DE on the BEC}: For a given degree distribution pair $(\lambda(x), \rho(x))$, the recursion formula of DE for iterative BP decoding over the BEC is given by
\begin{equation}
\label{equ:de_bec}
p_{e}^{(\ell)} = p_e^{(0)}\lambda\left(1 - \rho\left(1-p_e^{(\ell-1)}\right)\right), 
\end{equation}
where $p_e^{(0)} = \epsilon$ and $p_e^{(\ell)}$ denotes the probability of a variable to check node message in decoding iteration $\ell$ is erased.
The DE threshold $\epsilon^{*}$ of an LDPC ensemble over the BEC is defined as $\epsilon^{*} \triangleq\sup\{\epsilon>0|\lim_{\ell \rightarrow \infty}p_e^{(\ell)}=0\}$


\textit{1b) DE on the BI-AWGN channel:} The DE analysis of an LDPC ensemble defined by $\lambda(x)$ and $\rho(x)$ on the BI-AWGN channel is similar to that on the BEC.
The difference is that, during the recursion of the DE process, \emph{real-valued} space messages, known as the \emph{log-likelihood ratio} (LLR) of the probability of a given bit being 0 or 1, are passed among the edges of a Tanner graph.
Therefore, instead of a single probability, the probability densities of LLRs (viewed random variables) need to be tracked.
Let $P^{(\ell)}$ be the probability density function (PDF) of the LLR passed from VN to CN at iteration $\ell$.
The DE analysis over the BI-AWGN channel is given by the recursion formula
\begin{equation}
\label{equ:de_biawgn}
P^{(\ell)} = P^{(0)}\circledast \lambda^{\circledast}\left(\Gamma^{-1}\left( \rho^{\circledast}\left(\Gamma\left(P^{(\ell-1)}\right)\right)\right)\right), 
\end{equation}
where $\circledast$ denotes convolution and 
\begin{equation}
\lambda^{\circledast}(x) = \sum_{d=1}^{d_v}\lambda_dx^{\circledast(d-1)},
\end{equation}
\begin{equation}
\rho^{\circledast}(x) = \sum_{d=1}^{d_c}\rho_dx^{\circledast(d-1)}.
\end{equation}
The operator $\Gamma$ corresponds to the density change processed at the CNs.
An example of such a process is to let $\Gamma = \Phi(x) = -\log(\tanh(x/2))$, which is the CN processing function adopted in the sum-product algorithm (SPA). 
The density $P^{(0)}$ is the initial PDF of the LLR received from the channel.  
Assume that the all-zero codeword is transmitted, or the all $+1$ signal vector is transmitted according to the BPSK modulation mapping, i.e., $0\rightarrow +1$ and $1\rightarrow -1$. The probability of bit error after iteration $l$ is equal to the probability that a VN LLR is negative-valued, which is given by
\begin{align}
p_{\text{err}}^{(\ell)} = \int_{-\infty}^{0} P^{(\ell)}(x)dx.
\end{align}
The DE threshold of an LDPC ensemble over a BI-AWGN channel is defined as $\sigma^{*} \triangleq\sup\{\sigma>0|\lim_{\ell\rightarrow\infty}p_{\text{err}}^{(\ell)}=0\}$.

\subsubsection{Gaussian Approximation}
Although DE provides an exact decoding threshold for a given degree distribution pair, the computational complexity is very high.
To simplify the calculation process of DE, Gaussian approximation (GA) \cite{chung_analysis_2001} reduces the complexity of DE by assuming all messages are Gaussian distributed and that the mean of any message is equal to one-half of its variance. 
The approach greatly simplifies DE since only the mean value of the messages need to be tracked.  

\subsubsection{Extrinsic Information Transfer (EXIT) Chart}
Another approach for approximate decoding threshold is that of the EXIT chart \cite{ashikhmin_extrinsic_2004}.  
The EXIT chart was first introduced for turbo codes \cite{ten_brink_convergence_2001}, where the details can be found in Section \ref{sec:turbo}-\ref{ssec:turbo_tool}\ref{sec:exit_turbo}.
Then, the EXIT chart analysis was extended to design the degree distributions of LDPC codes \cite{ten_brink_design_2004}.
Unlike DE tracks the density of messages, the EXIT chart tracks the evolution of the average \emph{extrinsic mutual information} $I_{E,V}$ and $I_{E,C}$ passed from VNs to CNs and CNs to VNs, respectively.
Compared to DE, the EXIT chart analysis is simple to implement while only an approximation of the true decoding threshold of an ensemble is produced.

For an irregular LDPC code ensemble with degree distribution pair $(\lambda(x), \rho(x))$, the evolution of the average extrinsic mutual information is obtained by recursively averaging over the different node degrees, that is 
\begin{align}
\label{equ:exit_averaging_1}
I^{(\ell)}_{E,V} = \sum_{d=1}^{d_v}\lambda_d I^{(\ell-1)}_{E,V}(d),
\end{align}
and
\begin{align}
\label{equ:exit_averaging_2}
I^{(\ell)}_{E,C} = \sum_{d=1}^{d_c}\rho_d I^{(\ell-1)}_{E,C}(d),
\end{align}
where \cite{ryan_channel_2009}
\begin{equation}
I^{(\ell)}_{E,V}(d) = J\left(\sqrt{(d-1)\left[J^{-1}\left(1-I^{(\ell-1)}_{E,C}\right)\right]^2 + \sigma^2_{\text{ch}}}\right).
\end{equation}
and
\begin{equation}
I^{(\ell)}_{E,C}(d) = 1-J\left(\sqrt{(d-1)\left[J^{-1}\left(1-I^{(\ell-1)}_{E,V}\right)\right]^2}\right),
\end{equation}
Here, the variance of the LLR at the output of the channel $\sigma^2_{\text{ch}} = 8RE_b/N_0$ for a rate-$R$ code.
The function $J(\cdot)$ is defined by \cite{ryan_channel_2009}
\begin{equation}
J(\sigma) = 1 - \int_{-\infty}^{\infty} \frac{1}{\sqrt{2\pi}\sigma}e^{-\frac{(\eta-\sigma^2/2)^2}{2\sigma^2}}\log_2\left(1+e^{-\eta}\right)d\eta
\end{equation}
with $I^{(0)}_{E,V} = J(\sigma^2_{\text{ch}})$ is the initialization of the recursion.

\subsubsection{Protograph-Based EXIT Chart}
To identify good codes in the ensemble defined by $\lambda(x)$ and $\rho(x)$, an extension of the EXIT approach, namely the protograph-based EXIT (PEXIT) chart \cite{liva_protograph_2007}, has been proposed for LDPC ensembles defined by protographs, which takes into account the different edge connection properties.
%
%
The PEXIT chart analysis is facilitated by the relatively small size of protographs and permits the analysis of protograph code ensembles characterized by the presence of multiple parallel edges, degree-$1$ VNs, and punctured VNs.

The PEXIT chart analysis eliminates the average in (\ref{equ:exit_averaging_1}) and (\ref{equ:exit_averaging_2}) and considers the propagation of the messages on specific edges that are specified by the protograph of the ensemble. 
Let $I_{E,V}(i,j)$ be the extrinsic mutual information between code bits associated with type-$j$ VNs and the LLRs sent from these VNs to type-$i$ CNs.
Similarly, let $I_{E,C}(i,j)$ be the extrinsic mutual information between code bits associated with type-$j$ VNs and the LLRs sent from type-$i$ CNs to these VNs. Let $b_{i,j}$ be the element on the $i$-th row and the $j$-th element of the photograph base matrix. Then we have recursive formulas given by \cite{liva_protograph_2007}
\begin{multline}
I_{E,V}^{(\ell)} = J\Biggl(\sqrt{\sum_{s\neq i}b_{s,j}\left[J^{-1}\left(I^{(\ell-1)}_{E,C}(s,j)\right)\right]^2} \\ \notag
\overline{\rule{0pt}{3.0ex} +(b_{i,j}-1)\left[J^{-1}\left(I^{(\ell-1)}_{E,C}(i,j)\right)\right]^2+\sigma^2_{\text{ch}}(j)}\Biggl).
\end{multline}
and
\begin{multline}
I^{(\ell)}_{E,C} = 1 - J\Biggl( \sqrt{\sum_{s\neq j}b_{i,s} \left[J^{-1}\left(1-I^{(\ell-1)}_{E,V}(i,s)\right)\right]^2}\\ \notag
\overline{\rule{0pt}{3.0ex} + \left(b_{i,j}-1\right)\left[J^{-1}\left(1-I^{(\ell-1)}_{E,V}\left(i,j\right)\right)\right]^2}\Biggl).
\end{multline}
Note that if $b_{i,j} = 0$, $I_{E,V} = 0$ and $I_{E,C} = 0$ in VN and CN updates, respectively.
The recursion continues until either the maximum iteration is reached or the convergence condition $I_{APP}(j) = 1$ for $j = 0,1,\ldots,N-1$, where
\begin{equation}
I^{(\ell)}_{APP}(j) = J\left(\sqrt{\sum_{s}b_{s,j}\left[J^{-1}\left(I^{(\ell-1)}_{E,C}\left(i,j\right)\right)\right]^2 + \sigma^2_{\text{ch}}(j)}\right).
\end{equation}
Hence, the threshold is the lowest value of $E_b/N_0$ for which all $I_{APP}(j)$ converge to $1$.

\subsubsection{Reciprocal Channel Approximation }
The decoding threshold of Protograph LDPC ensembles can be analyzed and designed via GA, EXIT chart or PEXIT chart.
Alternatively, a simpler approach, namely the \emph{reciprocal channel approximation} (RCA) \cite{divsalar_capacity-approaching_2009}, is devised so that the decoding threshold analysis for protograph LDPC ensembles can be carried out on the associated protograph directly.
A single real-valued parameter, in this case, signal-to-noise ratio (SNR) $\mathcal{S}$, is used for full DE.
For every value of $\mathcal{S}$, a reciprocal of SNR $\bar{\mathcal{S}}$ is defined such that $\mathcal{C}(\mathcal{S}) + \mathcal{C}(\bar{\mathcal{S}}) = 1$, where $\mathcal{C}(x)$ denotes the capacity of the BI-AWGN channel with SNR $x$.
The parameters $\mathcal{S}$ and $\bar{\mathcal{S}}$ are interchangeable via $\mathcal{R}(x) = \mathcal{C}^{-1}(1-\mathcal{C}(x))$ for $\mathcal{S} = \mathcal{R}(\bar{\mathcal{S}})$ and $\bar{\mathcal{S}} = \mathcal{R}(\mathcal{S})$.

To apply RCA on a protograph, select an initial channel SNR $\mathcal{S}_{\text{ch}}$ and identify all transmitted variable nodes first. 
The transformation $\mathcal{S}_{v\rightarrow c}= \mathcal{R}(\bar{\mathcal{S}}_{c\rightarrow v})$ and $\bar{\mathcal{S}}_{c\rightarrow v}= \mathcal{R}({\mathcal{S}}_{v\leftarrow c})$ is applied and it refers to the message going out from a VN and CN, respectively.
At VN and CN nodes, the {extrinsic messages} is determined as $\mathcal{S}_{v\rightarrow c} = \mathcal{S}_{\text{ch}}+\sum_{c'\in \mathscr{C}\backslash c}\bar{\mathcal{S}}_{c'\rightarrow v}$ and $\bar{\mathcal{S}}_{c\rightarrow v} = \sum_{v'\in \mathscr{V}\backslash v}\mathcal{S}_{v'\rightarrow c}$, respectively.
The process continues and a threshold is determined by the smallest value of $\mathcal{S}_{\text{ch}}$ for which unbounded growth of all messages $\mathcal{S}_{v\rightarrow c}$ and $\bar{\mathcal{S}}_{c\rightarrow v}$ can be achieved.

\subsubsection{Finite Blocklength Scaling Law}
In \cite{amraoui_finite-length_2009}, the behavior of finite blocklength
LDPC codes over the BEC in terms of the waterfall performance were studied. It was observed that if an iterative decoding process goes through a phase transition as a channel parameter $\epsilon$ crosses the decoding threshold $\epsilon^{*}$, then around this transition point the decoding process obeys a very specific scaling law.
Let $P_{\text{BLER}}(N,\epsilon)$ be the block error probability as a function of the blocklength $N$ and the channel erasure probability $\epsilon$. 
According to DE, $P_{\text{BLER}}(N,\epsilon)$ exhibits a phase transition at the iterative decoding threshold $\epsilon^{*}$ as $N\rightarrow\infty$.
Then, the estimation of $P_{\text{BLER}}(N,\epsilon)$ for a finite length LDPC code is obtained by the method of the covariance evolution, also known as the \emph{scaling law} \cite{amraoui_finite-length_2009}
\begin{align}
P_{\text{BLER}}(N,\epsilon) = Q\left(\frac{\sqrt{N}(\epsilon^{*}-\epsilon)}{\alpha}\right),
\end{align}
where $\alpha$ is the \emph{scaling parameter} that only depends on the degree distributions $\lambda(x)$ and $\rho(x)$, and $Q(\cdot)$ is the Q-function.
The scaling behavior has been conjectured to more general settings and channels, and empirical evidence was shown to support the conjecture in \cite{amraoui_finite-length_2009}. 

\subsection{Finite-Length Construction of QC-LDPC Codes}
%
Once an optimal degree distribution pair is obtained via any of the design tools mentioned above for an LDPC code ensemble, the next step is to construct an LDPC code from the ensemble such that the error rate performance approaches the theoretical threshold of the ensemble.
The design of LDPC codes of finite lengths falls into two categories: \emph{pseudorandom} code design and \emph{structured} code design, where \emph{graph-theoretical} based approaches and \emph{algebraic-based or matrix-theoretical} based approaches are commonly adopted, respectively.
The well-known graph-theoretical based construction methods are the \emph{progressive edge-growth} (PEG) \cite{hu_progressive_2001,hu_regular_2005} and the \emph{protograph} methods \cite{thorpe_low-density_2003}.
The LDPC codes constructed using PEG algorithms are unstructured in the sense that they can only be described by specifying, for each VN, the indices of the CNs to which it is connected.
From the practical implementation perspective, this is equivalent to storing the row and column indices of all the non-zero elements of a parity-check matrix in a memory unit.
This significantly reduces the practicability of LDPC codes due to their high memory usage when implemented on high-speed hardware platforms such as field-programmable gate arrays (FPGA) or application-specific integrated circuits (ASIC).
In addition, unstructured LDPC codes usually are encoded only via the multiplication of the information sequence by a generator matrix of the code, an operation whose complexity is quadratic with the codeword length.

%
Alternatively, algebraic-based design approaches, which were first introduced in 2000 \cite{fan_array_2001}, are commonly adopted in the design of structured LDPC codes.
Since then, various algebraic methods for constructing LDPC codes, binary and non-binary, have been developed based on mathematical tools such as finite geometries, finite fields, difference sets and combinatorial designs, e.g., \cite{ryan_channel_2009, kang_quasi-cyclic_2010,zhang_quasi-cyclic_2010,kou_low-density_2001,nguyen_construction_2012,ammar_construction_2004,djurdjevic_class_2003,vasic_combinatorial_2004,chen_near-shannon-limit_2004,fossorier_quasi-cyclic_2004,xu_construction_2005,tang_codes_2005,lan_new_2008,sassatelli_nonbinary_2010,song_unified_2009,huang_cyclic_2012,li_algebraic_2014}. 
Moreover, the concepts of protograph can be further adopted in the algebraic-based design approaches, by limiting the permutation matrix to the cyclic shift of a $Z\times Z$ identity matrix.
Various early-stage studies on designing prograph-based LDPC codes have been conducted, \emph{e.g.,} \cite{divsalar_protograph_2005,divsalar_low-rate_2005,divsalar_construction_2006,divsalar_ensemble_2006,divsalar_capacity-approaching_2009,andrews_encoders_2005}.
Many of the construction techniques have been adopted in the construction of quantum LDPC codes, \emph{e.g.,} \cite{mackay_sparse-graph_2004,xie_protograph_2015,xie_reliable_2016,xie_design_2018}.
Furthermore, the PEG algorithm may be effectively used to construct QC-LDPC codes by adding randomness to the code design, e.g., \cite{li_class_2004,liu_construction_2009,xie_design_2010}. 
In practice, a QC-LDPC code is often designed using a hybrid of the two approaches so that the optimized pseudorandom construction performs well in the waterfall region, while code structure can help provide exemplary performance in the error floor region. In what follows, we focus on the construction of QC-LDPC codes. 

The parity-check matrix of a QC-LDPC code is commonly in the form of \emph{block-circulant}. 
Particularly, the parity-check matrix is an $m\times n$ array of $Z\times Z$ square \emph{circulant permutation matrices} (CPMs) defined as
%
%
 \begin{equation}
 \label{equ:CPM}
 \boldsymbol{P} :=
 \left[\begin{matrix}
 0&1&0& \cdots &0\\
 0&0& 1 & \cdots & 0 \\
  \smash{\vdots} & \smash{\ddots} & \smash{\ddots} & \smash{\ddots} &\smash{\vdots}\\
 0& 0 & \cdots & 0 &1\\
 1&0& \cdots & 0 & 0
 \end{matrix} \right].
\end{equation}
Note that $\boldsymbol{P}^i$ represents the $i$-th cyclic right shift of $\boldsymbol{P}$ for $0\leq i\leq Z-1$ and $\boldsymbol{P}^Z=\boldsymbol{P}^0 = \boldsymbol{I}_Z$ is the identity matrix of size $Z$.
Then the parity-check matrix of an $mZ\times nZ$ QC-LDPC code has the structure of
\begin{align}
\label{equ:qcldpc_structure}
\boldsymbol{H} = 
\left[\begin{smallmatrix}
\boldsymbol{P}^{b_{0,0}} & \boldsymbol{P}^{b_{0,1}} & \cdots & \boldsymbol{P}^{b_{0,n-1}} \\
\boldsymbol{P}^{b_{1,0}} & \boldsymbol{P}^{b_{1,1}} & \cdots & \boldsymbol{P}^{b_{1,n-1}} \\
\vdots & \vdots & \ddots & \vdots \\
\boldsymbol{P}^{b_{m-1,0}} & \boldsymbol{P}^{b_{m-1,1}} & \cdots &  \boldsymbol{P}^{b_{m-1,n-1}} \\
\end{smallmatrix}\right],
\end{align}
where $b_{i,j}<Z$ or $b_{i,j}=-1$ if the circulant is a zero-matrix for $0\leq i\leq m-1$ and $0\leq j\leq n-1$.
Such a parity-check matrix may be compactly described by means of its {exponent matrix}, also known as the \emph{base matrix},
\begin{align}
\label{equ:exponent_matrix}
\boldsymbol{H}_b =
\left[\begin{smallmatrix}
b_{0,0} & b_{0,1} & \cdots & b_{0,n-1} \\
b_{1,0} & b_{1,1} & \cdots & b_{1,n-1} \\
\vdots & \vdots & \ddots & \vdots \\
b_{m-1,0} & b_{m-1,1} & \cdots & b_{m-1,n-1} \\
\end{smallmatrix}\right].
\end{align}
The base matrix not only constitutes a compact description of the parity-check matrix $\boldsymbol{H}$, which yields efficient memory usage to describe the Tanner graph of the QC-LDPC code but also affects the performance of the code by maximizing the girth of the QC-LDPC constructed. 
For instance, in \cite{fossorier_quasi-cyclic_2004,smarandache_unifying_2022}  a necessary and sufficient condition was derived for the Tanner graph to be characterized by a girth not smaller than some value. 
Hence, constructing a good QC-LDPC code is equivalent to finding a good base matrix with optimized shift values for a given lifting size.

\subsubsection{QC-LDPC Codes Based on Protographs}
Unlike the base matrix $\boldsymbol{H}_b$ of QC-LDPC codes where each element represents a cyclic shift of the identity matrix, the protomatrix $\boldsymbol{B}_{p}$ is a constitute collection of edge types.
In this case, a QC-LDPC code can be constructed from its promatrix $\boldsymbol{B}_{p}$ via a two-step lifting process.
Recall that a non-binary entry of $\boldsymbol{B}_{p}$ represents the multiple edge connections between the same VN and CN types.
Thus, the first step is to break this multi-edge type connection.
This can be done by replacing element $b_{i,j}$, of $\boldsymbol{B}_{p}$, $0\leq i\leq m_p-1$ and $0\leq j\leq n_p-1$, by a $Z'\times Z'$ permutation matrix such that the summation of each row and column of each permutation equals to $b_{i,j}$.
The value of $Z'$ needs to be greater or equal to the maximum value in the $\boldsymbol{B}_{p}$ so that the resulting matrix $\boldsymbol{B'_p}$ is a $m_pZ'\times n_pZ'$ protomatrix contains elements of 1s and 0s only.
Note that the permutation matrix in the first step of the lifting process does not need to be a CPM.
An effective way of placing the 1s' inside each permutation matrix is to use PEG algorithms to add randomness to the code structure \cite{xie_design_2010}.
The second step is to obtain the $Zm_pZ'\times Zn_pZ'$ parity-check matrix of the derived graph by replacing each 1 in $\boldsymbol{B'_p}$ with a $Z\times Z$ CPM with a cyclic shift, and each 0 by a $Z\times Z$ zero matrix.
The cyclic shifts can be obtained either by algebraic design approaches or search-based optimization approaches.

\subsubsection{QC-LDPC Codes from Finite Geometry}
The geometric approach to constructing LDPC codes is based on lines and points of finite geometry. Well-known finite geometries are Euclidean and projective geometry over finite fields \cite{kou_low-density_2001}. 
It is known that finite geometry LDPC codes have relatively good minimum distances and their Tanner graphs do not contain cycles of length 4.
The LDPC codes constructed using finite geometries yield either cyclic or quasi-cyclic depending on the dimension of the Euclidean geometry and therefore possess a simple encoding structure.

Let $EG(m, 2^s)$ be an $m$-dimensional Euclidean geometry over the Galois Field $GF(2^s)$ where $m$ and $s$ are positive integers. 
Denote by $\boldsymbol{H}_{EG(m,2^s)}$ an $M\times N$ parity-check matrix over $GF(2)$  composed of $N=\left(2^{(m-1)s}-1\right) \left(2^{ms}-1\right) / \left(2^s - 1\right)$ lines in $EG(m, 2^s)$ that pass through $M=2^{ms} - 1$ non-origin points \cite{kou_low-density_2001}.
If $m>2$, the $N$ columns of $\boldsymbol{H}_{EG(m,2^s)}$ can be partitioned into $\mathscr{K} = \left(2^{(m-1)s} - 1\right)/\left(2^s - 1\right)$ cycle classes, and is denoted as $\boldsymbol{h^{i}}_{(m,2^s)}$, $i = 1,2,\ldots,\mathscr{K}$.
Each of these $\mathscr{K}$ cycle classes consists of $(2^{ms} - 1)$ lines, which are obtained by cyclically shifting (downwards) any line in the class $2^{ms}-1$ times.
Hence, 
\begin{align*}
\boldsymbol{H}_{EG(m,2^s)} =  \left[\boldsymbol{h^{1}}_{(m,2^s)},\boldsymbol{h^{2}}_{(m,2^s)},\ldots,\boldsymbol{h^{\mathscr{K}}}_{(m,2^s)}\right].
\end{align*}
Alternatively, if $m=2$, $\boldsymbol{H}_{EG(m,2^s)}$ is a square matrix with $\mathscr{K}=1$.
Let $\boldsymbol{g}^{i} = [g^i_0,g^i_1,\ldots,g^i_{2^{ms}-2}]$ be the first column of a cycle class $\boldsymbol{h}^{i}_{(m,2^s)}$ for $0\leq i,\leq \mathscr{K}$, and $\mathscr{S}_{\boldsymbol{g}^i}$ be the set of indices of the non-zero elements.
Note that $\boldsymbol{g}^{i}$ is also known as the \emph{generator} of the cycle class and the number of non-zero elements of $\boldsymbol{g}^{i}$ is $\vert\mathscr{S}_{\boldsymbol{g}^i}\vert = 2^s$.
Then 
\[\boldsymbol{h}^{i}_{(m,2^s)} = \sum_{j}^{2^s} \boldsymbol{P}^{q_j}, \forall q_j\in\mathscr{S}_{\boldsymbol{g}^i}\] is a weight-$2^s$ circulant block and $\boldsymbol{H}_{EG(m,2^s)}$ is a $1\times \mathscr{K}$ array of weight-$2^s$ circulant blocks.

One approach of constructing a QC-LDPC code of the form (\ref{equ:qcldpc_structure}) from $\boldsymbol{H}_{EG(m,2^s)}$ is to perform generator splitting   \cite{ryan_channel_2009}.
An example of such a method is the $(d_v,d_c)=(4,32)$ regular $(8176,7156)$ QC-LDPC code designed from $EG(3,2^3)$ Euclidean geometry.
The code has rate $0.8752>1 - d_v/d_c$ because the resulting parity-check matrix $\boldsymbol{H}_{(8176,7156)}$ is not in full rank.
The geometry $EG(3,2^3)$ has total of $M=2^{ms}-1=511$ non-origin points and $N=(2^{(m-1)s}-1)(2^{ms}-1)/2^s - 1 = 4599$ lines, and $\mathscr{K}=9$ cycle classes, each cycle class is a weight $2^s=8$ circulant block.
The parity-check matrix $\boldsymbol{H}_{(8176,7156)}$ is obtained by splitting the generator $\boldsymbol{g}^i$ of $\mathscr{K} = 8$ cycle classes into smaller subsets $\{\boldsymbol{g}^{i,1}, \boldsymbol{g}^{i,2}, \boldsymbol{g}^{i,3}, \boldsymbol{g}^{i,4}\}$.
Each $\boldsymbol{g}^{i,l}, l = 1,2,3,4, $ has $2$ elements. 
The resulting parity-check matrix $\boldsymbol{H}_{(8176,7156)}$ is given by 
\begin{align*}
\boldsymbol{H}_{(8176,7156)} = \left[\boldsymbol{h}^1_{(3,2^3)},\boldsymbol{h}^2_{(3,2^3)},\ldots,\boldsymbol{h}^8_{(3,2^3)}\right],
\end{align*}
where each 
\begin{align*}
\boldsymbol{h}^i_{(3,2^3)} = \left[\begin{smallmatrix}\boldsymbol{h}^{i,1} & \boldsymbol{h}^{i,2}\\ \boldsymbol{h}^{i,3} & \boldsymbol{h}^{i,4}\end{smallmatrix}\right]
\end{align*}
is a $2\times 2$ array of weight-$2$ circulant blocks, where $\boldsymbol{h}^{i,l} = \boldsymbol{P}^{q_1} + \boldsymbol{P}^{q_2}, l = 1,2,3,4$ for $q_1,q_2\in\boldsymbol{g}^{i,l}$.
The Tanner graph of the code has a girth of $6$, and at a bit error rate of $10^{-6}$ the code performs $1$dB from the Shannon limit under sum-product decoding with maximum $50$ iterations \cite{ryan_channel_2009}.
This code has been selected for use in the NASA Consultative Committee for Space Data Systems (CCSDS) telemetry system \cite{noauthor_tm_2022}.

\subsubsection{QC-LDPC Codes Based on Finite Fields}
In the early 1960s, finite fields were successfully used to construct linear block codes, especially cyclic codes, with large minimum distances for hard-decision algebraic decoding, such as BCH codes \cite{bose_class_1960, hocquenghem_a_codes_1959} and Reed-Solomon (RS) codes \cite{reed_polynomial_1960}.
In the past decades, there have been major developments in using finite fields to construct LDPC codes.
LDPC code constructions based on finite fields perform well over
the binary-input AWGN channel with iterative decoding based on belief propagation. 
Most importantly, these finite-field LDPC codes were shown to have low error floors. 
These codes are more suitable for wireless and optical communication systems and data-storage systems, for which very low bit and/or word error rates are required. 
Furthermore, most of the LDPC code construction/based on the basis of finite fields are quasi-cyclic and hence they can be efficiently encoded using simple shift registers with linear complexity.
The general construction of QC-LDPC codes using finite fields is that the elements of the base matrix $\boldsymbol{H}_b$ belong to the elements of a finite field of a certain size. 
The position of each element can be determined through different properties of finite fields, such as additive and multiplicative group properties, subgroups, primitive elements of prime fields, and extension fields.

\subsection{Spatially-Coupled/Convolutional LDPC Codes}
In recent years, spatially-coupled LDPC (SC-LDPC) codes have drawn a lot of attention to the research communities and industries. 
SC-LDPC codes can be viewed as a type of convolutional LDPC codes \cite{jimenez_periodic_1998,jimenez_felstrom_time-varying_1999} that have the ability to combine good features of regular and irregular LDPC block codes in a single design \cite{6852099}.
It has been proven, in \cite{richardson_modern_2008,kudekar_threshold_2011,kudekar_spatially_2013}, that SC-LDPC ensembles are capacity-achieving on binary-input memoryless output-symmetric (BMS) channels under BP decoding. 
To construct SC-LDPC codes from LDPC block codes, the well-known approach called \emph{matrix unwrapping} \cite{jimenez_felstrom_time-varying_1999} is commonly adopted. 
Good convolutional LDPC codes can be constructed from good LDPC codes, e.g., see \cite{pusane_deriving_2011} for guidelines.
Consider an $M\times N$ parity-check matrix $\boldsymbol{H}$ of an LDPC block code. 
Its design code rate is given by $R_{BC}=1-M/N$. 
In practice, SC-LDPC codes are \emph{terminated}, whose parity-check matrix can be represented by
\begin{align}
\label{equ:SCLDPC}
\setlength{\arraycolsep}{2.0pt}
\renewcommand{\arraystretch}{0.4}
\boldsymbol{H}^{SC} = \left[\overbrace {\begin{smallmatrix}
\begin{array}{*{20}{c}}
{{{\boldsymbol{H}}_0}}&{}&{}&{}\\
{{{\boldsymbol{H}}_1}}&{{{\boldsymbol{H}}_0}}&{}&{}\\
 \vdots &{{{\boldsymbol{H}}_1}}&{}&{}\\
{{{\boldsymbol{H}}_{{m_s}-1}}} & \vdots & \ddots &{}\\
{{{\boldsymbol{H}}_{{m_s}}}}&{{{\boldsymbol{H}}_{{m_s - 1}}}}& \ddots &{{{\boldsymbol{H}}_0}}\\
{}&{{{\boldsymbol{H}}_{{m_s}}}}&{\ddots}& {{{\boldsymbol{H}}_1}}\\
{}&{}&{}&\vdots \\
{}&{}&{}&{{{\boldsymbol{H}}_{{m_s}-1}}} \\
{}&{}&{}&{{{\boldsymbol{H}}_{{m_s}}}}
\end{array}
\end{smallmatrix}}^{L}\right],
\end{align}
where $m_s$ is often called the \emph{syndrome former memory} or coupling memory.
Each $\boldsymbol{H}_j$ of size $M\times N$, such that $\sum_{j=0}^{m_s}\boldsymbol{H}_j = \boldsymbol{H}$, is a \emph{descendent matrix} of the parity-check matrix $\boldsymbol{H}$. 
The set of descendent matrices is then repeated $L$ times as shown in (\ref{equ:SCLDPC}) to construct the parity-check matrix $\boldsymbol{H}^{SC}$ of the terminated SC-LDPC code, where $L$ is the \emph{termination length} or coupling length of the code.
Note that the process of termination results in a parity-check matrix $\boldsymbol{H}^{SC}$ that contains irregular row weights. The code rate of an SC-LDPC code is then a function of $L$, given by
$R_{SC} = 1 - (L+m_s)M/LN$.
It is obvious that if the termination length $L\rightarrow \infty$, the SC LDPC code has the same code rate as the underlying LDPC block code defined by $\boldsymbol{H}$, that is $R_{SC}\xrightarrow[]{L\rightarrow \infty} R_{BC}$. 
%
Different from conventional block codes, SC-LDPC codes possess some important properties over their uncoupled counterparts, including simple code construction approaches and sliding window decoding with high throughput.
\paragraph{Threshold Saturation}
The \emph{threshold saturation} phenomenon discovered in \cite{kudekar_threshold_2011} shows that the decoding threshold of regular SC-LDPC codes under sub-optimal BP decoding can reach that under the optimal maximum a posterior (MAP) decoding on the BEC.
Intuitively, this is the result of the termination, which introduces a slight structured irregularity in the graph.
The CNs with lower degrees at each end of the terminated graph pass more reliable messages to their neighboring VNs. This effect propagates throughout the graph as decoding iterations increase.
On top of that, it is also shown that the MAP decoding threshold of regular-$(d_v,d_c,L,m_s)$ SC-LDPC codes is the same as that of regular $(d_v,d_c)$-LDPC block codes, 
when $m_s\rightarrow \infty$ and $L\rightarrow\infty$.
In addition, for regular SC-LDPC codes, the terminated ensembles retain asymptotically good in the sense that their minimum distance grows linearly with blocklength $N$. Moreover, BP
decoding thresholds of regular SC-LDPC codes are extremely close to the BEC capacity, with gaps that diminished for larger $(d_v,d_r)$, in contrast with what happens for regular block LDPC code ensembles. On the other hand, the MAP decoding thresholds of regular-$(d_v,d_c)$ LDPC block code ensembles achieve capacity as $(d_v,d_c)$ become very large. Thanks to threshold saturation, one can simply increase $(d_v,d_c)$ to construct capacity-approaching SC-LDPC codes without tedious optimization steps as in the design of irregular LDPC block codes.
The proof of threshold saturation of SC-LDPC codes was generalized to the general BMS channel \cite{kudekar_spatially_2013}. A simplified proof was later given in \cite{6325197}, see Section \ref{sec:turbo}-\ref{ssec:sc_turbo}\ref{sssec:sc_turbo_prop}.

\paragraph{Universality} In \cite{kudekar_spatially_2013}, the universality of SC-LDPC code is investigated and proved that a single SC-LDPC ensemble is universally good for all BMS channels, without the need to customize the ensemble construction to a specific channel. It should be noted that this is not the case for many other types of codes such as irregular LDPC block codes, turbo codes, and the original polar codes. The universality of SC-LDPC codes was extended to other channel settings, e.g., Gaussian multiple access channel \cite{yedla_universal_2011}.


\paragraph{Sliding Window Decoding}
A terminated SC-LDPC code can be decoded in the same way as decoding a block code.
Alternatively, due to the diagonal structure of the parity-check matrix, SC-LDPC codes can be decoded using a much smaller window size, and slide across the matrix diagonally.
This type of decoding is known as the sliding window decoder \cite{iyengar_windowed_2012} and has the feature that the decoding process may start during the reception of the data frame, and hence, an advantage for streaming type of data transmission, e.g., \cite{jimenez_felstrom_time-varying_1999, iyengar_windowed_2012,pusane_implementation_2008, kang_reliability-based_2018}.

\begin{figure}
    \centering
    \includegraphics[width=1\linewidth]{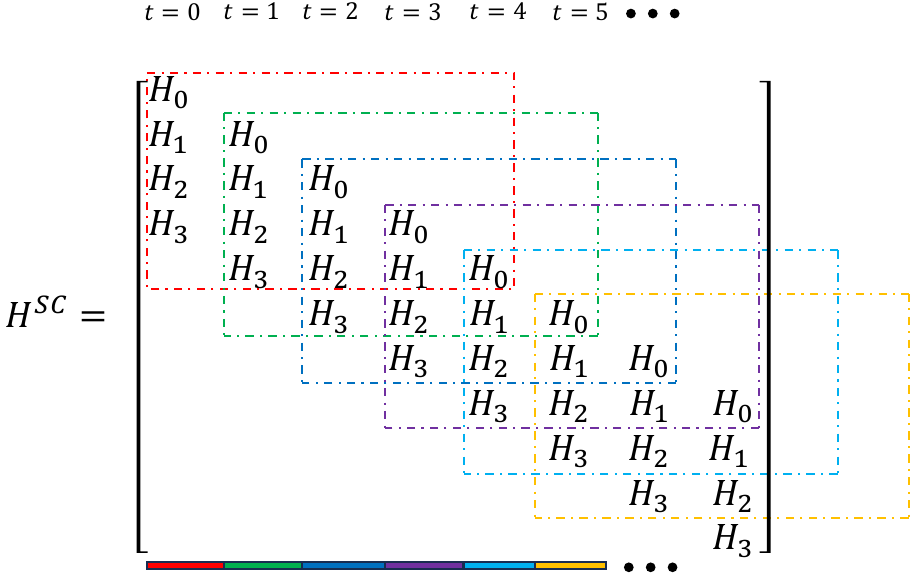}
    \caption{An example of sliding window decoder with window size $W=5$ at time index $t = 0,1,2,3,4,5$ (dash regions). The output of each decoding window is the target symbols shown at the bottom of the matrix.}
    \label{fig:sliding_window_decoder}
\end{figure}

Let $W$ be a window of size covering $WM$ CNs and $WN$ VNs.
The decoding window slides from time index $t = 0$ to time index $t = L-1$, which associates with different window positions in $\boldsymbol{H}^{SC}$ as shown in Fig. \ref{fig:sliding_window_decoder}.
At each time index, an iterative message-passing decoding is performed within the window in the same manner as decoding a block code. 
The decoding process stops if the syndrome check of the decoding window is satisfied or a predetermined maximum number of iterations is reached. 
Then the decoding window shifts by $M$ CNs vertically and $N$ VNs horizontally to the next time index $t$.
The first $N$ VNs shifted out of the decoding window are called target symbols.


%
The performance of window decoding degrades as $W$ becomes smaller.
This is because, in a decoding window, there exist VNs that have neighboring CNs outside the decoding window. 
Thus, the messages sent out from these VNs may not be reliable. 
These unreliable messages are propagated to the next window and deteriorate the error rate performance of the code.
Various research works have been conducted to improve the performance of sliding window decoding for SC-LDPC codes, e.g., \cite{jimenez_felstrom_time-varying_1999, lentmaier_efficient_2011, iyengar_windowed_2012,pusane_implementation_2008,ul_hassan_non-uniform_2017, kang_reliability-based_2018,ali_improving_2018,klaiber_avoiding_2018}. 
\subsubsection{Protograph SC-LDPC codes}
%

An insightful way of designing terminated SC-LDPC codes is to use a \emph{protograph} representation of a code ensemble.
%
%
Let $\boldsymbol{B}_{p}$ be the $m_p\times n_p$ protomatrix of a protograph LDPC block code ensemble. 
Then the protograph of a $(m_{p}, n_{p}, L)$ ensemble of SC-LDPC codes is obtained by performing \emph{edge-spreading} \cite{mitchell_spatially_2015,xie_euclidean_2016} to split the base matrix $\boldsymbol{B}_{p}$ into $(m_s+1)$ descendent protomatrices $\boldsymbol{B}_{p}^{(0)}, \boldsymbol{B}_{p}^{(1)},\ldots,\boldsymbol{B}_{p}^{(m_s)}$. 
Each of the descendent protomatrices has size $m_p\times n_p$ and $\sum_{i=0}^{m_s}\boldsymbol{B}_{p}^{(i)} = \boldsymbol{B}_{p}$. 
If $\boldsymbol{B}_{p}^{(i)}$ are identical for $0\leq i\leq m_s$, by arranging the set of descendent protomatrices $\boldsymbol{B}_{p}^{(0)}, \boldsymbol{B}_{p}^{(1)},\ldots,\boldsymbol{B}_{p}^{(m_s)}$ into the form as shown in (\ref{equ:SCLDPC}), the resulting terminated protomatrix of an SC-LDPC code is \emph{time-invariant}.
%
Otherwise, it is \emph{time-varying}, where each row of (\ref{equ:SCLDPC}) could start with a different $\boldsymbol{B}_{p}^{(i)}$. 
The design code rate of the protograph SC-LDPC codes is given by $R_{SC}=1 - (L+m_s)m_p/Ln_p$.

To construct a practical code from a protograph ensemble, the process of lifting is adopted to derive a large Tanner graph from the protograph of the ensemble. 
By lifting each element inside $\boldsymbol{B}_{p}^{(i)}$ by the $Z\times Z$ zero matrix or CPM defined in (\ref{equ:CPM}). 
The derived Tanner graph is a QC protograph SC-LDPC code that has $\left(m_s +L \right)Zm_p$ check nodes and $L Z n_p$ variable nodes.

\begin{figure}[h!]
\centering
\includegraphics[width=\linewidth]{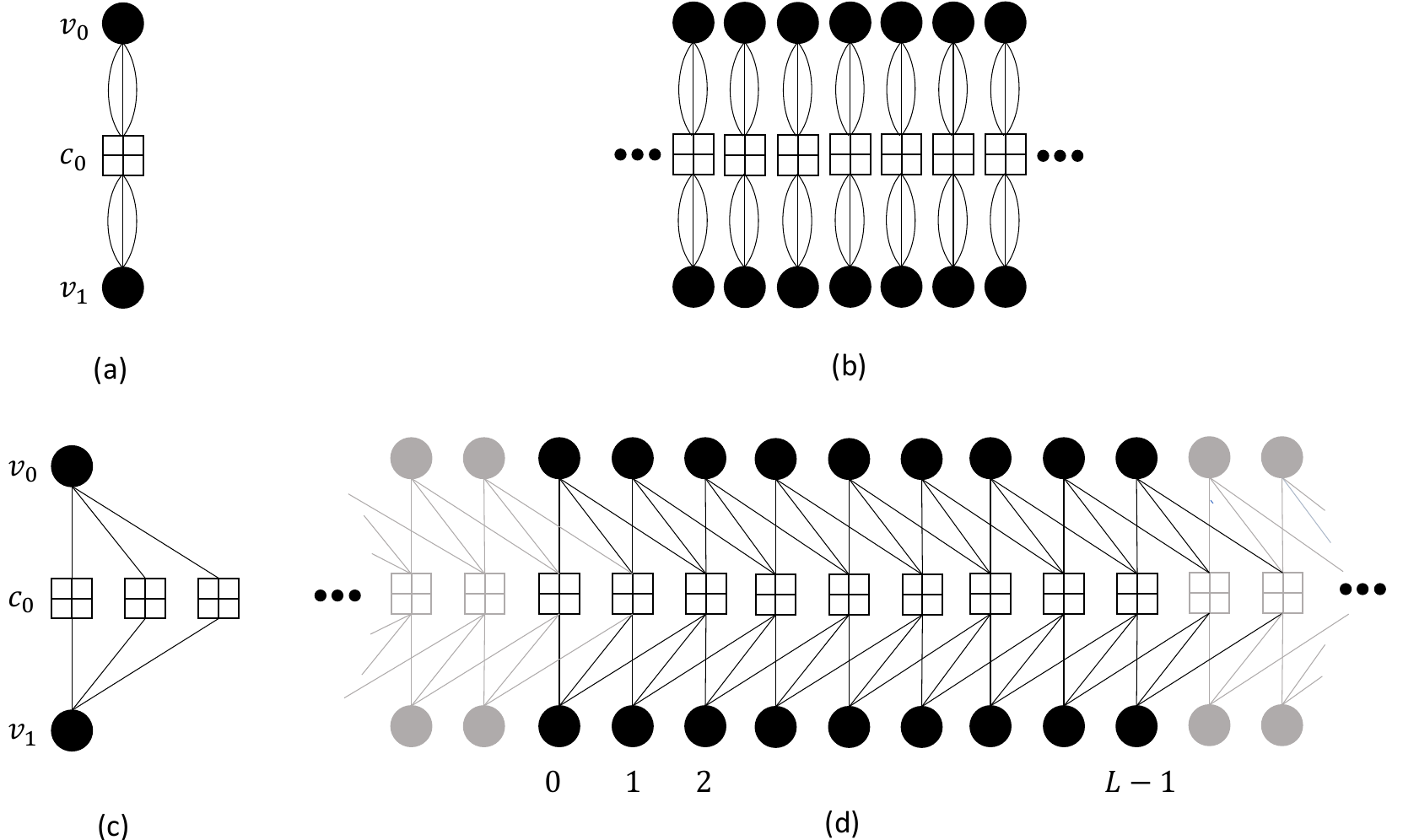}
\caption{(a) Protograph representation of a $(3,6)$ regular LDPC block code ensemble. (b) sequence of $(3,6)$ regular LDPC block code ensembles. (c) edge spreading for one $(3,6)$ regular LDPC block code ensemble with $m_s = 2$. (d) protograph representation of a terminated 
SC-LDPC code ensemble with coupling length $L$ and coupling memory $m_s = 2$.}
\label{fig:sc_LDPC_construction}
\end{figure}

Consider the example of $(d_v,d_c) = (3,6)$ regular protograph LDPC block code ensemble with protomatrix $\boldsymbol{B}_{p} = [3,3]$.  
Let $\boldsymbol{B}_{p}^{(0)} = \boldsymbol{B}_{p}^{(1)} = \boldsymbol{B}_{p}^{(2)} = [1,1]$.
Thus, $m_s = 2$ and $\sum_{i=0}^{m_s}\boldsymbol{B}_{p}^{(i)} = \boldsymbol{B}_{p}$.
The protomatrix of the corresponding SC-LDPC codes ensemble is 
\begin{align}
    \boldsymbol{B}_{p}^{SC} = \left[\begin{matrix}
\left[\begin{smallmatrix}1&1\end{smallmatrix}\right] & {} & {} & {} & {} \\
\left[\begin{smallmatrix}1&1\end{smallmatrix}\right] & \left[\begin{smallmatrix}1&1\end{smallmatrix}\right] & {} & {} & {} \\
\left[\begin{smallmatrix}1&1\end{smallmatrix}\right] & \left[\begin{smallmatrix}1&1\end{smallmatrix}\right] & \left[\begin{smallmatrix}1&1\end{smallmatrix}\right] & {} & {} \\
{} & \left[\begin{smallmatrix}1&1\end{smallmatrix}\right] & \left[\begin{smallmatrix}1&1\end{smallmatrix}\right] & \left[\begin{smallmatrix}1&1\end{smallmatrix}\right] & {} \\
{} & {} & \left[\begin{smallmatrix}1&1\end{smallmatrix}\right] & \left[\begin{smallmatrix}1&1\end{smallmatrix}\right] & \left[\begin{smallmatrix}1&1\end{smallmatrix}\right]\\
{} & {} & {} & \left[\begin{smallmatrix}1&1\end{smallmatrix}\right] & \left[\begin{smallmatrix}1&1\end{smallmatrix}\right]\\
{} & {} & {} & {} & \left[\begin{smallmatrix}1&1\end{smallmatrix}\right]
\end{matrix}\right]
\end{align}
for $L = 5$.
The design rate $R_{SC} = 1 - (5+2)/5\times 2 = 1 - 7/10<1/2$.
To construct the parity-check matrix of a QC SC-LDPC, each entry of $\boldsymbol{B}_{p}^{SC}$ is then replaced with a cyclic shift optimally design, followed by the lifting process with a lifting factor $Z$.
Fig. \ref{fig:sc_LDPC_construction} illustrated the construction of the SC-LDPC code ensemble from the $(3,6)$ regular protograph LDPC block code ensemble.

\begin{figure}[h!]
\centering
\includegraphics[width=\linewidth]{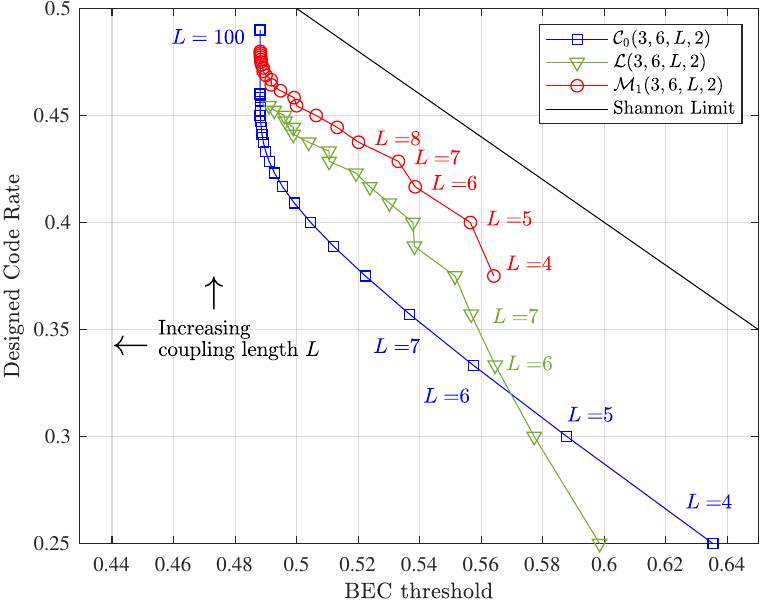}
\caption{Design rate versus BP threshold on the BEC for various coupled
ensembles under different chain lengths.}
\label{fig:SC_LDPC_threshold_vs_rate}
\end{figure}

The connected chain SC-LDPC codes were introduced in \cite{truhachev_code_2019}, which extends the spatial coupling phenomenon of individual graphs of LDPC block codes in a chained connection of multiple SC-LDPC graphs.
By adding additional edges to connect the terminated CN of one sub-chain to the VNs in another sub-chain, reliable information propagates from several directions rather than just from the ends of a single chain, leading to an improved BP decoding threshold when the coupling length $L$ is small.
The loop construction of connected chain SC-LDPC codes was proposed in \cite{dehaghani_improving_2022} with further improved BP decoding threshold over the BEC. 
In addition, the work in \cite{liao_self-connected_2023} not only improved the decoding threshold of connected chain SC-LDPC codes by introducing self-connected SC-LDPC code ensembles but also proposed a termination method to reduce the rate loss due to small $L$. As an example, the
designed code rates versus the BEC thresholds of the regular-$(3,6,L,2)$ SC-LDPC codes in \cite{mitchell_spatially_2015}, \cite{truhachev_code_2019}, and \cite{liao_self-connected_2023}, denoted by $\mathcal{C}_0$, $\mathcal{L}$, and $\mathcal{M}_1$, respectively, under different chain lengths are shown in Fig. \ref{fig:SC_LDPC_threshold_vs_rate}. The self-connected chain SC-LDPC code ensembles proposed in \cite{liao_self-connected_2023} achieve the best trade-off between rate loss and gap to capacity.

\subsubsection{Globally Coupled LDPC codes}
The globally-coupled LDPC (GC-LDPC) \cite{li_globally_2016} codes are another class of coupled LDPC codes.
Unlike spatially coupled Tanner graphs, where $m_s+1$ Tanner graphs are coupled by the common check node, the Tanner graph of GC-LDPC codes has $L$ disjoint LDPC block codes coupled by the additional globally-connecting check nodes.
As a result, the code length can be scaled up, which implies promising performance improvement, based on the component LDPC block codes, avoiding
construction of a completely new longer LDPC code.
An example of the parity-check matrix of a GC-LDPC code is given as
\begin{equation}
\label{equ:GCLDPC}
\setlength{\arraycolsep}{5.0pt}
\renewcommand{\arraystretch}{0.4}
\boldsymbol{H}^{GC} = \left[\begin{matrix}
\frac{\begin{array}{*{20}{c}}
{{{\boldsymbol{H}}_0}}&{}&{}&{}&{}&{}&{}\\
{}&{{{\boldsymbol{H}}_1}}&{}&{}&{}&{}&{}\\
{}&{}&{{{\boldsymbol{H}}_2}}&{}&{}&{}\\
{}&{}&{}&{\ddots}&{}&{}\\
{}&{}&{}&{}&{}&{{{\boldsymbol{H}}_{t-1}}}
\end{array}}{\boldsymbol{H}_{gc}}
\end{matrix}\right],
\end{equation}
where the upper submatrix of $\boldsymbol{H}^{GC}$ is a $t\times t$ diagonal array consisting of $t$ LDPC block codes $\boldsymbol{H}_i$, $0\leq i\leq t-1$, of size $M \times N$ on the main diagonal.
The upper submatrix is known as the \emph{local part} as each Tanner graph of the corresponding LDPC block is disjoint and independent. 
The lower part, also known as the \emph{global part}, is an $s\times tN$ matrix.
If each $M\times N$ submatrix $\boldsymbol{H}_i$ is replaced by a $m_p\times n_p$ protomatrix, the $(m_pt + s )\times n_pt$ protomatrix $\boldsymbol{B}_{p}^{GC}$ of a GC-LDPC code is obtained. 
Different constructions of the GC-LDPC codes have been investigated.
For instance, the GC-LDPC codes construction based on finite field and finite geometry \cite{li_globally_2016,nasseri_globally_2021}.
A family of non-binary GC-LDPC codes designed from RS codes is proposed in \cite{li_reed-solomon_2017} 
In \cite{zhang_construction_2018}, the construction of rate-compatible GC-LDPC codes is investigated. 

To effectively decode GC-LDPC codes, a two-phase local/global iterative decoding scheme for CN-GC-LDPC codes is proposed in \cite{li_globally_2016}.
Taking advantage of the cascading structure of the local part, a whole data frame can be split into $t$ independent sections, each section can be simultaneously decoded by an independent decoder.
If all sections of the local part are successfully decoded and the locally decoded codeword satisfies the parity-check constraints in the global part, the locally decoded codeword will be delivered to the user. 
If it does not, the global decoder starts to process the received codeword from the local decoder.
To reduce the decoding latency, sliding window decoders are commonly adopted in the BP decoding of SC-LDPC code.
The investigation of sliding window decoders for GC-LDPC codes is still ongoing. 

Due to that the rate loss caused by the additional global parity checks cannot be neglected for the finite coupling length $L$, free-ride coding \cite{wang_free-ride_2023} is proposed to construct coupled LDPC codes without any rate loss. 
The basic idea is to transmit some extra bits over the original coded link in a superposition (XOR) manner, where the coded length remains unchanged in comparison with the original coded link.
At the receiver side, a successive cancellation decoder can be used. 
It is shown in \cite {wang_free-ride_2023} that the proposed GC-LDPC codes can improve the performance of the
component LDPC codes, yielding an extra coding gain of up to $0.8$ dB, but without any code rate reduction.

\subsubsection{Partially Information Coupled LDPC Codes}
In the current 5G networks, the effective user data rate is approximately increased by 100 folds compared to 4G, and the maximum transport block (TB) size is over $1.2$ million bits. 
For peak throughput scenarios, the highest number of code blocks (CBs) in a TB reaches 151, and hence, the TB-level HARQ protocol will need more spectrum resources to send feedback information to the transmitter.
To improve the spectrum efficiency and the error rate performance of the transport block (TB), partially information-coupled LDPC (PIC-LDPC) codes were proposed \cite{yang_chained_2018}. 
The idea is to share a few information bits between every two adjacent CBs during the encoding.
In addition, by adding dummy bits to the first and the last CBs, there is a considerable and consistent SNR gain since the reliable messages from these two CBs spread out across other CBs with the aid of the coupled information bits when performing iterative decoding.

\begin{figure}[h!]
\centering
\includegraphics[width=3.3in]{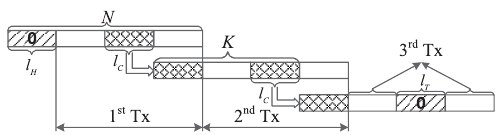}
\caption{The block diagram of encoding a TB with three CBs.}
\label{fig:PIC_LDPC}
\end{figure}

The frame structure of the PIC-LDPC codes with three CBs is illustrated in Fig. \ref{fig:PIC_LDPC}.
The CBs are component codewords of a systematic $(N,K)$ LDPC code of rate $R = K/N$. To obtain the CB at time instant $t$, the LDPC encoder takes $l_c$ information bits from the information bits at time $t-1$ and $K-l_c$ information bits at time $t$ as its input. The encoder outputs a length-$N$ LDPC component codeword. In other words, the CB at time $t$ shares $l_c$ information bits with the CB at time $t-1$. These shared information bits are called coupled bits and are only transmitted once, such that those $l_c$ bits are punctured from the $t$-th CB. For the first CB, $l_H$ information bits are set to zero for initialization. Moreover, $l_T$ information bits in the last CB are set to zero for terminating the coupled code chain. As the coupled bits and the zero bits are not transmitted, the length of the transmitted TB is $N_{TB} = (N-l_H) + (N-l_c) + (N-l_c-l_T )$ and the total information bits in the transmitted TB is $K_{TB} = (K-l_H) + (K-l_c) + (K-l_c-l_T )$.
Let $l_H=l_T=l_c = l$. For coupling length $L$, 
the code rate of PIC-LDPC codes is
\begin{align}
R_{\text{PIC}} =\frac{K_{TB}}{N_{TB}}= \frac{L(K-l)-l}{L(N-l)-l}.  \notag
\end{align}
When $L\rightarrow \infty$, the code rate becomes
$R_{\text{PIC}}\rightarrow \frac{K-l}{N-l}$. \color{black}

The PIC is performed on the encoder of LDPC codes, which is different from SC-LDPC codes whose coupling is performed based on the parity-check matrix.
As a result, PIC-LDPC codes can directly employ conventional encoders and decoders for LDPC block codes as their component code encoding and decoding.
The decoding of PIC-LDPC codes can be done at the TB level using the feed-forward and feed-backward (FF-FB) window decoding \cite{yang_chained_2018}.
The FF-FB decoding exploits the correlation, as well as the coupled bits, between every two consecutive CBs to achieve a good TB error rate.
Simulation results show that the PIC-LDPC codes yield at least $0.5$ dB gain over the LDPC block codes counterpart from IEEE 802.16e.

\subsection{LDPC Codes in Communications Standards} 
\subsubsection{IEEE 802 Communication Standards}
%


\paragraph{802.3an} The IEEE 802.3an \cite{noauthor_standard_2006} standard adds a physical layer for 10 Gigabit Ethernet over unshielded twisted pair cabling (10GBASE-T) for distances of up to 100 meters. 
The LDPC code is not quasi-cyclic and is designed from the $[N,K,d_{min}]=[32,2,31]$ shortened RS base code over $GF(2^6)$, where $d_{min}$ denotes the minimum distance of the code.
The resulting LDPC code is a rate $0.84$ binary $[N,K]=[2048,1723]$ code that guarantees no cycles of length four within the Tanner graph. 
%
%
The matrix has a constant column weight of 6 and a constant row weight of 32.
Thus, the LDPC code is a $(6,32)$-regular LDPC code.
Detail of the code designs is described by Djurdjevic \emph{et al.} in \cite{djurdjevic_class_2003}.

\paragraph{802.11n Wireless LAN}
%
IEEE 802.11n \cite{noauthor_ieee_2009} is an amendment to the previous 802.11 a/b/g standards in the 2.4GHz and 5GHz bands.
The amendment adds a high throughput physical layer specification that encodes data fields using either a convolutional code with a memory length of $7$ or a QC-LDPC code.
%
%
There are in total of $12$ independent QC-LDPC codes with $3$ code lengths and $4$ code rates.
Each of the 12 codes is derived from a protograph with one of the lifting sizes $27, 54$, and $81$, which corresponds to the three blocklengths $N = 648,1296$ and $1944$.
For rates $1/2, 2/3, 3/4,$ and $5/6$, the size of the base matrix is $12\times 24$, $8\times24$, $6\times24$, and $4\times 24$, respectively. 
The exponent matrix \cite{noauthor_ieee_2009} for each code is optimized with respect to the error rate performance and the girth property. 
The parameters of the codes are summarized in Table \ref{tab:802.11n_ldpc}. 
%
%
The parity-check matrix of the code has a dual-diagonal structure.
Such a code structure enables efficient encoding 
\cite{li_efficient_2006} to be performed on the parity-check matrix, and hence, no generator matrix is needed.

\begin{table}[h!]
\centering
\caption{IEEE 802.11n LDPC code parameters}
\begin{tabular}{ |c|c|c| } 
 \hline
  \textbf{Code length} $N$ & \textbf{Lifting size} $Z$ & \textbf{Code rates} $R$ \\
 \hline
  648 & 27 & 1/2, 2/3, 3/4, 5/6 \\ \hline
 1296 & 54 & 1/2, 2/3, 3/4, 5/6 \\ \hline
 1944 & 81 & 1/2, 2/3, 3/4, 5/6 \\
 \hline
\end{tabular}
\label{tab:802.11n_ldpc}
\end{table}

\paragraph{802.11ad Wireless LAN at 60GHz}
IEEE 802.11ad \cite{noauthor_ieee_2012} extend the previous wireless LAN standards into the $60$ GHz band, i.e., millimeter wavelength. 
This version contains a directional multigigabit physical layer specification utilizing four QC-LDPC codes of rates $1/2$, $5/8$, $3/4$, and $13/16$ to send control and data.
The base matrix is of size $8\times 16$, $6\times 16$, $4\times 16$, and $3\times 16$, respectively, corresponding to each one of the code rates.
In all cases, the length of the codes is $672$ and the lifting size is $42$. 
%
%
The code has the structure of a lower triangular form, which is different from the one designed for 802.11n.
LDPC codes with a lower triangular code structure also yield efficient encoding using the parity-check matrix \cite{richardson_efficient_2001}.

\paragraph{802.16e Mobile WiMAX}
Mobile WiMAX (Worldwide Interoperability for Microwave Access), IEEE 802.16e \cite{noauthor_ieee_2006}, was one of the first standards to adopt LDPC codes for forward error correction. 
The standard added mobility to the metropolitan area network standards.
For rates $1/2$, $2/3$ \footnote{Two different base matrix as were designed for rates $2/3$ and $3/4$.}, $3/4$, and $5/6$, the base matrix is of size $12\times 24$, $8\times 24$, $6\times 24$ and $4\times 24$, respectively.
%
%
Moreover, the code lengths vary from 576 to 2304 with a step size of 96 bits, and the corresponding lifting sizes vary from 24 to 96 with a step size of 4. 
%

\paragraph{802.22 Cognitive Wireless}

The IEEE 802.22 \cite{noauthor_ieee_2011} standard is for cognitive wireless regional area networks that operate in TV bands between 54MHz and 862MHz.
The aim of the standard is to bring broadband access to low-population-density areas. The maximum data rate is about 20 Mbit/s. Due to its cognitive radio techniques, it has the potential to be applied in many regions worldwide.					
The standard adopted the same LDPC codes as in 802.16e with minor modifications. 
Two shorter code lengths, 384 and 480, were added, and only one base matrix is used for rates $2/3$ and $3/4$.

\paragraph{802.15.3c Millimeter WPAN}
IEEE 802.15.3c \cite{noauthor_ieee_2009} is a standard for high data rate wireless personal area networks (WPAN).
The standard adds a new mmWave PHY layer that operates in the 60 GHz band (57 - 64 GHz) and allows for air throughput of up to 5 Gbit/s. 
The standard defines 5 LDPC codes with two blocklengths. 
For rates $1/2$, $5/8$, $3/4$, and $7/8$, the base matix is of size $16\times 32$, $12\times 32$, $8\times 32$, and $4\times 32$, respectively. 
Moreover, for these code rates, the lifting size is 21, which yields codes with a length of 672.
For a rate of $14/15$, the base matrix is of size $1\times 15$ with a lifting size of 96. 
The resulting code length is $1440$. 
%
%
The code structure for this code is neither a dual-diagonal nor a lower triangular form.
Instead, the code structure can be considered as an approximate lower triangular form, which is able to perform efficient encoding in a similar way as for codes in a lower triangular form with an encoding complexity 'gap'.
Such a gap is defined as the distance of the given parity-check matrix to a lower triangular matrix \cite{richardson_efficient_2001}. 


\subsubsection{Consultative Committee for Space Data Systems (CCSDS)}

The CCSDS has included LDPC codes in its recommendation for near-Earth and deep-space telemetry \cite{noauthor_tm_2022}.
A total of ten QC-LDPC codes have been included in the recommendation. 
An $[8160, 7136]$ code with a rate $R=223/255$ is recommended for near-Earth telemetry applications.
The rest nine codes, whose information lengths and code rates are the combination of $K=1024, 4096, 16384$ and $R=1/2, 2/3, 4/5$, are recommended for deep-space telemetry. 

The $[8160, 7136]$ LDPC code is constructed from the $[8176, 7156]$ QC-LDPC code via shortening and extention.
The parity-check matrix of the $[8176, 7156]$ QC-LDPC is a $2\times 16$ array of $511\times 511$  circulant blocks constructed based on Euclidean geometry.
Moreover, for rates $1/2, 2/3, 4/5$, the nine QC-LDPCs for deep-space telemetry are represented by a $3\times 5$, $3\times 7$, or $3\times 11$ array of $Z\times Z$ circulant blocks, respectively, where $Z = 2^b$, $b=7,8,\cdots,13$.
The parameters of the nine QC-LDPC codes are summarized in Table \ref{tab:ccsds_ldpc}.

\begin{table}[!ht]
\centering
\caption{Submatrix size for nine CCSDS QC-LDPC codes}
\label{tab:ccsds_ldpc}
\begin{tabular}{|c|c|c|c|} \hline
{} & \multicolumn{3}{|c|}{\textbf{Lifting size} $Z$} \\ \hline
\textbf{Information length} $K$ & rate $1/2$ & rate $2/3$ & rate $4/5$ \\ \hline
1024 & 512 & 256 & 128 \\ \hline
4096 & 2048 & 1024 & 512 \\ \hline
16384 & 8192 & 4096 & 2048 \\ \hline
\end{tabular}
\end{table}

\begin{figure}
    \centering
    \includegraphics[width=0.5\linewidth]{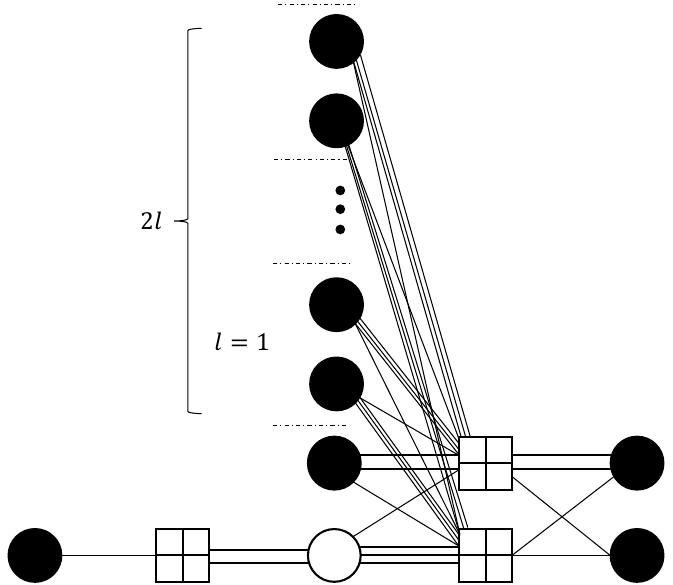}
    \caption{Protograph for AR4JA code family.}
    \label{fig:ar4ja_family}
\end{figure}

The nine QC-LDPC codes for deep-space telemetry are constructed by lifting the AR4JA LDPC code family \cite{noauthor_low_2007}.  
Note that for the AR4JA protograph LDPC code family, one variable node with the highest degree is punctured.
This refers to the last column of the protomatrix.    
The lifting procedure is performed in two stages.
The first stage uses an expansion factor equal to 4 and the second expansion factor $Z'$ equal to the power of $2$ leading to $Z = 4Z'$.
In each stage, the lifting procedure employs CPM whose exponents are selected based on the extrinsic message degree (EMD) or approximate cycle EMD (ACE) metrics \cite{tian_selective_2004}.
The protomatrix for the rate $1/2$ protograph code is given by

\begin{align*}
\centering
\label{circH}
\boldsymbol{H}_{\frac{1}{2}}=\left[\begin{array}{ccccc}
0_{Z} & 0_{Z} & I_{Z} & 0_{Z} & I_{Z} \oplus \prod_{1}\\
I_{Z} & I_{Z} & 0_{Z} & I_{Z} & \prod_{2} \oplus \prod_{3} \oplus \prod_{4} \\
I_{Z} & \prod_{5} \oplus \prod_{6} & 0_{Z} & \prod_{7} \oplus \prod_{8} & I_{Z} \\
\end{array}\right],
\end{align*}
where $n_p = 5$, $m_p = 3$.
The optimized permutation matrices $\prod_{i}$, $i = 1,2,\cdots,8$, are in \cite{noauthor_tm_2022}.
Since the last column is punctured during the transmission, the code rate $R = (n_p - m_p)/(n_p - \hat{n}_{pu}) = 2/4 = 1/2$. 
By adding two columns at a time to $\boldsymbol{H}_{\frac{1}{2}}$, the AR4JA protograph LDPC code family, shown in Fig. \ref{fig:ar4ja_family}, is obtained with the code rate defined as
\begin{equation}
R_{\frac{1+l}{2+l}} = \frac{2l+(n_p - 3)}{2l+(n_p - 1)},
\end{equation}
where $l = 0,1,2,3,\cdots$ is an integer meaning the number of additional double columns added to $\boldsymbol{H_{\frac{1}{2}}}$, and $n_p = 5$ is the number of columns in $\boldsymbol{H}_{\frac{1}{2}}$.
The optimized permutation matrices $\prod_{i}$, $i = 9,10,\cdots$ for rates $2/3,$ and $4/5$ can be found in \cite{noauthor_tm_2022}.

\subsubsection{DVB/DVB-S2}
In 2005, the 2nd generation of Digital Video Broadcasting (DVB-S2) became the first standard to adopt LDPC codes \cite{noauthor_etsi_nodate-1}.
The standard defines four different system configurations and applications: broadcasting, interactive services, digital satellite news gathering (DSNG), and professional services. 
The systematic non-QC-LDPC codes in this standard are constructed via the accumulation of information bits, which the addresses of the parity bits are specified in Annex B in \cite{noauthor_etsi_nodate-1}.
There are $21$ LDPC codes in total included in the standard, $10$ supported code rates for the short frame ($N=16200$) and $11$ code rates are supported for the normal frame ($N=64800$).
Moreover, the DVB-S2X \cite{noauthor_etsi_nodate} is the next-generation satellite transmission standard which is an extended version of its predecessor DVB-S2. 
Apart from the $21$ LDPC codes in DVB-S2, the new specification introduced additional $34$ LDPC codes with $3$ new codes specified for the new frame length of $32400$.
The new specifications in DVB-S2X allow for spectral efficiency gains of up to $50\%$ by offering lower roll-off factors, higher modulation orders, and finer code rate granularity compared to DVB-S2.
Table \ref{tab:dvb_s2_ldpcs} summarized the LDPC codes in the DVB-S2/DVB-S2X standards.

\begin{table*}[h!]
\begin{centering}
\caption{DVB-S2/DVB-S2X LDPC code parameters}
\label{tab:dvb_s2_ldpcs}
\begin{tabular}{ |c|c|c| } 
 \hline
 \textbf{Blocklength} $N$ & \textbf{Submatrix size} $Z$ & \textbf{Code rates} $R$ \\
 \hline
 16200 & 360 & 1/4, 1/3, 2/5, 1/2, 3/5, 2/3, 3/4, 4/5, 5/6, 8/9 (DVB-S2)	 \\ 
 \null & \null & 11/45, 4/15, 14/45, 7/15, 8/15, 26/45, 32/45 (DVB-S2X$^{*}$) \\ \hline
 32400 & 360 & 1/5, 11/45, 1/3 (DVB-S2X)	\\ \hline
 64800 & 360 & 1/4, 1/3, 2/5, 1/2, 3/5, 2/3, 3/4, 4/5, 5/6, 8/9, 9/10 (DVB-S2)	 \\
 \null & \null & 2/9, 13/45, 9/20, 11/20, 26/45, 28/45, 23/36, 25/36, 13/18, 7/9,  90/180, 96/180,  100/180,  104/180,     116/180, ... \\
 \null & \null &124/180, 128/180, 132/180, 135/180, 140/180, 154/180, 18/30, 20/30,   22/30    (DVB-S2X$^{*}$) \\  \hline
\end{tabular}
 \end{centering}

$*$ denotes additional codes added on top of DVB-S2.
\end{table*}

\subsubsection{3GPP 5G NR}
%
According to the 5G NR specifications, URLLC and mMTC services are sensitive to latency and hence used for high-reliability short data transmissions, while eMBB targets transmitting large blocks of data with high throughput.
The 5G deployment scenarios of eMBB require the support of not only a high throughput of up to $20$ Gbps, but also a wider range of code rates, code lengths, and modulation formats than 4G LTE.
Hence, QC-LDPC codes are recommended as the channel coding scheme for eMBB \cite{3gpp_chairmans_nodate-1}.
In particular, the recommended code lengths for eMBB ranged from $100$ bits to $8000$ bits, and code rates ranged from $1/5$ to $8/9$.
Furthermore, the promising transport block error rate (BLER) performance at $10^{-2}$ and invisible error floor down to BLER of below $10^{-4}$ for code blocks are required for the QC-LDPC codes in 5G standard. 

Besides QC structure, the 5G NR LDPC codes simultaneously possess rate compatibility and support multiple lifting factors.
The rate compatibility of 5G LDPC codes is effectively implemented with the aid of a raptor-like code structure.
Moreover, as multiple lifting sizes are supported, a vast range of information lengths and code rates can be easily adapted.
Furthermore, 5G LDPC codes support code rate adjustment up to bit-level granularity.
This is accomplished by performing puncturing, shortening, and repetition of coded bits after the lifting process of the base matrix.
Such a process is called the \emph{rate-matching}, which is one of the important modules in a practical 5G LDPC coding/decoding chain. 


\paragraph{Code Structure}

The LDPC codes for 5G New Radio (NR) are ensembles of PBRL-LDPC codes.
The base matrix structure can be represented as
\begin{equation}
\begin{centering}
\label{equ:5g_bg_structure}
\boldsymbol{B_{\text{5G}}} = 
\left[\begin{matrix}
\boldsymbol{B}_{\text{core}} & \boldsymbol{0} \\
 \boldsymbol{B}_{\text{ex}} &  \boldsymbol{I}
 \end{matrix}\right],
\end{centering}
\end{equation}
where $\boldsymbol{B}_{\text{core}}$ denotes the dense core matrix with dual-diagonal code structure as LDPC codes in IEEE 802.11n and 802.16e and $\boldsymbol{B}_{\text{ex}}$ denotes the sparse matrix indicating the connection in the single parity-check (SPC) extension of the 5G LDPC codes from high rate to low rate. 
The $\boldsymbol{0}$ and $\boldsymbol{I}$ indicate the zeros and identity matrices, respectively. 
%
%

Two base matrices, namely base graph 1 (BG1) and base graph 2 (BG2), are adopted for 5G LDPC codes.
Both BG1 and BG2 have similar code structures as shown in (\ref{equ:5g_bg_structure}), but BG1 supports information blocks up to $8448$ bits and code rates from $1/3$ to $8/9$, while BG2 supports information blocks up to $3840$ bits and code rates from $1/5$ to $2/3$ \cite{3gpp_etsi_nodate}.

As mentioned earlier, QC-LDPC codes with dual-diagonal structure yield efficient encoding on its parity-check matrix and hence no generator matrix is required. 
5G LDPC codes also benefit from this property due to its dual-diagonal structure in the core matrix $\boldsymbol{B}_{\text{core}}$.  
The base matrix format of both BGs in (\ref{equ:5g_bg_structure}) can be sub-divided and represented as
\begin{align}
\begin{centering}
\label{equ:BG_matrix}
\boldsymbol{B_{\text{5G}}} = 
\left[\begin{matrix}
\boldsymbol{A} & \boldsymbol{B} & \boldsymbol{0} \\
 \boldsymbol{C} & \boldsymbol{D} & \boldsymbol{I}
 \end{matrix}\right],
\end{centering}
\end{align}
where the size of each sub-divided matrix is given by:
\begin{itemize}
  \item $\boldsymbol{A}$ : $g \times k_b$.
  \item $\boldsymbol{B}$ : $g\times g$.
  \item $\boldsymbol{C}$ : $(m_p - g)\times k_b$.
  \item $\boldsymbol{D}$ : $(m_p - g)\times g$.
  \item $\boldsymbol{I}$ : $(m_p-g)\times (m_p - g)$ identity matrix.
  \item $\boldsymbol{0}$ : $g \times (m_p-g)$ all-zero matrix.
\end{itemize}
The parameter $g = 4$ is set for both BGs, whereas $n_p = 68, m_p = 46$ for BG1 and $n_p=52, m_p=42$ for BG2.
The information block, denoted by $k_b$, is $22$ for BG1 and $10$ for BG2.
The base matrix $\boldsymbol{B_{\text{5G}}}$ is lifted using CPMs of one of the supported lifting factors $Z$ to obtain a derived Tanner graph of desired length and rate.
Hence, the resulting LDPC code is in the QC structure.
%


\paragraph{Encoding of 5G NR LDPC Codes}
%
The encoding of 5G NR LDPC codes is performed on the parity-check matrix by solving the equation
\begin{align}
\label{equ:5g_encoding_equation}
\boldsymbol{x}\cdot\boldsymbol{H}^T = \boldsymbol{0},
\end{align}
where $\boldsymbol{x}$ is a systematic codeword and $T$ denotes the matrix transpose.
%
In the following, the encoding process is performed on a base matrix $\boldsymbol{B_{5G}}$ rather than on a derived graph $\boldsymbol{H}$. 
The encoding process on a derived graph will be the same except that the complexity is up-scaled by a factor of $Z$.
In the following, a two-stage encoding process for 5G LDPC codes using the BG1 represented in the form of  (\ref{equ:BG_matrix}) is introduced: 1) encoding of the core check $\boldsymbol{B}_{\text{core}}$ and 2) encoding of the SPC $\boldsymbol{B}_{\text{ex}}$.

Let $\boldsymbol{x} = [\boldsymbol{s}, \boldsymbol{\hat{p}}, \boldsymbol{\bar{p}}]$ be a codeword, where $\boldsymbol{s} = [s_0,s_1,\ldots,s_{k_b-1}]$ denotes the information block of size $k_b$.
Let $\boldsymbol{p} = [\boldsymbol{\hat{p}}, \boldsymbol{\bar{p}}]$ be the $m_p$ parity blocks, where $\boldsymbol{\hat{p}} = [\hat{p}_0, \hat{p}_{1}, \ldots, \hat{p}_{{g-1}}]$ and $\boldsymbol{\bar{p}} = [\bar{p}_{{0}}, \bar{p}_{1}, \ldots, \bar{p}_{{m_p-g-1}}]$ and $g = 4$.
Then, the precise representation of codeword $\boldsymbol{x}$ is 
\begin{align}
\label{equ:cw_equation}
\boldsymbol{x} = [s_0, s_1, \ldots, s_{k_b-1}, \hat{p}_{0}, \hat{p}_{1}, \ldots, \hat{p}_{g-1}, \bar{p}_{0}, \bar{p}_{1}, \ldots, \bar{p}_{{m_p-g-1}}].
\end{align}
The encoding of 5G NR LDPC codes is carried out by solving 
%
\begin{align}
\label{equ:parity_multiplication}
\boldsymbol{B_{5G}}\cdot\boldsymbol{x}^T = \left[\begin{matrix}
                       \boldsymbol{A} & \boldsymbol{B} & \boldsymbol{0} \\
                       \boldsymbol{C} & \boldsymbol{D} & \boldsymbol{I}
                     \end{matrix}\right] \left[\begin{matrix}
                             \boldsymbol{s}^T \\
                             \boldsymbol{\hat{p}}^T \\
                             \boldsymbol{\bar{p}}^T
                           \end{matrix}\right] = \boldsymbol{0}.
\end{align}
Equation (\ref{equ:parity_multiplication}) can be naturally split into two equations, as follows:
\begin{subequations}
\label{equ:p22}
\begin{align}
\boldsymbol{A}\boldsymbol{s}^T + \boldsymbol{B}\boldsymbol{\hat{p}}^T &= \boldsymbol{0},  \label{equ:p22_a}\\
\boldsymbol{C}\boldsymbol{s}^T + \boldsymbol{D}\boldsymbol{\hat{p}}^T + \boldsymbol{I}\boldsymbol{\bar{p}}^T &= \boldsymbol{0}. \label{equ:p22_b}
\end{align}
\end{subequations}
From these two equations, it can be seen that the parity bits can be computed in two steps. 
The first step is to calculate $\boldsymbol{\hat{p}}$ from equation (\ref{equ:p22_a}), and the second step is to compute $\boldsymbol{\bar{p}}$ from (\ref{equ:p22_b}) using the computed $\boldsymbol{\hat{p}}$.

Consider the first step, rewrite (\ref{equ:p22_a}) into the form
\begin{align}
\label{equ:5g_encoding_1}
\boldsymbol{A}\left[\begin{smallmatrix}
                                 s_0 \\
                                 s_1 \\
                                 \vdots \\
                                 s_{k_b-1}
                               \end{smallmatrix}\right] + \boldsymbol{B}
                                     \left[\begin{smallmatrix}
                                                                \hat{p}_{0} \\
                                                                \hat{p}_{1} \\
                                                                \hat{p}_{2} \\
                                                                \hat{p}_{3}
                                                              \end{smallmatrix}\right] = \boldsymbol{0},
\end{align}
where $\boldsymbol{A} = \left[a_{i,j}\right]^{g\times k_b}, 0\leq i\leq g-1, 0\leq j\leq k_b - 1$ and
\begin{align}
\boldsymbol{B} = \left[
\begin{matrix}
1 & 1 & 0 & 0 \\
1 & 1 & 1 & 0 \\
0 & 0 & 1 & 1 \\
1 & 0 & 0 & 1
\end{matrix}\right].
\end{align}
%
%
By expanding (\ref{equ:5g_encoding_1}) and perform matrix multiplication,
the following is obtained due to modulo-$2$ addition:
\begin{align}
\sum\limits_{i = 0}^{3}\sum\limits_{j=0}^{k_b-1} a_{i,j} s_j + \hat{p}_{0} = 0.
\end{align}
Hence, the first parity block $\hat{p}_{0}$ of $\boldsymbol{\hat{p}}$ is computed by accumulating all the results from $\boldsymbol{A}\boldsymbol{s}^T$.
Let
\begin{align}
  \theta_i = \sum\limits_{j=0}^{k_b-1} a_{i,j} s_j \hspace{0.5cm} \text{for} \hspace{0.5cm} i = 0,1,2,3.
\end{align}
The individual parity blocks in $\boldsymbol{\hat{p}} = \left[\hat{p}_{0}, \hat{p}_{1}, \hat{p}_{2}, \hat{p}_{3}\right]$ can be expressed as the following set of equations:
\begin{subequations}
\begin{align}
  \hat{p}_{0} &= \sum\limits_{i=0}^{3} \theta_i, \label{equ:parity_calculate_eqs1}\\
  \hat{p}_{1} &= \theta_0 + \hat{p}_{0}, \label{equ:parity_calculate_eqs2}\\
  \hat{p}_{2} &= \theta_2 + \hat{p}_{3},\label{equ:parity_calculate_eqs3}\\
  \hat{p}_{3} &= \theta_3 + \hat{p}_{0}. \label{equ:parity_calculate_eqs4}
\end{align}
\end{subequations}
The above equations compute each $\theta_i$ value by accumulating all the $a_{i,j}s_j$ values.
In modulo-$2$ operation, $\theta_i$ is obtained by carrying out XOR operations on all the elements of $a_{i,j}s_j$.
%
%
%
%
%
Furthermore, it is worth to mention that the submatrix $\boldsymbol{B}$ in (\ref{equ:5g_encoding_1}) is different for BG2\footnote{For BG2, the first column of $\boldsymbol{B}$ is $\left[\begin{matrix}1& 0& 1& 1\end{matrix}\right]^T$.}. 
Hence, the equations in (\ref{equ:parity_calculate_eqs1}) - (\ref{equ:parity_calculate_eqs4}) are different when performing encoding for BG2.

In the second step, the $\boldsymbol{\bar{p}}$ portion can be easily determined based on (\ref{equ:p22_b}). 
By performing rearrange and expansion, each of the parity blocks in $\boldsymbol{\bar{p}} = [\bar{p}_{0}, \bar{p}_{1},\cdots,\bar{p}_{{m_p-5}}]$ can be computed using the following equations:
\begin{align}
\label{equ:pc_parity_bits_a}
  \bar{p}_{i} = \sum\limits_{j=0}^{k_b-1} c_{i,j} s_j + \sum\limits_{l=0}^{3} \hat{p}_{l} d_{i,k_b+l} , i= 0,1,\ldots,m_p-5.  
\end{align}
%
In the actual encoding process, the entries $a_{i,j}$, $c_{i,j}$, and $d_{i,j}$ represent a CPM with a circular shift value.
Thus, the multiplication of $a_{i,j}s_j$, $c_{i,j}s_j$, and $\hat{p}_{l}d_{i,k_b+l}$ in the above calculations are multiplication between a CPM matrix and a vectors sequence. 
Finally, combining the information block $\boldsymbol{s}$ and the parity block $\boldsymbol{p} = [\boldsymbol{\hat{p}}, \boldsymbol{\bar{p}}]$, the encoded systematic codeword $\boldsymbol{x}$ is obtained.

\subsection{Iterative Message Passing Decoding of LDPC Codes}
\label{sec:ldpc_dec_MP_process}
Decoding of LDPC codes happens on Tanner graph using {iterative message passing} decoding algorithms.
There are two types of LDPC decoders have been investigated and researched in the past decades: hard-decision decoders and soft-decision decoders.
The difference between the two types of decoders is the input to the decoder and the format of the message passed along each edge during the iterative decoding process. 
The input to a hard-decision decoder is a binary sequence representing the sign of each bit and is passed along edges for the iterative decoding process.
A majority-based decision rule is made after each decoding iteration to determine the sign of each bit.  
On the other hand, the input to a soft-decision decoder is a sequence of log-likelihood ratio (LLR), which is in the range of $[+\infty,-\infty]$.
The magnitude of an LLR represents the reliability of the bit, and the sign represents the polarity of the bit.
The sequence of LLRs is passed along the edges of a Tanner graph to update the reliability of each bit iteratively. 
The final decision is made according to the sign of the LLR of each bit. 
Alternatively, there are decoders proposed in the literature which combine the features of both hard-decision and soft-decision decoders.
In this case, the input of a decoder is a sequence of LLR, whereas only the sign of each variable node is passed along the edges to perform the iterative decoding process.
A majority-based decision rule together with weighting is used to determine the sign of each bit at the end of each iteration.
One of the well-known examples of such a type is the \emph{weighted bit-flipping} (WBF) decoder \cite{ryan_channel_2009, kou_low-density_2001}.

For the sake of simplicity, consider that binary codeword $\boldsymbol{x}=(x_0,x_1,\cdots,x_{N-1})$ is transmitted over a binary-input memoryless channel, and denoted by $\boldsymbol{y}=(y_0,y_1,\cdots,y_{N-1})$ the received sequence at the output of the channel. 
The nature of $\boldsymbol{y}$ depends on the channel:
\begin{itemize}
\item For the Binary Symmetric Channel (BSC), $\boldsymbol{y}\in \{0,1\}^{N}$ is a binary sequence of length $N$, obtained by flipping each bit of $\boldsymbol{x}$ with crossover probability $p$.

\item For the Binary Erasure Channel (BEC), $\boldsymbol{y}\in\{0,1,\mathscr{X}\}^N$, where $\mathscr{X}$ denotes an erasure. Each bit of $\boldsymbol{x}$ is either erased ($y_j=\mathscr{X}$) with probability of $\epsilon$ or perfectly received ($y_j=x_j$) with probability of $1-\epsilon$.

\item For the Binary-input Additive White Gaussian Noise (BI-AWGN) channel, $\boldsymbol{y}\in\mathbb{R}^N$ is a length $N$ real vector, obtained by $y_j=(1-2x_j)+w_j$, where $(1-2x_j)\in\{\pm 1\}$ is the BPSK modulation of the bit $x_j$, and $w_j$ is the white Gaussian noise with zero mean and variance $\sigma^2$.
\end{itemize}
The commonly adopted message format in BP decoding is the LLR, which is represented as a ratio of the \emph{a posterior propabaility} (APP) of the transmitted bits and of the channel output, that is,
\begin{equation}
    r_j=\log\left(\frac{\Pr(x_j=0\vert y_j)}{\Pr(x_j=1\vert y_j)}\right).
\end{equation}
For the BSC channel with crossover probability $p$, the LLR of the $j$-th variable node is computed as
\begin{align}
r_j 
=  (1-2y_j)\log\left(\frac{1-p}{p}\right), \text{for} \hspace{0.2cm}y_j = 0, 1.
\end{align}
For the BEC channel with erasure probability $\epsilon$, the LLR of the $j$-th variable node is computed as
\begin{equation}
r_j = \left\{\begin{matrix}
 +\infty & \text{if} \hspace{0.1cm} y_j = 0, \\ 
 -\infty & \text{if} \hspace{0.1cm} y_j = 1, \\ 
 0& \text{if} \hspace{0.1cm} y_j = \mathscr{X} 
\end{matrix}\right.
\end{equation}
For the BI-AWGN channel with noise variance $\sigma^2$, the LLR of the $j$-th bit is computed as 
\begin{align}
r_j = \frac{2y_j}{\sigma^2}.
\end{align}
%
Denoted by $\mathcal{A}(v)$ the set of indices with the corresponding CN connected to the VN $v$.
Similarly, let $\mathcal{B}(c)$ be the set of indices with the corresponding VN connected to the CN $c$.
Furthermore, let $V_{i,j}$ be the variable-to-check (V2C) message sent from the $j$-th VN to the $i$-th CN, and $E_{i,j}$ be the check-to-variable (C2V) message sent from the $i$-th CN to the $j$-th VN.
Then the iterative decoding process is summarized in the following steps:

\begin{itemize}

\item The decoder takes $\boldsymbol{r}=\{r_0,r_1,\cdots,r_{N-1}\}$ as the input \emph{a priori} information of the VNs, and it is computed from $\boldsymbol{y}$ depends on the nature of the channel.

\item For each VN, the new V2C message $V^{(u)}_{i,j}$ sent out at iteration $u$ is computed as a function $\mathcal{F}_1(*)$ of all the incoming C2V messages $E^{(u-1)}_{i',j}$ of the previous iteration and the initial channel LLR $r_j$, that is,
        \begin{equation}
        \label{equ:mp_v2c_calculation}
            V^{(u)}_{i,j} = \mathcal{F}_1\left(r_j, E^{(u-1)}_{i',j}\right),i'\in\mathcal{A}(v_j),{i'\neq i}.
        \end{equation}
When $u=0$, $E^{(u-1)}_{i,j} = 0$ for $0\leq i\leq M-1$ and $0\leq j\leq N-1$.
\item For each CN, the new C2V message $E^{(u)}_{i,j}$ sent out at iteration $u$ is computed as a function $\mathcal{F}_2(*)$ of the incoming V2C messages $V^{(u-1)}_{i,j'}$ of the previous iteration, that is,
        \begin{equation}
        \label{equ:mp_c2v_calculation}
            E^{(u)}_{i,j} = \mathcal{F}_2\left(V^{(u-1)}_{i,j'} \right), j'\in{\mathcal{B}(c_i),{j'\neq j}}.
        \end{equation}
\item For each VN, the updated APP $\hat{r}_j$ at iteration $u$ is computed as a function of $r_j$ and all the incoming C2V messages $E^{(u)}_{i,j}$, that is, 
    \begin{equation}
        \label{equ:mp_app_update}
        \hat{r}^{(u)}_j = \mathcal{F}_1\left(r_j, E^{(u)}_{i,j}\right), i\in\mathcal{A}(v_j). 
    \end{equation}
The tentative decision of the bit $\hat{x}_j, 0\leq j\leq N-1,$ is decoded to $0$ if $\hat{r}^{(u)}_j\geq 0$ and is decoded to $1$ otherwise. 
\item The iterative decoding process terminates and outputs the decoded sequence $\hat{\boldsymbol{x}} = \left(\hat{x}_0,\hat{x}_1,\cdots,\hat{x}_{N-1}\right)$ if 
    \begin{equation}
        \label{equ:mp_tentative_decision}
        \boldsymbol{H}\hat{\boldsymbol{x}}^T = \boldsymbol{0} \mod 2.
    \end{equation}
Otherwise, the iterative decoding process
continuous until the maximum iteration number $I_{\max}$ is reached.  
\end{itemize}
In the following, various iterative message-passing decoding algorithms are introduced.

\subsection{LDPC Decoding Algorithms}
\label{sec:ldpc_dec_alg}
\subsubsection{ Sum-Product Algorithm}
Along with the introduction of LDPC codes in Gallager's seminal work in 1960, a near-optimal probabilistic decoding algorithm for LDPC codes was introduced that is now called the \emph{Sum-Product algorithm} \cite{tanner_recursive_1981,kschischang_factor_2001}.
%

In the SPA decoding, the function to compute the C2V messages in (\ref{equ:mp_c2v_calculation}) is explicitly given by
\begin{align}
\label{equ:spa_c2v}
E_{i,j} = \prod_{j'\in\mathcal{B}(c_i),j'\neq j}\alpha_{i,j'} \cdot \Phi\left(\sum_{j'\in\mathcal{B}(c_i),j'\neq j}\Phi\left(\beta_{i,j'}\right)\right),\end{align}
where $\alpha_{i,j'} = \text{sign}(V_{i,j'})$, $\beta_{i,j'} = \vert V_{i,j'}\vert$ and $\Phi(x) = -\log(\tanh(x/2))$.
Next, the function to compute the V2C messages in (\ref{equ:mp_v2c_calculation}) is explicitely given by
\begin{align}
\label{equ:spa_v2c}
V_{i,j} = r_j + \sum_{i'\in\mathcal{A}(v_j),i'\neq i}E_{i',j}.
\end{align}
The updated APP of each VN in (\ref{equ:mp_app_update}) is given by
\begin{align}
\label{equ:spa_app}
\hat{r}_j = r_j + \sum_{i\in\mathcal{A}(v_j)} E_{i,j}.
\end{align}

\subsubsection{Min-Sum Algorithm (MSA)}
While SPA has near-optimal error rate performance, it also has some drawbacks which may limit its use in practical applications.
One of the drawbacks is the use of the computationally expensive $\Phi(*)$ function.
The second drawback is the sensitivity of the BP decoding to the accuracy of the channel parameter estimate used in the initialization step to compute the a priori information $\boldsymbol{r}$.

Both drawbacks can be addressed by using an approximation to compute C2V message \cite{fossorier_reduced_1999,eleftheriou_reduced-complexity_2001}. 
It can be easily seen that the function $\Phi(*)$ in (\ref{equ:spa_c2v}) is a decreasing function satisfying $\Phi\left(\Phi\left(x\right)\right) = x$. 
Therefore, for any set of $b$ real values $(a_0,a_1,a_2\ldots,a_{b-1})$, we have $\Phi\left(\sum_{i=0}^{b-1}\Phi(a_i)\right)\leq \Phi(\Phi(a_{i'})) = a_{i'}$ for $0\leq i'\leq b-1$.
Thus,
\begin{align}
\label{equ:ms_approx}
\Phi\left(\sum_{i=0}^{b-1}\Phi(a_i)\right) \leq \custommin_{0\leq i\leq b-1}(a_i).
\end{align}
Based on (\ref{equ:ms_approx}), the min-sum decoding was proposed to simplify the calculation (\ref{equ:spa_c2v}) in SPA. 
More specifically, the computation of C2V messages $E_{i,j}$ is approximated as 
%
%
%
\begin{align}
 E_{i,j} = \left(\prod_{j'\in\mathcal{B}(c_i),j'\neq j}\text{sign}(V_{i,j'}) \right)\cdot \left(\custommin_{j\in\mathcal{B}(c_i),j'\neq j}\left(\vert V_{i,j'}\vert\right)\right). 
\end{align}
Compared to the $\Phi(*)$ function in the SPA, MSA is commonly adopted in practice as it has low implementation cost because the check node operation is simplified to compare operations.

\subsubsection{Normalized and Offset MSA}
The MSA approximates SPA by assigning the upper bound value in (\ref{equ:ms_approx}) in each decoding iteration. Hence, the computed C2V messages are an overestimation of the true value computed via SPA. 
Due to this overestimation, the MS algorithm suffers from performance degradation compared to SPA. 
To reduce this performance gap, various improved MS-based decoding algorithms have been studied, e.g., \cite{hu_efficient_2001,chen_reduced-complexity_2005,chen_near_2002}. 
The normalized min-sum (NMS) \cite{chen_near_2002,chen_reduced-complexity_2005} and the offset min-sum (OMS) \cite{chen_reduced-complexity_2005} are probably the most popular ones, due to their simplicity.
The NMS and OMS decoding algorithms rely on a scaling factor or an offset factor to compensate for the overestimation of the C2V messages.
Hence, the computation of C2V messages $E_{i,j}$ is modified to 
\begin{align}
\label{equ:NMS}
&E_{i,j}= \\\notag &\left(\prod_{j'\in \mathcal{B}(c_i),j'\neq j}\text{sign}(V_{i,j'})\right)\cdot \Big(\alpha \custommin_{j\in\mathcal{B}(c_i),j'\neq j}\left(\vert V_{i,j'}\vert\right) \Big),
\end{align}
or
\small
\begin{align}
\label{equ:OMS}
&E_{i,j}=\\\notag &\left(\prod_{j'\in \mathcal{B}(c_i),j'\neq j}\text{sign}(V_{i,j'})\right) \cdot \max\left(\custommin_{j\in\mathcal{B}(c_i),j'\neq j}\left(\vert V_{i,j'}\vert\right)-\beta,0\right),
\end{align}
\normalsize
where $\alpha$ is the scale factor adopted in the NMS algorithm, and $\beta$ is the offset factor used in the OMS algorithm.
For a range of SNR, the optimal values of $\alpha$ and $\beta$ can be determined by Monte-Carlo simulation for regular LDPC codes, or through DE analysis for irregular LDPC codes \cite{chen_improved_2005,chen_density_2002}.

Although the performance of MS decoding can be improved by using a normalization factor or an offset factor, it is also important to properly tune these factors to avoid creating artificial error floors, particularly, in the case where only a limited room for optimization is provided such as fixed-point decoders implemented on a small number of bits.
Consequently, the performance of the OMS and NMS may exhibit high error floors \cite{zhao_implementation_2005}.
To overcome these drawbacks, two-dimensional (2-D) NMS and OMS decoding algorithms have been provided \cite{zhang_two-dimensional_2006} \cite{kang_enhanced_2019,kang_enhanced_2020}.
The 2D correction schemes rely on normalization factors (resp. offset factors) used to normalize (resp. offset) both V2C and C2V messages and whose values can be optimized as a function of the variable node or the check node degree. 

\subsubsection{Self-Corrected MS}
Another MS-based decoding algorithm, referred to as self-corrected MS (SCMS) decoding, was proposed in \cite{savin_self-corrected_2008}.
The main idea is to detect unreliable V2C messages during the iterative decoding process, and to erase them by setting their values to zero.
More specifically, a V2C message $V^{(u)}_{i,j}$ in iteration $u$ is assigned $0$ if its sign changed with respect to the V2C message $V^{(u-1)}_{i,j}$ in iteration $u-1$.
Hence, the computation of V2C messages $V_{i,j}$ is modified to 

\small
\begin{align}
\label{equ:scms}
V^{(u)}_{tmp} &=r_j + \sum_{i'\in\mathcal{A}(v_j), i'\neq i}E^{(u-1)}_{i',j}; \\ \notag
V^{(u)}_{i,j} &= \left\{\begin{matrix}
0 & \text{if sign}\left(V^{(u-1)}_{i,j}\right) \neq \text{sign}(V^{(u)}_{tmp}), V^{(u-1)}_{i,j}\neq 0,\\ 
V^{(u)}_{tmp} & \text{else}.
\end{matrix}\right.
\end{align}
\normalsize
The performance of SCMS decoding is very close to BP in the error floor region \cite{savin_self-corrected_2008, andrade_near-lspa_2013}.
Moreover, its built-in feature of erasing unreliable messages can also be advantageously exploited for energy-efficient implementations \cite{amador_energy_2009,amador_hybrid_2010}.

\subsubsection{Approximate-Min}
Approximate-min (A-min*) \cite{jones_approximate-min_2003} decoding algorithm is an approximation of the SPA decoder by applying the Jacobian logarithmic identity to change the way of computing the C2V messages $E_{i,j}$ using a recursive method. 
Let $\exteriorLambdaNoboldfont$ be the recursive function
$$\exteriorLambdaNoboldfont(a,b)=\text{sign}(a)\text{sign}(b)\cdot\mathcal{J}(a,b),$$
where
\begin{align*}&\mathcal{J}(\vert a\vert,\vert b\vert) = \\\notag &\left(\min(\vert a\vert,\vert b\vert)+\ln(1+e^{-|\vert a\vert+\vert b\vert|}) \newline
-\ln(1+e^{-||a|-|b||})\right).\end{align*}
The C2V message of a CN $c_i$ to VN $v_j$ is computed recursively as
\begin{align}\label{equ:a_min}
&E_{i,j} = \\ \notag &\exteriorLambdaNoboldfont\left(V_{i,j'_{\vert\mathcal{B}(c_i)\vert}},\cdots,\exteriorLambdaNoboldfont\left(V_{i,j'_4},\exteriorLambdaNoboldfont\left(V_{i,j'_3},\exteriorLambdaNoboldfont\left(V_{i,j'_1}, V_{i,j'_2}\right)\right)\right)\right),
\end{align}
where $\{j'_1,j'_2,\cdots,j'_{\vert\mathcal{B}(c_i)\vert}\}\subset\mathcal{B}(c_i)$ which excludes $j$.
Let 
\begin{align}
    E_{i,j} = \exteriorLambdaboldfont_{j'\in\mathcal{B}(c_i),j'\neq j} V_{i,j'}
\end{align}
be the recursive operation given in (\ref{equ:a_min}). 
Then the check node computation is performed in the following steps:
\begin{itemize}
\item{} For each CN $c_i$, find the incoming V2C message with the minimum magnitude and label its source VN as $v_{j_{min}}$ and the V2C message as $V_{i,j_{min}}$.
\item{} The C2V message $E_{i,j_{min}}$ send to the VN $v_{j_{min}}$ is the calculation (\ref{equ:a_min}) with $j = j_{min}$.
\item{} For all other variable node $j\in\mathcal{B}(c_i), j\neq j_{min}$, 
\begin{align*}
E_{i,j} = \left(\prod_{j'\in \mathcal{B}(c_i),j'\neq j}\text{sign}(V_{i,j'})\right)\cdot\exteriorLambdaNoboldfont\left(E_{i,j_{min}},V_{i,j_{min}}\right)
\end{align*}
\end{itemize}

Observe that in the A-min* approximation of SPA, only two magnitudes are computed at each check node, requiring only $d_c - 1$ computation of $\exteriorLambdaNoboldfont$ function to compute both, where $d_c$ is the check node degree.
Moreover, compared to MS-based decoding algorithms, the message sent to the least reliable variable node $v_{j_{min}}$, in the A-min* decoder is exactly that of the SPA decoder.
This explains the performance improvement of the A-min* decoder over the MS decoder and its negligible
loss relative to the SPA decoder.


\subsubsection{Adjusted MS}
The Adjusted min-sum (AdjMS) algorithm of LDPC codes is proposed in \cite{richardson_adjusted_2018}, and the C2V approximation function  $\hat{\Lambda}^{*}(a,b)$ is given by
\begin{align}
\hat{\Lambda}^{*}(a,b) \approx \min(a,b) - f\left(\vert a- b\vert + h(M)\right),
\end{align}
where $f(x) = \log(1+\exp^{-x}) $ and $h(x)$ may be defined as $\log(\text{coth}(x))$ or $-\log(1-\exp^{-2x})$.

Unlike the conventional implementation of MS-based algorithms, where the incoming V2C messages with minimum two magnitudes are found to update the outgoing C2V messages, the C2V messages in the AdjMS decoding rely on the maximum and minimum value of the incoming $V_{i,j}$.
Similar to A-min*, AdjMS applies the approximation function $\hat{\Lambda}^{*}$ to recursively calculate two values assigned to $E_{i,j}$. 
Let $j_{\max}$ and $j_{min}$ be the index of the incoming V2C messages with the maximum and the minimum magnitude.
Then the C2V messages computed for check node $i$ are given by
\begin{align}
    E_{tmp} = \exteriorLambdahatboldfont_{j'\in\mathcal{B}(c_i),j'\neq j_{min}, j'\neq j_{\max}} V_{i,j'}
\end{align}
and 
\begin{align}&E_{i,j\neq j_{min}}=\hat{\Lambda}^{*}(V_{i,j_{min}},E_{tmp}), \\\notag &E_{i,j_{min}}=\hat{\Lambda}^{*}(V_{i,j_{\max}},E_{tmp}).\end{align}
In \cite{3gpp_ldpc_2016}, the error rate performance of AdjMS over a range of code lengths and code rates is shown. 
The results show that the AdjMS algorithm with layered scheduling\footnote{Decoder scheduling will be further discussed in the next section.} can achieve SPA decoding performance with flood scheduling using only half of the required iterations.

\subsection{Quasi-Maximum Likelihood (QML) decoding of LDPC Codes}
It is shown in \cite{etzion_which_1999} that the SPA is sub-optimal and has a considerable performance gap to the maximum likelihood (ML) decoding with short blocklength codes, which is due to the existence of small cycles in their associated Tanner graphs.
To further improve code performance, the concept of QML decoding has been further investigated in recent years. 
One of the first QML decodings is the ordered statistic decoding (OSD) \cite{fossorier_soft-decision_1995}, and later the idea was adopted in the decoding of LDPC codes by the same author in \cite{fossorier_iterative_2001}.
Various research has been conducted in this direction to reduce the complexity of QML decoders 
\emph{e.g.,} \cite{varnica_augmented_2007,scholl_saturated_2016,kudekar_efficient_nodate,kang_enhanced_2020,liu_iterative_2018}.
The most common strategy adopted in these
works are to introduce multiple rounds of \emph{decoding tests},
so-called \emph{reprocessing}, after the failure of the conventional
BP decoding. 
More specifically, the decoder is reinitialized with a list of different decoder inputs during the reprocessing, where each input sequence is generated by substituting the channel outputs of the selected unreliable VNs with the maximum or minimum values. 
The conventional BP decoding is conducted with each input sequence, and the decoding output is stored if it generates a valid codeword. 
The ‘best’ codeword is chosen from the list of valid codewords as the decoder output according to a certain decision metric.
This type of decoder is referred to as the \emph{list} decoder.

\subsection{Machine Learning-Based LDPC Decoders}
In a communication system, a large number of signal-processing tasks, such as detection and decoding, can be formulated as optimization problems. 
Conventionally, these optimization problems are typically solved using numerical algorithms that iteratively refine the solution.
However, in practice, the iterative process can only be afforded with only a small number of iterations. 
To accurately find the solution with a small number of iterations, the numerical solver requires parameter tuning.
The most straightforward way to tune the algorithm parameters is by using heuristic approaches based on simulation results.  
However, such conventional approaches are prone to result in suboptimal performance and may cause stability issues if the system conditions change. 
Following the idea of introducing intelligence to the future B5G or 6G wireless networks, various machine
learning-aided approaches have been developed to reinforce the design of next-generation wireless communication systems, such as signal detection and channel encoding and decoding \cite{borgerding_amp-inspired_2017,zappone_wireless_2019,he_model-driven_2020, liu_channel_2021, liu_novel_2021}. 

Among all the deep learning-based data detection or channel decoding algorithms, “deep unfolding” \cite{hershey_deep_2014} is an efficient method to improve an existing algorithm’s performance.
More specifically, deep unfolding takes an iterative algorithm with a fixed number of iterations, unfolds its structure onto several hidden layers of a \emph{neural network} system, and introduces a number of trainable parameters such as multiplicative weights and bias. 
Hence, deep learning has become one of the promising optimization tools for iterative decoding of linear block codes, and much research has been dedicated to this direction in recent years. 
For instance, in \cite{nachmani_learning_2016,nachmani_deep_2018, beery_active_2020, liu_deep_2020, buchberger_learned_2021}, an off-line training model for BP decoding of linear block codes with high-density (also sometimes referred to as high-density parity-check codes) codes were investigated.
Moreover, research on machine learning-aided LDPC decoding has also attracted a lot of attention. 
In \cite{wu_decoding_2018}, unfolding MS algorithm to decode LDPC codes is proposed by introducing additional parameters tuning the scales and offset of the standard NMS and OMS algorithms for 5G LDPC codes, respectively.
The work of \cite{vasic_learning_2018} proposes the idea of unfolding to learn finite-alphabet (FA) decoding of LDPC codes.
Simulation results show that by unfolding and learning FA decoders, gains of up to $0.3$ dB can be achieved for a $(1296,972)$ QC-LDPC code over conventional MS decoder when using $3$ quantization bits.
The authors in \cite{wang_neural-network-optimized_2021} proposed a neural 2D normalized MS decoder, together with various weight-sharing techniques to reduce the number of parameters that must be trained.
Furthermore, the machine learning-aided decoding for protogrpah LDPC codes has been investigated in \cite{dai_learning_2021}, along with a trajectory-based extrinsic information transfer (T-EXIT) chart developed for various decoders.
Various other machine learning-related research on LDPC decoding have also been conducted. 
For instance, in \cite{sandell_machine_2021} a syndrome-based neural network is used to estimate the LLR for flash memory controller with LDPC codes. 
In \cite{geiselhart_learning_2022}, the optimal quantization bits are estimated using machine learning, and the works in \cite{stark_decoding_2020} proposed a machine learning-based quantization decoding for 5G LDPC codes. 
In \cite{buchberger_pruning_2021}, a pruning-based neural BP decoder is proposed, resulting in a different parity-check matrix in each decoding iteration.


\subsection{Decoding of Non-Binary LDPC Codes} \label{sec:NB_LDPC_DEC}
The non-binary (NB) formulation of LDPC codes over $GF(q)$, $q>2$, was proposed by Davey and MacKay in the late 90s \cite{davey_low_1998}. 
Let $GF(q)$ be a Galois field with $q$ elements, where $q$ is a power of a prime.
A $q$-ary LDPC code of length $N$ is given by the null space over $GF(q)$ of the sparse parity-check matrix $\boldsymbol{H}_{NB}$ over $GF(q)$.
The non-zero elements of $\boldsymbol{H}_{NB}$ are represented by symbols
contained in the respective $GF(q)$ or the binary $m$-tuple in the extension field $GF(2^m)$, such that for any valid codeword $\boldsymbol{x}\in GF(q)$
\begin{equation}
\boldsymbol{H}_{NB}\otimes\boldsymbol{x}^T = \boldsymbol{0},
\end{equation}
where $\otimes$ denotes the multiplication over $GF(q)$.

The BP decoding of non-binary LDPC codes, known as the $q$-ary SPA, is introduced in \cite{davey_low_1998}, and later on improved by using fast Fourier transform (FFT) \cite{mackay_evaluation_2001,barnault_fast_2003,declercq_decoding_2007}.
Denoted by $\boldsymbol{z} = \{z_0,z_1,\cdots,z_{N-1}\}$ the hard-decision of the received sequence $\boldsymbol{y}$.
In binary cases, only two prior probabilities of the $j$-th received symbol are needed, $\Pr(z_j=0)$ and $\Pr(z_j=1)$, and the two probabilities can be compactly represented in the form of LLR when passing along edges in a Tanner graph.
In $q$-ary SPA decoding of a non-binary LDPC code, a set of $q$ prior probabilities represent the $j$-th received symbol, that is $P^{\alpha_k}_j = \Pr(z_j=\alpha_k)$ for all elements $\alpha_k, 1\leq k\leq q $, in the  $GF(q)$.
Denoted by $\boldsymbol{\omega^{\alpha}}_{i,j} = (\omega^{\alpha_1}_{i,j}, \omega^{\alpha_2}_{i,j},\ldots,\omega^{\alpha_q}_{i,j})$ and $\boldsymbol{\theta^{\alpha}}_{i,j} = (\theta^{\alpha_1}_{i,j},\theta^{\alpha_2}_{i,j},\ldots,\theta^{\alpha_q}_{i,j})$ the set of probability message sent from VN $v_j$ to CN $c_i$ and CN $c_i$ to VN $v_j$, respectively.
Upon receiving these incoming prior probabilities, the CN computes the outgoing C2V messages as
\begin{equation}
\theta^{\alpha_k}_{i,j} = \sum_{\boldsymbol{z}:z_j=\alpha_k}\Pr\left(s_i = 0\vert \boldsymbol{z}, z_j = \alpha_k\right) \cdot \prod_{j'\in\mathcal{B}(c_i),j'\neq j}\omega^{z_{j'}}_{i,j'},
\end{equation}
where $\boldsymbol{s} = \{s_0,s_1,\cdots,s_{M-1}\}$ is the syndrome vector obtained from \[\boldsymbol{H}_{NB}\otimes \boldsymbol{z}^T = \boldsymbol{s}.\] 
The probability $\Pr\left(s_i = 0\vert \boldsymbol{z}, z_j = \alpha_{k}\right) = 1$ when $z_j=\alpha_k$ and $\boldsymbol{z}$ satisfies the $i$-th checksum, i.e., $s_i = 0$; otherwise the probability equals the zero. 
The computed C2V messages $\boldsymbol{\theta^{\alpha}}_{i,j}$ are then used to update the V2C probability $\boldsymbol{\omega^{\alpha}}_{i,j}$ for the next decoding iteration 
\begin{align}
\omega^{\alpha_k}_{i,j} = f_{i,j}\cdot P^{\alpha_k}_j\prod_{i'\in\mathcal{A}(v_j),i'\neq i} \theta^{\alpha_k}_{i',j},
\end{align}
where $f_{i,j}$ is the normalization term to ensure that $\sum_{k=1}^{q}\omega_{i,j}^{\alpha_k} = 1$.  
The tentative decision of $\boldsymbol{z} =\{\hat{z}_1,\hat{z}_2,\cdots,\hat{z}_N\}$ is estimated based on \begin{align}\hat{z}_j = \arg\max_{\alpha_k} P^{\alpha_k}_j \prod_{i\in\mathcal{A}(v_j)}\theta_{i,j}^{\alpha_{k}}.\end{align}
The decoding process continues if $\boldsymbol{H}_{NB}\otimes\hat{\boldsymbol{z}}^T\neq 0$ or until the maximum decoding iteration is reached.

For each nonzero entry in $\boldsymbol{H}_{NB}$, the number of computations required to compute the probability messages passing between CN $c_i$ to VN $v_j$ in each decoding iteration is on the order of $\mathcal{O}(q^2)$. 
For large $q$, the computational complexity may become prohibitively large.
The FFT-based $q-$ary SPA has complexity reduced to $\mathcal{O}(q\log q)$, or $\mathcal{O}(pq)$ for $q = 2^p$.
Furthermore, the work in \cite{declercq_decoding_2007} proposed the \emph{extended min-sum} (EMS) algorithms to simplify message
update at the check-node output by computing suboptimal reliability measures with low computational complexity rather than the extract reliability in $q$-ary SPA.
To make this happen, the concept of the \emph{configuration set} is introduced such that only a subset $n_m$ of the most significant values of each
vector at a check node is used to compute the output reliability, resulting in a complexity order of $n_m q$.
%
Since then, low-complexity decoding of non-binary LDPC code become one of the toughest problems in the field for many years, and many related works to non-binary LDPC codes have been investigated.
In \cite{savin_min-max_2008}, the ‘Min-max’ algorithm was introduced with the same complexity order as $q$-ary SPA, but using only addition and comparison operations.
The symbol-flipping decoding (SFD) \cite{liu_weighted_2010,garcia-herrero_non-binary_2014} for non-binary LDPC codes was explored intensively by using flipping metrics to determine which unreliable symbols should be flipped during the iterations.
Improved SFD based on prediction (SFDP) has been investigated in \cite{wang_symbol_2017,zhao_momentum-based_2023}.
In \cite{li_efficient_2019}, the early termination architecture of a modified Trellis Min-Max (T-MM) decoding algorithm for non-binary LDPC codes is presented, and the maximum achievable throughput is $4.68$ Gb/s.
The work in \cite {wijekoon_decoding_2021} investigated the decoding of non-binary LDPC codes over subfields by adopting the method of expanding a non-binary Tanner graph over a finite field into a graph over a subfield, resulting in a reduction of decoding complexity.
Moreover, the analysis of the trapping set and absorbing set for non-binary protograph LDPC codes is investigated in \cite{ben_yacoub_trapping_2023}.

\subsection{Performance Comparison}
\subsubsection{Code Construction Comparison}

\begin{figure}[ht!]
\centering
\includegraphics[width=3.4in]{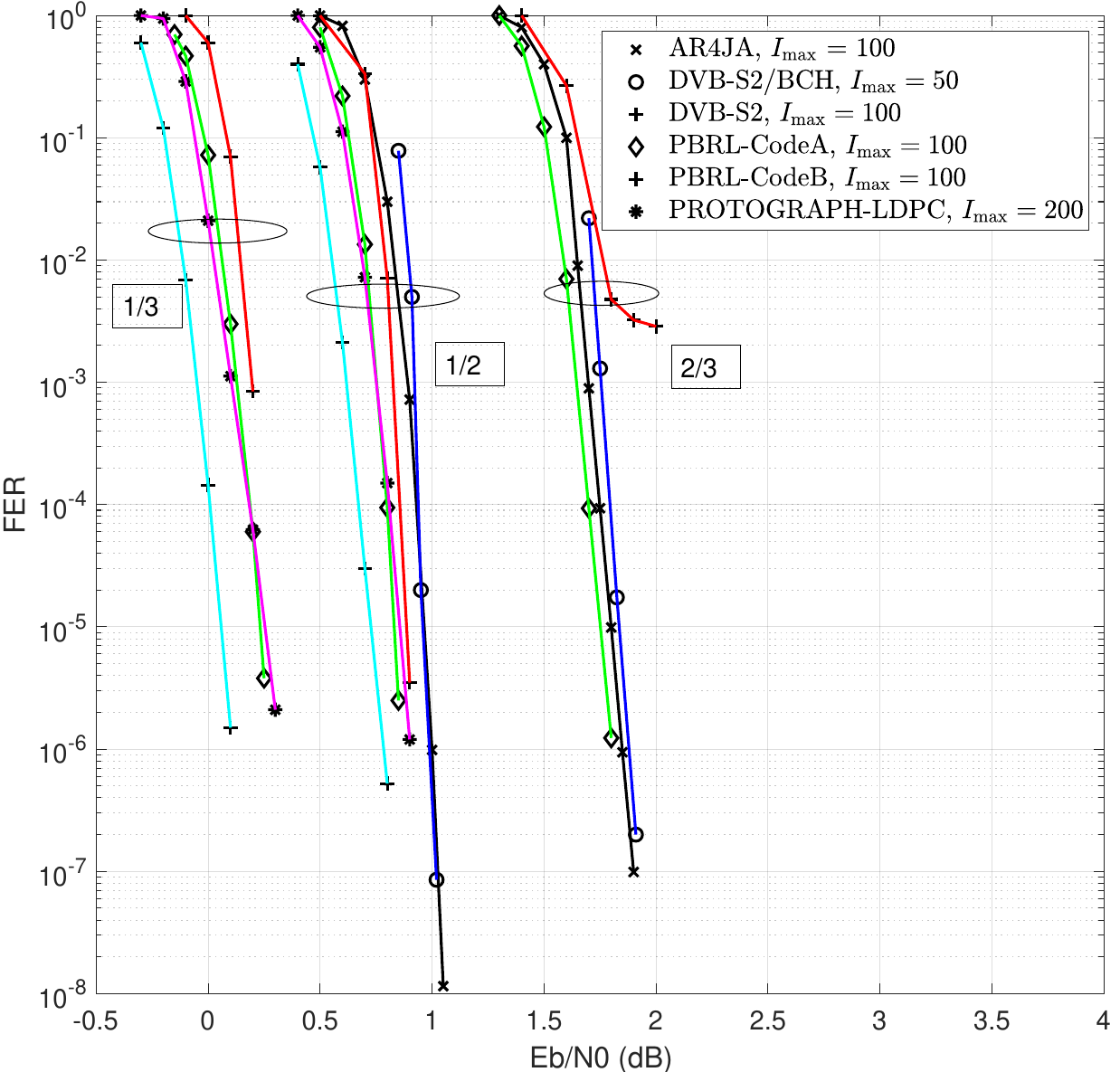}
\caption{FER performance of rate-compatible protograph-based LDPC codes of different structures.}
\label{fig:proto_ldpc_performance}
\end{figure}

Fig. \ref{fig:proto_ldpc_performance} compares the FER performance of LDPC codes of different structures at rates 1/3, 1/2, and 2/3.
More specifically, the PBRL LDPC codes, denoted as `PBRL-CodeA' and 'PBRL-CodeB' \cite{chen_protograph-based_2015}, and the rate-compatible protograph LDPC code, denoted as 'PROTOGRAPH-LDPC' \cite{nguyen_design_2012}, is compared to the AR4JA code from the CCSDS standard \cite{noauthor_tm_2022} and the LDPC code from the DVB-S2 standard (with and without an outer BCH code) \cite{noauthor_etsi_nodate-1}.
The block lengths of the DVB-S2 codes are fixed to 64800 bits, whereas the PBRL and AR4JA codes have a fixed information length of 16368 bits and block lengths of 32736 bits and 24552 bits for rate 1/2 and rate 2/3, respectively.
Moreover, the block lengths of the rate-compatible protograph LDPC code PROTOGRAPH-LDPC are 32736 bits and 49104 bits for rate 1/2 and rate 1/3, respectively.

From the figure, in the waterfall region, both of the PBRL codes outperform both the AR4JA codes and the DVB-S2 codes even though the DVB-S2 codes have longer code lengths and benefit from concatenation with a BCH code.
Further, at a FER around $10^{-6}$, the PBRL code PBRL-CodeB is $0.2$ dB outperforms the PBRL-CodeA and the rate-compatible PROTOGRAPH-LDPC code at rate $1/3$, and about $0.1$ dB better at rate $1/2$. 

\begin{figure}[ht!]
\centering
\includegraphics[width=3.4in]{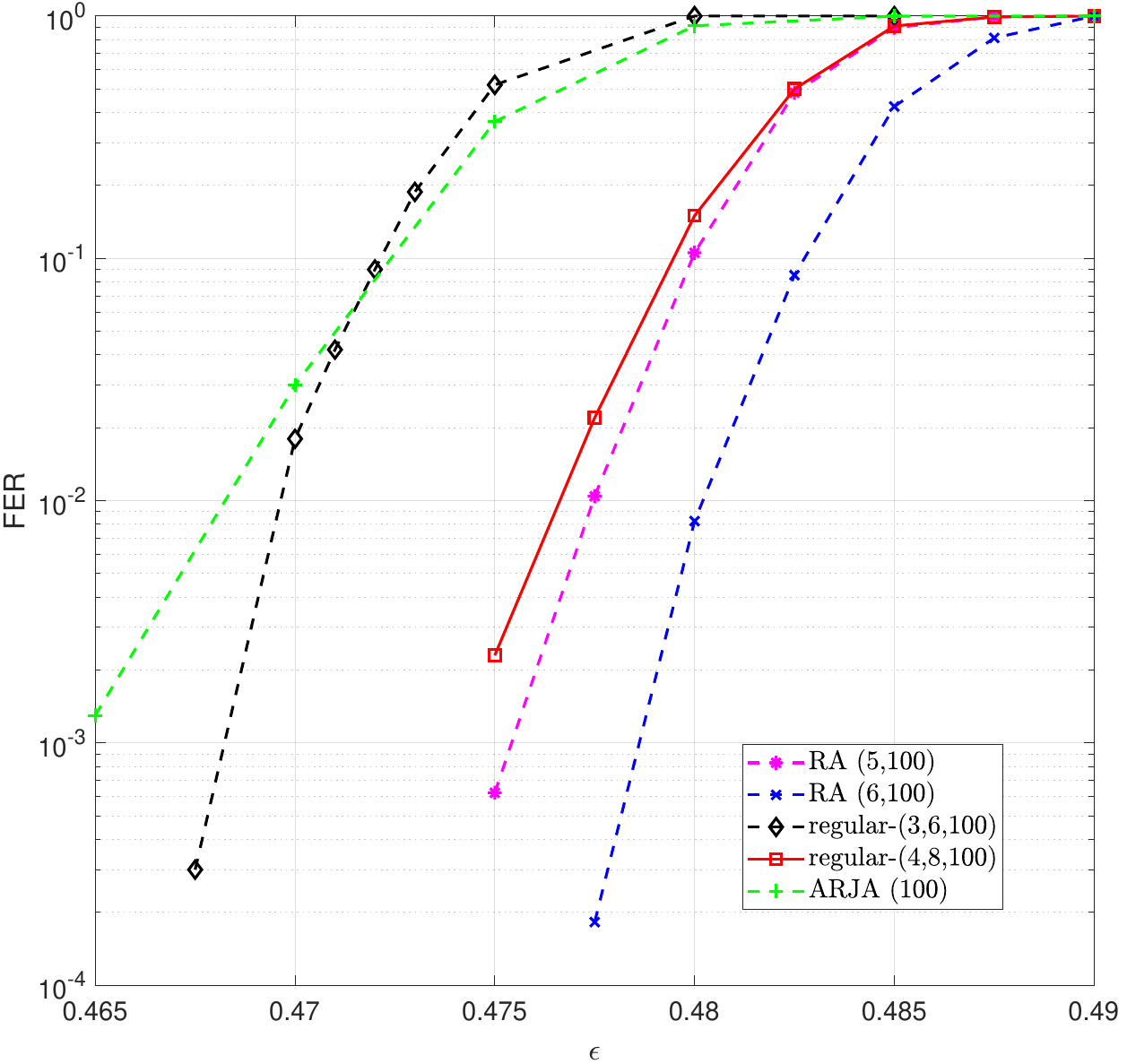}
\caption{FER for different protograph SC-LDPC constructions with $L = 100$ and $N = 2000$ bits. }
\label{fig:SC_ldpc_performance_comparison}
\end{figure}

Fig. \ref{fig:SC_ldpc_performance_comparison} shows the frame error rate performance for different protograph SC-LDPC codes constructed from the following component codes: 1) regular $(3,6)$ and $(4,8)$ LDPC block codes \cite{mitchell_spatially_2015} with coupling memory $m_s=2$ and $m_s=3$, respectively; 2) rate-$1/2$ ARJA code \cite{5513633} with $m_s=1$; 3) regular RA$(q, L)$ codes \cite{johnson_spatially_2013} with coupling memory $m_s=q$ and repetition factor $q=5$ and $q=6$, respectively, over the BEC channel. 
The coupling length of each code is $L = 100$, while the protograph lifting factor $M=2000$.
As shown in the figure, the ARJA($L$) code has a poor finite-length performance, due to the high structure of the ARJA protograph with 5
variable nodes in the uncoupled protograph and the small block length $N$ compared to the rest of the codes. 
It is also interesting to note that both spatially coupled RA codes outperform the regular $(4,8,100)$ SC-LDPC code in finite blocklength performance, although they have worse BP decoding threshold than SC-LDPC codes \cite[Table III]{stinner_waterfall_2016}.

\subsubsection{Decoding Algorithm Comparison}
\begin{figure}[ht!]
\centering
\includegraphics[width=3.4in]{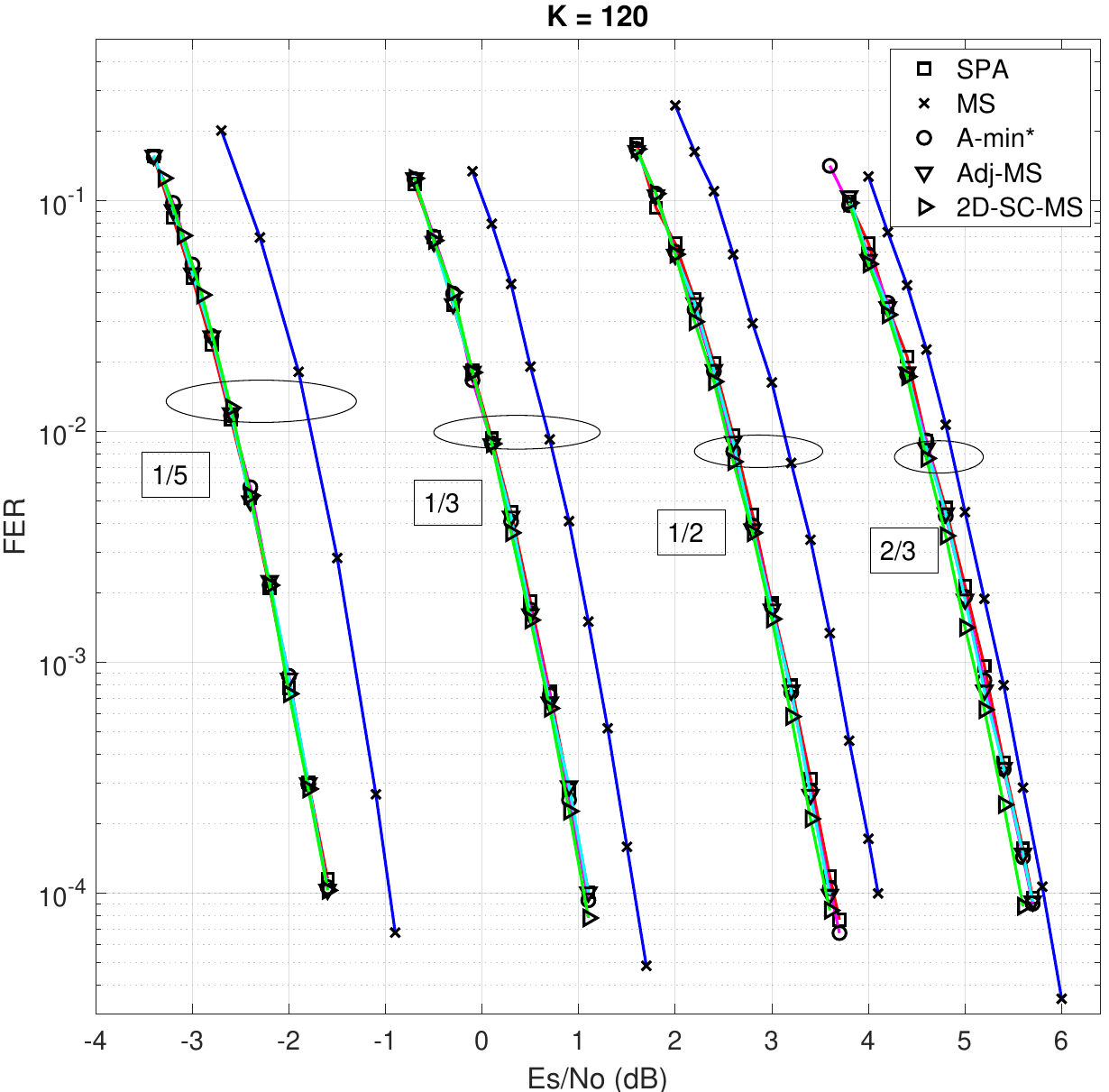}
\caption{FER performance of 5G NR LDPC with $K=120$, $R = 1/5$, $1/3$, $1/2$, $2/3$. }
\label{fig:k120_diff_alg}
\end{figure}

\begin{figure}[ht!]
\centering
\includegraphics[width=3.4in]{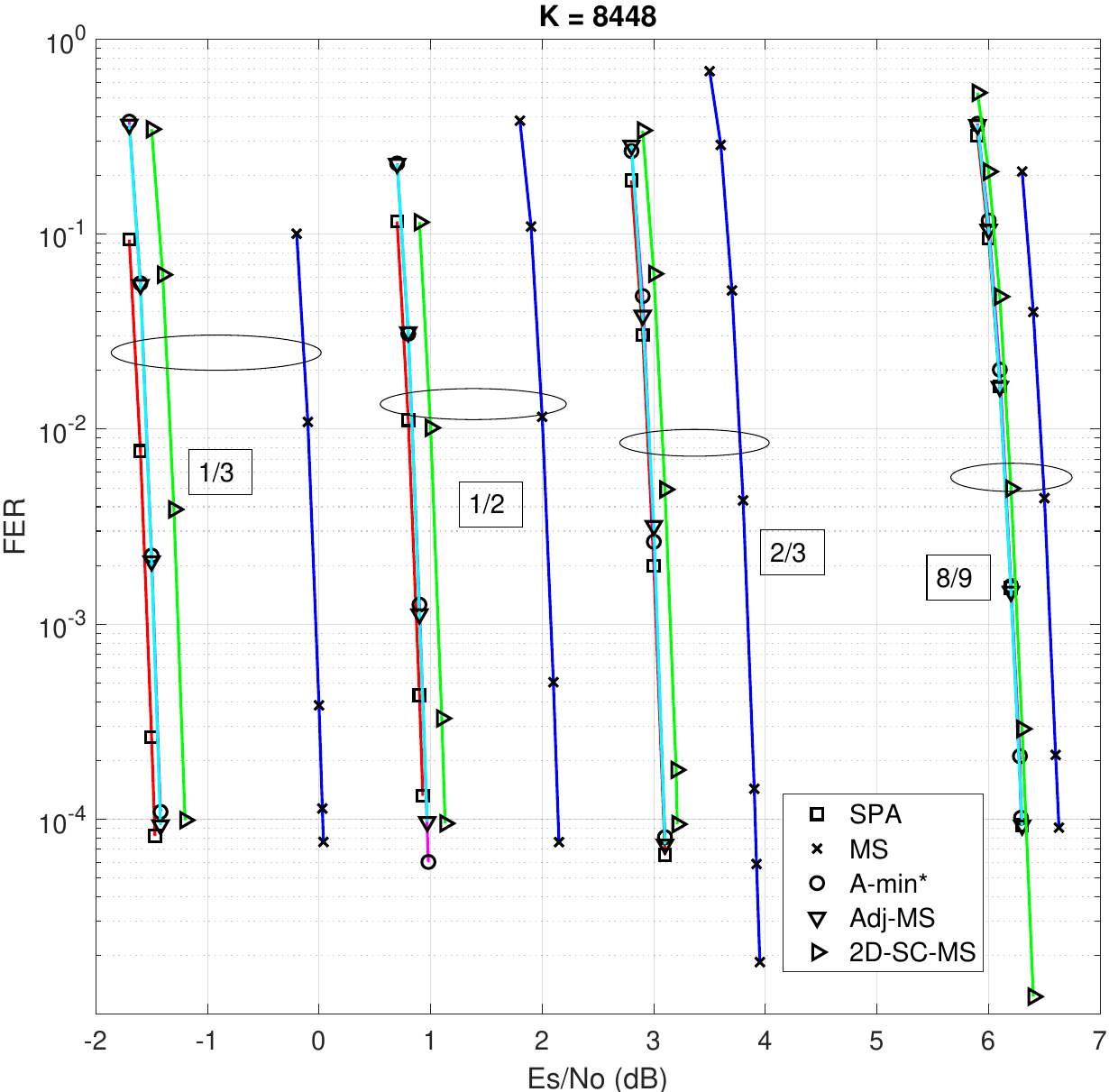}
\caption{FER performance of 5G NR LDPC with $K=8448$, $R = 1/3$, $1/2$, $2/3$, $8/9$.}
\label{fig:k8448_diff_alg}
\end{figure}

Fig. \ref{fig:k120_diff_alg} and Fig. \ref{fig:k8448_diff_alg} show the frame error rate (FER) performance of 5G NR LDPC codes with information length $K=120$ and $K=8448$ simulated over an AWGN channel with $4$-QAM modulation and maximum iteration number $I_{\max} = 50$. 
The codes are decoded using the SPA, the MS, the A-min*, the AdjMS, and the 2D-SCMS \cite{kang_enhanced_2020} algorithms. 
From Fig. \ref{fig:k120_diff_alg}, the performance of A-min*, AdjMS, and 2D-SCMS are similar to SPA when the information block size $K$ is small.
The performance gap of MS decoding from SPA reduces as the code rate increases.
However, when $K$ is very large, the performance of AdjMS and A-min* is similar and has a negligible performance gap from the performance of SPA when the code rate reduces.
On the other hand, the performance of 2D-SCMS decoding shows a slightly larger gap from SPA compared to AdjMS and A-min*, and the performance gap of MS decoding from SPA is much larger compared to AdjMS, A-min*, and 2D-SCMS.

\subsubsection{Quasi ML decoding comparison}
\begin{figure}[ht!]
\centering
\includegraphics[width=3.4in]{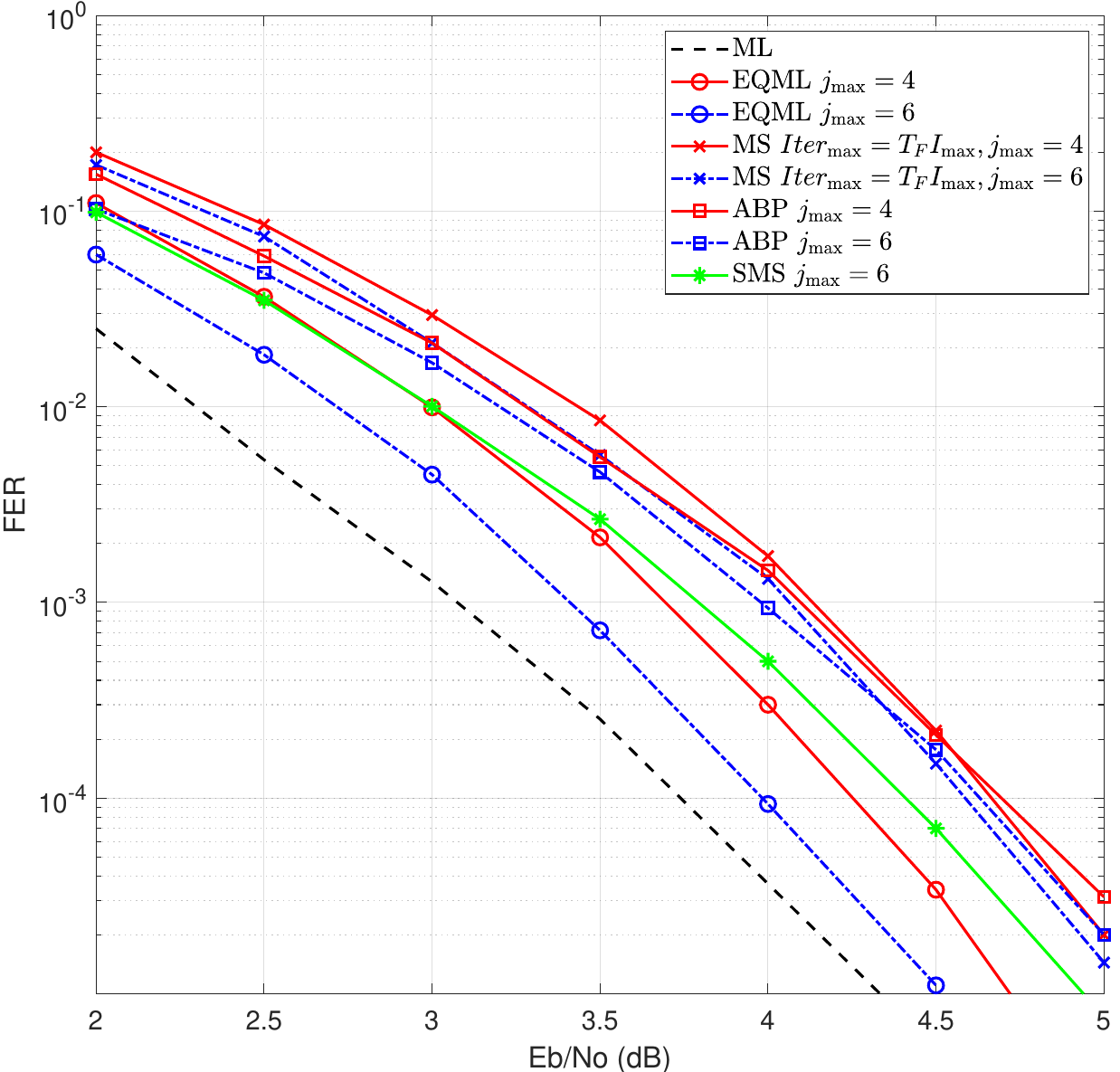}
\caption{FER performance of the regular $(4,7)$ LDPC code.}
\label{fig:qml_1}
\end{figure}

Fig. \ref{fig:qml_1} illustrated the FER performance of the regular $(4,7)$ LDPC code with $K = 48$ and $R=1/2$ \cite{noauthor_ml_nodate} decoded over the AWGN channel with BPSK modulation and maximum decoding iteration of $I_{\max} = 30$.  
The parameter $j_{\max}$ is defined such that $T_{F} = 2^{j_{\max}+1} -1 $ is the number of the decoding tests that a QML decoder runs.
Moreover, the conventional MS decoder is performed with the same number of decoding iterations as the QML decoders run, that is $Iter_{\max} = T_F I_{\max}$.  
From the figure, by performing reprocessing, the enhanced QML (EQML) \cite{kang_enhanced_2019,kang_enhanced_2020} decoder outperforms the MS decoder by about $0.5$ dB and $0.6$ dB for $j_{\max} = 4$ and $j_{\max}=6$, respectively. 
The ML performance of the code \cite{noauthor_ml_nodate}, the FER performance of the augmented
belief propagation (ABP) decoder \cite{varnica_augmented_2007} and the Saturated MS decoder (SMS) \cite{scholl_saturated_2016} are also shown in the figure.
For $j_{\max} = 6$, the EQML decoder outperforms the ABP decoder and SMS decoder by about $0.6$ dB and $0.4$ dB, respectively, and it can approach the performance of the ML decoder within $0.3$ dB at FER=$10^{-4}$.

\begin{figure}[ht!]
\centering
\includegraphics[width=3.4in]{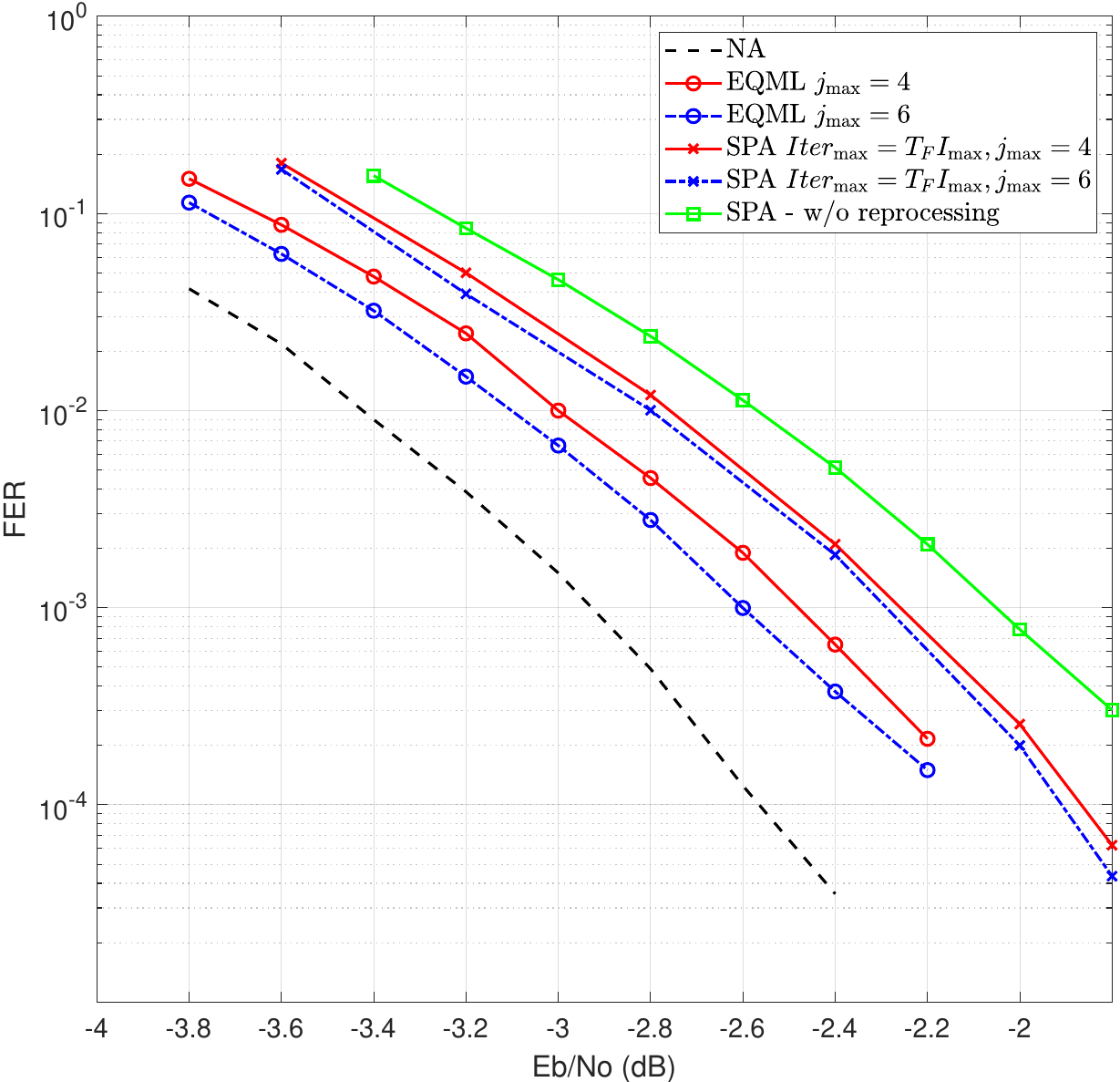}
\caption{FER performance of 5G NR LDPC with $K=120$, $R=1/5$.}
\label{fig:qml_2_K120}
\end{figure}

Fig. \ref{fig:qml_2_K120} shows the FER performance of the 5G LDPC code with $K = 120$ and $R=1/5$ decoded over AWGN channel with $4$-QAM modulation and $I_{max}=50$.
By performing reprocessing, the enhanced QML (EQML) decoder outperforms the SPA decoder with the same number of decoding iterations as the EQML decoder by about $0.2$ dB and $0.3$ dB for $j_{\max} = 4$ and $j_{\max}=6$, respectively.
In addition, the normal approximation (NA) \cite{polyanskiy_channel_2010} bound for blocklength $N=600$ is shown.
The performance gap between the EQML decoder with $j_{\max} = 6$ and the NA is within $0.6$ dB at FER=$10^{-4}$.

\subsection{Implementation of LDPC Decoders}
\subsubsection{Decoder Scheduling}
\begin{figure}[h!]
\centering
\includegraphics[width=1.7in]{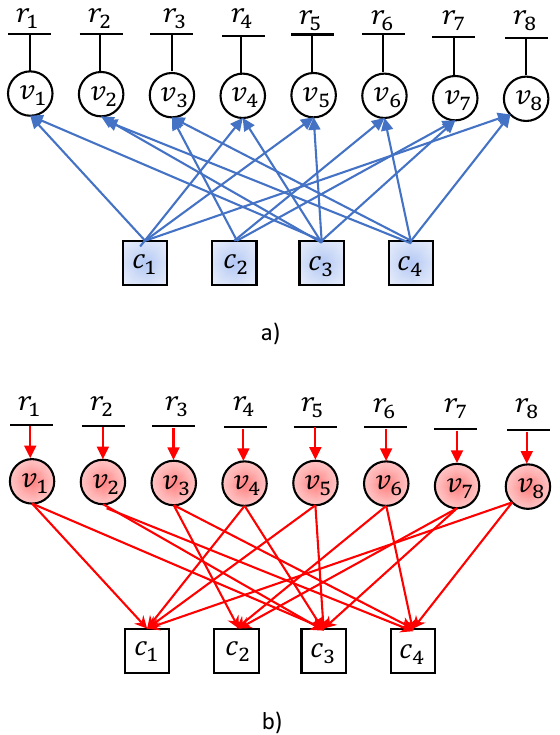}
\caption{An example of the flooding schedule.} 
\label{fig:LDPC_flooded}
\end{figure}

The scheduling of the LDPC decoding process determines the order in which VNs and CNs are processed.
The three most common scheduling methods, namely flooding \cite{kschischang_iterative_1998}, layered belief propagation (LBP) \cite{hocevar_reduced_2004} and informed dynamic scheduling (IDS) \cite{casado_informed_2007}, are described in the following.  
\paragraph{Flooding}
In a flooded LDPC decoder, all check nodes update the C2V messages simultaneously in the first half of the iteration, followed by all variable nodes updating the V2C messages simultaneously in the second half of the iteration. 
During each half of the iteration, only one side of the Tanner graph is activated and performs message calculation. 
An example of the flooded decoder is illustrated in Fig. \ref{fig:LDPC_flooded}. It can be seen in Fig. \ref{fig:LDPC_flooded}-a) that in the first half of the iteration, the CNs $c_1, c_2, c_3, c_4$ update their C2V messages, which are then sent to its neighboring VNs, while all the VNs are not computing. 
Fig. \ref{fig:LDPC_flooded}-b) illustrates the second half of the iteration, where VNs $v_1,v_2,\cdots,v_8$ update their V2C messages and send them to their neighboring CNs. 

\begin{figure}[h!]
\centering
\includegraphics[width=3.5in]{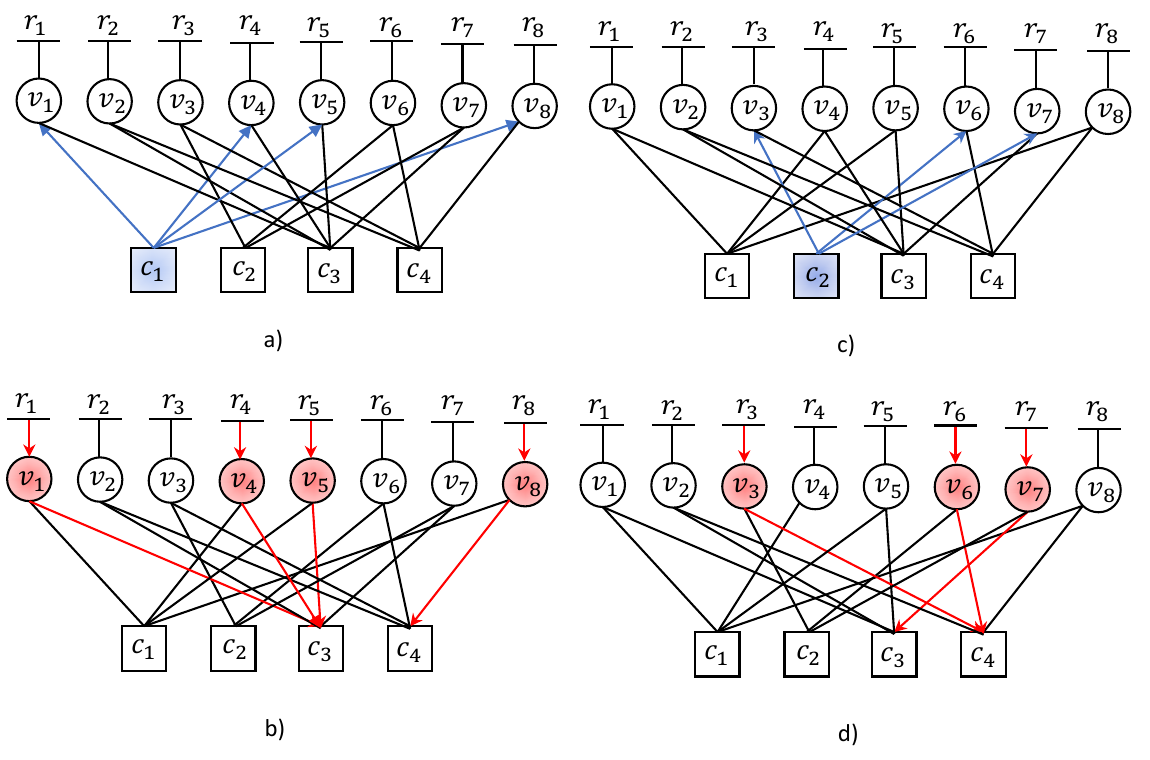}
\caption{An example of the layered schedule.}
\label{fig:LDPC_layered}
\end{figure}

\paragraph{Layered}
Layered BP (LBP) decoder, on the other hand, is operated in a sequential manner within each iteration. 
It sequentially processes each CN in turn.
Once a CN has sent the updated C2V message to its neighboring VNs, these VNs perform V2C message calculations before moving on to the next CN.   
The iteration is complete when all CNs have been processed.
Using Fig. \ref{fig:LDPC_layered} as an example, a layered LDPC decoder may commence each decoding iteration by activating CN $c_1$ first, sending the updated C2V messages to VNs $v_1,v_4,v_5$ and $v_8$ as shown in Fig. \ref{fig:LDPC_layered}-a). 
Each of these VNs then activated, sending updated V2C messages to its neighboring CNs except $c_1$, as illustrated in Fig. \ref{fig:LDPC_layered}-b). 
Fig. \ref{fig:LDPC_layered}-c) and Fig. \ref{fig:LDPC_layered}-d) show the similar process for CN $c_2$. 
The iteration completes until the rest of CNs $c_3$ and $c_4$ complete the process above. 

The advantage of the layered LDPC decoder is that the information obtained during an iteration is available to aid the remainder of the iteration. 
In addition, the layered decoder tends to have a faster convergence speed than the flooded decoder, and hence less number of iterations are needed to converge to the correct codeword.
However, the drawback of the layered decoder is that it does not have the same high level of parallelism as the flooded decoder, possibly resulting in low processing throughput and higher processing latency due to its sequential processing.

\paragraph{Informed Dynamic Scheduling}\color{black}
Informed dynamic scheduling (IDS) inspects the messages that are passed between the nodes, selecting to activate whichever node is expected to offer the greatest improvement in message belief.
During the iterative decoding process, the inspection of messages requires additional calculations to determine which node to activate.
The additional calculation takes place at the CNs, where the difference between the previous C2V message $E_{i,j}^{(u-1)}$ sent over an edge and the C2V $E_{i,j}^{(u)}$ that is obtained using recently-updated information in the current iteration.
This difference is termed the \emph{residual} and represents the improvement in the belief of the new C2V message $E_{i,j}$. 
At the start of the decoding iteration, the residual of each outgoing C2V message of each CN is calculated.
As $E_{i,j}^{(-1)} = 0$, the residual equals the magnitude of the C2V message to be sent over that edge. 
The message with the greatest residual is identified and the receiving VN is then activated, sending updated V2C messages $V_{i,j}$ to each of its neighboring CNs.
These CNs are then activated and calculate the residual for each of their edges. 
The new maximum residual is then obtained among all the residuals in the graph before the process is repeated.

\begin{figure}
    \centering
    \includegraphics[width=1.0\linewidth]{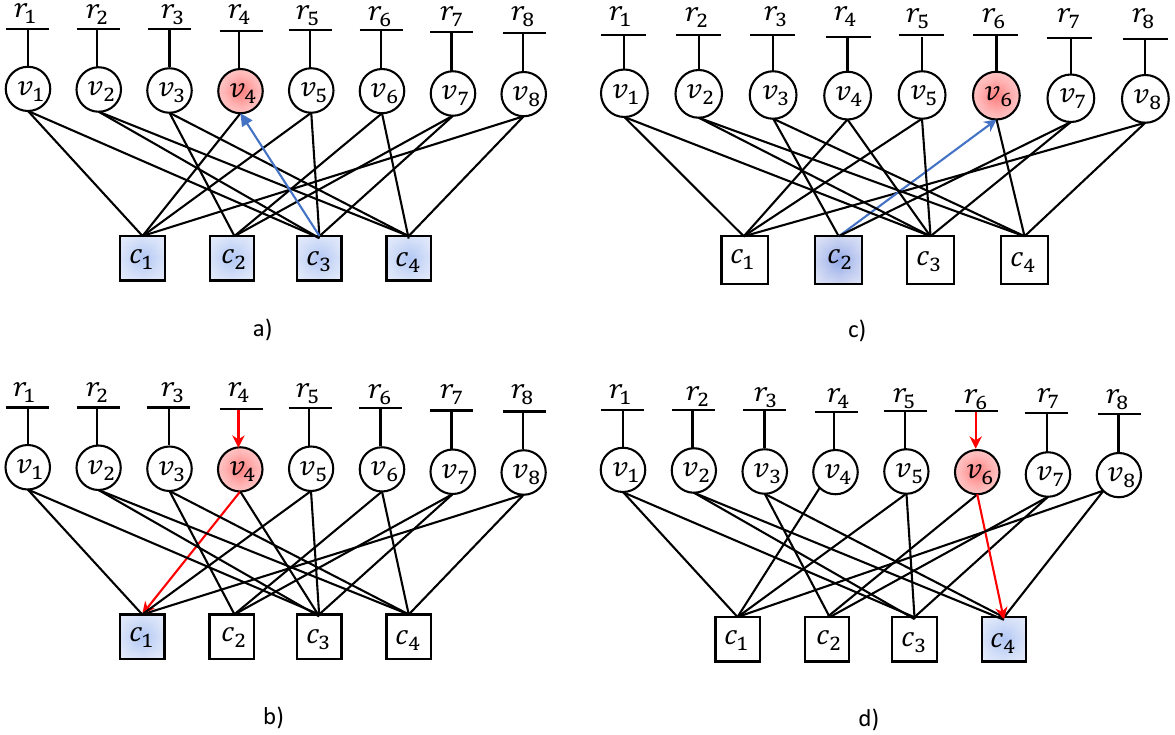}
    \caption{An example of informed dynamic scheduling}
    \label{fig:scheduling_IDS}
\end{figure}

Using the example code in Fig. \ref{fig:scheduling_IDS}, at the start of the decoding iteration, the C2V message $E_{3,4}$ from CN $c_3$ to VN $v_4$ is identified as having the highest magnitude of all the C2V messages in the Tanner graph.
The VN $v_4$, after receiving the message from CN $c_3$, is then activated and calculates the updated V2C $V_{1,4}$, which is then passed to CN $c_1$. 
The CN $c_1$ can then be activated to calculate new residuals for its other three edges.
The new residual is compared with others from the previous step, allowing a new maximum to be identified.
As shown in Fig. \ref{fig:scheduling_IDS}-c), the next highest residual identified is the message from CN $c_2$ to VN $v_6$. 
Thus, VN $v_6$ is activated to calculate the updated V2C message sent to its neighboring CN $c_4$ as illustrated in Fig. \ref{fig:scheduling_IDS}-d).
This implies that in the decoding of IDS, the next highest residual does not necessarily have to originate from the most recently updated CN $c_1$.
Hence, a particular CN can be updated several times before another one is updated once.
\subsubsection{Decoder Architecture}

The implementation of a practical LDPC decoder can be varied.
The well-known architectures are fully parallel, unrolled fully parallel, and partially parallel. 
%
The requirements of a decoder are measured in several aspects, including area efficiency, energy efficiency, error performance, and throughput. 
A simple and effective model to estimate the throughput of a decoder architecture for LDPC block codes is based on the average number of edges the decoder architecture processes in one clock cycle, denoted as $\textbf{Proc(}\mathscr{E}(\boldsymbol{H})\textbf{)}$.  
Let $\mathscr{E}(\boldsymbol{H}) = Z\mathscr{E}(\boldsymbol{H}_b)$ be the total number of edges in a Tanner graph, where $Z$ is the lifting size, and $\boldsymbol{H}_b$ is the protomatrix.
The information throughput of an architecture for one iteration is estimated by
\begin{equation}
\label{equ:throughput_model}
\hat{\mathscr{T}}(\boldsymbol{H}) = \frac{\textbf{Proc(}\mathscr{E}(\boldsymbol{H})\textbf{)}}{\mathscr{E}(\boldsymbol{H})}\cdot n_p\cdot Z\cdot f_{\max} \cdot R,
\end{equation}
where $n_p$ is the column number of a protomatrix, $f_{\max}$ is the maximum operating frequency of the decoder and $R$ is the code rate. 
In practice, decoding of an LDPC code is an iterative process and hence the information throughput is estimated as $\mathscr{T}(\boldsymbol{H}) = \hat{\mathscr{T}}(\boldsymbol{H})/I_{\max}$, where $I_{\max}$ denotes the preset iteration number decoder performs\footnote{If early stop option is adopted in the decoder, then the average iteration number $I_{avg\_iter}\leq I_{\max}$ is often considered.}.
Based on the model in (\ref{equ:throughput_model}), the three well-known architectures can be defined:
\begin{itemize}
\item Fully parallel: \textbf{Proc(}$\mathscr{E}(\boldsymbol{H})\textbf{)} = \mathscr{E}(\boldsymbol{H})$,
\item Unrolled fully parallel: \textbf{Proc(}$\mathscr{E}(\boldsymbol{H})\textbf{)} > \mathscr{E}(\boldsymbol{H})$, and
\item Partially parallel: \textbf{Proc(}$\mathscr{E}(\boldsymbol{H})\textbf{)} < \mathscr{E}(\boldsymbol{H})$.
\end{itemize}

\begin{figure}[h!]
\centering
\includegraphics[width=3.0in]{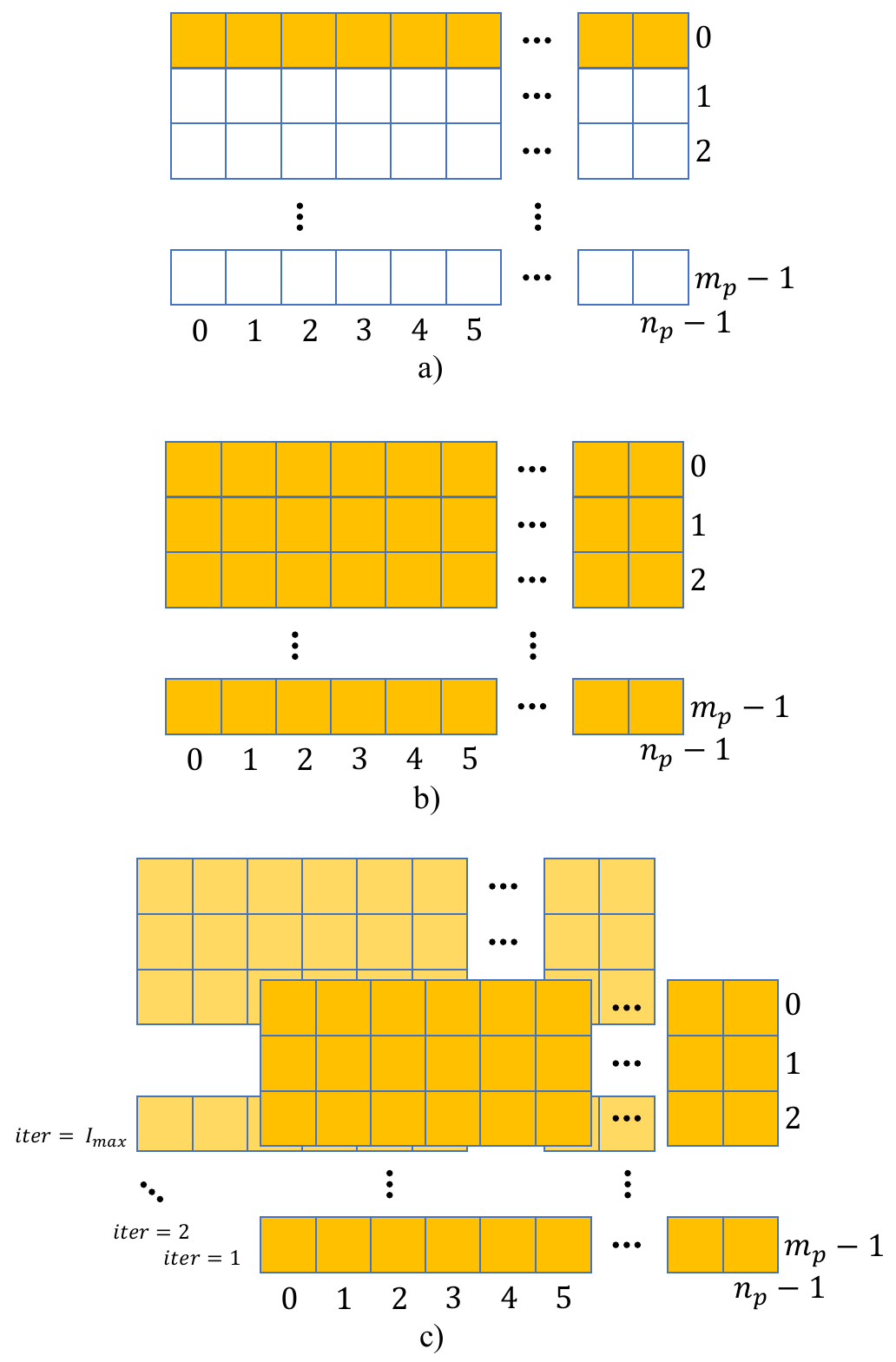}
\caption{Decoder architectures for LDPC codes: a) partially parallel processing; b) fully parallel processing; c) unrolled fully parallel processing.}
\label{fig:architecture_overview}
\end{figure}

An illustration of three types of decoder architecture is given in Fig. \ref{fig:architecture_overview}, where rows that are processed in parallel are marked in yellow.

\paragraph{Fully Parallel} 
A fully parallel LDPC decoder is a realization of a flooded LDPC decoder described in the previous section.
A fully parallel decoder consists of:
\begin{itemize}
\item Node processors (NP): the number of VN processors (VNP) is equal to the number of columns, $N=Zn_p$, in the parity-check matrix $\boldsymbol{H}$ and the number of CN processors (CNP) is equal to the number of rows $M=Zn_p$ in $\boldsymbol{H}$.
\item Routing network (RNW): the routing network is represented by wires connecting the VNPs and CNPs according to $\boldsymbol{H}$.
\end{itemize}
As all VNs or CNs are activated simultaneously to update messages, only a small number of clock cycles is required for an iteration, resulting in high processing throughput, which turns out to be the biggest advantage for a fully parallel decoder.
%
%
However, routing congestion, especially for large block sizes, is a major challenge in implementing the RNW due to the large number of wires needed to describe the connections between VNs and CNs.
For LDPC codes that have thousands of VNs and CNs, the routing network involves tens of thousands of connections between VNPs and CNPs.
Moreover, if $\boldsymbol{H}$ is an irregular structure, the interconnections of the RNW are highly irregular, which will further contribute to the increase in cost, as well as the reduction in the maximum operating frequency due to a high routing delay.
Another drawback of a fully parallel LDPC decoder is its low flexibility.
Although the architecture has no limitations on the structure of $\boldsymbol{H}$, a decoder is often specific to an LDPC with fixed interconnections in the RNW.
A redesign of the entire decoder is needed if the LDPC code is modified. 
Hence, this type of architecture cannot easily accommodate features such as reconfigurable decoders.


%
%
%
%
To reduce the complexity of these fully parallel decoders, the straightforward way is to reduce the wires in the RNW unit. 
The \emph{bit-serial decoder} \cite{darabiha_bit-serial_2006} uses a single wire to send the message from a VN to a CN or vice versa.
Thus, the connection between a VNP and a CNP consists of only two wires, instead of $\mathcal{Q}(V_{i,j})$ wires and $\mathcal{Q}(E_{i,j})$ wires for the V2C and C2V messages, respectively, where $\mathcal{Q}(*)$ represents quantization function.
Such a decoder trades throughput for cost reduction since each VNP or CNP only becomes activated when all  $\mathcal{Q}(E_{i,j})$ or  $\mathcal{Q}(V_{i,j})$ bits of the message are received. 
Hence, the throughput reduction is related to the quantization level.
The reduced quantization of messages leads to a reduced number of wire interconnects between VNPs and CNPs, however at a cost of degradation of the error correction capability.

\begin{figure}[h!]%
\centering
\includegraphics[width=3.45in]{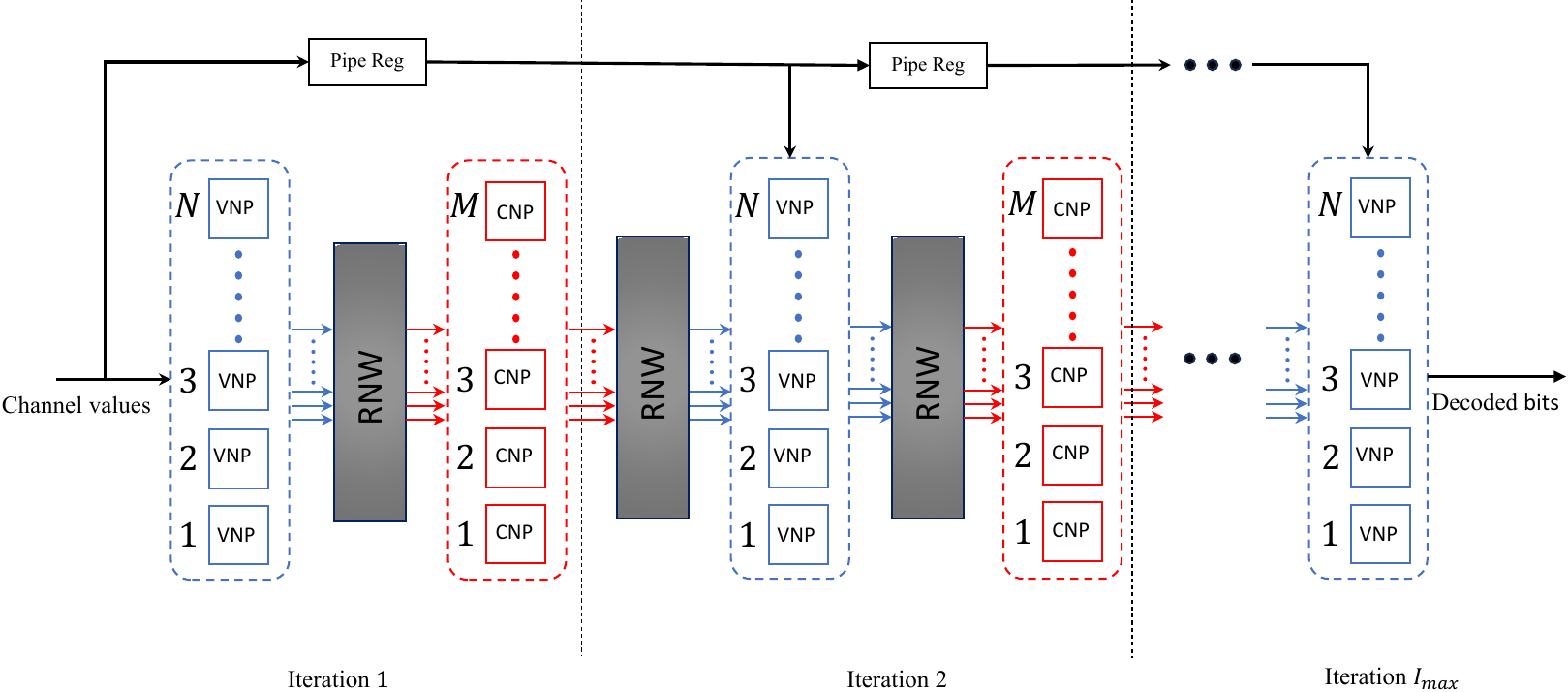}
\caption{Unrolled LDPC decoder.}
\label{fig:ldpc_unrolled}
\end{figure}

\paragraph{Unrolled Fully Parallel}
Unrolled fully parallel architecture is applicable for extremely high throughputs (\emph{e.g.,} hundreds to thousands of Gb/s), which would be considered as the desired architecture for broadband data transmission in 6G.
The basic idea behind this type of architecture is to introduce another level of parallelism by unrolling the decoding iterations.
%
%
%
Consider the Tanner graph in Fig. \ref{fig:LDPC_flooded} as an example.
There are in total $N=8$ VNs and $M=4$ CNs, which can be referred to as $8$ VNPs and $4$ CNPs in a fully parallel decoder.
In an unrolled decoder, the total number of VNPs and CNPs is $8I_{\max}$ and $4I_{\max}$, respectively.
Fig. \ref{fig:ldpc_unrolled} shows an example of the unrolled LDPC decoder.
The latency of the decoder is determined by the number of iterations, and hence, the number of pipeline stages (referring to the `Pipe Reg' in Fig. \ref{fig:ldpc_unrolled}), but the throughput is fixed by the cycle duration. 
While such a fully unrolled decoder architecture requires significant hardware resources, it also has very high throughput since one codeword can be decoded in each clock cycle.
Thus, the hardware efficiency (\emph{i.e.,} throughput per unit area) of the fully unrolled decoder often turns out to be significantly better than the hardware efficiency of the fully parallel (non-unrolled) decoders and partially parallel decoders \cite{schlafer_new_2013}. However, the flexibility is limited similar to the fully parallel architecture.
However, the unrolled architecture implies mainly local wires, which reduces the routing
congestions.

\begin{figure}[h!]%
\centering
\includegraphics[width=3.5in]{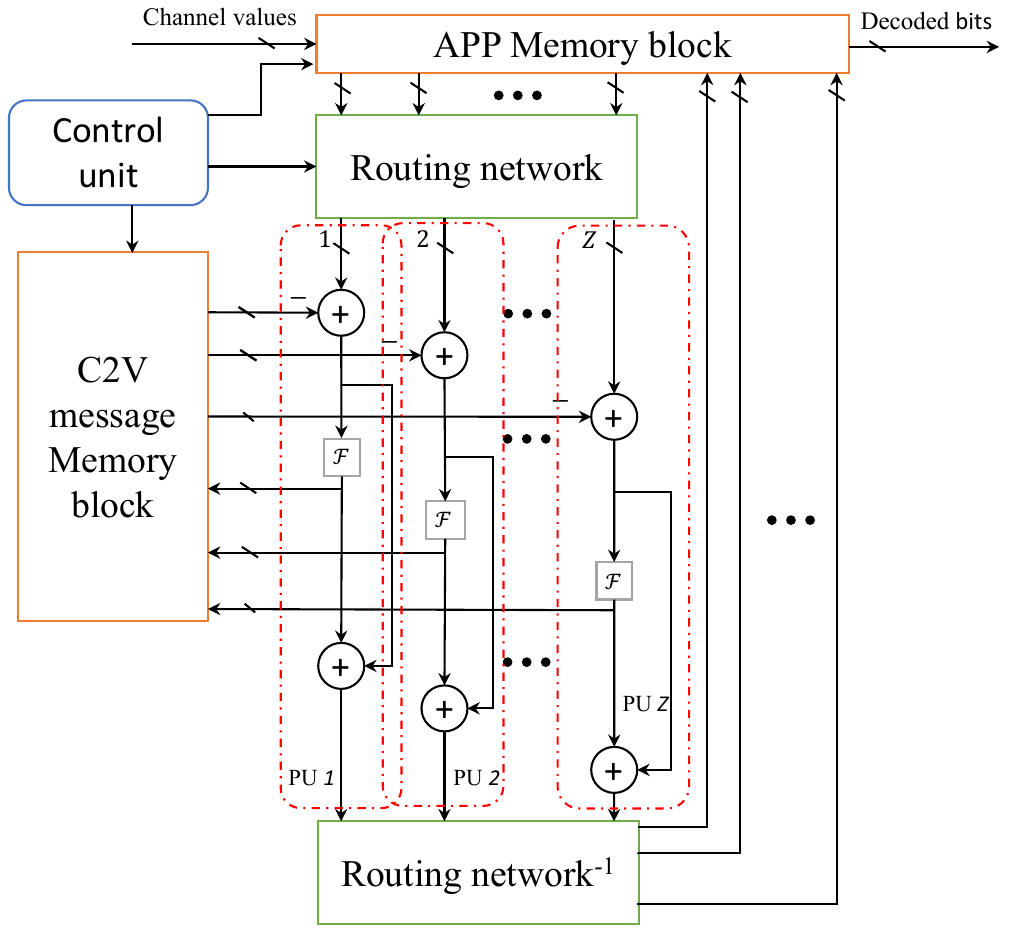}
\caption{Pipelined QC-LDPC decoder.}
\label{fig:ldpc_pipelined}
\end{figure}

\paragraph{Partially Parallel} 
Another type of decoder architecture, namely \emph{partially parallel} \cite{chen_overlapped_2004,wang_memory_2007,chen_fpga-based_2009,chen_memory_2011,nimara_fpga_2016}, has been considered to reduce the complexity and cost of the (unrolled) fully paralleled decoder.
The main feature of such an architecture is to introduce some level of serialization of the CN and VN operations by processing only a subset of edges in parallel.
This can be done in either a row-based or a column-based manner.
The sequential processing of rows or columns allows layered decoding, i.e. taking advantage of intermediate node updates, which accelerates
convergence and thus reduces the number of iterations.
Fig. \ref{fig:ldpc_pipelined} shows an example of a partially parallel architecture for layered decoding of QC-LDPC codes. The key components included are:

{\emph{Processing unit (PU):} }
The PUs associated with each VN consist of the computations of:
\begin{itemize}
\item V2C messages of the current iteration: this is obtained by subtracting the C2V message of the previous iteration from the stored APP value, where $\hat{r}^{-1}_j = {r}_j$, 
\begin{align}
\label{equ:compute_current_V2C}
V^{(u)}_{i,j} = \hat{r}^{(u-1)}_j - E^{(u-1)}_{i,j}. 
\end{align}
\item C2V messages of the current iteration: this is obtained via the function $\mathcal{F}_2(*)$ with certain decoding algorithms described in Section \ref{sec:ldpc_dec_alg},
\begin{align}
E^{(u)}_{i,j} = \mathcal{F}_2\left(V^{(u)}_{i,j'}\right), j'\in\mathcal{B}(c_i),j'\neq j.
\end{align}
\item Updated APP message of the current iteration by adding the C2V and V2C messages of the current iteration,
\begin{align}
\hat{r}^{(u)}_j = E^{(u)}_{i,j} + V^{(u)}_{i,j}.
\end{align}
\end{itemize}
The number of PU is equal to the circulant size $Z$, and hence, for each layer (row) of the base matrix $\boldsymbol{H}_b$, $d_c$ clock cycles are needed to load all $d_c$ circulants of APP values to the PUs for pipeline processing.   
Let $\delta_{pipe}$ be the pipeline depth of PU, that is, the number of pipeline stages for an input APP message to complete all the calculation steps in the PU.  
Then the total number of cycles required to complete a layer of decoding is $d_c + \delta_{pipe}$.  
Let $\mathscr{E}(\boldsymbol{H}_b)$ be the total number of non-zero elements in the base matrix.
The total number of clock cycles to complete a decoding iteration is $\mathscr{E}(\boldsymbol{H}_b)+\delta_{pipe}$. 

{\emph{Block RAM: }} Three blocks of memory are needed in this kind of architecture.
The initial channel LLR values need to be stored in a block RAM, and this memory block will be constantly updated by new APP messages computed during decoding iterations. 
This is illustrated as the `APP Memory block' in the Fig. \ref{fig:ldpc_pipelined}.
%
The size of this BRAM is $n_p\times Z\mathcal{Q}(r)$, where $n_p$ and $Z\mathcal{Q}(r)$ are the depth and width of the memory block, respectively, $\mathcal{Q}(r)$ denotes the number of quantization bits to represent the initial channel LLR $r$. 
The second block RAM unit of size $\mathscr{E}(\boldsymbol{H}_b)\times Z\mathcal{Q}(E_{i,j})$ is needed to store the computed C2V messages as this message needs to be used in the computation of V2C message in the next iteration according to (\ref{equ:compute_current_V2C}).
An additional code description memory (usually a ROM) is needed to store the information of the parity-check matrix $\boldsymbol{H}$. 
This information includes the column indices of each row, the shift positions for each circulant block, as well as other useful information determined during the design stage.
%

{\emph{Routing network: }} The routing network is needed for this type of architecture because each circulant block is loaded in the decoder sequentially, and the shifting value corresponding to each circulant block varies.
Furthermore, the updated APP message in one decoding iteration needs to be routed in the reverse direction before being stored back in the APP memory for use in the next iteration.  
In this case, the routing network must be reconfigurable to handle all possible shifting values smaller than the lifting size  $Z$, while also being capable of both forward and reverse routing networks.
Examples of well-known routing networks are Benes network \cite{benes_optimal_1964}, Oh-Parhi network (OPN)\cite{oh_low-complexity_2010} and QC-LDPC shift network (QSN) \cite{chen_qsnsimple_2010}. 

\subsubsection{Comparison Between Decoders}

The channel codec implementation is critical in terms of power consumption and silicon area, particularly at data rates approaching $1$ Tb/s, which is one of the important KPIs foreseen in 6G.
The recent results of the implementation of the LDPC decoder are briefly reviewed in Table \ref{tab:decoder_implementation}.
Note that for a comparison the results collected in the table have a similar code rate.
Results of other code rates can be found in the referenced works. 
In addition, more results in LDPC decoder implementations can be found in \cite{shao_survey_2019}.

To achieve ultra-high throughput of $1$ TB/s, there are many possible optimization directions: increasing the decoding parallelism, reducing the decoding latency, increasing the clock speed, and increasing the code length. 
From the table, it can be seen that iteration unrolled fully parallel architecture is promising, however at the expense of $I_{max}$ copies of the circuit.
Alternatively, row-layered scheduling with multi-core instantiated architecture could also approach a throughput of $1$ TB/s.
For instance, the $1$ Tb/s throughput achieved in \cite{li_high-speed_2021} instantiated $12$ single core and is configured for maximum $4$ iterations.
Moreover, extended pipelined design, by inserting additional flip-flops (or register units) in between each operation during the decoding, significantly boosts the operating frequency, so that the overall throughput is improved. 
For instance, the implementations in \cite{lopacinski_ultra_2022} and \cite{lopacinski_hardware_2022} add a pipeline with seven stages for each iteration to increase the frequency up to $1.88$ GHz for the $(648,540)$ LDPC code in IEEE 802.11n, and up to $1.511$GHz for the $(1944,1620)$ LDPC code in IEEE 802.11n.
The resulting peak throughput is $1.218$Gb/s and $2.937$Gb/s, respectively.
The throughput of the $(1944,1620)$ code is $3$ times the throughput of the $(648,540)$ code due to the fact that the information length is $3$ times different.
Hence, increasing the information block size would also increase the throughput for the same decoding architecture and configurations.

\begin{table*}
\begin{centering}
\caption{Comparison of ASIC implementation of LDPC decoders}
\label{tab:decoder_implementation}
\begin{tabular}{ |c|p{1.8cm}|p{1.8cm}|p{1.5cm}|p{1.5cm}|p{1.5cm}|p{1.5cm}|p{1.5cm}|} 
 \hline
 \textbf{Ref.} & \cite{schlafer_new_2013} & \cite{cushon_low-power_2016} & \cite{ghanaatian_588-gbs_2018} & \cite{li_high-speed_2021} & \cite{lopacinski_ultra_2022} & \cite{lopacinski_hardware_2022} & \cite{lee_multi-mode_2022}\\\hline
 \textbf{Technology}& $65$nm & $28$nm & $28$nm & $16$nm & $28$nm & $28$nm & $65$nm\\\hline
 \textbf{Algorithm} & Min-sum & Adaptive degeneration  &  Finite alphabet & Min-sum & Min-sum & Min-sum & Offset min-sum\\\hline
 \textbf{Architecture}  & Unrolled fully parallel & Fully parallel & Unrolled fully parallel & Layered & Unrolled fully parallel & Unrolled fully parallel & Layered\\\hline
 \textbf{LDPC code} $(N,K)$ & $(672, 546)$& $(60000,53570)$ & $(2048,1723)$ & $(1027,856)$ & IEEE 802,11n $(648,540)$ & IEEE 802.11n $(1944,1620)$ & 3GPP 5G NR $(10368, 8448)$ \\\hline
 \textbf{Code rate} $R$ & $0.8125$ & $0.88$ & $0.84$ & $0.833$ & $0.833$ & $0.833$ & $0.88$ \\\hline
 \textbf{Iterations} &$9$ & $49$ & $5$ & $4$ & $5$ & $5$ & $3$\\\hline
 \textbf{SNR} $@$BER=$10^{-7}$ (dB) & $-$ & $4.55$ & $4.95$ & $5.3$ & $5$ & $5.05$ & $4.25^{***}$\\\hline
 \textbf{Throughput}$^{*}$ (Gb/s)& $131(161)$ & $400 (455)$ & $494(588)$ & $833(1000)$ & $1015(1218)$ & $2446(2937)$ & $19.2(21.8)$ \\\hline
 \textbf{Clock speed} $f_{max}$&$257$MHz &$373$MHz & $862$MHz & $1$GHz & $1.88$GHz & $1.511$GHz & $500$MHz\\\hline
 \textbf{Latency} &$105$ns & $134$ns & $69.6$ns & $38$ns & $19.68$ns & $27.8$ns & $-$\\\hline
 \textbf{Core Area} mm$^{2}$ & $12.09$ & $7.46$ & $16.2$ & $2.24$ & $5.49$ & $16.46$ & $5.74$\\\hline
 \textbf{Area efficiency}$^{**}$ (Gb/s/mm$^2$)& $13$Gb/s/mm$^2$ & $61$Gb/s/mm$^2$ &  $36$Gb/s/mm$^2$ & $446$Gb/s/mm$^2$ & $222$Gb/s/mm$^2$ & $178$Gb/s/mm$^2$ & $3.8$Gb/s/mm$^2$\\\hline
 \textbf{Core power} (mW)& $-$ & $624$ & $13350$ & $3.19$ & $-$ & $-$ & $413$\\\hline  
 \textbf{Energy efficiency} (pJ/bit) & $3.61$ & $1.56$ & $22.7$ & $3.82$@$6$dB & $12.74$@$4$dB & $10.31$@$4$dB & $-$ \\
 \hline
\end{tabular}
 \end{centering}
\vspace{0.1cm}

$*$: Here, the information throughput is shown. The corresponding coded throughput is given inside the bracket.\\
$**$: Area efficiency is calculated based on the coded throughput.\\
$***$: Estimated value at BER = $10^{-6}$.
\end{table*}

\subsection{Future Directions}
\subsubsection{New Code Structures}
For future-generation wireless communication systems, such as B5G and 6G, the required transmission data rate will be extremely high due to demanding, data-hungry applications and technologies such as streaming multimedia, augmented reality (AR), virtual reality (VR), the metaverse and more.
For LDPC codes, it may be necessary to introduce a lifting size larger than $384$, which is currently the largest lifting size used in the 5G NR, while the protograph remains unchanged because the 5G NR LDPC code's performance has already been pushed to the limit using $16$nm and $7$nm technologies \cite{noauthor_enabling_nodate}.
As shown in (\ref{equ:throughput_model}) the throughput of an LDPC decoder is directly proportional to the size of the information block; increasing the lifting size will enhance the decoder's throughput. 
Hence, new base matrices with optimized shift values need to be designed based on the new or existing 5G NR PBRL LDPC code structure.  
Furthermore, as the performance of 5G NR LDPC codes has been pushed to the limit, trade-offs need to be made between parallelization, pipelining, iterations, and unrolling, while linking them with the decoder architecture.

\subsubsection{Spectrum Efficient LDPC Coding Schemes}
Recent standards exploit the collaborative use of re-transmission protocol and channel coding schemes for lowering end-to-end delay essential for high-speed applications. 
However, this requires the need for additional bits in sending the acknowledgment of the received ensemble data. 
Further, each transportation data block consists of several code blocks. 
If any of the code blocks get corrupted, the retransmissions can improve the spectrum efficiency. 
As 5G trends focus on increasing the data rate a hundred times, there are more code blocks in one transport block.
%
Several techniques have been implemented to overcome this challenge that re-transmit selective code blocks of corrupted data but at the cost of increased overhead due to the additional cyclic redundancy check (CRC) bits needed for each code block. 
The work \cite{yang_chained_2018} proposed the PIC method, which involves the use of some interlinked CBs of encoded LDPC codewords which improves the problem of spectrum efficiency and reduces delay too. 
Also, Non-Binary-LDPC Codes (NBLC) \cite{peng_wlc45-2_2006,chen_high-throughput_2012,feng_nonbinary_2018} have been proposed to escalate spectrum efficiency. 
Furthermore, due to the superior error rate performance and simple code construction approach, SC-LDPC codes with windowed decoder would be an option for a spectrum-efficient channel coding scheme.
It is known that the maximum size of a transportation block in the 5G NR standard is over $1.2$ million bits, and is expected to be larger in future communication proposals.
The rate-loss of SC-LDPC codes due to the termination length $L$ and coupling memory $m_s$ can be minimized with parameters of SC-LDPC codes, such as $L$ and $m_s$, properly chosen. 
However, one of the key challenges for SC-LDPC codes to be considered in practice is the efficient encoding methods, which is one of the research directions that has very limited reports.

\subsubsection{Unified Reconfigurable Decoders}
As one of the key technologies for 6G, the unified design of channel coding schemes at the circuit level of different types of codes is important \cite{wang_road_2023}, such as Turbo/LDPC decoders \cite{noauthor_vlsi_nodate} and LDPC/Polar decoders \cite{cao_reconfigurable_2021}.  
The unified decoder architecture would significantly benefit the decoder implementation in chipset design.
In \cite{wang_unified_2018}, a deep-learning-based unified polar-LDPC decoder for 5G communication systems is proposed.
Moreover, the authors in \cite{cao_reconfigurable_2021} proposed a joint LDPC/polar decoding algorithm based on the BP decoding algorithm. 
Based on the proposed decoding algorithm, a reconfigurable decoding architecture is proposed for standard-compatible decoding of LDPC codes and polar codes.
Furthermore, it is known that the biggest challenge in the design of a unified BP decoder for LDPC and Polar codes is the significant performance disparity between Polar codes decoded using belief propagation (BP) and those decoded under the successive cancellation list (SCL) and CRC-aided SCL decoders. 
The performance of Polar codes under BP decoding is notably worse which is one of the potential barriers that need to be overcome in the design of a unified Polar-LDPC decoder.

\section{Polar Codes}\label{sec:polar}
Since their introduction in 2008, polar codes have garnered significant attention. They were the first family of binary linear codes with explicit construction that provably achieve symmetric capacity
of arbitrary binary-input discrete memoryless channels (B-DMCs) \cite{arikan2009channel}. Their construction is explicit as implied by the transformations that lead to channel polarization (CP).  Due to their structural similarity with binary Reed-Muller codes, many concepts and decoding schemes can be borrowed from RM codes and adopted for use with polar codes. 

In this section, while briefly reviewing the polar coding, we discuss promising advances that could potentially reduce the polar code decoding latency or improve their reliability. Toward this goal, we first describe the basis of polar coding, which is channel polarization, the origin of the idea, the properties of polar codes, and their relation with Reed-Muller codes. Then, we review the code construction methods and commonly used code concatenation schemes, such as CRC-polar coding. This is followed by the known decoding algorithms used for polar codes and their variants along with their hardware implementations. This section is completed by discussing the puncturing and shortening techniques used for rate-compatible polar codes for practical applications and polar-coded modulation schemes for high-order modulation. We end this section with a comparison of the performance and complexity of various polar coding and decoding schemes. 
\subsection{Channel Polarization Effect}\label{ssec:polarization}
\begin{figure}[h]
    \centering
    \includegraphics[width=0.7\columnwidth]{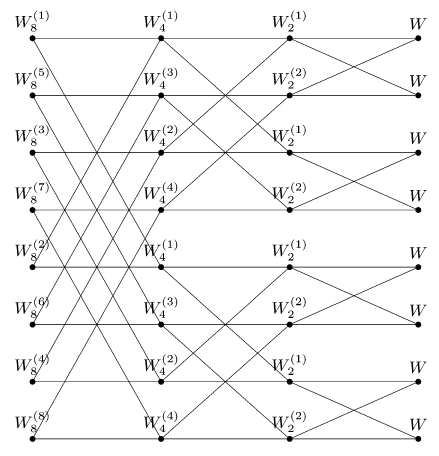}
    \caption{The channel transformation for $N=8$ \cite{arikan2009channel}.}
    \label{fig:ch_transform_8}
\end{figure}

The polarization effect is realized by the transformation of $N=2^n,n\geq1$ identical and independent copies of a physical/raw B-DMC $W: \X \rightarrow \Y$ into a set of $N$ correlated virtual channels or a vector channel with $N$ sub-channels as
\begin{equation}
    W_{N}^{(i)}: \mathcal{X} \rightarrow \mathcal{Y}^N \times \mathcal{X}^{i-1},\;\; i\in [1, N],
\end{equation}
which are either better or worse than the original channel $W$. As $N$ grows large, the channels perceived by individual bits start to polarize, that is, they approach the status of either a perfect channel or an unreliable channel. Note that the polarization is not restricted to a particular transformation, but is considered a general phenomenon. 
A single-step (local) channel transform is performed for two raw channels based on the linear map $\boldsymbol{G}_2 \overset{\Delta}{=} \begin{bmatrix}
1 & 0 \\
1 & 1
\end{bmatrix}$, 
which is equivalently represented by two channel transformations of $W \boxast W$ and $W \circledast W$ resulting in $W \mapsto\left(W^{-}, W^{+}\right)$. Hence, we denote the basic channel transformation by
\begin{equation}\label{eq:basic_ch_trans}
    W^0=W^-=W \boxast W,\quad W^1=W^+=W \circledast W. 
\end{equation}
The output alphabet of $W \boxast W$ is $\Y^2$, the output alphabet of $W \circledast W$ is $\Y^2 \times \X$, and their transition probabilities are given by
\begin{equation}\label{eq:w0}
W^{0}\left(y_1, y_2 \mid u_1\right)\!\triangleq\!
\frac{1}{2} \sum_{u_2 \in \mathcal{X}} W\left(y_1 \mid u_1 \oplus u_2\right) W\left(y_2 \mid u_2\right)
\end{equation}
and
\begin{equation}\label{eq:w1}
W^{1}\left(y_1, y_2, u_1 \mid u_2\right) \triangleq\!
 \frac{1}{2}W\left(y_1 \mid u_1 \oplus u_2\right) W\left(y_2 \mid u_2\right).
\end{equation}
The channel transformation can then be recursively expanded to any power of two $N=2^n$ channels for an integer $n\geq 1$. 
The corresponding transformation matrix can be obtained by the Kronecker power of $\boldsymbol{G}_2$ as $\boldsymbol{G}_N=\boldsymbol{G}_2^{\otimes n}$. Starting from $N=2^n$ raw channels $W = W_1^{(1)}$, then in each channel transformation stage $j\in[1,n]$, every two sub-channels $W_{2^{j-1}}^{(i)}$ are transformed into two child sub-channels as
\begin{equation}\label{eq:ch_trans}
\begin{aligned}
& W_{2^{j}}^{(2i-1)}=W_{2^{j-1}}^{(i)} \boxast W_{2^{j-1}}^{(i)}, \\
& W_{2^{j}}^{(2i)}=W_{2^{j-1}}^{(i)} \circledast W_{2^{j-1}}^{(i)}. \\
&
\end{aligned}
\end{equation}
Fig. \ref{fig:ch_transform_8} demonstrates the transformation of eight raw channels $W$ to sub-channels $\{W^{(i)}_8 : i\in[1,8]\}$ from right to left.  
Let us denote the binary expansion of integer $i-1,\text{ for }i \in[1, N]$ by $\left(b_{n-1}, \ldots, b_1, b_0\right)$, for $b_k \in\{0,1\}$ with the most significant bit on the left. Then, the sub-channels can be defined as
\begin{equation}
W_N^{(i)}=\left(\left(W^{b_0}\right)^{b_1} \ldots\right)^{b_{n-1}}, i\in[1,N]
\end{equation}
where $(\cdot)^{b_k}$ is obtained from equations \eqref{eq:w0} and \eqref{eq:w1}.

To proceed further, let us define the notions of symmetric capacity and the Bhattacharyya parameter as rate and reliability parameters. For a B-DMC $W:\{0,1\} \rightarrow \Y$, the channel transition probabilities are denoted by $W(y \mid x)$, where $y \in \mathcal{Y}, x \in\mathcal{X}=\{0,1\}$. The channel $W$ is said to be symmetric if for every $y \in \Y$ and a permutation $\pi$ where $\pi=\pi^{-1}$, we have $W(y \mid 1)=W(\pi(y)\mid 0)$. Then, the symmetric capacity and the Bhattacharyya parameter of $W$ are defined as
$$
I(W) \triangleq \sum_{y \in \mathcal{Y}} \sum_{x \in \mathcal{X}} \frac{1}{2} W(y \mid x) \log \frac{W(y \mid x)}{\frac{1}{2} W(y \mid 0)+\frac{1}{2} W(y \mid 1)},
$$
and
$$
Z(W) \triangleq \sum_{y \in \mathcal{Y}} \sqrt{W(y \mid 0) W(y \mid 1)}.
$$


For the basic transformation where there exist only two channels, as in \eqref{eq:w0} and \eqref{eq:w1}, the capacities of these symmetric channels are related by 
\begin{equation}\label{eq:capacity_ch_trans}
I\left(W^{-}\right)+I\left(W^{+}\right)=2 I(W).
\end{equation}
As can be seen, the capacities of the raw channels are preserved in the channel transformation. 
This can be expanded to any $N$ channels of power of two as follows. Given the input vector $u^N_1$, the transformed vector $x_1^N=u_1^N\boldsymbol{G}_N$ to be transmitted through the raw channel, and the corresponding vector $y^N_1$ representing the output of the raw channel through $N$ uses, the transition probabilities from the input of the synthesized vector channel $W_N$ to the output of the underlying $N$ raw channels $W^N$ are related by $W_N(y^N_1|u^N_1)=W^N(y^N_1|u^N_1\boldsymbol{G}_N)$. From $W_N(y^N_1|u^N_1)$, the transition probability of the bit-channel (a.k.a sub-channel or synthetic channel) $i\in[1,N]$ is implicitly defined as
\begin{equation}\label{eq:Wi} 
W_{N}^{(i)}\left ({y_{1}^{N}, u_{1}^{i-1}|u_{i}}\right ) = \sum _{u_{i+1}^{N}\in\{0,1\}^{N-i}} \frac {1}{2^{N-1}}W_{N}\left ({y_{1}^{N}|u_{1}^{N}}\right )\!. 
\end{equation}

\begin{figure}
    \centering
    \includegraphics[width=0.7\columnwidth]{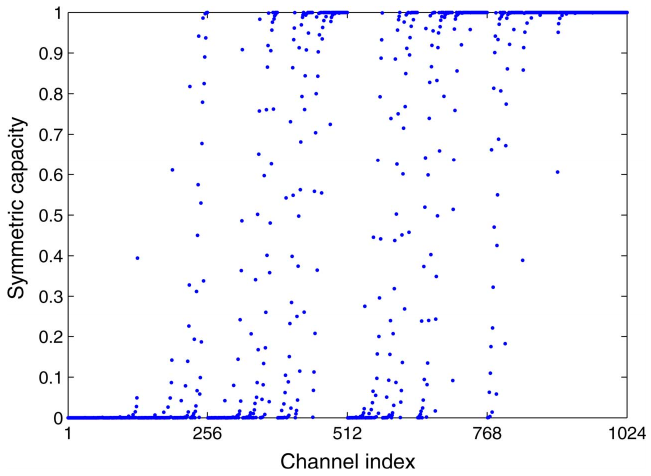}
    \caption{The channel transformation for $N=2^{10}$ \cite{arikan2009channel}.}
    \label{fig:polarization_1kb}
\end{figure}

Let us now investigate how the rate (the sub-channel capacity) and the reliability change as a result of transformation. For any $i\in[1,N]$, $N=2^n,n\geq0$ and B-DMC $W$, the transformation $\left(W_N^{(i)}, W_N^{(i)}\right) \mapsto\left(W_{2 N}^{(2 i-1)}, W_{2 N}^{(2 i)}\right)$ is rate-preserving and reliability-improving as \cite[Prop. 7]{arikan2009channel}
$$
\begin{aligned}
& I\left(W_{2 N}^{(2 i-1)}\right)+I\left(W_{2 N}^{(2 i)}\right)=2 I\left(W_N^{(i)}\right), \\
& Z\left(W_{2 N}^{(2 i-1)}\right)+Z\left(W_{2 N}^{(2 i)}\right) \leq 2 Z\left(W_N^{(i)}\right),
\end{aligned}
$$
where the Bhattachariya parameter for sub-channel $W_N^{(i)}$ given the channel observation $y_1^N$ is
\begin{equation}
    Z(W_N^{(i)})\!\triangleq\!\sum_{y_1^N, u_1^{i-1}}\! \sqrt{W_N^{(i)}\!\left(y_1^N, u_1^{i-1} \mid 0\right) W_N^{(i)}\!\left(y_1^N, u_1^{i-1} \mid 1\right)}.
\end{equation}
If $I(W_N^{(i)})>I(W_N^{(j)})$, or equivalently $Z(W_N^{(i)})<Z(W_N^{(j)})$, it is said that the sub-channel $W_N^{(i)}$ is more reliable than $W_N^{(j)}$, denoted as $W_N^{(i)} \succ W_N^{(j)}$, or simply $i \succ j$. 
The channel polarization theorem \cite{arikan2009channel} states that the mutual information of the $i$-th sub-channel, $I(W^{(i)}_N)$ for every $i\in[1,N]$, converges to 0 or 1 as $N$ approaches infinity. That is, the channels obtained after $n$ levels of transformation by \eqref{eq:basic_ch_trans} are either almost perfect, $I\left(W_{2^n}^{(i)}\right) \geq$ $1-\delta$ where $i\in[0,2^n-1]$ and $\delta>0$, or almost unreliable, $I\left(W_{2^n}^{(i)}\right) \leq \delta$. Note that the fraction of channels with $I\left(W_{2^n}^{(i)}\right) \in(\delta, 1-\delta)$ diminishes:
$$
\lim _{n \rightarrow \infty} \frac{\left|\left\{i: I\left(W_{2^n}^{(i)}\right) \in(\delta, 1-\delta)\right\}\right|}{2^n}=0.
$$
Given \eqref{eq:capacity_ch_trans}, by induction to 
$$\sum_{i\in[0,2^n-1]} I\left(W_{2^n}^{(i)}\right)=2^n I(W),$$ 
we can conclude that the fraction of almost perfect channels approaches the symmetric capacity. Therefore, 
we would have $N\cdot I(W)$ perfect sub-channels to use for $K$ information bits. 
Conversely, the fraction of indices $i\in[1,N]$ for which the sub-channels become extremely bad sub-channels approaches ($1-I(W)$). Fig. \ref{fig:polarization_1kb} illustrates the polarization effect for the binary erasure channel (BEC) $W$ where the erasure probability is 0.5. In the case of BEC $W$, the rate $I(W_N^{(i)})$ can be computed using the following recursive relations: 
\begin{equation*}
\begin{aligned}
I\left(W_N^{(2 i-1)}\right) & =I\left(W_{N / 2}^{(i)}\right)^2 \\
I\left(W_N^{(2 i)}\right) & =2 I\left(W_{N / 2}^{(i)}\right)-I\left(W_{N / 2}^{(i)}\right)^2
\end{aligned}
\end{equation*}

Polar codes with rate $R=K/N$ are constructed by selecting $K$ indices with the highest $I(W^{(i)}_N)$ for $i\in[1,N]$. These are dedicated to information bits and called the {\em non-frozen set}, $\mathcal{A}$. The input bits corresponding to {\em frozen set} $\mathcal{A}^c$ are usually set to zero. We further discuss the construction of polar codes in the next section. Note that this construction method is optimal for the original decoding algorithm proposed for polar codes called successive cancellation (SC) decoding. 


\subsection{Origin of Polar Codes}\label{ssec:origin_pc}
The idea of building synthetic channels \cite{arikan2015origin} originated from the concatenated schemes for convolutional codes under sequential decoding by Massey \cite{massey1981capacity} and Pinsker \cite{pinsker1965complexity} in order to boost the cutoff rate. The cutoff rate is said to be ``boosted" when the sum of the cutoff rates of the synthetic channels is greater than the sum of the cutoff rates of the raw channels. The key idea to achieve that was to correlate the independent copies of raw channels through concatenation. 
In Pinsker's scheme, the identical outer convolutional transforms were employed while the inner block code (with length $N$) was suggested to be chosen at random. This requires maximum likelihood (ML) decoding with prohibitive complexity. 
Different from Pinsker's scheme, in multi-level coding and multi-stage decoding (MLC/MSD), originally proposed in \cite{imai1977new} as an efficient coded-modulation technique, $N$ convolutional codes at different rates $\{R_i\}$ are used, which consequently require a chain of $N$ outer convolutional decoders. 
In contrast to the aforementioned schemes, polar coding was originally designed as a low-complexity recursive channel combining and splitting operations, where the polarization effect constrains the rates $R_i$ to 0 or 1. This method of building synthetic channels turned out to be so effective that no outer code was employed to achieve the original aim of boosting the cutoff rate. 

\subsection{Properties of Polar Codes}\label{ssec:properties}
The generator matrix of the polar code $(K,N)$ is a $K\times N$ submatrix of polar transform $\boldsymbol{G}_N=\bB\bG_2^{\otimes n}=[\boldsymbol{g}_1\,\cdots\boldsymbol{g}_N]^T$ consisting of rows $\boldsymbol{g}_i$ with indices $i\in\mathcal{A}$, denoted by $\boldsymbol{G}$, where $\bB$ is a bit-reversal permutation matrix defined in \cite{arikan2009channel}. The matrix $\bB$ is symmetric, that is, the $(i, j)$th entry is equal to the $(j, i)$th entry for all $i$ and $j$. Hence, $\bB^T=\bB$. As the inverse of a symmetric matrix is the matrix that reverses the permutation that the original matrix performs, we have $\bB^{-1}=\bB$. 
On the other hand, $\bG_2^{\otimes n}$ is invariant under bit-reversal, we have $\boldsymbol{G}_N=\bB^T\bG_2^{\otimes n}\bB$. Since $\bB^T=\bB^{-1}$, hence $\bB_N$ commutes with the bit-reversal operator, that is, $\bB \bG_2^{\otimes n}=\bG_2^{\otimes n}\bB$.  

As $\bG_2^{\otimes n}$ is a lower-triangular matrix with 1s on the diagonal, it is invertible. In fact, the inverse of $\bG_2^{\otimes n}$ is itself \cite{arikan2009channel}. Given $\bG_2^{-1}=\bG_2$ and $\bB$ commutes $\bG_2^{\otimes n}$, we have 
\begin{equation}
    \bG_N^{-1}=
(\bG_2^{\otimes n})^{-1} \bB^{-1}=(\bG_2^{-1})^{\otimes n} \bB=\bB\bG_2=\bG_N. 
\end{equation}

Let us denote the elements of the row $\bg_i$ of the matrix $\bG_N$ by $\{g_{i,j}\},j\in[N]$. Then, the submatrix $\bG_{\A}$ includes the rows $\bg_i,i\in\A$ while the submatrix $\bG_{\A\B}$ consists of $g_{i,j},i\in\A,j\in\B$. Also, let $\bu_{\A}$ denote the subvector of $\bu$ consists of $u_i, i\in\A$. 

For encoding, the information bits $\boldsymbol{d}=[d_1\,\cdots\, d_K]$  are inserted into the input vector $\boldsymbol{u}=[u_1\,\cdots\, u_N]$ at the coordinates $i\in\A$, while $u_i=0$ for $i\in\A^c$, that is, the bad subchannels are used for the transmission of known values (by default 0). Therefore, we can encode an information sequence $\boldsymbol{d}=\bu_{\A}$ using $\boldsymbol{d}\bG_{\A}=\boldsymbol{u}\boldsymbol{G}_N=\boldsymbol{x}$.

The parity check matrix $\bH$ of a polar code is characterized as follows \cite{goela2010lp}: Given $\bG_N^{-1}=\bG_N$ and $\bx=\bu\bG_N$, we have $\bu=\bx\bG_N$. To impose parity check constraints $\bx\cdot\bH^T=\bzero$ on $\bx$, it suffices to select the columns of $\bG_N$ corresponding to $u_j=0$. Therefore, the parity check matrix $\bH$ of polar codes is a submatrix of $\bG_N$ consisting of columns $j\in\A^c$. 

\subsubsection{Systematic Polar Codes}
A systematic code allows the original message to be recovered directly from the received codeword. That is, the systematic coding results in $\bx_{\A}=\bu_{\A}$. To achieve this, we obtain coded bits $\bx=\{\bx_{\A}, \bx_{\A^c}\}$ using submatrix $\bG_{\A\A}$ as 
\begin{equation}\label{eq:sys_enc_xA}
    \bx_{\A}=\bu_{\A} \bG_{\A\A}^{-1} \bG_{\A\A}=\bu_A \boldsymbol{I}_K=\bu_{\A},
\end{equation}
while the parity bits $\bx_{\A^c}$ are obtained by  
\begin{equation}\label{eq:sys_enc_xAc}
    \bx_{\A^c}=\bu_A \bG_{\A\A}^{-1} \bG_{\A\A^c}=\bu_{\A} \bP,
\end{equation}
where $\bP=\bG_{\A\A}^{-1} \bG_{\A\A^c}$. 
\subsubsection{Minimum Distance of Polar Codes}
\begin{figure}
    \centering
    \includegraphics[width=0.7\columnwidth]{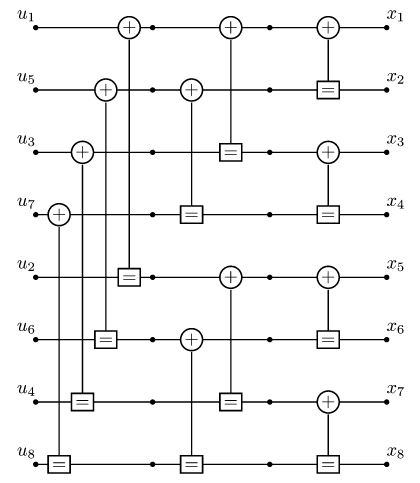}
    \caption{Factor graph for $N=8$ \cite{arikan2009channel}.}
    \label{fig:FG_8}
\end{figure}

The minimum distance $d$ of polar codes is equal to the minimum weight of the rows of the generator matrix, that is, $d=\min \{w(\boldsymbol{g}_i),i\in\mathcal{A}\}$. Fig. \ref{fig:FG_8} illustrates the factor graph representation of $\boldsymbol{G}_2^{\otimes3}$ proposed in \cite{forney2001codes} for the BP decoding of Reed-Muller codes. Observe that if we pass the input vector $\boldsymbol{u}=[u_1\,\cdots\, u_8]$ from the left-hand side through the factor graph consisting of an addition operator $\oplus$ in the Galois field $F_2$ (we call it $f$-node) and the passing operator $=$ (we call it $g$-node), we get the coded sequence $\boldsymbol{x}=[x_1\,\cdots\, x_8]$ on the right-hand side. 

We can divide all codewords of a polar code $\mathcal{C}(\mathcal{A})$ (excluding the all-zero codeword) into cosets  $\mathcal{C}_i(\mathcal{A})$ as 
\begin{equation}
    \Ci(\A) \triangleq \left\{\bgi\oplus\bigoplus_{h\in \H_i} \bgh \colon \H_i \subseteq \A \setminus [0,i]\right\}\subseteq \C(\A).
\end{equation}
Then, the minimum distance of the code $\mathcal{C}(\mathcal{A})$,  is $d_{\min}  = \min_{i\in\A} \w(\bgi)$ \cite{hussami2009performance},\cite[Lemma 3]{rowshan2021error,rowshan2023formation}. 
\subsubsection{Formation of Min-weight Codewords}
The weight of any codeword in the coset $\mathcal{C}_i(\mathcal{A})$ follows \cite[Corollary 3]{rowshan_error_2021}
\begin{equation}\label{eq:geq_wi}
    \w(\bgi\oplus\bigoplus_{j\in\H_i}\boldsymbol{g}_h)\geq \w(\bgi),
\end{equation}
where $\H_i\subseteq [i+1,N-1]$. Then, the minimum weight codewords in each coset $\Ci$ is decomposed into $\bGN$-rows as \cite[Lemma 6]{rowshan_error_2021}
\begin{equation}\label{eq:wt_decomp}
    \w\big(\bgi\oplus \bigoplus_{j\in\J}\bgj \oplus \bigoplus_{m\in\M(\J)}\bgm\big) =  \wm,
\end{equation}
where $\bgi$ is the leading row, $\bg_j, j\in\J$ are the core rows, and $\bg_m, m\in\M(\J)$ are the balancing rows. The indices of the core rows in $\J$ are a subset of the set $\Ki \triangleq \{j \in \I\backslash[0,i]\colon |\supp(j)\backslash\supp(i)|=1\}$ \cite[Defnition 4, Lemma 5]{rowshan_error_2021}. As a result, every subset of $\Ki$ along with the other rows in \eqref{eq:wt_decomp} forms a codeword of minimum weight. 
The number of subsets of $\Ki$ is given by $2^{|\Ki|}$. Given $\B\triangleq\{i\in\I:\w(\bgi)=\dm\}$, the total number of minimum weight codewords of the polar code will be $\sum_{i\in\B}2^{|\Ki|}$. 
The set $M(\J)$ is a function of the set $\J$ and every $m\in M(\J)$ has the property $|\supp(m)\backslash\supp(i)|>1$ (see \cite[Equations (33),(34)]{rowshan_error_2021} for a detailed definition of $M(\J)$). The number of codewords with minimum weight generated by the leading row $\bgi$ is denoted by $A_{i, \dm}$. 


The elements of the set $\mathcal{A}$ can be represented by monomials depending on the binary representation of the indices. For example, given the index $i=i_{0}, i_1, \dots, i_{n-1}=10101$, the corresponding monomial $f$ is a product of variables $f=\prod_{j=0}^{m-1}x_{j}^{i_{j}}=x_0x_2x_4$ of degree 3. Now, we collect all these monomials associated with the elements of the set $\mathcal{A}$ of a certain polar code and form the set $\mathcal{I}$, then we can call it the monomial code. 

In \cite{bardet2016cryptanalysis} it was shown that there is a decreasing order relation between the elements of the set $\mathcal{I}$, as described in \cite[Definition 3]{bardet2016algebraic}, equivalent to the partial order. The decreasing property induces new algebraic properties that can give a slightly better understanding of the algebraic structure of polar codes. 
\subsubsection{Rate of Polarization}
For matrix $\bG_2$, block-length $N=2^n$, and rate $R<I(W)$, the probability of block error $P_e(N,R)$ for polar coding and successive cancellation decoding can be bounded as $P_e(N,R)\leq 2^{-N^{\beta}}$, or $P_e(N,R)=O(2^{-N^{\beta}})$, for any $\beta<\frac{1}{2}$ \cite{arikan2009rate}. The parameter $\frac{1}{2}$ is called the \emph{exponent} of $G_2$, denoted by $E(\bG_2)=\frac{1}{2}$ and is a performance measure. 
This implies that when $n$ is sufficiently large, there exists a set $\A$ of size $N\cdot R$ such that $\sum_{i \in \mathcal{A}} Z\left(W_{2^n}^{(i)}\right) \leq 2^{-N^\beta}$. 
Let us define the \emph{partial distances} $D_i, i=1, \ldots, \ell$ of an $\ell \times \ell$ matrix $G=\left[g_1 \ldots, g_{\ell}\right]^T$ as \cite{korada2010polar}
$$
\begin{aligned}
& D_i \triangleq d_H\left(g_i,\left\langle g_{i+1}, \ldots, g_{\ell}\right\rangle\right), \quad i=1, \ldots, \ell-1 \\
& D_{\ell} \triangleq d_H\left(g_{\ell}, 0\right),
\end{aligned}
$$
where $d_H$ and $\langle\cdot\rangle$ denote the minimum Hamming distance and row span, then the \emph{rate of polarization} $\mathrm{E}(G)$ is 
$$
\mathrm{E}(G)=\frac{1}{\ell} \sum_{i=1}^{\ell} \log _{\ell} D_i.
$$
The partial distances of a polarizing matrix constructed from the Kronecker
product can be expressed as a product of those of its component matrices \cite{lee2014exponent}. As a result, the exponent of the polarizing matrix is a weighted sum of the exponents of its component matrices. 

The rate of polarization of an $\ell\times\ell$ kernel matrix $\boldsymbol{G}$  indicates how fast the error probability $P_e$ decays in the code length $N$ assuming that the rate $R$ is fixed. Therefore, this rate can be used to characterize asymptotically the performance of polar codes based on underlying kernels. It was shown in \cite{korada2010polar} that by constructing polar codes based on larger matrices $\bG$, the exponent can improve, i.e., becomes larger. We will discuss this in Section \ref{ssec:large_kern}. 


\subsection{Reed-Muller Codes versus Polar Codes}\label{ssec:polar_vs_rm}
A binary Reed-Muller (RM) code of length $N=2^n$, order $r$, dimension $K=\sum_{i=0}^r {n\choose i}$ and the minimum distance $d=2^{n-r}$ can be represented based on the transformation matrix $\boldsymbol{G}_N$ where the generator matrix is a $K\times N$ submatrix of $\boldsymbol{G}_N=[\boldsymbol{g}_1\,\cdots\boldsymbol{g}_N]^T$ consisting of rows $\boldsymbol{g}_i$ with indices $i\in\{j\in[N] \mid w(\boldsymbol{g}_j)\geq2^{n-r}\}$. We refer to this approach of selecting rows based on their weights as \emph{RM rule}. As can be seen, the differences between RM codes and polar codes are mainly in the choice of rows of the transformation matrix $\boldsymbol{G}_N$ and the flexibility in the choice of the code dimension $K$. 
Furthermore, the decreasing property described in Section \ref{ssec:properties} also applies to RM codes. Therefore, this family of monomial codes is called decreasing monomial codes. 

From a decoding perspective, successive cancellation decoding (see Section \ref{sssec:sc}) closely resembles recursive decoding of RM codes \cite{arikan2010survey}. 

\subsection{Code Construction}\label{ssec:construct}
In this section, we discuss the main approaches to designing polar codes. We focus mainly on the reliability-based construction, which is optimal for SC decoding. We also consider error-coefficient, the number approach for ML and near-ML decoding algorithms. 

The construction (a.k.a. rate profile) defines the set $\mathcal{A}\subset\{1,2,\dots,N\}$ of rows of $\boldsymbol{G}_N$ corresponding to the $K$ best sub-channels $W_N^{(i)}$ whose mutual information $\{I(W_N^{(i)}),i\in\mathcal{A}\}$ is the largest among $\{I(W_N^{(i)}),i\in[N]\}$. We call this approach as \emph{Ar\i kan rule}, which is different from the RM rule used in the construction of RM codes.
While the construction of polar codes is explicitly defined, the exact evaluation of $I(W_N^{(i)})$ for an additive white Gaussian noise channel is intractable, as it depends on the calculation of the parameters of channels whose output alphabets grow exponentially in code length. 

As discussed in Section \ref{ssec:polarization}, the seminal work on polar codes \cite{arikan2009channel} proposed using the Bhattacharyya parameter $Z(W_N^{(i)})$ as a measure of the probability of error (a reliability metric) over the binary erasure channel (BEC) and then choosing $K$ sub-channels with the smallest $Z(W_N^{(i)}),i\in[N]$. 
The block error event, $\E$, of the code resulting from the set $\A$ under SC decoding is a union over $\A$ of the events $\B_{i}$ in which the first bit error occurs at the $i$-th bit, expressed as $\E=\bigcup_{i\in\A} \B_i$. The set is defined as \cite{arikan2009channel}
\begin{equation}
\begin{aligned}
    \mathcal{B}_{i} & =\{(u_1^N, y_1^N):\hat{u}_1^{i-1}=u_1^{i-1}, \hat{u}_i(y_1^N, u_1^{i-1}) \neq u_i\} \\
    & \subseteq\{(u_1^N, y_1^N):\hat{U}_i(y_1^N, u_1^{i-1}) \neq u_i\} \triangleq \A_i.
\end{aligned}
\end{equation}
Then, the block error probability can be upper bounded by 
\begin{equation}
    P(\E)=\sum_{i \in \A} P(\B_{i}) \leq \sum_{i \in \A} P(\A_i) \leq \sum_{i \in \A} Z(W_N^{(i)}).
\end{equation}

\subsubsection{Density Evolution}
In \cite{mori2009performance}, $P(\A_i)$ was considered as decoding error probabilities in belief propagation (BP) decoding on the tree graph corresponding to the $i$th bit while eliminating other edges of the Tanner graph. The root and leaves of the tree correspond to the variables nodes of $u_i$ and $y_1^N$, respectively. This helps to evaluate the probability of error at the root node using \emph{ density evolution} as a known approach for LDPC codes \cite{richardson2008modern}. 

The log-likelihood ratio (LLR) of the $i$-th bit, defined as 
\begin{equation}\label{eq:LLRi}
L_N^{(i)}(y_1^N, \hat{u}_i^{i-1}) \triangleq \log\frac{W_N^{(i)}(y_1^N, \hat{u}_1^{i-1} \mid 0)}{ W_N^{(i)}(y_1^N, \hat{u}_1^{i-1} \mid 1)},
\end{equation}
is calculated recursively the intermediate LLRs from leaves to the root of the tree by the following updating rules \cite{hartmann1976optimum,battail1979replication,hagenauer1996iterative,luby1998analysis,forney2001codes,arikan2009channel,mori2009performance}:
\begin{equation}\label{eq:var_LLR}
\begin{aligned}
&L_N^{(2 i-1)}\left(y_1^N, \hat{u}_1^{2 i-2}\right) \\
& =2 \tanh ^{-1}\left(\tanh \left(L_{N / 2}^{(i)}\left(y_1^{N / 2}, \hat{u}_{1, e}^{2 i-2} \oplus \hat{u}_{1, o}^{2 i-2}\right) / 2\right)\right. \\
& \left.\times \tanh \left(L_{N / 2}^{(i)}\left(y_{N / 2+1}^N, \hat{u}_{1, e}^{2 i-2}\right) / 2\right)\right) 
\end{aligned}
\end{equation}
\begin{equation}\label{eq:check_LLR}
\begin{aligned}
L_N^{(2 i)}(y_1^N,& \hat{u}_1^{2 i-1})  =L_{N / 2}^{(i)}\left(y_{N / 2+1}^N, \hat{u}_{1, e}^{2 i-2}\right) \\
+ & (-1)^{\hat{u}_{2 i-1}} L_{N / 2}^{(i)}\left(y_1^{N / 2}, \hat{u}_{1, e}^{2 i-2} \oplus \hat{u}_{1, o}^{2 i-2}\right)
\end{aligned}
\end{equation}
where $\hat{u}_{1, e}^i$ and $\hat{u}_{1, o}^i$ are subvectors which consist of elements of $\hat{u}_1^i$ with even and odd indices, respectively, and channel LLR: 
\begin{equation}\label{eq:ch_LLR}
    L_1^{(i)}(y_i)=\log \frac{W(y_i \mid 0)}{W(y_i \mid 1)}.
\end{equation}

According to \cite{richardson2008modern,mori2009performance}, for a symmetric B-DMC with the probability density functions (PDFs) $a_w(x)$ of the channel LLRs in \eqref{eq:ch_LLR} for all-zero codeword, we have  
\[
P\left(\A_i\right)=\mathfrak{E}\left(\mathrm{a}_N^i\right)
\]
where
\begin{equation}
\mathfrak{E}(\mathrm{a}):=\lim _{\epsilon \rightarrow+0}\left(\int_{-\infty}^{-\epsilon} \mathrm{a}(x) \mathrm{d} x+\frac{1}{2} \int_{-\epsilon}^{+\epsilon} \mathrm{a}(x) \mathrm{d} x\right),
\end{equation}
and $\mathrm{a}_N^i,i\in[1,N]$ are recursively obtained using
\begin{equation}\label{eq:conv}
\mathrm{a}_{2 N}^{2 i}=\mathrm{a}_N^i \circledast \mathrm{a}_N^i, \quad \mathrm{a}_{2 N}^{2 i-1}=\mathrm{a}_N^i \boxast \mathrm{a}_N^i, \quad \mathrm{a}_1^1=\mathrm{a}_w,
\end{equation}
and $\circledast$ and $\boxast$ denote the convolutions of LLR density functions corresponding to \eqref{eq:var_LLR} and \eqref{eq:check_LLR}, respectively. Recall that the addition of independent random variables, here LLRs, implies the convolution of their densities. 
Then, a polar code can be constructed by choosing a set $\A$, where $|\A|=N\cdot R$, which minimizes 

\[
\sum_{i\in\A} P(\A_i)
\]

\subsubsection{Approximate Density Evolution}
To reduce the complexity of DE, several approximate methods have been proposed. In \cite{tal2013construct}, an approach was proposed based on the upper bound and the lower bound on the error probability of the sub-channels.  

The success of the Gaussian approximation \cite{chung2001analysis} for sparse graph codes suggests the use of a similar approach for polar codes to approximate the density evolution of LLRs throughout the tree graph \cite{korada2010empirical,trifonov2012efficient}. The intuition behind this approximation is that the channels $W^{(i)}_N$ behave like binary AWGN channels with varying noise levels when $N$ is sufficiently large. Therefore, we only need to monitor the noise variances of these channels. 
For AWGN channels, the distribution of the channel LLRs in \eqref{eq:ch_LLR} is Gaussian, that is, $L_1^{(i)}\left(y_i\right) \sim \mathcal{N}(m, 2m)$ with mean $m=\frac{2}{\sigma^2}$ and variance $\frac{4}{\sigma^2}$, where $\frac{1}{\sigma^2}$ is the SNR of the channel. 
In \cite{chung2001analysis}, it was suggested that the intermediate LLRs in \eqref{eq:check_LLR} and \eqref{eq:var_LLR} be approximated by Gaussian random variables where the relationship between the mean (expected value) $E$ and the variance $V$ is $V[L^{(i)}_N]=2 E[L^{(i)}_N]$. 
As a result, considering the check and variable nodes of degree $d_c=d_v=3$ in the factor graph of polar codes (Fig. \ref{fig:FG_8}), the convolution operations in \eqref{eq:conv} are reduced to \cite[Equations (6),(4)]{chung2001analysis}:
\begin{equation}
E\left[L_N^{(2 i-1)}\right]=\phi^{-1}\left(1-\left(1-\phi\left(E\left[L_{N / 2}^{(i)}\right]\right)\right)^2\right)
\end{equation}
\begin{equation}
E\left[L_N^{(2 i)}\right]=2 E\left[L_{N / 2}^{(i)}\right]
\end{equation}
where function $\phi(x)$ for $x \in[0, \infty)$ is defined as \cite[Definition 1]{chung2001analysis}
\begin{equation}\
\phi(x)= \begin{cases}1-\frac{1}{\sqrt{4 \pi x}} \int_{\mathbb{R}} \tanh \frac{u}{2} e^{-\frac{(u-x)^2}{4 x}} d u, & \text { if } x>0 \\ 1, & \text { if } x=0\end{cases}
\end{equation}
To reduce the complexity of calculating $\phi(x)$, an approximation was suggested with acceptable accuracy in \cite{chung2001analysis}. For $x<10$, one can use the following curve fitting:
$$
\phi(x) \sim e^{\alpha x^\gamma+\beta}
$$
where $\alpha=-0.4527, \beta=0.0218$, and $\gamma=0.86$. For $x\geq 10$, we can use the average of the upper and lower bounds for $\phi(x)$ as \cite[Lemma 1]{chung2001analysis}:
\begin{equation}
\sqrt{\frac{\pi}{x}} e^{-\frac{x}{4}}\left(1-\frac{3}{x}\right)<\phi(x)<\sqrt{\frac{\pi}{x}} e^{-\frac{x}{4}}\left(1+\frac{1}{7 x}\right), x>0.
\end{equation}
\subsubsection{Partial Order \& Polarization Weight}
There are also SNR-independent low-complexity methods for reliability evaluation. 
It was shown in \cite{bardet2016cryptanalysis,bardet2016algebraic,schurch2016partial} that there exists a certain partial order relation between the reliabilities of the sub-channels in polar coding. They are called ``partial" because the available relations cannot form a fully ordered integer sequence corresponding to the indices of all $N$ sub-channels. 
The partial orders are deterministic and universal; that is, the relation holds for any transmission channel. Hence, they are called universal partial order (UPO). 

Given $i,j\in [0,2^n-1]$, we denote the partial order $i \preceq j$ if they satisfy one of the following conditions \cite[Def. 1]{rowshan2021error,rowshan2023formation}: a) $\text{supp}(i) \subseteq \text{supp}(j)$, b) 
 $\text{supp}(j) = (\text{supp}(i)\setminus \{a\})\cup \{b\}$ for some $a\in \text{supp}(i)$, $b \notin \text{supp}(i)$ and $a < b$, and c) there exists $k\in [0,2^n-1]$ that satisfy $i\preceq k$ and $k\preceq j$. 
In \cite[Section V]{mori2009performance}, it was shown that for all $i\in \mathcal{A}$ and $j\in [0,N-1]$, if $j \succeq i$ then $j\in \mathcal{A}$. 

The universal partial order of polar codes has two main properties \cite [Prop. 1]{he2017beta}: (a) The orders determined for a code of block-length $N$ remain unchanged for block-length of $2 N$ (nested property), 
(b) the order of $x \prec y$ and the order of $(N-$ $1-x) \succ(N-1-y)$ are twin pairs for a polar code with block-length $N$. 

Later, a closed-form algorithm based on binary expansion known as \emph{polarization weight} (PW) was proposed in \cite{huawei2016polar,he2017beta} to characterize the reliability order of the subchannels for AWGN channels. 
The polarization weight of the sub-chnnel $x$ with binary expansion $\left(b_{n-1}, \ldots, b_1, b_0\right)$ is defined as \cite[Def. 3]{he2017beta}
\begin{equation}\label{eq:pw}
f^{\mathrm{PW}}: x \mapsto \sum_{i=0}^{n-1} b_i \beta^i
\end{equation}
where $\beta$ is an optimization parameter (the suggested value is $\beta=2^{1 / 4}$). 
Computing the PW for all sub-channels gives a reliability measure, and subsequently a reliability order. 
The PW algorithm preserves the nested structure (or nested frozen sets) for polar codes \cite[Prop. 2]{he2017beta}. This property helps in using one long reliability sequence that is found offline, for shorter block-lengths. 
Due to the low complexity, this method was adopted in the 5G standard \cite{noauthor_5g_2020} in the form of a reliability sequence, which gives the indices of sub-channels in ascending reliability order. 
\subsubsection{RM-Polar Construction}
\begin{figure}[ht]
	\begin{center}
    	\includegraphics[width=1\columnwidth]{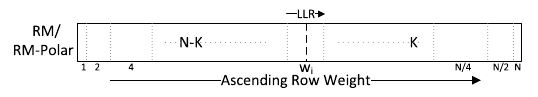}
    \end{center}
    \caption{RM-Polar Construction \cite{rowshan2021polarization}, a construction based on RM rule and Ar\i kan rule.}
    \label{fig:rm_polar}
\end{figure}

Alternatively, we can employ both the weight-based RM rule and the reliability-based Ar\i kan rule to construct a polar-like code as demonstrated in \cite[Section VI.1]{rowshan2021polarization}. That is, the set $\A$ is divided into two;
\begin{equation}\label{eq:rm_polar}
    \A=\A^{\text{RM}} \cup \A^{\text{Polar}}.
\end{equation}
Given the blocklength $N=2^n$ and code dimension $K$,  we find $r'=
{\arg\max}_r \sum_{j=0}^{r}{n \choose j}\leq K$ and denote $K' = \sum_{j=0}^{r'}{n \choose j}$. Then, we find the indices of $K'$ rows of $\bGN$ with weights $\w(\bg_i)\geq 2^{n-r'}$. That is,
\[
\A^{\text{RM}}=\{i\in[0,N-1]:2^{\operatorname{w}(\operatorname{bin}(i))}\geq2^{n-r'}\},
\]
where $K'=|\A^{\text{RM}}|$. To find $\A^{\text{Polar}}$, 
we exclude the indices in $\A^{\text{RM}}$ from the reliability sequence of length $N$ and select the remaining $K-K'$ indices from the most reliable indices remaining in the sequence with row weight $2^{n-(r'+1)}$. That is, 
\[
\A^{\text{Polar}}=\underset{\substack{\mathcal{S}\subseteq\left([0,N-1]\setminus\mathcal{A}^{\text{RM}}\right)\\ |\mathcal{S}|=K-|\mathcal{A}^{\text{RM}}|}}{\arg\max}\sum_{\substack{i\in\mathcal{S}\\ {\operatorname{w}(\operatorname{bin}(i))}={n-r'-1}}}L_N^{(i)},
\]
where $L_N^{(i)}$ is the average LLR of sub-channel $i$ which can be found from the methods discussed in the previous sections. Observe that this approach gives $r$-th order RM code $(r,n)$ when $K=\sum_{j=0}^{r}{n \choose j}$ and therefore $\A^{\text{Polar}}=\emptyset$. 
If we use the 5G reliability sequence in this process, we call it \emph{5G-RM} construction \cite{gu2023rate}, otherwise (crossed) \emph{RM-polar} construction, can be denoted by RMxPolar codes, due to intersection with RM codes. Fig. \ref{fig:rm_polar} illustrates the selection of $K$ indices with this approach. 

A different approach for constructing RM-Polar codes is to constrain the selection of most reliable sub-channels 
with row weights of $\w(\bg_i)\geq2^{n-r'-1}$ \cite{li_rm-polar_2014} where set $\A$ is defined as below: 
\[
\A=\underset{\substack{\mathcal{S}\subset[0,N-1]\\ |\mathcal{S}|=K}}{\arg\max}\sum_{\substack{i\in\mathcal{S}\\ {\operatorname{w}(\operatorname{bin}(i))}\geq{n-r'-1}}}L_N^{(i)},
\]
To differentiate from the RMxPolar codes, we refer to this code as \emph{constrained RM-Polar} codes. 
Note that the set $\A$ in this approach is most likely different from \eqref{eq:rm_polar}, because there might be sub-channels with row weight(s) larger than the minimum weight $\w_{min}$ excluded from $\A$ and instead include more sub-channels with row weight $\w_{min}$. This results in a code with a higher number of cosets $\C_i(\A)$, where $\w(\bin(i))=\log_2(\w_{min})$ and therefore consequently containing more minimum weight codewords. Due to this improvement, relative to the constrained RM-polar codes in \cite{li_rm-polar_2014}, crossed RM-polar codes are also called improved RM-polar codes \cite{cai2023modified}. 


According to the Union bound, the block error rate is a function of the minimum distance $d_{min}$ and the number of small-weight codewords, in particular, the number of minimum weight $A_{\wm}$, a.k.a. the error coefficient. Hence, a code design approach that considers both the sub-channel reliability and the error coefficient is of interest; however, optimizing both would be highly complex and requires performing a search for every code. An approach in \cite{rowshan2023formation,rowshan2022improving} was suggested that simply improves the error coefficient of an available reliability-based construction, which was discussed above. Also, as shown in \cite{rowshan2019modify}, we can modify the reliability-based construction by bit-swapping to reduce the possibility of eliminating the correct path in the list decoding. This shows that one can tailor the construction of polar code for a specific decoding algorithm. This insight was also used in \cite{rowshan2018stepped} to partition the code block and to use different list sizes for each partition (depending on the possibility of correct path elimination in order to reduce computational complexity. 


\subsection{Code Concatenation and Pre-transformation}\label{ssec:concat_pretrans}
A polar code can be used as an inner code, while an outer code such as a cyclic code can be used for its detection capability. In this case, the systematic cyclic coded bits are called cyclic redundancy check (CRC) bits, which are appended to the information bit sequence $\boldsymbol{d}$. Observe that the CRC bits as additional parity bits are placed on sub-channels in $\mathcal{A}^c$. 
As the probability of false detection increases with the length of the sequence $\boldsymbol{d}$, so-called distributed CRCs are suggested to be used. In the distributed CRCs scheme, the information sequence is divided into segments, and each segment is outer encoded separately. 

Another pre-transformation scheme considers parity bits $u_i$ resulting from a linear combination of bits in the sequence $\boldsymbol{d}$ such that $u_i=\sum_{j\in\J}u_j$ where $\J$ is a subset of $\A$ and for every $j\in\J$, 
 we have $j<i$. These parity bits are carried similarly by sub-channels in $\mathcal{A}^c$, hence sometimes called dynamic frozen bits \cite{trifonov2013polar}. The choice of the linear combination of bits has not been explicitly structured except in a recently introduced variant of polar codes called polarization-adjusted convolutional (PAC) codes. We discuss them in detail in Section \ref{ssec:pac}. 

\subsection{Information (or Spatially) Coupled Polar Codes}\label{ssec:specially-coupled}
\begin{figure}[ht]
	\begin{center}
    	\includegraphics[width=0.9\columnwidth]{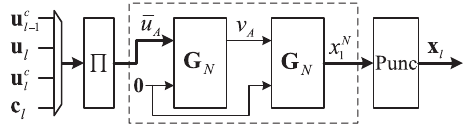}
    \end{center}
    \caption{Encoding scheme for a single CB \cite{wu2018partially}.}
    \label{fig:CB_enc}
\end{figure}

To improve the efficiency of the transport block (TB) in communication standards, consecutive systematic code blocks (CBs) in a TB are coupled by sharing a portion of the information bits, hence called \emph{partially information coupled} (PIC) polar codes \cite{wu2018information,wu2018partially,wu2021partially}. 
In this scheme, a $\bu$ corresponding to the $l$-th CB, consists of the coupled bits $\bu^c_{l-1}$ with the previous CB, the code bits $\bu_{l}$, the coupled bits $\bu^c_{l}$ with the next CB, and the CRC bits $\bc_{l}$ of the CB. Fig. \ref{fig:CB_enc} illustrates the coding process that includes permutation $\Pi$, systematic encoding $\bx=[\bar{\bu}_{\A},\bar{\bu}_{\A}\bG_{\A\A}^{-1} \bG_{\A\A^c}]$ according to \eqref{eq:sys_enc_xA} and \eqref{eq:sys_enc_xAc}, followed by puncturing of the coded bits corresponding to $\bu^c_{l-1}$. 
The partially coupled information bits are punctured in the next CB. Given that the coupled bits are correctly decoded in the previous CB, the error propagation resulting from these bits is avoided in successive cancellation-based decoding algorithms. The inter-CB decoding scheme, which realizes a windowed decoder with variable window size, achieves a trade-off between the decoding performance and complexity. 
In \cite{wang2018improving}, the coupled portions of the information bits are transmitted as frozen bits or added to the information bits, both in the next CB in a row. 

\subsection{PAC Codes}\label{ssec:pac}
Polarization-adjusted convolutional (PAC) codes are pre-transformed polar codes suggested initially in the Shannon lecture at the International Symposium on Information Theory (ISIT), 2019. As mentioned in Section \ref{ssec:origin_pc}, although the motivation behind polar codes was to build a synthetic channel to boost the cutoff rate inspired by classical concatenated schemes, the concatenation was not employed at that stage due to the performance of polar codes and the availability of the low complexity decoding procedure. Nevertheless, it was later demonstrated that a rate-1 convolutional pre-transformation can improve the block error rate (BLER) for short codes under sequential decoding \cite{arikan2019sequential}. 

In PAC coding, the information bits $\boldsymbol{d}=(d_0,d_1,...,d_{K-1})$ are first mapped to a vector $\boldsymbol{v}=(v_0,v_1,...,v_{N-1})$ using a rate-profile defined by set $\mathcal{A}$. 
This set includes the indices of the positions where the information bits are placed in the input vector to the convolutional transform (CT). The rule to form this set does not have to be the same as polar codes. 
The bit values in the remaining positions in $\boldsymbol{v}$ are set to 0. 
The constraint $v_{\mathcal{A}^c}=0$ simply leads to an irregular decoding tree. 
Note that 1) the outputs of CT for indices in $\mathcal{A}^c$, which enter the polar transform, are no longer fixed, i.e., known a priori - unlike in polar coding. 
2) In contrast to convolutional coding, in which usually $R_c<1$, here we use a one-to-one transform $\boldsymbol{G}$, hence the vectors $\boldsymbol{v}$ and $\boldsymbol{u}$ in $\boldsymbol{u}=\boldsymbol{v}\boldsymbol{G}$ have the same dimension.


The relation between the input bits $d_{i-m,i}$ and the output bit $u_i$, at time-step $i$, is  obtained as a binary convolution by
\begin{equation}
     u_i = \sum_{j=0}^m g_j d_{i-j}, 
\end{equation}
where $g_i\in\{0,1\}$ and $m$ is the number of previous input bits stored in a shift register ($m$ is also known as {\em memory size}). By representing bit sequences as polynomials in the delay variable $D$ representing a time step in the encoder, an output sequence $x(D)$ is obtained as $g(D) d(D)$, where $g(D) = \sum_{j=0}^m g_j D^j$ is the {\em generator polynomial}. The coefficients of the generator polynomial in the context of convolutional codes are represented in octal notation. For example, the commonly used $\bg=[1\;0\;1\;1\;0\;1\;1]$ is represented by $133$.  

\begin{figure}[ht]
	\begin{center}
    	\includegraphics[width=0.9\columnwidth]{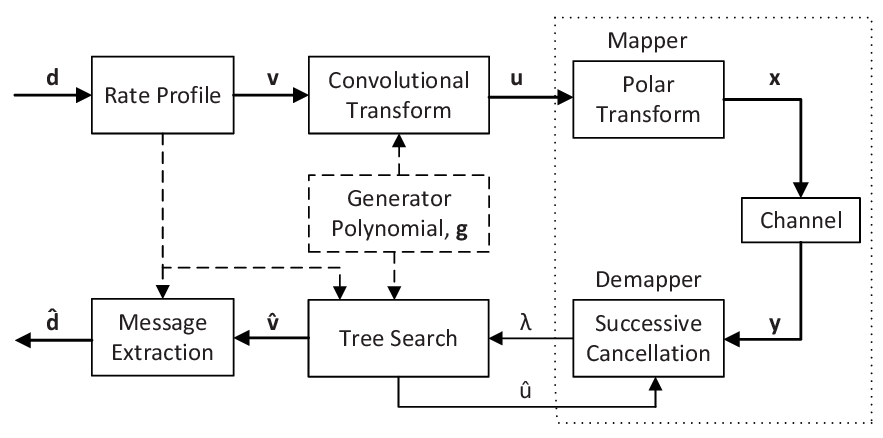}
    \end{center}
    \caption{PAC Coding Scheme \cite{rowshan_polarization-adjusted_2021}.}
   \label{fig:PAC_scheme}
\end{figure}

After the convolutional transform, the vector $\boldsymbol{u}$ is mapped to $\boldsymbol{x}$, as Fig. \ref{fig:PAC_scheme} shows, employing the polar transform $\boldsymbol{G}_N$; therefore, $\boldsymbol{x}=\boldsymbol{u}\boldsymbol{G}_N$. 

The readers can refer to Section \ref{sec:turbo}-\ref{ssec:conv} on Convolutional codes, a component code used in Turbo coding. 
In \cite{rowshan_convolutional_2021,rowshan_fast_2022,rowshan2023minimum} it was shown that the precoding stage can reduce the number of codewords of minimum weight as a result of the inclusion of frozen rows in $\bG_N$. Hence, precoding can potentially improve the block error rate of polar codes. Various decoding algorithms have been adapted to PAC codes, such as sequential decoding \cite{arikan2019sequential,rowshan2020complexity,rowshan2021towards,mozammel2021hardware}, list decoding \cite{rowshan2021polarization,yao2021list}, list Viterbi decoding \cite{rowshan2021list}, belief propagation decoding \cite{zhang2022crc}, GRAND \cite{rowshan2022constrained,rowshan2023segmented,10355824}, stack decoding \cite{rowshan2021polarization,jiang2023soft,wu2023multi,10139571}, fast and simplified list decoding \cite{saber2024simplified,dai2024fast}, and other decoding algorithms \cite{yu2021threshold,zhang2023novel}. 

Furthermore, reverse PAC coding has recently been proposed \cite{gu2022selective,gu2023improved} that can overcome the limitations of forward precoding to further reduce the number of minimum weight codewords. Among other precoding schemes and variants of PAC codes, we can name tail-biting PAC codes \cite{feng2022tail}, row-merged polar codes \cite{gelincik2022pre,zunker2023rowmerged}, spatially coupled PAC codes \cite{ying2022spatially}, modified polar codes \cite{cai2023modified}, parity check PAC codes \cite{yu2024parity}, and other schemes \cite{mishra2020selectively,choi2023deep,liu2023novel,10465170}. 
Numerous construction methods for PAC codes, predominantly based on search, have also been suggested in \cite{abdullah2023new,mishra2023heuristic,9900337,10087278}. It was proposed in \cite{rowshan2023formation} to use the reliability-based rate-profile of polar codes as it is easy to construct, and instead to modify it to drastically reduce the number of minimum weight codewords. Furthermore, rate-matching techniques, such as puncturing and shortening, for systematic and non-systematic PAC codes have also been investigated in \cite{sun2021optimized,e22111301,gu2023rate,zhang2023rate,qiu2023construction,jiang2023construction,10464959}. Finally, PAC coding has been considered for source coding and joint source-channel coding in \cite{9682079,10206454,kann2023source,zheng2023pac}. 

\subsection{Large Kernel Polar Codes (binary and non-binary)}\label{ssec:large_kern}
Ar\i kan's pioneering work \cite{arikan2009channel} was based on the linear binary \emph{kernel matrix} $\bG_2$ of dimension two and exponent $\frac{1}{2}$ (see Section \ref{ssec:properties}). 
To improve the exponent and, consequently, the performance of polar codes under SC decoding, large non-singular kernel matrices $\bG$ were suggested in \cite{korada2010polar}. The polarization effect over BI-DMC still holds for such an $\ell \times \ell$ kernel matrix $\bG$ provided that none of its column permutations is an upper triangle matrix. The block error probability is found to be $O\left(2^{-\ell^{n \beta}}\right)$ if $\beta$ is less than the exponent of the kernel matrix. 

The upper and lower bounds of the achievable exponents for the kernels showed that there are no matrices of size smaller than $15 \times 15$ with exponents $\mathrm{E}(\bG)$ exceeding $\frac{1}{2}$. Furthermore, it was shown that a general construction based on BCH codes for large $\ell$ can achieve exponents arbitrarily close to 1. For example, at size $16 \times 16$, this construction yields an exponent $\mathrm{E}(\bG)>\frac{1}{2}$. 

Different linear non-binary kernels have also been proposed. 
In \cite{mori2010non}, non-binary kernel matrices were constructed based on Reed-Solomon codes and Hermitian codes that were later expanded to other algebraic geometry codes, including Suzuki codes, in \cite{anderson2014exponents}, concatenated algebraic geometry codes, due to their large minimum distance and often nested structure, in \cite{eid2013using}. These kernels have a larger exponent than any linear binary kernel of the same dimension. Furthermore, non-linear binary kernels were studied in \cite{presman2015binary,lin2015linear} which provide exponents superior to any linear binary kernel of the same number of dimensions. 

The SC decoding algorithm for large kernel polar codes has the complexity of order $O(2^{\ell} N \log N)$ operations, where $O(2^{\ell})$ is the complexity of kernel processing \cite{trofimiuk2021window} based on trivial implementation. Although this complexity can be slightly reduced, direct calculation remains prohibitive for kernels with relatively large sizes, such as $\ell=16,32$. More efficient algorithms, such as window processing \cite{trifonov2014binary,abbasi2020large} and recursive trellis processing \cite{trifonov2014binary} have been proposed. The complexity of these algorithms is still exponential with respect to kernel size $\ell$. 
Moreover, some of these approaches, such as window processing which is based on the idea of expressing LLRs for a large kernel via LLRs for the dimension-2 kernel $\bG_2$, are not admitted by all kernels in an efficient way. 

\subsection{Mixed/Multi-kernel Polar Codes}\label{ssec:multi_kern}
As we observed in Section \ref{ssec:polarization}, the code construction based on the $n$-fold Kronecker product of $\boldsymbol{G}_2$ restricts the length $N$ of polar codes to be powers of two, i.e., $N=2^n$. In the case of larger kernels, the code length becomes powers of integers, i.e., $N=l^n$, where $l$ is the dimension of the kernel. One can employ puncturing, shortening, and repetition techniques to adjust the blocklength and have an arbitrary blocklength. However, these techniques come with disadvantages. In addition to degradation in performance, as the punctured and shortened codes are decoded based on their mother polar codes,  the decoding latency is proportional to the actual code length. Furthermore, the coordinates of dummy bits representing shortened or punctured bits affect the polarization of the codes and improper selection of these coordinates may result in catastrophic error correction performance. 

In this section, we consider the multi-kernel polar codes \cite{presman2015mixed} in which different kernel sizes over the binary alphabet \cite{bioglio2020multi} are mixed, while the polarization effect is preserved \cite{benammar2017multi}. Encoding of multi-kernel polar codes follows the general structure of polar codes, and decoding can be performed by any SC-based decoding discussed earlier. 

\begin{figure}[ht]
	\begin{center}
    	\includegraphics[width=0.9\columnwidth]{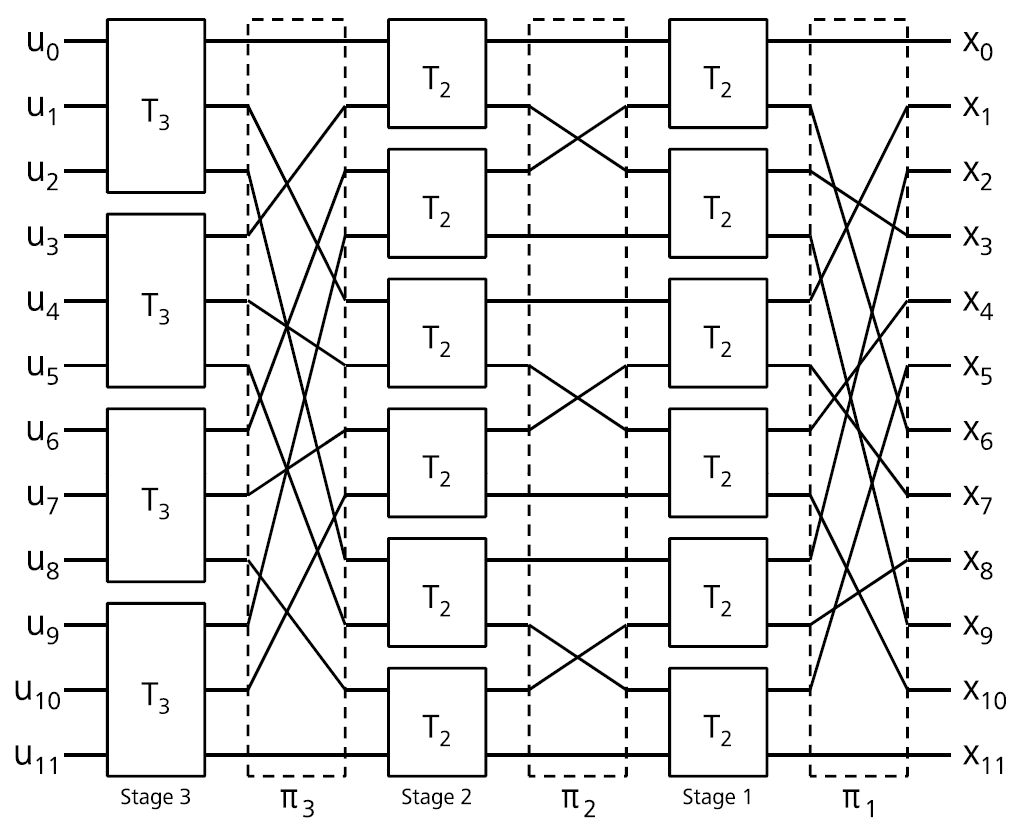}
    \end{center}
    \caption{Tanner graph of the multi-kernel polar code with 
the transformation matrix $T_{12} = T_2 \times T_2 \times T_3$ \cite{bioglio2020multi}.}
    \label{fig:factor_graph_multi}
\end{figure}

An $(N,K)$ multi-kernel polar code is defined by an $N \times N$ transformation matrix $T_N$ as
\begin{equation}\label{eq:multi-kern-transform}
T_N=T_{p_1} \otimes T_{p_2} \otimes \cdots \otimes T_{p_s},
\end{equation}
where the code length is $N=p_1p_2\cdots p_s$ and the binary polarizing matrices $T_{p_i}$ of size $p_i \times p_i$ for $i=1,\dots,s$, called kernels of dimension $p_i$. Note that the Kronecker product in \eqref{eq:multi-kern-transform} is not commutative. 
Fig. \ref{fig:factor_graph_multi} illustrates the Tanner graph of a polar code with length $N=12=2\times2\times3$ composed of two kernels of dimension 2 and one kernel of dimension 3. 
Observe that when all kernels are composed of kernel $T_2=\boldsymbol{G}_2$, we will have the polar codes discussed in Section \ref{ssec:polarization}. The permutations of edges between kernels are implicitly defined by the Kronecker product, similarly to polar codes \cite{arikan2009channel}. Permutation $\pi_i$ connects the output $j,j=0,1,\dots,p_{i-1}$ of stages $i-1, i=2,3,\dots,s$ to the input $\pi_i(j)$ of stage $i$.  

A multi-kernel polar code can be designed based on three principles of reliability, Hamming distance, and a combination of both.  Let us review these code design approaches: 1) The reliability approach is based on the polarization effect, with the aim of minimizing the BLER under SC decoding.


\subsection{Decoding Algorithms}\label{ssec:decoding}
Since Ar\i kan's seminal work, many decoding algorithms have been adapted for decoding polar codes and their variants. In this section, we review the major decoders and their recent improvements and compare them in terms of complexity and block error rate performance. We also consider the realization of these decoders as hardware architectures and provide an assessment of their compatibility with the KPIs of the potential scenarios in 6G. 

\subsubsection{Successive Cancellation (SC) Decoding}\label{sssec:sc}
As we observed, channel combining introduces a correlation between the source bits. As a result, each coded bit with a given index relies on all its preceding source bits. 
This correlation can be conceptually treated as {\em interference} in the source-bit domain and can be exploited in the decoding process. Therefore, the bits are decoded one at a time in a specific order. The bit decision $\hat{u}_i$ is made before the calculations start to find the next bit $\hat{u}_{i+1}$, and the already decided bits influence the decision of the following bit decisions. The successive cancelation of the "interference" caused by the previous bits improves the reliability of retrieving the source bits. Apparently, the name of this decoding process has been borrowed from the successive interference cancellation (SIC) decoding technique used by a receiver in wide-band wireless communications where two (or more) packets arrive simultaneously (which otherwise cause a collision). SIC is an iterative algorithm in which received data are decoded on the order of decreasing power levels. That is, in the case of two signals, the stronger signal is first decoded and then subtracted from the combined signal, and then the weaker signal is extracted from the residue. 

In each use of the system, a codeword is transmitted and a channel output vector $y\in \mathcal{Y}^N$ is received. 
In SC decoding, the source bits corresponding to the frozen bits are set to zero, $\hu_i=0,i\in\A^c$. The information bits $\hu_i,i\in\A$ are decoded sequentially through maximum likelihood (ML) decoding of the channel $W_N^{(i)}$ as
\begin{equation}
    \hat{u}_i\left(y_1^N, \hat{u}_1^{i-1}\right)=\underset{u_i=0,1}{\operatorname{argmax}} \;W_N^{(i)}\left(y_1^N, \hat{u}_1^{i-1} \mid u_i\right).
\end{equation}
Practically, similar to \eqref{eq:ch_LLR}, the receiver first calculates the vector of logarithmic likelihood ratios (LLRs) 
with 
\begin{equation}\label{eq:ch_LLR_sc}
L_n^{(i)} = \log\frac{W(y_i|x_i=0)}{W(y_i|x_i=1)},
\end{equation}
for each element of the channel output vector and feeds it into a decoder, here the SC decoder. 

In SC decoding, information bits are estimated by a hard decision based on the final evolved LLRs $\lambda_i^0$. When decoding the $i$-th bit, if $i \notin \mathcal{A}$, regardless of the final LLR value $\lambda_i^0$, $\hat{u}_i$ is set as a frozen bit, that is, $\hat{u}_i=0$. Otherwise, $u_i$ is decided by a maximum likelihood (ML) rule as equation \eqref{eq:sc_hard_decision_ch2} based on the previously estimated vector $(\hat{u}_1, ..., \hat{u}_{i-1})$.
\begin{equation}
\label{eq:sc_hard_decision_ch2}
\hat{u}_i = h(L^{(i_0)}) = \begin{dcases*}
        0 & $L^{(i)}_0 = \log \frac{W_N^{(i)}(y_1^N,\hat{u}_0^{i-1}|\hat{u}_i=0)}{W_N^{(i)}(y_1^N,\hat{u}_0^{i-1}|\hat{u}_i=1)}>0$,\\
        1 & otherwise\\
\end{dcases*}
\end{equation}

The decoder estimates the transmitted bits successively by computing LLRs of the indexed edges. The LLR of edge $(i,j)$ is computed by
\begin{equation}
\label{eq:edge_llr}
L^i_j = \begin{dcases*}
        f(L^i_{(j+1)},L^{i+2^j}_{j+1})  & if $B(i,j)=0$\\
        g(L^{(i-2^j)}_{j+1},L^i_{(j+1)},\hat{\beta}^{i-2^j}_j)  & if $ B(i,j)=1$\\
\end{dcases*}
\end{equation}
where $0\leq i<N$, $0\leq j\leq n$, $B(i,j)=\lfloor \frac{i}{2^j}\rfloor\mod 2$, and $\hat{\beta}^{i}_j$ denote the partial sum, which corresponds to the propagation of estimated bits $\hat{u}_i$ backward into the factor graph, from right to left in Fig. \ref{fig:FG_8}. Note that $i$ and $j$ denote the bit index and the stage index, respectively, and the channel/intermediate LLRs are sometimes denoted by $\lambda$ (we use it in the rest of this subsection) or $\alpha$. The reason is that depending on the decoding algorithm, $L$ might be used for other purposes. 

\begin{figure}[ht]
	\begin{center}
    	\includegraphics[width=0.8\columnwidth]{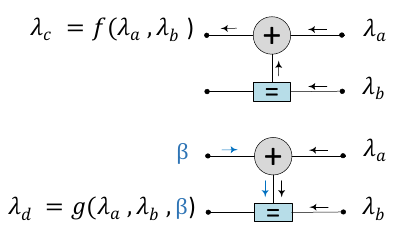}
    \end{center}
    \caption{Internal LLR ($\lambda$) calculations \cite{rowshan2021towards}}
    \label{fig:sc_decoder_2b}
\end{figure}

The functions $f$ and $g$ in \eqref{eq:edge_llr} as illustrated in Fig. \ref{fig:sc_decoder_2b}, equivalent to \eqref{eq:check_node} and \eqref{eq:var_node}, respectively, can be obtained by min-sum approximations, similar to decoding algorithm used for Reed-Muller codes in \cite{schnabl1995soft}, by
\begin{equation}
\label{eq:check_node}
 f(\lambda_a,\lambda_b)\approx\sgn(\lambda_a)\cdot \sgn(\lambda_b)\cdot \min(|\lambda_a|,|\lambda_b|)
\end{equation}
\begin{equation}
\label{eq:var_node}
 g(\lambda_a,\lambda_b,\hat{\beta})=(-1)^{\hat{\beta}}\lambda_a+\lambda_b
\end{equation}
where $\lambda_a$ and $\lambda_b$ are the incoming LLRs to a node and $\hat{\beta}$ is the partial sum of previously decided bits.


\subsubsection{Belief propagation (BP) Decoding}\label{sssec:bp}
\begin{figure}
    \centering
    \includegraphics[width=0.7\columnwidth]{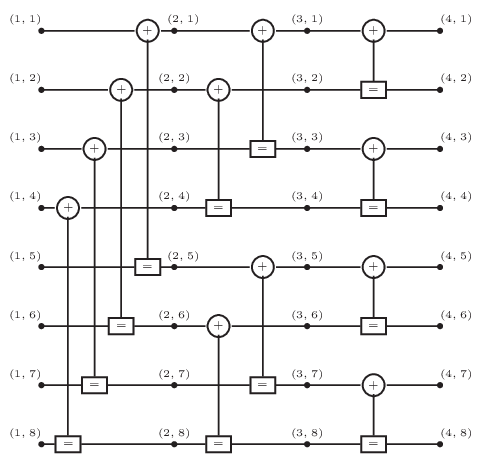}
    \caption{The factor graph in Fig. \ref{fig:FG_8} labeled for BP decoding \cite{arikan2010polar}.}
    \label{fig:fact_graph_bp}
\end{figure}

Belief propagation decoding uses graphical models and message passing to iteratively update the beliefs or probabilities of the transmitted codeword symbols based on received channel observations. It is commonly used for low-density parity-check (LDPC) codes (see Section \ref{sec:ldpc}) and turbo codes (see Section \ref{sec:turbo}). 
The sparseness of the graph representation of the $\boldsymbol{G}_N$ transformation in Fig. \ref{fig:FG_8} suggests that Gallager's belief propagation
algorithm \cite{gallager1962low} can be effective in decoding polar codes. In Fig. \ref{fig:fact_graph_bp}, the nodes are labeled with pairs of integers $(i, j), 1 \leq i \leq n+1,1 \leq j \leq N$. The leftmost nodes, $(1, j)$, correspond to the source data $u_j$ that are to be estimated, while the rightmost nodes $(n+1, j)$ are associated with channel input variables $x_j$ that are transmitted through a noisy channel. 
The BP decoder associates two messages to each node $(i, j)$: a right-propagating message $R_{i, j}^{(t)}$ and a left-propagating message $L_{i, j}^{(t)}$, where $t\geq0$ denotes the time index. These messages correspond to log-likelihood ratios (LLR) at time $t$ and are set as
$$
\begin{aligned}
L_{n+1, j}^{(0)} & =\log\frac{P\left(x_j=0 \mid y_j\right)}{P\left(x_j=1 \mid y_j\right)} \\
R_{\text {1, j }}^{(0)} & =\log\frac{P\left(u_j=0\right)}{P\left(u_j=1\right)} 
 = \begin{cases}0 & \text { if } j\in\mathcal{A}  \\
\infty & \text { otherwise }\end{cases}
\end{aligned}
$$
Note that a frozen coordinate is by default $u_j=0,j\in\mathcal{A}^c$, hence $R_{1, j}^{(0)}=\infty$. However, the $u_j=0 \text{ and } 1$ are equally likely values for information coordinates $\j\in\mathcal{A}$, i.e., $P\left(u_j=1\right)=P\left(u_j=0\right)$, thus $R_{1, j}^{(0)}=0$. The rest of $R_{i, j}^{(0)}$ and $L_{i, j}^{(0)}$ are initialized to 0. 

\begin{figure}
    \centering
    \includegraphics[width=0.7\columnwidth]{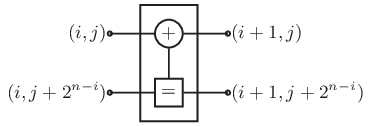}
    \caption{The basic computational block in factor graphs  \cite{arikan2010polar}.}
    \label{fig:basic_block}
\end{figure}

In each stage of Fig. \ref{fig:fact_graph_bp}, there are $\frac{1}{2}N \log N=4, \text{ for }N=8$ basic computational blocks shown in Fig. \ref{fig:basic_block}. The upper node is a check node representing the parity-check constraint, while the lower node is a variable node. This computational block computes the left/right-propagating  messages based on three inputs as 
$$
\begin{aligned}
L_{i, j}^{(t+1)} & =f\left(L_{i+1, j}^{(t)}, L_{i+1, j+N_i}^{(t)} R_{i, j+N_i}^{(t)}\right) \\
L_{i, j+N_i}^{(t+1)} & =L_{i+1, j+N_i}^{(t)} f\left(L_{i+1, j}^{(t)}, R_{i, j}^{(t)}\right) \\
R_{i+1, j}^{(t+1)} & =f\left(R_{i, j}^{(t)}, L_{i+1, j+N_i}^{(t)} R_{i+1, j+N_i}^{(t)}\right) \\
R_{i+1, j+N_{i}}^{(t+1)} & =R_{i, j+N_{i}}^{(t)} f\left(R_{i, j}^{(t)}, L_{i+1, j}^{(t)}\right)
\end{aligned}
$$
where $f(x, y)=(1+x y) /(x+y)$ for real values $x, y$ and $N_i=2^{n-i}$.

The messages carry information about the reliability of each symbol or constraint and are exchanged between the variable and the check nodes. At each iteration represented by time index $t$, a variable node calculates its outgoing messages based on the incoming messages from the connected check nodes. Similarly, check nodes update their outgoing messages based on the incoming messages from the connected variable nodes. These message updates are repeated until a stopping criterion is met or until convergence is achieved. The stopping criterion is satisfying $\boldsymbol{x}=\boldsymbol{u}\boldsymbol{G}_N$ where $\boldsymbol{u}$ and $\boldsymbol{x}$ are the hard decision (similar to \eqref{eq:sc_hard_decision_ch2}) made on the left-most and right-most messages, respectively. 

Belief propagation decoding is known for its effectiveness in achieving near-optimal decoding performance for low-density parity-check codes due to possessing a sparse parity-check matrix. However, it may suffer from convergence issues in the presence of high noise levels or certain code structures.

\subsubsection{SC List Decoding}\label{sssec:scl}
As we discussed in Section \ref{sssec:sc}, the successive cancellation decoder locally makes a hard decision for each bit value $u_i$  and proceeds with decoding the subsequent bits. Although the decision at each decoding step is locally optimal, given that the previous bits have been decoded correctly, it is not necessarily globally optimal. To overcome this shortcoming, one can consider both options for the value of each bit, $u_i=0,1$, and form a binary tree \cite{tal2015list,dumer2006soft,lucas1998improved}. A path in the tree from the origin to a leaf is a solution to decoding. 
Obviously, the path with the highest likelihood will be selected from the list for optimal decoding. That is, the probability to be maximized is
$$
P\left(\hat{u}_0^{N-1} \mid y_0^{N-1}\right)=\prod_{t=0}^{N-1} P\left(\hat{u}_t \mid \hat{u}_0^{i-1}, y_0^{N-1}\right). 
$$
However, we cannot explore the entire decoding tree to examine all paths and we constrain the traversal to $L$ paths. Therefore, the solution obtained from the decoding may be sub-optimal. In SCL decoding, the probability of the partial path $l\in[1,L]$ representing the sequence $\hat{u}_0^{i-1}=\left(\hat{u}_0, \hat{u}_1, \ldots, \hat{u}_{i-1}\right)$ is calculated by
\begin{equation}\label{eq:sc_metric_prob}
    P\left(\hat{u}_0^i[l] \mid y_0^{N-1}\right)=\prod_{t=0}^i P\left(\hat{u}_i[l] \mid \hat{u}_0^{i-1}[l], y_0^{N-1}\right).
\end{equation}
In practice, the logarithm of \eqref{eq:sc_metric_prob} is used. Since $\log (x)<0$ for $x<1$, we multiply the resulting logarithm by -1 to have a positive metric. Therefore, we get the following logarithmic path metric for the sequence $\hat{u}_0^{i-1}$ on path $l$ \cite[Section 2.5.2]{rowshan2021towards}:
\begin{equation}\label{eq:scl_metric_prob}
\begin{aligned}
& P M_l^{(i-1)}=-\log P\left(\hat{u}_0^{i-1}[l] \mid y_0^{N-1}\right) \\
& \quad=-\sum_{j=0}^{i-1} \log P\left(\hat{u}_j[l] \mid \hat{u}_0^{j-1}[l], y_0^{N-1}\right)\\
& \quad=P M_l^{(i-1)}+\mu_l^{(i)}.
\end{aligned}
\end{equation}
where $\mu_l^{(i)}=-\log P\left(\hat{u}_i[l] \mid \hat{u}_0^{i-1}[l], y_0^{N-1}\right)$ denotes the branch metric and $P M_i^{(-1)}=0$. 
A genie such as CRC bits (see Section \ref{ssec:concat_pretrans}) can also help detect the correct path \cite{niu2012crc}, that is, the transmitted data. The same list decoding procedure can be used with other concatenated codes, such as PAC codes or PC-polar codes.  

\begin{figure}
    \centering
    \includegraphics[width=1\columnwidth]{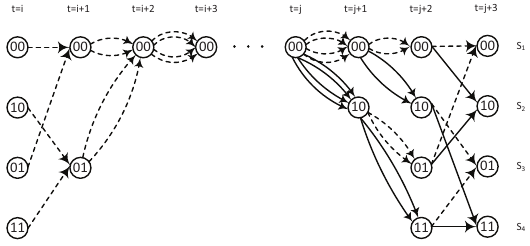}
    \caption{Irregular trellis in list Viterbi decoding where $v_t=0$ for $t=[i+1, \ldots, j]$ and $t \in \mathcal{A}^c$. Note that the paths from $t=i+1$ to $t=j$ are not pruned. \cite{rowshan2021list}.}
    \label{fig:viterbi_trellis}
\end{figure}

For PAC codes, one can use SC-based list Viterbi decoding \cite{rowshan2021list}. As a result of the convolutional transformation, each path is associated with a single state at each decoding step. Fig. \ref{fig:viterbi_trellis} illustrates the trellis as a graphical representation of the Viterbi list decoding with list size 4 (the number of surviving paths), equal to the number of states. 

\subsubsection{Simplified/Fast SC-based Decoding}\label{sssec:symp_sc}
\begin{figure}
    \centering
    \includegraphics[width=1\columnwidth]{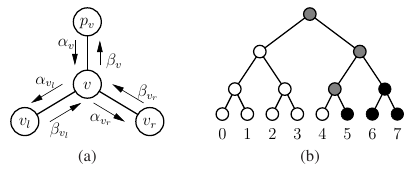}
    \caption{(a) Local decoder, (b) Labelling the leaf nodes for $2^n=8$. The white circles represent rate-0 nodes, while the rate-1 nodes are shown as black circles. \cite{alamdar2011simplified}.}
    \label{fig:simplified_SC}
\end{figure}

The distribution of frozen coordinates in sequence $[0,N-1]$ forms sub-sequences of length $2^s,s=0,\dots,n-1$ with specific patterns. The two trivial ones are the sub-sequences with all frozen coordinates and no frozen coordinates. As shown in Fig. \ref{fig:simplified_SC}, the nodes (in the tree that represents the factor graph) associated with these leaves representing these sub-sequences are called rate-0 and rate-1 nodes, respectively.  The update rule in \eqref{eq:var_LLR}, that is, for LLRs values of $x$ and $y$ we have $x \boxast y \triangleq2 \operatorname{atanh}(\tanh (x / 2) \tanh (y / 2))$, has the following property that can be exploited to simplify the message passing process:
\begin{equation}
    h(x \boxast y)=h(x) \oplus h(y) \text { if } x\cdot y \neq 0.
\end{equation}
Let $V_v$ denote the set of nodes of the subtree rooted at node $v$, and $\ell(u)$ denote the index of a leaf node $u$. Furthermore, we define the set $\I_v$ for each node $v$ that contains the indices of all leaf nodes that are descendants of node $v$, as follows:
$$
\mathcal{I}_v=\left\{\ell(u): u \in V_v \text { and } u \text { is a leaf node }\right\}.
$$
Now, when a rate one-1 $v$ is activated, it immediately calculates $\beta_v$ via
$$
\beta_v=h\left(\lambda_v\right),
$$
and all bits with indices in $\mathcal{I}_v$ can be immediately decoded using 
$$
\left(\hat{u}\left[\min \mathcal{I}_v\right], \ldots, \hat{u}\left[\max \mathcal{I}_v\right]\right)=\beta_v G_{n-d_v},
$$
where $d_v$ indicate the depth of node $v$. Effectively, the decoding of a rate-1 node is simply finding the constituent code by hard-decision $h(\cdot)$, see \eqref{eq:sc_hard_decision_ch2}, of soft-input values $\lambda_v$ and computing the inverse transform which is $G_{n-d_v}^{-1}=G_{n-d_v}$ (see Section \ref{ssec:properties}). 
The decoding of rate-0 nodes is simpler. Since all descendants of a rate zero node $v$ are themselves rate-0, node $v$ can immediately set the constituent code $\beta_v$ to an all-zero sequence. 

Later, these special nodes were extended to single parity check (SPC) nodes and repetition (REP) nodes \cite{sarkis2014fast}, type-I,II,III,IV nodes,  \cite{hanif2017fast} and sequence repetition node \cite{ren2022sequence}, and adapted to other decoding algorithms such as SC list decoding \cite{hashemi2015list}. 

\subsubsection{Iterative SC-based decoding}\label{sssec:iter_sc}
Using CRC bits for block error detection, one can recover the correct path (transmitted data) in additional $T$ attempts by making different decisions, i.e., flipping the initial estimate of the bit (s), in SC decoding \cite{afisiadis2014low,8259253}, or effectively following different paths in the decoding tree in SC list decoding \cite{rowshan2019improved,rowshan2022efficient}. The earlier decoding is called SC-flip (SCF) decoding, and the latter is referred to as SC list decoding with shifted-pruning (SP), SCL-flip (SCLF) decoding, or other names such as SCL decoding with list-flipping and path flipping \cite{rowshan2022sc,rowshan2021shifted,8705213,9294144,9622756,9684773,10180029,9742711,10279400,e24121806,10038849,rowshan2019repetition}. Furthermore, there is an efficient way to start re-decoding in the additional iterations by partial rewinding, that is, the re-decoding process is performed by partially rewinding the SC process, not necessarily from the first bit. This can significantly reduce the complexity of decoding in the SCF and SCLF decoding algorithms. Recently, a different approach called SCL-perturbation decoding has been developed \cite{10464952}, which is based on the idea of adding noise or perturbation to the received sequence in the literature, where small random perturbations are introduced to the received symbols before each SCL decoding attempt. Numerical results show that the perturbation yields a higher coding gain than flipping at a large block length with the same number of decoding attempts.

\subsubsection{Automorphism Ensemble Decoding}\label{sssec:ensemble}
\begin{figure}
    \centering
    \includegraphics[width=1\columnwidth]{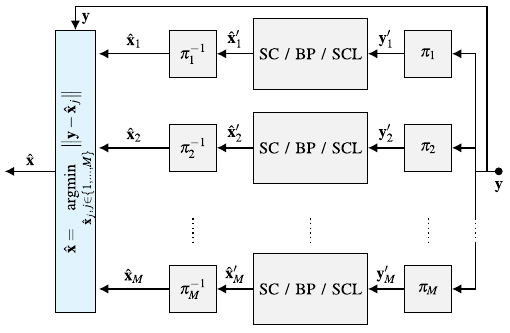}
    \caption{Automorphism/Permutation Ensemble Decoding scheme \cite{geiselhart2021automorphism}.}
    \label{fig:AEDec}
\end{figure}

A different approach to decoding is to exploit the symmetries of the codes and to have a set of decoders, where each decodes a distinct permuted received sequence $\pi_i(\by),i\in[1,M]$, as shown in Fig. \ref{fig:AEDec}. The output of every decoder as an estimated transmitted codeword is then de-permuted and the one with maximum likelihood is chosen \cite{geiselhart2021automorphism}. That is, 
\begin{equation}
\hat{\boldsymbol{x}}=\underset{\hat{\boldsymbol{x}}_j, j \in\{1, \ldots, M\}}{\operatorname{argmin}}\left\|\hat{\boldsymbol{x}}_j-\boldsymbol{y}\right\|^2=\underset{\hat{\boldsymbol{x}}_j, j \in\{1, \ldots, M\}}{\operatorname{argmax}} \sum_{i=0}^{N-1} \hat{x}_{j, i} \cdot y_i,
\end{equation}
where
\begin{equation}    \hat{\boldsymbol{x}}_j=\pi_j^{-1}\left(\operatorname{Dec}\left(\pi_j(\boldsymbol{y})\right)\right).
\end{equation}
An automorphism group of a code $\C$ is the set of permutations that map codewords to other codewords and leave the code globally invariant. This approach, called automorphism group decoding, has been employed for Bose-Chaudhuri-Hocquenghem (BCH) codes in \cite{hehn2008permutation}. 
The automorphism group of decreasing monomial codes, including polar codes, was shown in \cite{bardet2016algebraic} to be at least the lower-triangular affine group (LTA), which relies on the partial order of sub-channels. This was later extended to block LTA \cite{li2021complete}. 

This approach can decode $M$ permuted received sequences in parallel with potentially the latency of the component decoding, as it does not require the path selection stages used in SCL decoding. However, the choice of permutations to achieve a performance close to SCL decoding is challenging despite an effort in \cite{pillet2022classification} to classify the permutations. 

\subsubsection{Other Near-ML and ML Decoding Algorithms}\label{sssec:ML} 
Among other decoding algorithms adapted to polar codes and their variants, we can name the following decoding algorithms: 
\begin{itemize}
    \item Soft CANcellation (SCAN) decoding \cite{fayyaz2014low} is an iterative decoder that allows one to reduce the number of iterations of the BP decoder (see Section \ref{sssec:bp}) by adopting the SC schedule (see Section \ref{sssec:sc}). 
    Unlike SC decoding, SCAN decoding propagates soft information in both directions. 
    \item Fano decoding \cite{massey1972variable,jeong2019sc,rowshan2020complexity,TIMOKHIN2023104008} is a variable-complexity ML decoding algorithm where the decoder examines a sequence of adjacent non-frozen bits and explores the partial paths of the code tree by moving forward and backward, aiming to find the maximum likelihood path. 
    \item Stack decoding \cite{jelinek1969fast,niu2012stack,rowshan2021polarization} keeps a stack of size/depth $D$ of partial paths sorted with respect to the path metric. The algorithm extends the path with the best metric at the top of the stack. The stack decoding is a memory-intensive algorithm with a variable time complexity that, instead of backtracking as in the Fano decoding, selects to extend the best partial path in the stack at each time step. 
    \item Sphere decoding \cite{pohst1981computation,kahraman2012code,gu2022selective}  based on the depth-first approach first finds a candidate solution close to the received sequence in terms of the Euclidean distance (ED). Then, it searches for a closer candidate to the received sequence by tree pruning, if there exists. 
    \item Generic/universal decoding algorithms such as ordered statistics decoding (OSD) \cite{fossorier1995soft,wu2016ordered} and guessing random additive noise decoding (GRAND) \cite{duffy2022ordered,rowshan2022constrained,rowshan2023segmented} are reliability-based algorithms that guess the error pattern introduced through the channel in a specific order and subsequently check them using a systematic generator matrix or a parity check matrix, respectively, to find the candidate closest to the received sequence. They have shown remarkable performance of ML/near-ML decoding, given enough trials for search. Among them, GRAND best suits high-rate codes. 
\end{itemize}
The algorithms above predominantly provide excellent performance at the cost of high and variable complexity. We compare the hardware implementation of these algorithms in Section \ref{ssec:hardware}. 

\subsection{Rate-Compatible Polar Codes}\label{ssec:rate-compatible}
As discussed in Section \ref{ssec:polarization}, the length of polar codes is restricted to the powers of two, $N=2^n, n\geq 1$. 
To accommodate the various practical $E$-length requirements or equivalently rate requirements, a mother code $(N,K)$ is punctured or shortened according to the employed index pattern, by $P$ bits or $S$ bits, respectively. Note that the focus of this section is on non-systematic polar codes. 

The punctured code $(N-P,K)$ with length $E_p=N-P$, where $P<N-K$, clearly has a higher rate, $K/(N-P) > K/N$. The $P$ coded bits are not transmitted, and the decoding is performed on the mother code of length $2^{\left\lceil\log _2 E_p\right\rceil}$ considering the punctured bits as erased, that is, the corresponding LLRs are set to zero. The LLRs of these unreliable bits in the decoder are evolved to zero at the output of the decoder, resulting in \emph{incapable bits} \cite{shin2013design}. Therefore, these indices should be included in the frozen set $\A^c$ to avoid a drastic error rate. Note that the number of incapable bits is equal to the number of punctured bits \cite{shin2013design}.  
In the 5G standard, the puncturing set $\P$ is given by $\P = \{0, \dots , P-1\}$, that is, the first $P$ indices of $[0,N-1]$ in natural order (alternatively in bit-reversal order \cite{bioglio_low-complexity_2017}) are considered frozen. The rest of the $N-K$ frozen bits are chosen from the reliability sequence, from the least reliable indices, in a circular buffer configuration \cite{noauthor_5g_2020}. 

\begin{figure}
    \centering
    \includegraphics[width=1\columnwidth]{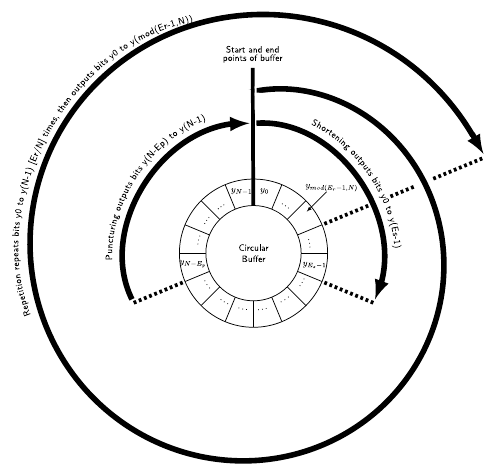}
    \caption{The Circular buffer in 5G for rate-matching \cite{egilmez2019development}.}
    \label{fig:circ_buffer}
\end{figure}

On the other hand, for the shortened polar code $(N-S,K)$ with length $E_s=N-S$, there are $S$ coded bits restricted to zero in all codewords of its mother code $(N,K)$. Therefore, these known bits are not transmitted and the decoding is performed on the mother code of length $2^{\left\lceil\log _2 M\right\rceil}$ considering the known shortened bits, That is, the corresponding LLRs are set to $+\infty$ (a large value). The LLRs of these known bits in the decoder are evolved to large values at the output of the decoder, resulting in \emph{overcapable bits}. Therefore, these indices can be included in the frozen set $\A^c$, since the source bits $\bu_{\S}$ corresponding to the coded bits $\bx_{\S}$ are also zero, i.e. $\bu_{\S}=\bx_{\S}=\bzeros$. Note that similarly to incapable bits, the number of overcapable bits is equal to the number of shortened bits. 
The shortening set $\S$ in the 5G standard is given by $\S=(N-S, \dots, N-1)$, the last $S$ indices of $[0,N-1]$ in natural order (alternatively in bit-reversal order \cite{bioglio_low-complexity_2017}). 
Fig. \ref{fig:circ_buffer} illustrates the circular buffer used in 5G based on the natural order of the bit indices. 

\subsection{Polar Coded Modulation (PCM)} \label{ssec:polar_coded_modulation}
Let $W: \S \rightarrow \mathcal{Y}$ be a $2^m$-ary channel with input symbol from the constellation set $\S$ of order $|\S|=2^m$, and output alphabet $\Y$. 
Each symbol in the constellation is labelled with a binary $m$-tuple, and we say that the symbols in this constellation have $m$ bit levels. A binary labeling rule, denoted by $\mathcal{L}:\{0,1\}^m \rightarrow \S$, maps binary $m$-tuples bijectively to the $2^m$ input symbols $s\in\S$. The popular labelling rules are gray labelling and set partitioning. In the following, we review the two major polar coded modulation schemes: 
\subsubsection{Multilevel Polar Coded Modulation (ML-PCM)} \label{sssec:multilevel_polar_coded_modulation}
In an ML-PCM, there exist $m$ component polar codes, each of length $N$. 
At the encoding stage, a binary vector of length $mN$ consisting of both data bits and frozen bits that are placed together according to an overall rate profile is divided into $m$ vectors of equal length. Then, each $N$-length subvector is polarly encoded as a component polar code. 
Let $\bx_j = (x_{j,1}, x_{j,2}, \dots, x_{j,N})$ denote the coded bits of the $j$-th component code for $j = 1, 2, \dots ,m$. Then, at the modulation stage, the $m$-tuple $(x_{1,i}, x_{2,i}, \dots , x_{m,i})$ for $i = 1, 2, \dots, N$ is mapped to a constellation symbol for transmission. In this way, in every channel use, the bits of each component code only appear at the corresponding single-bit level. For example, in channel use $i=1,\dots, N$ modulated with 4QAM constellation that maps every 2 bits to one symbol point, we will have two component polar codes $\bx_m,m=1,2$ where every bit pair of $x_{1,i}x_{2,i}$ is mapped to a symbol. 

In the receiver, the demodulation and decoding are performed for each bit level sequentially, as shown in Fig. \ref{fig:coded_mod} (right). In the previous example, the reliability information corresponding to the first bit level ($m=1$) of all received symbols is computed as $y_{1,1}y_{1,2}$ and the associated component code is decoded. Then the decoding results are used for demapping $y_{2,1}y_{2,2}$ corresponding to the bit level $m=2$ and, consequently, decoding. The decoding results are passed for demapping and decoding the next bit levels. 
Observe that set partitioning better suits this approach, which is called \emph{multi-stage} demodulation and decoding. 

As observed above, in a multilevel coding approach, the raw channel $W$ is effectively decomposed into $m$ binary sub-channels while preserving their mutual information. Let us denote this channel decomposition, which is called the sequential binary partition (SBP) \cite{seidl2013polar}, as
$$
\psi: W \rightarrow\left\{B_\psi^{(1)}, B_\psi^{(2)}, \cdots, B_\psi^{(m)}\right\},
$$
where $B_\psi^{(j)}:\{0,1\} \rightarrow \mathcal{Y} \times\{0,1\}^{j-1}$ is the binary sub-channel corresponding to the bit level $j=1,2, \dots, m$. Therefore, an $M$-ary channel $W$ is divided into $m$ bit levels (or sub-channels) $\mathrm{B}_\psi^{(i)}(0 \leq i<m)$ that are B-DMCs as long as $W$ is a DMC. In SBP, each sub-channel $B_\psi^{(j)}$ has the knowledge of both the channel output $y \in \mathcal{Y}$, and their previous bit levels obtained through decoding. 
Mutual information between channel input $\S$ and channel output $\Y$ of $W$, assuming equiprobable source symbols, is referred to as the coded modulation capacity $C_{\mathrm{cm}} (\mathrm{W})$ \cite{wachsmann1999multilevel}: 
$$
C_{\mathrm{cm}}(\mathrm{W}):=I(X ; Y)=\sum_{i=0}^{m-1} I\left(\mathrm{~B}_\psi^{(i)}\right).
$$
This capacity does not depend on a specific labeling rule $\mathcal{L}$. Note that according to \cite[Theorem 1]{arikan2009channel}, as the blocklength of component codes increases, the polar component codes approach the bit level capacities $I\left(\mathrm{~B}_\psi^{(i)}\right)$. 

A drawback of the MLC approach for practical use lies in the need to use several relatively short component codes with different rates for various bit levels. Given the length requirement of the power of twos in polar coding, this constraint adds to the disadvantages of the multilevel coding approach. 

\subsubsection{Bit-Interleaved Polar Coded Modulation (BI-PCM)} \label{sssec:bit_interleaved_polar_coded_modulation}
\begin{figure}
    \centering
    \includegraphics[width=0.7\columnwidth]{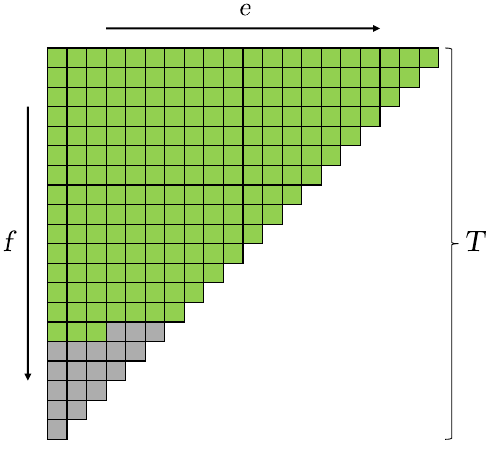}
    \caption{Triangular structure for channel interleaving in 5G.}
    \label{fig:ch_interleav_5G}
\end{figure}

The rate-matched binary coded bits $\boldsymbol{e}$ of length $E$ are permuted by an interleaver. The channel interleaver in the 5G standard is formed by an isosceles triangular structure of length $T$ bits, where the interleaver depth $T$, the maximum separation between two consecutive bits, is obtained by $T=\left\lceil\frac{\sqrt{8 E+1}-1}{2}\right\rceil$. The encoded bits in the vector $\boldsymbol{e}$ are written in the rows of the triangular structure shown in Fig. \ref{fig:ch_interleav_5G}, while the interleaved vector denoted by $\boldsymbol{f}$ is obtained by reading the bits column-wise, skipping the NULL entries. 
The triangular structure can be represented as a matrix $\boldsymbol{V}$ of size $T \times T$, where the entries are formed as
$$
V_{i, j}= \begin{cases}\text { NULL } & \text { if } i+j \geq T \text { or } r(i)+j \geq E \\ e_{r(i)+j} & \text { otherwise }\end{cases}
$$
where $r(i)=\frac{i(2 T-i+1)}{2}$. 

Then, each subblock of $m$ bits of vector $\boldsymbol{f}$ is mapped to a constellation symbol in $\S$ for channel transmission. When a symbol is received, the demodulator on the other end disregards the correlation between bit levels and calculates soft information for all bit levels based only on the channel observation. 

In a BI-PCM scheme, the channel $W$ is decomposed into $m$ binary sub-channels that are viewed as independent channels by the receiver. This channel transform is called \emph{parallel binary partition} (PBP) \cite{seidl2013polar} denoted as
$$
\varphi: W \rightarrow\left\{B_{\varphi}^{(1)}, B_{\varphi}^{(2)}, \cdots, B_{\varphi}^{(m)}\right\},
$$
where the channel $\mathrm{W}$ is mapped to a set of mutually independent binary sub-channels $B_{\varphi}^{(j)}:\{0,1\} \rightarrow \mathcal{Y},j=1,2, \cdots, m$ for the $j$-th bit level. 
These sub-channels have symmetric capacities of
$$
I\left(\mathrm{~B}_{{\varphi}}^{(i)}\right):=I\left(B_i ; Y\right).
$$
Obviously, when compared to the corresponding sub-channel in SBP, for all pairs of sub-channels, the following holds: 
\begin{equation}
\begin{aligned}
I\left(\mathrm{~B}_{{\varphi}}^{(i)}\right) & =I\left(B_i ; Y\right) \\
& \leq I\left(B_i ; Y \mid B_0, \ldots, B_{i-1}\right)=I\left(\mathrm{~B}_{\varphi}^{(i)}\right)
\end{aligned}
\end{equation}

In PBP, each sub-channel $B_{\varphi}^{(j)}$ only uses the channel output $y \in \mathcal{Y}$, while SBP requires knowledge of previous bits $\{0,1\}^{j-1}$. 
The commonly used labeling rule for sub-channels in BI-PCM is gray labeling which generates bit levels that are as independent as possible, assuming the sub-channels correspond to bit levels with symmetric capacities that do not differ significantly.

\begin{figure}
    \centering
    \includegraphics[width=1\columnwidth]{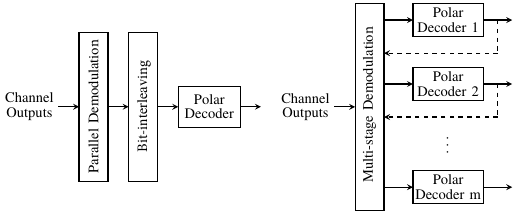}
    \caption{Multi-stage decoding vs the decoding of bit-interleaved codewords.}
    \label{fig:coded_mod}
\end{figure}

After demodulation, as shown in Fig. \ref{fig:coded_mod} (left), the soft information of all $mN$ bits is de-interleaved, and fed to the de-rate-matcher. Note that to use a single polar decoder for BI-PCM, 
the order $m$ of the constellation has to be a power of 2.

Unlike BICM where each codeword is independently modulated and transmitted in a single time slot, delayed BICM (D-BICM) \cite{wu2021delayed} divides each codeword into $m$ subblocks and modulates the subblocks from both the previous time slots and the current time slot onto the same signal sequence. The receiver starts decoding when all sub-blocks of a codeword are received. Next, the decoded delayed subblocks in the current time slot are used as a-priori information to improve the detection of the undelayed subblocks in the succeeding time slots.


\subsection{Error Correction Performance}\label{ssec:coparison} 
In this section, we compare known variants of polar codes, namely CRC-polar codes and PAC codes (with and without CRC concatenation). These results are obtained for AWGN channel and binary phase shift keying (BPSK) modulation. 
For this purpose, we choose two codes; the short code (128, 64) and the medium-length code (1024, 512). They are within the length range used for 5G, as mentioned in Table \ref{tbl:5G_coding_params}. The CRC and convolutional polynomials used for precoding in CRC-polar and PAC codes are $g_{\mathrm{CRC11}}(D)=D^{11}+D^{10}+D^9+D^5+1$ and $g_{\mathrm{Conv6}}(D)=1+D^{2}+D^{3}+D^5+D^6$, respectively.
To have a thorough comparison, we consider three sets of setups where in each setup, only one coding factor is changing. These factors are: 
\begin{itemize}
    \item coding scheme and code construction: As Table \ref{tab:fer_snr_all_constructions} shows, we choose CRC-polar, PAC and CRC-PAC constructed using 5G reliability sequence and its modified version called 5G-RM (see Section \ref{ssec:construct}), 
    \item code rate: $1/4, 1/2, 3/4$, as Table \ref{tab:fer_snr_all_rates} states, and
    \item decoding algorithms according to the setup in Table \ref{tab:fer_snr_all_decoders}.
\end{itemize}
Note that there are quite a number of methods to construct a code. However, they are either not explicit or they have a high complexity, which makes them impractical and not a good candidate to be adapted to a standard. Similarly, there are other decoding algorithms that we do not use in this comparison; however, we have discussed them in Section \ref{ssec:decoding}. In the following, we discuss each scenario in the same order as above.

\begin{figure}
    \centering
    \includegraphics[width=1\columnwidth]{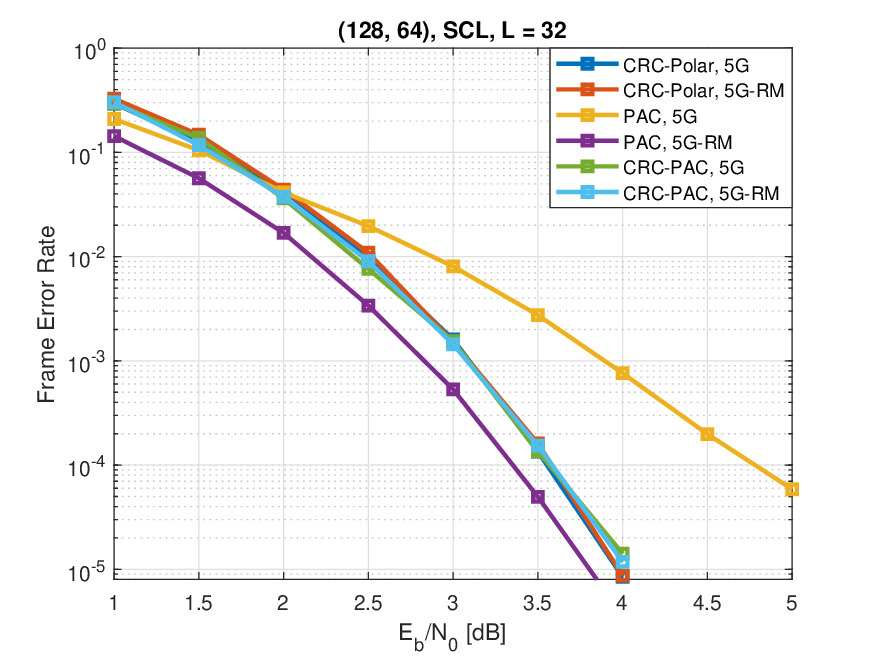}
    \caption{FER comparison between code constructions of polar codes and PAC codes with $N = 128, K = 64$.}
    \label{fig:polar_N_128_different_code_constructions}
\end{figure}

\begin{figure}
    \centering
    \includegraphics[width=1\columnwidth]{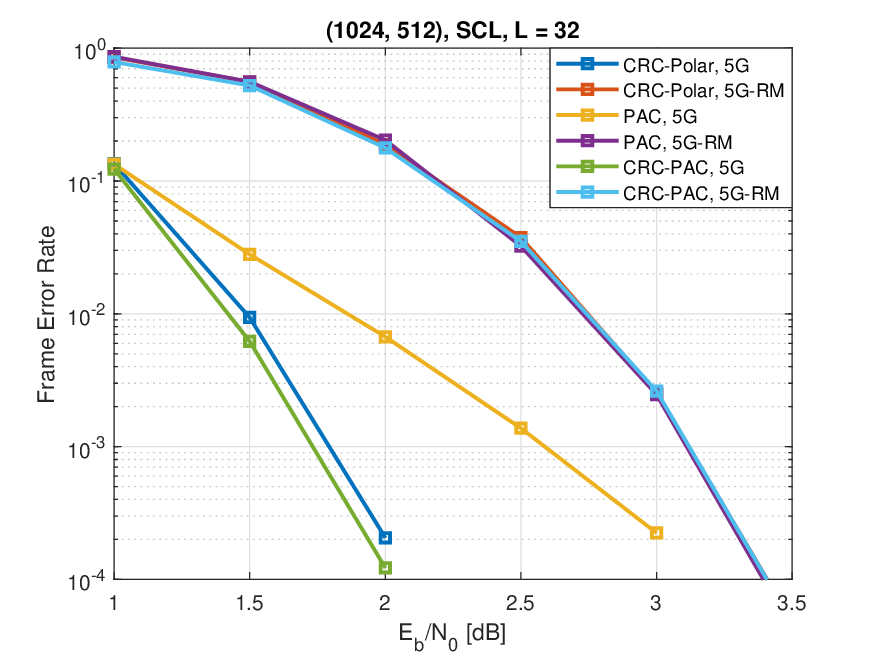}
    \caption{FER comparison between code constructions of polar codes and PAC codes with $N = 1024, K = 512$.} 
    \label{fig:polar_N_1024_different_code_constructions}
\end{figure}

\begin{table}[h]
    \centering
     \caption{Evaluation of the frame error rate performance versus $E_b/N_0$ for variants of polar codes with two different constructions.}
    \label{tab:fer_snr_all_constructions}
    \begin{tabular}{|c|c|c|c|}
        \hline Channel & \multicolumn{3}{|c|}{ AWGN } \\
        \hline Modulation & \multicolumn{3}{|c|}{ BPSK } \\
        \hline Coding variant & CRC-Polar & PAC & CRC-PAC \\
        \hline Constructions
         & \multicolumn{3}{|c|}{5G, 5G-RM} \\
        \hline Code rate, $R$ & \multicolumn{3}{|c|}{$1/2$} \\
        \hline Info. length, $K$ (bits w/o CRC)
         & \multicolumn{3}{|c|}{
        $64, 512$ } \\
        \hline Decoding Algorithm & \multicolumn{3}{|c|}{SCL, $L=32$}\\
        \hline
    \end{tabular}
\end{table}
Figs. \ref{fig:polar_N_128_different_code_constructions} and \ref{fig:polar_N_1024_different_code_constructions} illustrate the frame (or block) error rate for variants of the polar codes of (128, 64) and (1024, 512), respectively. In this scenario, summarized in Table \ref{tab:fer_snr_all_constructions}, we use the SCL decoder with the list size $L=32$ for all codes. 
As can be seen in Fig. \ref{fig:polar_N_128_different_code_constructions}, for the relatively short code (128, 64), the PAC code constructed with the 5G-RM construction outperforms the PAC codes with the 5G construction and the CRC-polar codes. Note that the 5G construction gives the minimum distance of $d=8$ for both PAC and CRC-polar codes, while the 5G-RM's minimum distance is $d=16$. Furthermore, CRC concatenation with short codes is punishing as it may decrease the minimum distance and utilizes more low-reliability sub-channels. When it comes to medium-length codes such as (1024, 512) in Fig. \ref{fig:polar_N_1024_different_code_constructions}, the 5G-RM construction performs poorly. The reason lies in employing severely bad sub-channels though the corresponding rows of $\bGN$ have weight that is larger than minimum weight/distance. Therefore, it is recommended to use the 5G construction for PAC coding in medium blocklength. To obtain a comparable result with CRC-polar codes in this range of blocklength, we need to concatenate PAC codes with CRC.

\begin{table}[ht]
    \centering
     \caption{Evaluation of the frame error rate performance versus $E_b/N_0$ for variants of polar codes at three different rates.}
    \label{tab:fer_snr_all_rates}
    \begin{tabular}{|c|c|c|c|}
        \hline Channel & \multicolumn{3}{|c|}{ AWGN } \\
        \hline Modulation & \multicolumn{3}{|c|}{ BPSK } \\
        \hline Coding variant & CRC-Polar & PAC & CRC-PAC \\
        \hline Constructions
         & \multicolumn{3}{|c|}{5G, 5G-RM} \\
        \hline Code rate, $R$ & \multicolumn{3}{|c|}{$1/4, 1/2, 3/4$} \\
        \hline Info. length, $K$ (bits w/o CRC)
         & \multicolumn{3}{|c|}{
        $64, 512 $ } \\
        \hline Decoding Algorithm & \multicolumn{3}{|c|}{SCL, $L=8$}\\
        \hline
    \end{tabular}
\end{table}

\begin{figure}
    \centering
    \includegraphics[width=1\columnwidth]{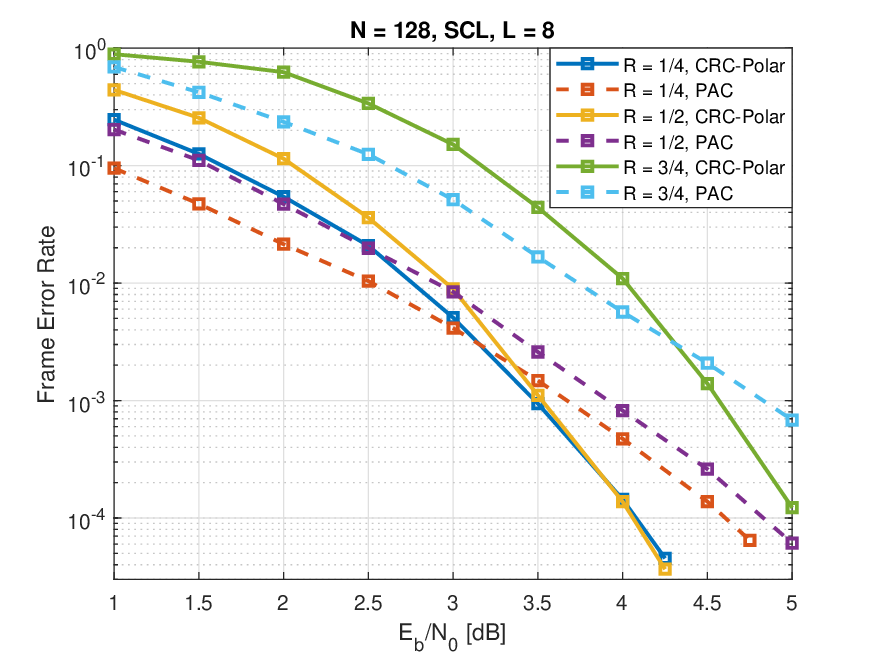}
    \caption{FER comparison between rates of polar codes with $N = 128$ under 5G construction.} 
    \label{fig:polar_N_128_different_rates}
\end{figure}

\begin{figure}
    \centering
    \includegraphics[width=1\columnwidth]{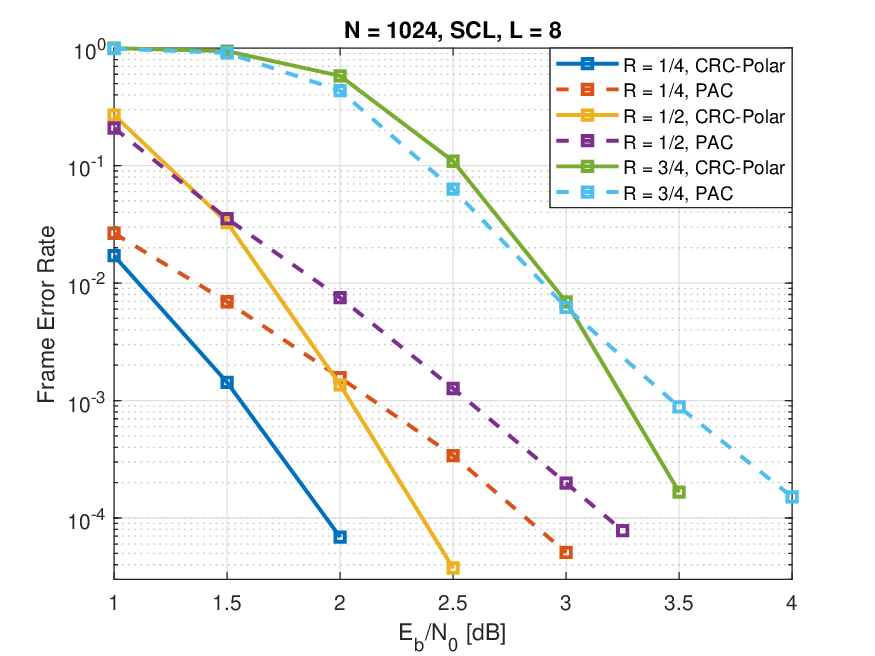}
    \caption{FER comparison between rates of polar codes with $N = 1024$ under 5G construction.}
    \label{fig:polar_N_1024_different_rates}
\end{figure}

Now, let us compare the considered codes under different code rates. 
Figs. \ref{fig:polar_N_128_different_rates} and \ref{fig:polar_N_1024_different_rates} demonstrate the FER performance of the CRC-polar and PAC codes for the two codes of (128, 64) and (1024, 512), respectively. In this scenario, summarized in Table \ref{tab:fer_snr_all_rates}, we use the SCL decoder with list size 8 and 5G construction for all codes. As can be seen, PAC codes constructed with the 5G sequence behave differently for short codes and medium-length codes. A short PAC code can outperform the corresponding CRC-polar code at low SNR regimes. At high code rates, this is also the case at medium SNR regimes. Note that short PAC codes can outperform CRC-polar codes when constructed with 5G-RM, as we observed in Fig. \ref{fig:polar_N_128_different_code_constructions}; however, here we use 5G for all PAC codes. On the contrary, at medium blocklengths, PAC codes cannot compete with CRC-polar and we need a CRC concatenation to improve it, as we observed in Fig. \ref{fig:polar_N_1024_different_code_constructions}. However, there is one exception for high-rate PAC codes.

Note that there is an approach \cite{rowshan2021error,rowshan2023formation} based on removing many minimum-weight codewords by a simple modification of the 5G-based construction of PAC codes so that it can outperform CRC-polar codes in low and medium SNR regimes for medium blocklengths and all rates. 


\setlength{\tabcolsep}{0.44em} 
\begin{table}[h]
    \centering
     \caption{Evaluation of the frame error rate performance versus $E_b/N_0$ for variants of polar codes under different decoding algorithms.}
    \label{tab:fer_snr_all_decoders}
    \begin{tabular}{|c|c|c|c|c|c|}
        \hline Channel & \multicolumn{5}{|c|}{ AWGN } \\
        \hline Modulation & \multicolumn{5}{|c|}{ BPSK } \\
        \hline Coding variant & \multicolumn{4}{|c|}{CRC-Polar} & PAC \\
        \hline Constructions
         & \multicolumn{5}{|c|}{5G} \\
        \hline Code rate, $R$ & \multicolumn{5}{|c|}{1/2} \\
        \hline Info. length, $K$ (bits w/o CRC)
         & \multicolumn{5}{|c|}{
        $64, 512 $ } \\
        \hline Decoding Algorithm & SC & BP & SCL & SCLF & SCL\\
        \hline
    \end{tabular}
\end{table}

\begin{figure}
    \centering
    \includegraphics[width=1\columnwidth]{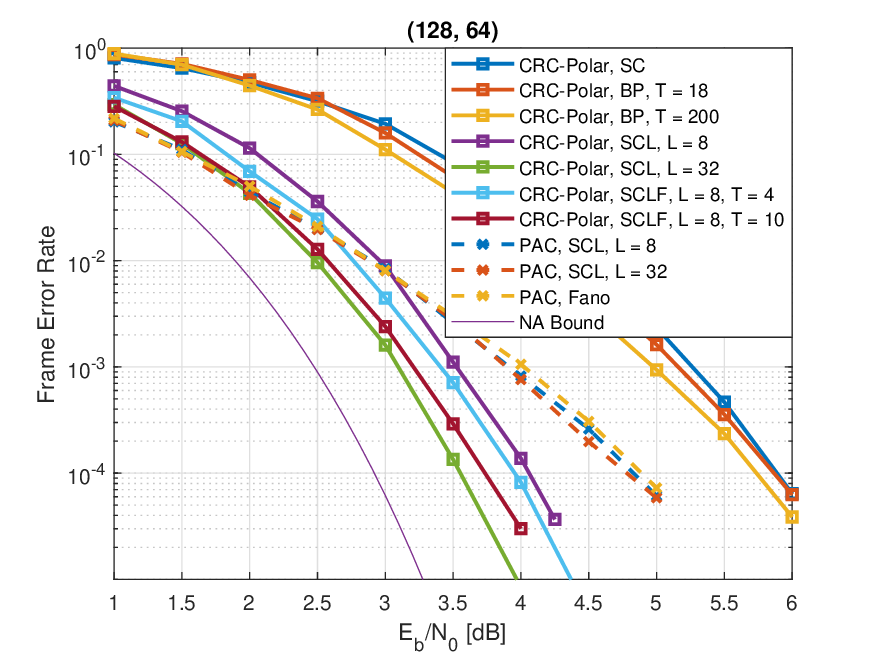}
    \caption{FER comparison between decoders for polar codes and PAC codes with $N = 128, K = 64$ under 5G construction.}
    \label{fig:polar_N_128_different_decoders}
\end{figure}

\begin{figure}
    \centering
    \includegraphics[width=1\columnwidth]{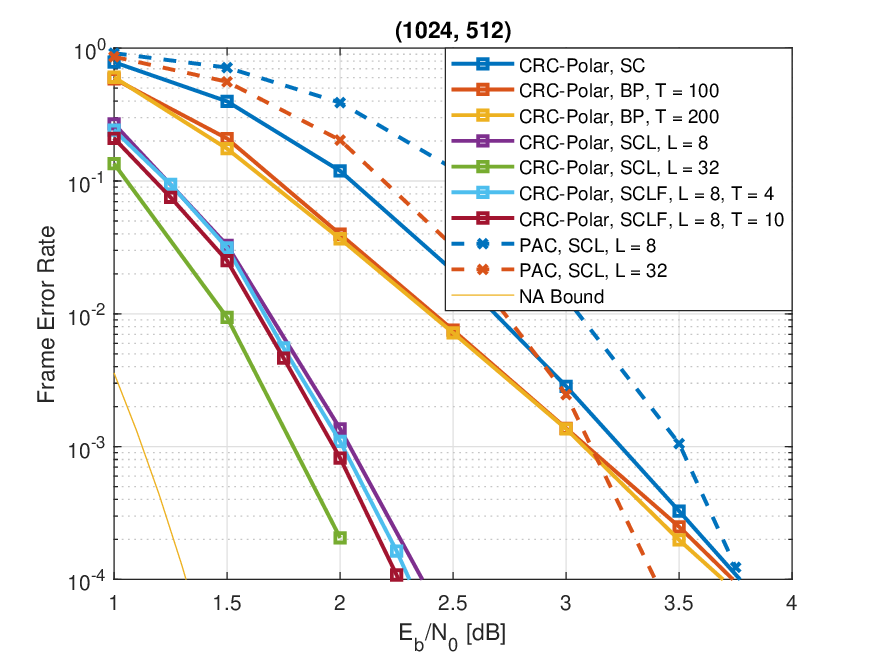}
    \caption{FER comparison between decoders for polar codes and PAC codes with $N = 1024, K = 512$ under 5G construction.} 
    \label{fig:polar_N_1024_different_decoders}
\end{figure}

Finally, we compare the error correction performance of the CRC-polar codes and PAC codes under various decoding algorithms. 
Figs. \ref{fig:polar_N_128_different_decoders} and \ref{fig:polar_N_1024_different_decoders} compare the block (or frame) error rate of the CRC-polar code (128, 64) under the SC, BP, SCL, and flipping SCL decoding algorithms with different decoding parameters of list sizes ($L=8,32$), iterations ($T$), or both. In this scenario, summarized in Table \ref{tab:fer_snr_all_decoders}, we use the 5G construction for all codes. As the figures demonstrate, the SCL decoding algorithm with a large list $l=32$ seems to be the best option as it outperforms the rest. Under resource constraints such as limited silicon area, SCLF with list size $L=8$ could alternatively be used at the cost of longer latency due to multiple iterations ($T=4,10$). Note that the PAC codes demonstrated here are constructed with 5G reliability sequences, not 5G-RM. As a result, it does not outperform a short CRC-polar code.


\subsection{Performance of Decoding Hardware Architectures}\label{ssec:hardware}
In this section, we review the hardware architectures designed for the decoders discussed in Section \ref{ssec:decoding}, and compare them in terms of their throughput, latency, energy efficiency, scalability, etc. Although the most popular decoder for polar codes is SC list decoding as it provides a competitive error correction performance, due to its sequential nature that limits the throughput and high computational complexity, alternative decoders have been adapted to polar codes, which admit parallelism in design, computationally simpler, etc. Taking into account the tradeoff between performance and other parameters, these alternative decoders can be used in different scenarios depending on their requirements. 

\begin{figure*}
    \centering
    \includegraphics[width=1.7\columnwidth]{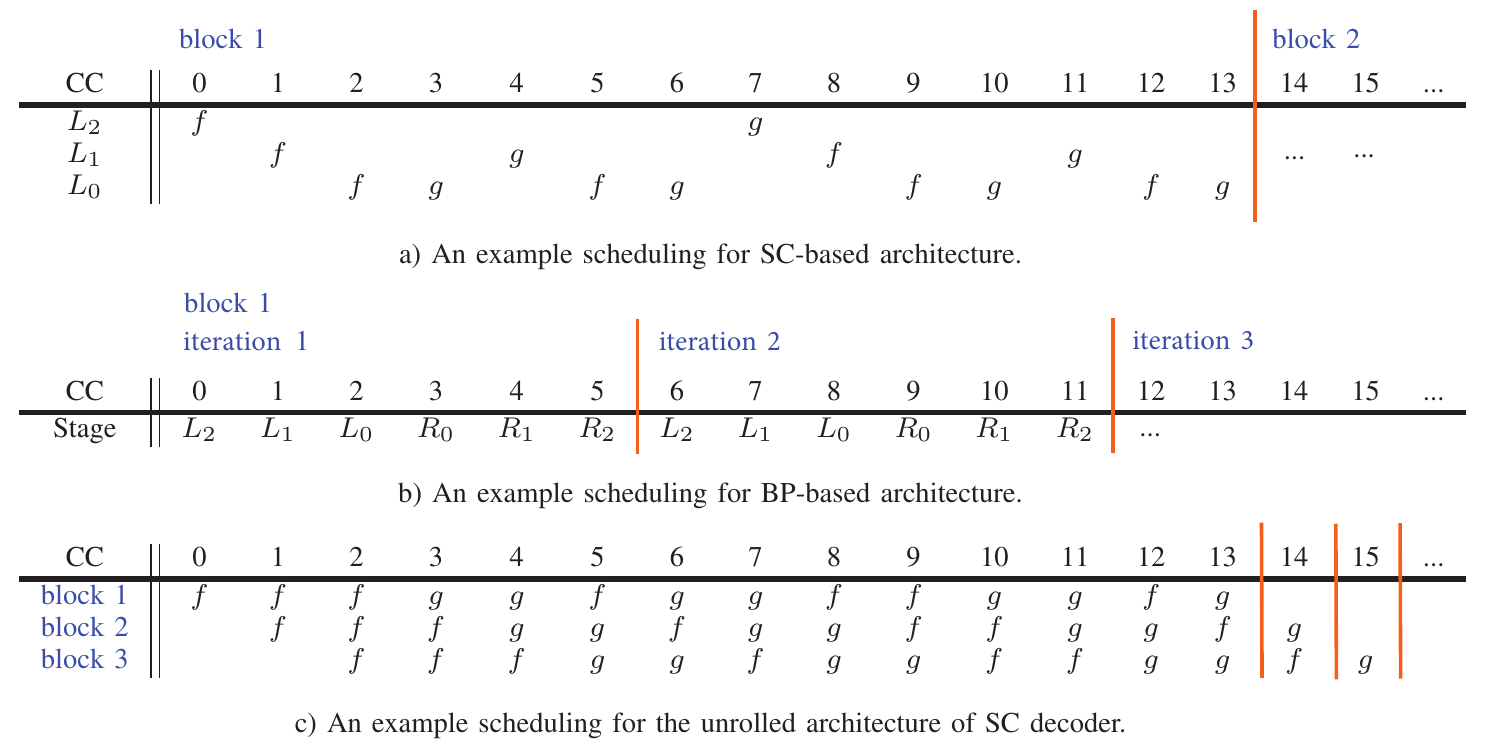}
    \caption{Examples of scheduling based on clock cycle (CC) for an underlying polar code of length N = 8.}
    \label{fig:hardware_polar_scheduling}
\end{figure*}

The architectures of hardware implementations can be categorized on the basis of various scheduling structures. 
Fig. \ref{fig:hardware_polar_scheduling} provides scheduling examples for decoding a single block of a polar code with length $N$, considering clock cycle (CC) based execution. The depicted scheduling covers SC-based and BP-based architectures, and unrolled architecture (used for SC decoder), all applied to a polar code with $N = 8$, where $L_i, 0\le i\le n$ represent the stages in the factor graph, while $f$ and $g$ are the computations within the SC decoder.
For the SC-based decoder shown in Fig. \ref{fig:hardware_polar_scheduling}.a, the determination of each bit relies on the LLR propagated from the preceding node. The scheduling is influenced by the time complexity of the decoder, resulting in a successive architecture that requires $2N-2$ clock cycles (CC) to decode a single block.
As described in Section \ref{sssec:bp}, the BP-based decoders operate in stages. Consequently, the scheduling entails $2n \cdot t$ CC, as illustrated in Fig. \ref{fig:hardware_polar_scheduling}.b, where $t$ represents the iteration number. 
In the unrolled architecture, all the necessary processing elements (PEs) for the decoding process are listed. The scheduling of this architecture is illustrated in Fig. \ref{fig:hardware_polar_scheduling}. For decoding the first block, it has the same latency as the structure shown in Fig. \ref{fig:hardware_polar_scheduling}.a. However, all the PEs are deeply pipelined. Once a PE finishes the computation of one block, it will pass the information to the next PE and start processing the information of the next block. As a result, following the initial block, the unrolled architecture can generate the result of decoding one block for every CC, leading to notably higher throughput than the other two architectures at the cost of much higher demand of the resources.

\subsubsection{Successive Cancellation-based Architecture}

\begin{figure}
    \centering
    \includegraphics[width=1\columnwidth]{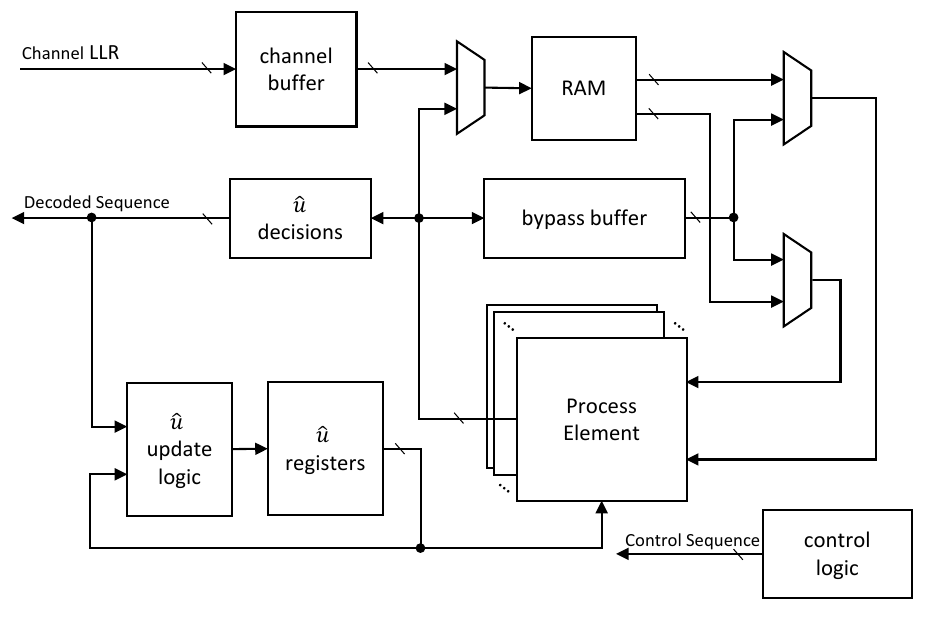}
    \caption{An example of a top-level architecture for the SC decoder \cite{leroux2012semi}.}
    \label{fig:hardware_SC_top}
\end{figure}

In \cite{leroux2012hardware}, three hardware structures were developed to implement the original successive cancellation (SC) decoder. An example of a top-level architecture for the SC-based decoder is shown in Fig. \ref{fig:hardware_SC_top}, which includes the main logic modules as follows:
\begin{itemize}
    \item \textit{Control logic}: The controller generates the control signals for the sub-modules to initiate their operations.

    \item \textit{RAM}: The memory block stores values of the propagating LLRs for subsequent usage by the decoder. The RAM addresses are acquired from the control logic.

    \item \textit{Process element (PE)}: Each PE computes a pair of functions (i.e. $f$ and $g$ in \eqref{eq:check_node} and \eqref{eq:var_node}) of the decoder. The number of PEs employed depends on the specific architecture design and scheduling.

    \item \textit{Buffers}: Read-and-write operations of the RAM follow the clock schedule. Therefore, in situations where certain data require immediate input or reuse, buffers are employed to act as an intermediary.

    \item \textit{$\hat{u}$ logics}: The values of $\boldsymbol{u}$ are determined by hard decisions of the LLRs as computed in equation \eqref{eq:sc_hard_decision_ch2}. Moreover, they are designed to update the value of the bit for the node,  i.e. update the $\hat{\beta}$ in equation \eqref{eq:var_node}, so that it has the value of the partial sum of the previously decided bits.
    
\end{itemize}

A semi-parallel structure \cite{leroux2012semi} was introduced to enhance resource utilization of the PEs, while a combinational logic structure \cite{dizdar2016high} focused on reducing the energy consumption of intermediate LLR calculations in the SC decoder. 
The simplified SC (SSC) \cite{alamdar2011simplified} and 2-bit-SC \cite{yuan2013low} were introduced to minimize latency by optimizing the calculations of frozen bits. These approaches aimed to reduce the decoding stages, resulting in lower latency and improved throughput. Building upon this idea, Fast-SSC decoders were proposed in \cite{sarkis2014fast, 8110014, 8010821, 9814456, 9195264}, which further optimized the decoding tree by categorizing special nodes, leading to improved throughput.  Additionally, SC flip (SCF) decoders were implemented in \cite{8854992, 9195767, 8839446}, achieving better error correction performance at the cost of lower throughput. 

Based on iterative message passing following the successive architecture, hardware implementations for the soft-output cancellation (SCAN) \cite{berhault2015hardware, lin2016efficient} and Fast-SCAN decoders \cite{zhang2021efficient} were designed. These decoders showed better error correction performance than the fundamental SC decoders.  However, because of their combined successive and iterative architectures, they exhibited increased latency and reduced throughput.

Hardware architectures of the SCL and CRC-aided SCL (CA-SCL) decoders were designed in \cite{zhang2014hardware, balatsoukas2015llr, liu20185}. 
The implementations of multi-paths in the SCL decoder either require large resources or slow down the throughput to trade with its competitive error correction performance. The sorters were designed to accelerate the path-pruning of the SCL decoder to improve throughput \cite{balatsoukas2015metric}. Node classification schemes and latency reduction schemes were adopted and improved for the SCL decoder in \cite{sarkis2015fast, hashemi2017fast,7339658, 7337462, 9928370, lee2020node}, showing significant improvements in latency, throughput, and area efficiency. 

An architecture for automorphism (or permutation) ensemble decoding was presented in \cite{kestel2023automorphism}. The AED employs an ensemble size of SC decoders implemented in parallel, similar to the SCL decoder, but without the path-sorting and pruning procedures, resulting in lower latency and higher throughput compared to SCL-based decoders.

In \cite{coppolino2018multi}, an SC-based decoder was adopted and designed for multi-kernel polar codes to meet specific requirements in practical applications, offering flexible code lengths, code rates, and kernel sequences. Subsequent improvements in latency and throughput were achieved by implementing the Fast-SSC decoder \cite{rezaei2022low} and a combinational logic structure \cite{rezaei2023high}. These advancements further enhanced the performance and efficiency of hardware implementations of polar codes.

Combining the SC decoder with the Fano decoder, a hardware implementation was developed for PAC codes in \cite{mozammel2021hardware}. While the Fano decoder offers good error correction performance and a small required area for short codes, its high-complexity iterative decoding processes result in extremely high latency and low throughput.

\subsubsection{Belief Propagation-based Architecture}

BP decoders offer advantages over SC-based decoders as the operations on the factor graphs can be executed in parallel, providing the potential for higher throughput and lower latency \cite{pamuk2011fpga}. The reduction in latency compared to the SC-based decoder becomes more pronounced as the code length increases. However, one of the drawbacks of BP-based decoders is the uncertainty in the required number of iterations. In worst-case scenarios, when BP-based decoding reaches the maximum number of iterations, it can result in significantly low latency and throughput. 
Various algorithms and hardware architectures, such as bi-directional, folding, double-column structures, and early stopping criteria \cite{zhang2014simplified, yuan2013architecture, 8552674, yuan2014architectures, park20144, sha2015memory} have been proposed to reduce latency and improve the throughput of BP decoders. Also, it was shown in \cite{gu2022nonuniform,rowshan2021logarithmic} that non-uniform quantization of messages can improve the BLER performance over uniform quantization in hardware implementation. Additionally, special node designs have been adopted to further enhance throughput \cite{abbas2016high}.
The BP-flipping (BPF) \cite{ji2020hardware, 9051990} and BP list (BPL) \cite{ren2022high} were introduced, demonstrating error correction performance comparable to SCL decoders, but with higher throughput.

\subsubsection{Unrolled Architecture}


Unrolled architectures deploy all the PEs for individual nodes within the decoder tree of a given polar code. This approach maintains consistent latency while achieving the highest throughput compared to other architectures, albeit with notable resource requirements, allowing each PE to handle a distinct block.
Unrolled architectures for the fast SSC decoder and the combined SC-majority logic (MJL) decoder were designed in \cite{giard2015237, sural2019terabits} to boost the throughput with larger area requirements, addressing the challenges of low throughput in successive structures. 
Unrolled structures of the SCL decoder were implemented in \cite{giard2016multi, giard2017high, kestel2020506gbit}. 
Moreover, unrolled BP decoders with a fixed number of iterations have been implemented to achieve ultra-high throughput in \cite{lopacinski2022ultra}.

\subsubsection{Comparison}
The performance of various polar decoders with state-of-the-art hardware architectures in \cite{hashemi2017fast, kestel2020506gbit, 9051990, ren2022high, zhang2021efficient, coppolino2018multi, kestel2023automorphism, xiao2023low, mozammel2021hardware} is presented and compared in Table \ref{tb:polar_hardware_imple_sums}. The results are presented relative to the 28nm technology for fair comparisons. 
The unrolled structure of the CA-SCL decoder \cite{kestel2020506gbit} achieves an impressive throughput of about 500Gb/s at the cost of a proportionally larger area requirement compared to other structures. 
For polar codes with a mother code length of $N = 1024$ bits, the parallel decoder in the BPL decoder \cite{ren2022high} allows it to achieve a power gain of 0.6 dB in error correction performance over the EBPF decoder \cite{9051990} with higher average throughput and lower average latency. However, the parallel structure comes with a larger required area.
Fast-SSCL \cite{hashemi2017fast} and BPL decoders demonstrate the best error correction performance. The implementations of the BP-based decoders show higher throughput than the Fast-SSCL decoder due to their highly paralleled decoding structure, requiring fewer average stages than the successive structure of the SC-based decoder.  However, the iterative decoding structure leads to much lower throughput in worst-case scenarios.
The Fast-SCANF decoder \cite{zhang2021efficient}, based on iterative message passing and a successive decoding structure, achieves lower throughput while requiring a smaller area compared to the SCL and BPL decoders.
The implementation of the multi-kernel polar codes shows inferior error correction performance and throughput compared to the state-of-the-art SCL-based and BP-based decoders, while it provides smaller area requirements and flexibility in the code rates and code lengths for the polar codes.
For hardware implementations that focus on shorter codes, the Fano decoder \cite{mozammel2021hardware} demonstrates the best error correction performance at the cost of significantly low throughput and high latency due to its high complexity. 
The multiple paralleled SC decoders enable the AED \cite{kestel2023automorphism} to have a reasonable error correction performance and throughput, but with a larger required area compared to 
Fano decoder.

\begin{table*}[h]
\setlength{\tabcolsep}{0.45em} 
\renewcommand{\arraystretch}{1.2} 

\caption{Hardware implementation results for polar and PAC codes decoders.} \label{tb:polar_hardware_imple_sums}
\centering
\begin{tabular}{|c c||c c c c c c c c c|}
\whline
\multicolumn{2}{|c||}{Implementation} 
& \cite{hashemi2017fast} & \cite{kestel2020506gbit} & \cite{sural2019terabits} & \cite{9051990} & \cite{ren2022high} & \cite{zhang2021efficient} & \cite{coppolino2018multi} & \cite{kestel2023automorphism} 
& \cite{mozammel2021hardware}  \\

\hline
\multicolumn{2}{|c||}{Algorithm} 
& Fast-SSCL & CA-SCL$^{\ast}$ & SC-MJLL$^{\ast}$ & EBPF & BPL & Fast-SCANF  & Multi-Kernel & AED 
& Fano \\

\whline
\multicolumn{2}{|c||}{Process [nm]} 
& 65 & 28 & 45 & 65 & 28 & 40 & 65 & 12 
& 28  \\

\hline
\multicolumn{2}{|c||}{Code} 
& (1024, 512) & (1024, 512)$^{\top_1}$ & (1024, 854) & (1024, 512) & (1024, 512) & (1024, 512) & (768, 384) & (128, 60) 
& (128, 64)$^{+}$  \\

\hline
\multicolumn{2}{|c||}{Quantization} 
& 6 & 6 & 5-to-1 & 7 & 7 & 6 & 6 & - 
& 7 \\

\hline
\multicolumn{2}{|c||}{List/Attempt} 
& 4 & 2 & 1 & 20 & 32/50 & 10 & 1 & 8 
& $2^{18}$   \\


\hline
\multicolumn{2}{|c||}{SNR@BLER=$10^{-4}$} 
& 2.65 & 2.94 & 5.54 & 3.25 & 2.65 & 3.07 & 3.48 & 3.71 
& 3.14  \\


\whline
\multicolumn{2}{|c||}{Area [mm$^2$] }
& 1.822 & 7.89 & 2.4 & 3.11 & 0.87 & 0.44 & 0.46 & 0.17 
& 0.059  \\

\hline
\multicolumn{2}{|c||}{Frequency [MHz]} 
& 840 & 494 & 500 & 319 & 1333 & 980 & 1110 & 498 
& 500 \\

\hline
\multicolumn{2}{|c||}{W.C. Latency [$\mu$s]} 
& - & - & - & 51.2 & 0.34 & - & - & - 
& 524  \\

\hline
\multicolumn{2}{|c||}{Avg. Latency [$\mu$s]} 
& 0.64 & 0.31 & 0.08 & 0.28$^{\diamond_1}$ & 0.04$^{\diamond_1}$ & 0.46$^{\diamond_2}$ & 2.09 & 0.022 
& 1.68$^{\diamond_4}$  \\

\hline
\multicolumn{2}{|c||}{W.C. T/P [Gb/s]} 
& - & - & - & 0.02 & 0.09 & - & - & - 
& 0.074  \\

\hline
\multicolumn{2}{|c||}{Coded T/P [Gb/s]} 
& 1.61 & 506 & 512 & 3.72$^{\diamond_1}$ & 25.63$^{\diamond_1}$ & 2.04$^{\diamond_2}$ & 0.358 & 63.7 
& 0.037$^{\diamond_4}$ \\

\hline
\multicolumn{2}{|c||}{Area Eff. [Gbps/mm$^2$] }
& 0.883 & 64.13 & 213.33 & 1.2 & 29.46 & 4.63 & 0.78 & 375.1 
& 0.646   \\


\whline
\multicolumn{11}{l}{Normalized to 28nm$^{\star}$}
    \\ 

\whline
\multicolumn{2}{|c||}{Coded T/P [Gb/s]}
& 3.74 & 506 & 823 & 8.64 & 25.63 & 2.91 & 0.83 & 27.3 
& 0.037 
   \\

\hline
\multicolumn{2}{|c||}{Area [mm$^2$] }
& 0.338 & 7.89 & 0.94 & 0.577 & 0.87 & 0.216 & 0.085 & 0.926 
& 0.059  \\

\hline
\multicolumn{2}{|c||}{Area Eff. [Gbps/mm$^2$]} 
& 11.07 & 64.13 & 872 & 14.97 & 29.46 & 13.47 & 9.76 & 29.48 
& 646
\\ 
\whline

\multicolumn{1}{l}{$\ast \,$} & \multicolumn{9}{l}{  These work design unrolled architectures.} \\
\multicolumn{1}{l}{$\top_{1,2}$} & \multicolumn{9}{l}{  These works employ 6-bit and 11-bit CRC codes for the polar codes respectively.}   \\
\multicolumn{1}{l}{$+$} & \multicolumn{9}{l}{   This work is developed for PAC codes.} \\
\multicolumn{1}{l}{${\diamond_{1,2,3,4}}$} & \multicolumn{9}{l}{  Average results reported at SNR = 4, 3, 7.5, 3.5 dB repectively. Worse case is not discussed for ${\diamond_{1}}$.} \\
\multicolumn{1}{l}{ $\star \;$} & \multicolumn{9}{l}{ Normalized to 28nm technology: Area $\propto \alpha^2$ and frequency $\propto 1/\alpha$, where $\alpha$ is the scaling factor to 28nm. } \\

\end{tabular}

\end{table*}


\subsection{Future Directions for Polar Codes}\label{ssec:future}
Polar coding was proposed about 1.5 decades ago. However, throughout this period, new ideas and initiatives have kept this scheme the focus of research attention in the field of channel coding. Among the major coding schemes, perhaps this is the only scheme that still is gaining popularity in research, albeit at a slower rate, as Fig. \ref{fig:number_of_articles} indicates. 
The popularity of polar codes is mainly due to their superior error correction performance for short codes at a reasonable complexity cost. Nevertheless, there is still room to further improve the performance through pre-transformation and code construction, and reduce the complexity of the decoding algorithms. In the following, we suggest three general directions that researchers can follow.  

Concatenation and pre-transformation of polar codes, such as CRC-polar codes and PAC codes, have remarkably improved the performance of polar codes. The available concatenation and transformation schemes can be further improved towards improving the distance properties of underlying codes. For instance, an attempt to improve pre-transformation in PAC codes was made in \cite{gu2022selective}. 

\begin{figure}[ht] 
    \centering
    \includegraphics[width=1\columnwidth]{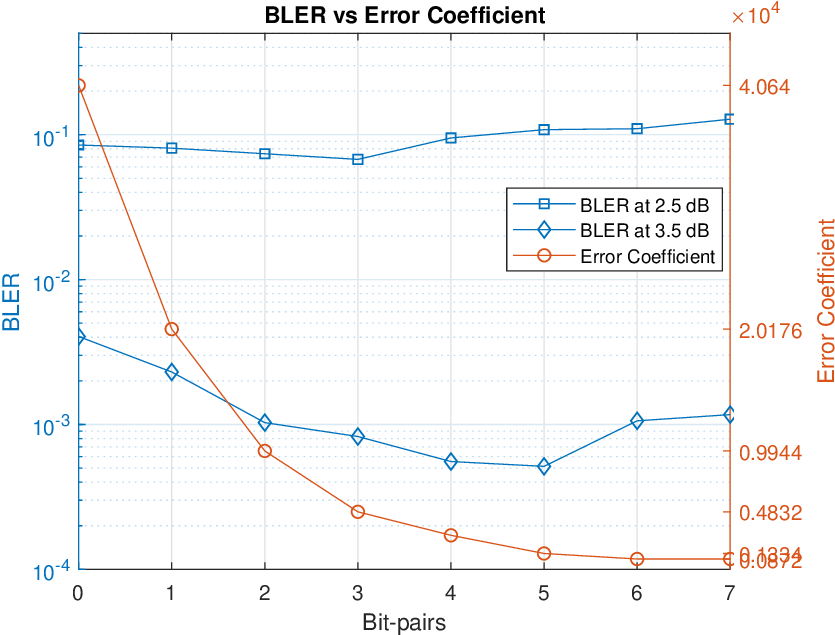}
    \caption{BLER at two $E_b/N_0$'s versus error coefficient $A_{d_{min}}$ of PAC code of (512,384) under SCL decoding with $L=16$. Bit-pairs are the number of bit indices added to/removed from $\A$ which are the indices of the most reliable sub-channels \cite{rowshan2023formation}.} 
    \vspace{-5pt}
    \label{fig:BLER_Admin}
\end{figure}

Construction of polar codes and their variants considering both the error coefficient (or weight distribution in general, based on the weight structure of polar codes studied extensively in \cite{bardet_algebraic_2016,rowshan2023formation,rowshan2023closed,rowshan2024weight}) and sub-channels reliability with a low-complexity approach is desired. 
As discussed in Section \ref{ssec:construct}, reliability-based code construction is optimal for SC decoding. However, (near) ML decoding performance depends on the weight distribution of the code. Fig. \ref{fig:BLER_Admin} shows that improving/reducing the number of minimum weight codewords, or error coefficient $A_{d_{min}}$, improves the block error rate, in particular at high SNR regimes. However, reducing $A_{d_{min}}$ beyond some extent deteriorates the performance due to excessive utilization of bad sub-channels. Hence, we need to consider the trade-off between the weight distribution and the selection of the most reliable sub-channels. 
An attempt to achieve this goal was made in \cite{rowshan2023formation,rowshan2022improving}. However, this direction requires further attention. 

The complexity of SC-based decoding algorithms has been a concern in achieving low latency, despite the fact that polar codes and their variants make high-reliability communication possible. Improving the existing algorithms and designing novel decoding still remain a direction to explore. 

As mentioned above and as we can see from the numerical result in Section \ref{compare_schemes}, polar codes cannot outperform other coding schemes at long block lengths. This is one of the directions that can be investigated to improve the performance of polar codes for longer codes and reduce their decoding complexity, which is another concern for longer codes. This improvement can be achieved through different code construction for longer codes or by a different decoding strategy. However, further studies are required on the properties and differences between long codes and short codes. 
\section{Performance Comparison of Turbo, LDPC, and Polar Codes}\label{compare_schemes}
We evaluated the performance of various coding schemes under different decoders, constructions, interleavers, component codes, etc., in Sections \ref{sec:turbo}, \ref{sec:ldpc}, and \ref{sec:polar}. We consider the information lengths $K=100, 400, 1000, 4000, 8000$ bits and the code rates $R=1 / 5,1 / 3,1 / 2,2 / 3,8 / 9$. Results are obtained from 3GPP report \cite{R1-1613029, R1-1611071}. Note that information block lengths $K = 96, 992$ are employed for turbo codes, whereas $K=100, 1000$ are used for LDPC and polar codes.
For LDPC and polar codes, we use the 5G standard's constructions. The turbo codes in the comparison are enhanced turbo codes with tail-biting termination from \cite{R1-167413,R1-1612938}, where interleavers and puncturers therein are adopted. The extrinsic information of the Max-log-MAP decoder is scaled as $s^{(1)}=0.6$, $s^{(\ell_{\max})}=1$, and $s^{(\ell)}=0.7,\forall \ell\in\{2,\ldots,\ell_{\max}-1\}$, where $\ell_{\max}$ denoting the maximum number of iterations. The setup of the numerical evaluations is summarized in Table \ref{tab:fer_snr_all_codes}. Note that puncturing, shortening, and repetition techniques are used to have the same rate (or the same block lengths) for all three codes.

\setlength{\tabcolsep}{0.25em} 
\begin{table}[ht]
    \centering
     \caption{Evaluation of the frame error rate performance versus SNR of  coding schemes.}
    \label{tab:fer_snr_all_codes}

    \begin{tabular}{|c|c|c|c|}
        \hline Channel & \multicolumn{3}{|c|}{ AWGN } \\
        \hline Modulation & \multicolumn{3}{|c|}{ QPSK } \\
        \hline Coding Scheme & Turbo & LDPC & Polar \\
        \hline Code rate $R$ & \multicolumn{3}{|c|}{$1 / 5,1 / 3,1 / 2,2 / 3,8 / 9$} \\
        \hline \begin{tabular}{l}
        Decoding \\
        Algorithm
        \end{tabular} & \begin{tabular}{l}
        Max-Log-MAP, \\
        Max Iter = 8
        \end{tabular} & \begin{tabular}{l}
        Adjusted Min-Sum, \\
        Max Iter = 25
        \end{tabular} & \begin{tabular}{l}
        SCL, \\
        L = 32
        \end{tabular} \\
        \hline \begin{tabular}{c}
        Info. length $K$\\
         (bits w/o CRC)
        \end{tabular} & \multicolumn{3}{|c|}{\begin{tabular}{c}
        $100,400,1000,4000,8000$ \\
        \end{tabular}} \\
        \hline
    \end{tabular}
\end{table}

\begin{figure}
    \centering
    \includegraphics[width=1\columnwidth]{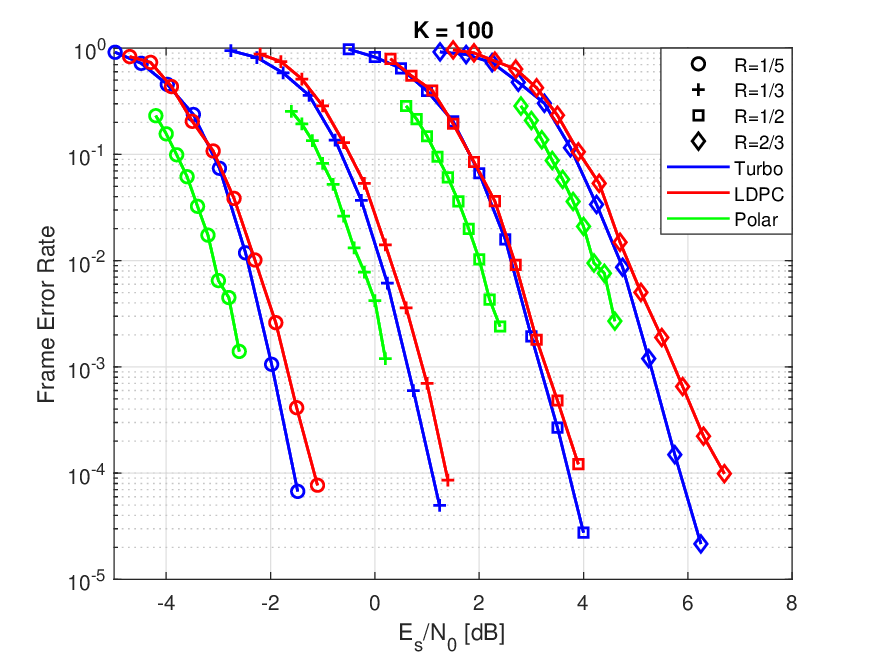}
    \caption{FER comparison between codes for information block $K = 100$ at different code rates.}
    \label{fig:compare_Codes_K_100_rates}
\end{figure}

\begin{figure}
    \centering
    \includegraphics[width=1\columnwidth]{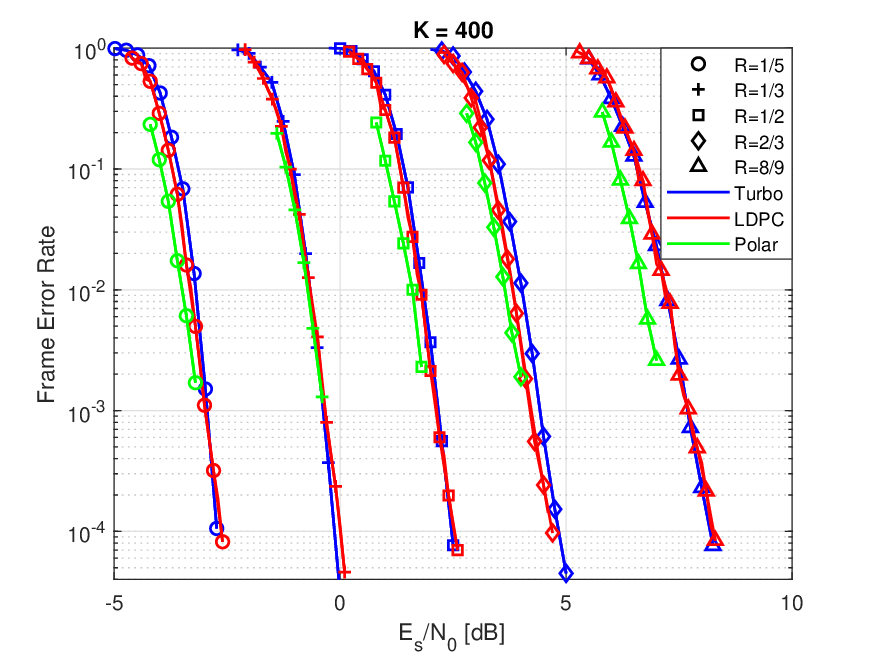}
    \caption{FER comparison between codes for information block $K = 400$ at different code rates.}
    \label{fig:compare_Codes_K_400_rates}
\end{figure}

\begin{figure}
    \centering
    \includegraphics[width=1\columnwidth]{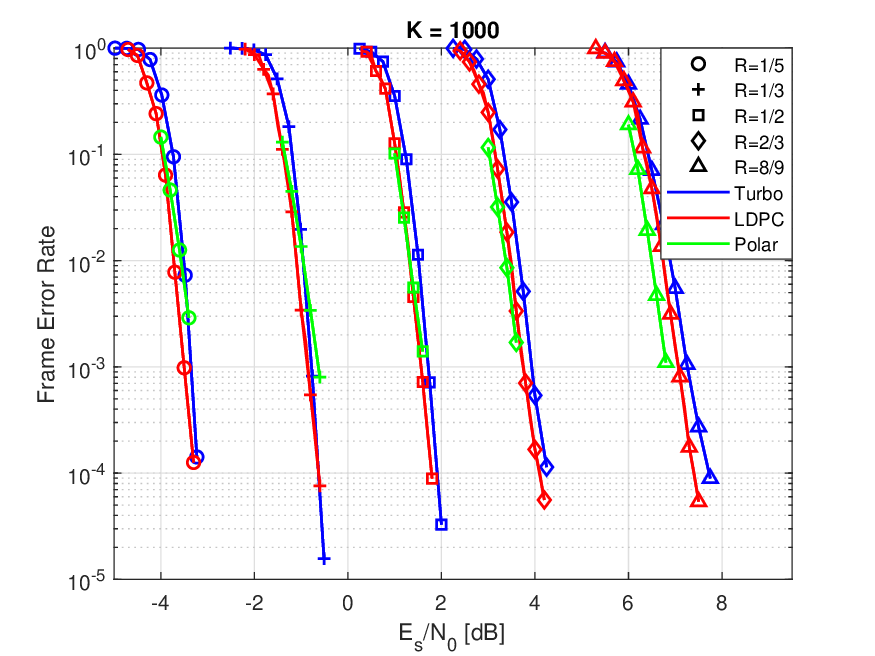}
    \caption{FER comparison between codes for information block $K = 1000$ at different code rates.}
    \label{fig:compare_Codes_K_1000_rates}
\end{figure}

As can be seen in Figs. \ref{fig:compare_Codes_K_100_rates}, \ref{fig:compare_Codes_K_400_rates}, and \ref{fig:compare_Codes_K_1000_rates}, polar codes outperform LDPC and Turbo codes when the length of the information block is $K\leq400$ bits. For $K=1000$ bits, the performance of polar codes is inferior compared to that of LDPC codes at rates $R\leq1/3$. As the length of the information block increases to $K=4000$ and 8000 bits, LDPC codes outperform Turbo and polar codes at all rates and the power gain becomes more significant as $K$ increases.

\begin{figure}
    \centering
    \includegraphics[width=1\columnwidth]{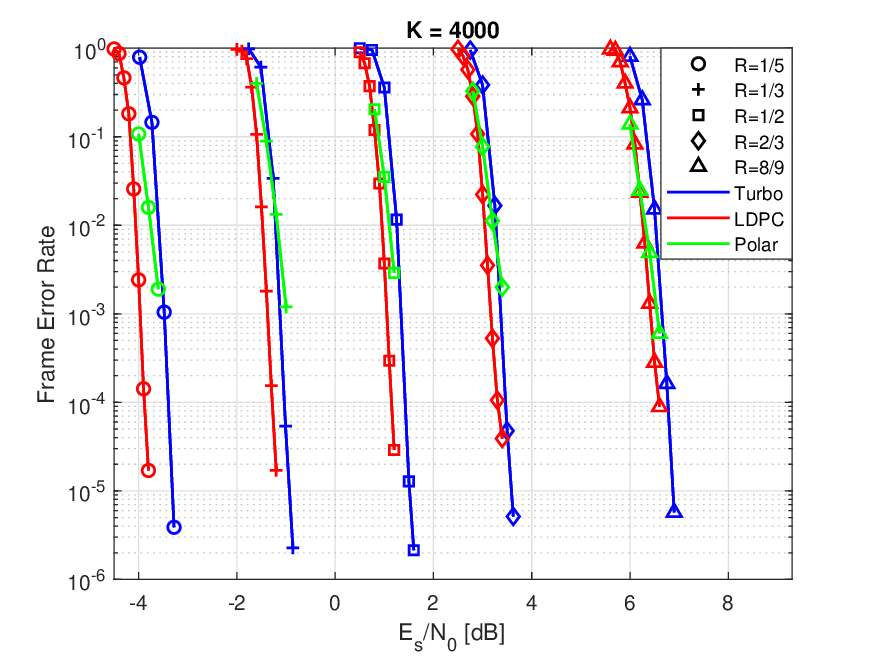}
    \caption{FER comparison between codes for information block $K = 4000$ at different code rates.}
    \label{fig:compare_Codes_K_4000_rates}
\end{figure}

\begin{figure}
    \centering
    \includegraphics[width=1\columnwidth]{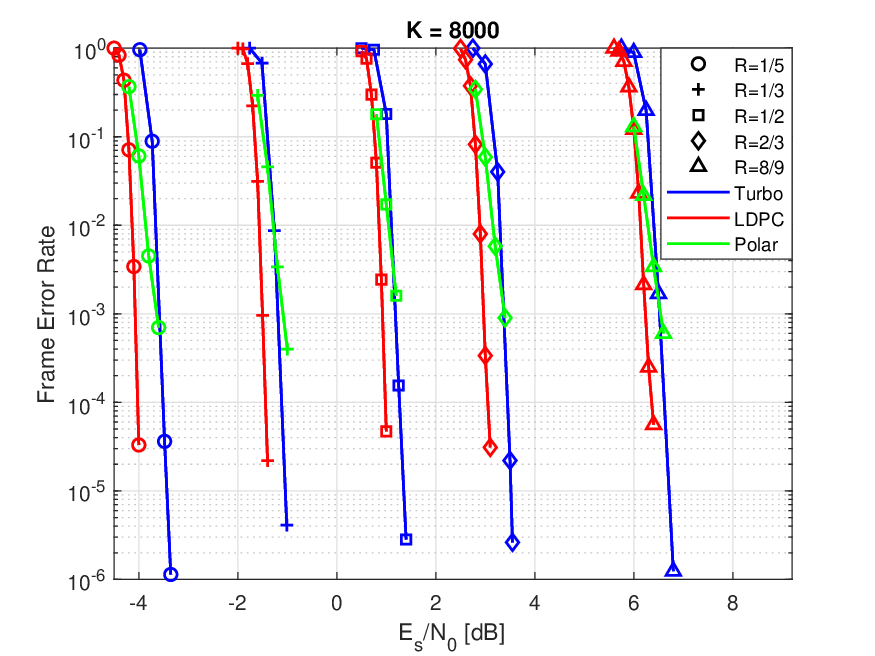}
    \caption{FER comparison between codes for information block $K = 8000$ at different code rates.}
    \label{fig:compare_Codes_K_8000_rates}
\end{figure}

Note that we did consider recent improvements of the coding schemes discussed earlier, in particular with regard to the polar coding scheme. Recent developments are still the subject of research and need to be investigated and improved further to the level needed for adaptation to a standard. Nevertheless, we have discussed and evaluated them in their own section in this paper. 
\section{Other Coding Schemes}\label{sec:other}
\begin{figure}[h]
    \centering
    \includegraphics[width=1\columnwidth]{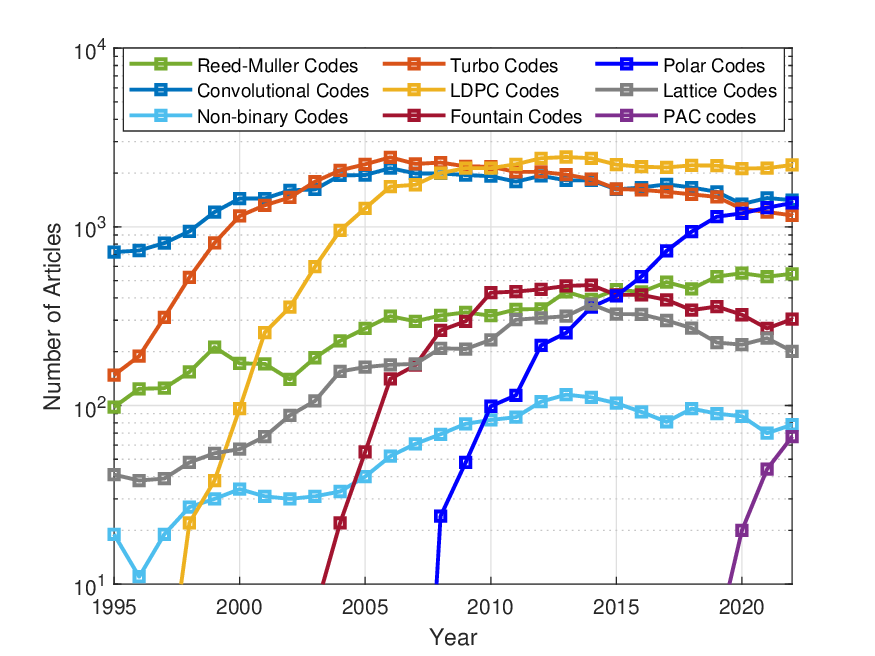}
    \caption{The trend of number of articles published annually (extracted from Google Scholar hits).}
    \label{fig:number_of_articles}
\end{figure}

The binary codes discussed in the previous sections, i.e., LDPC codes, polar codes, convolutional codes, and turbo codes are the coding schemes that have been adapted in standards, and hence have proved their merits. However, other coding schemes can also be considered. Fig. \ref{fig:number_of_articles} shows the trend of the number of publications throughout the last 30 years\footnote{This figure is inspired by a similar figure in \cite{koike2021evolution}}. As can be seen, while publications on polar codes have exponentially increased over the first decade of their invention, other codes such as LDPC codes have retained their popularity.

In this section, we briefly review other potential coding candidates which are included in Fig. \ref{fig:number_of_articles}  and discuss the major challenges to adapt them to a mobile communications standard, including decoding complexity, real-time requirements, power consumption, scalability in finite block length, and/or compatibility and interoperability with existing systems. Additionally, while the focus of this paper is on physical layer coding, we consider the option of application layer channel coding in the following.

\subsection{Application Layer Channel Coding}
Application layer (AL) channel coding, predominantly based on fountain codes and Reed-Solomon codes, has been considered in the literature for a variety of applications such as multimedia broadcasting and streaming \cite{stockhammer2008application,gomez2009application}, IoT \cite{marcelis2017dare,sandell2018application}, deep space communications \cite{hou2017application}, etc. In AL channel coding, additional packets of application-specific data are produced as redundant packets for packet recovery at the application layer, while physical layer channel coding involves mechanisms for error detection and correction in raw data bits,  specific to the characteristics of the transmission medium. 

One major advantage of this scheme is that operating at a higher layer in the protocol stack makes it easier to implement and manage the coding scheme, particularly in software-based systems, whereas physical layer coding may involve specialized hardware components and signal processing. 
However, a primary challenge of the scheme is the increase in latency. Additionally, larger memory spaces are required to handle larger source packets. Other challenges are as follows. 1) It is difficult for AL channel coding to fully exploit channel characteristics and potential error correction capabilities available in the physical layer, which may lead to suboptimal performance. 2) Additional headers or metadata may be required to incorporate AL coding schemes, resulting in increased overhead and reduced overall efficiency. 3) AL coding may not have the same level of error correction capability as physical layer coding techniques, especially in environments with high levels of noise or interference. Lastly, since AL channel coding is application-dependent, it may be difficult for standardization, while physical layer coding must adhere to standardized protocols and modulation techniques to ensure interoperability with different network devices and technologies. 

\subsection{Non-Binary Codes}
The trivial alternatives for the well-known coding schemes are non-binary variants of these codes over Galois field GF($q$). For binary codes, $q=2$, whereas for non-binary codes $q$ can be any prime number or a power of a prime number. The specific choice of $q$ depends on the design requirements and characteristics of the code. In the previous sections, we have briefly discussed the non-binary variants of turbo codes (see Sections \ref{sec:turbo}-\ref{sec:turbo_CM}), LDPC codes (see Section \ref{sec:ldpc}-\ref{sec:NB_LDPC_DEC}), and polar codes (see Section \ref{sec:polar}-\ref{ssec:large_kern}). Hereafter, we provide a summary of the benefits and challenges of non-binary codes.

\begin{itemize}
\item Advantages: 1) Encoding is directly over the $q$-ary alphabet corresponding to the signal constellation. This saves the mapping and de-mapping processes by converting between binary bits and non-binary modulation symbols. Most importantly, non-binary codes do not have information loss as in the de-mapping of binary codes (unless with costly iteration between de-mapper and decoder or using multilevel coding). This makes non-binary codes appealing for high spectral efficiency coding over higher order constellations; 2) Non-binary codes such as non-binary turbo codes and non-binary LDPC codes suffer from less performance loss in short blocklength compared to their binary counterparts.

\item Challenges: 1) The decoding complexity of non-binary codes increases with the alphabet size $q$. Consequently, the implementation complexity of non-binary codes is higher than that of binary codes; 2) Existing communication systems and standards are often designed with binary codes in mind. Introducing non-binary codes may raise compatibility and interoperability issues with legacy systems and standards. Transitioning from binary codes to non-binary codes would require significant infrastructure changes.
\end{itemize}





\subsection{Fountain Codes}
Fountain codes are a class of erasure codes designed for robust communication over unreliable channels, particularly in scenarios with frequent packet loss. They fall into the category of rateless codes capable of generating an infinite number of encoded symbols. In this coding scheme, a source block is partitioned into $k$ equal length sub-blocks, called source symbols. The basic idea is that the transmitter encodes the source symbols and sends the encoded symbols in packets. A receiver uses the encoded symbols received in the packets to recover the original source block. Ideally, fountain codes should possess the following properties \cite{CIT-060}:
\begin{itemize}
\item Ratelessness: The encoder of a fountain code should be able to generate as many encoded symbols as required for each receiver from the $k$ source symbols of a source block. 
\item Erasure resilience: Given any subset of $k$ encoded symbols, the fountain code decoder should be able to recover the original source block with high probability, regardless of which subset of the generated encoded symbols is received and independent of whether the subset was generated by one fountain encoder or generated by more than one encoder from the original block of source data. 
\item Linear-time complexity: The computation time for encoding and decoding should only scale linearly with respect to the size of a source block.
\end{itemize}

These properties illustrate the concept of a digital fountain, an analogy drawn from a fountain of water. Any receiver who aims to receive the source block holds a bucket under the fountain. It does not matter which particular drops fill the bucket. Only the amount of water, i.e., the number of encoded source symbols received that have been collected in the bucket is important to recover the original data. With a digital fountain, data packets may be obtained from one or more servers, and once enough packets are obtained (from whatever source), file transmission can end. The main figure of merit to assess the performance of a fountain code is the \emph{reception overhead} $\epsilon$, where $(1+\epsilon) k$ received encoded symbols are required to recover the $k$ source symbols. It is also worth pointing out that the underlying characteristic of fountain codes can be exploited to realize secure wireless delivery. That is, secure communication is achieved if the legitimate receiver has received enough fountain-coded packets for decoding before the eavesdropper does \cite{6777406}. Compared to existing physical layer security strategies, fountain-coding-based secure transmission strategies can significantly increase the transmission rate, which is bounded by Shannon capacity rather than secrecy capacity.


Moving forward, we introduce several popular classes of fountain codes. 

\subsubsection{Luby Transform (LT) codes}
The LT code was the first practical realization of fountain codes \cite{luby2002lt}. It is defined from its degree distribution 
\begin{equation}\label{eq:LT_code_degree}
\Omega(x)=\sum_d\Omega_dx^d,
\end{equation}
where $\Omega_d$ is the probability to assign a degree $d$ to an encoded symbol such that $d$ is the number of source symbols participating in the encoding of this symbol. LT codes are a class of low-density generator-matrix (LDGM) codes, enabling the use of BP decoding. It was shown that by using the ideal soliton distribution \eqref{eq:soliton}, only $k$ encoded symbols are sufficient to recover the $k$ source symbols by using the ideal soliton distribution.
\begin{equation}\label{eq:soliton}
\Omega_k(x)=\frac{x}{k}+\sum^k_{d=2}\frac{x^d}{d(d-1)}.
\end{equation}
However, this distribution works poorly in practice, as it is likely that at some decoding step there will be no available degree-one check node in the graph, leading to decoding failure. To overcome this issue, a robust soliton distribution was then introduced to allow the BP decoder to work well in practice, resulting in an overhead of $O(\log^2(k/\delta)\cdot \sqrt{k})$ and an average degree of $O(\log(k/\delta))$, where $\delta$ is the probability of decoding failure. The average number of symbol operations per encoded symbol generated and to decode the $k$ source symbols are is $O(\log(k/\delta))$ and $k\cdot O(\log(k/\delta))$, respectively. It can be seen that on average every encoded symbol needs $O(\log k)$ operations and that the decoding algorithm needs $O(k\log k)$ symbol operations, which are not linear encoding and decoding time. 

\subsubsection{Raptor codes}
\begin{figure}[ht]
				\centering
				\includegraphics[width=\linewidth]{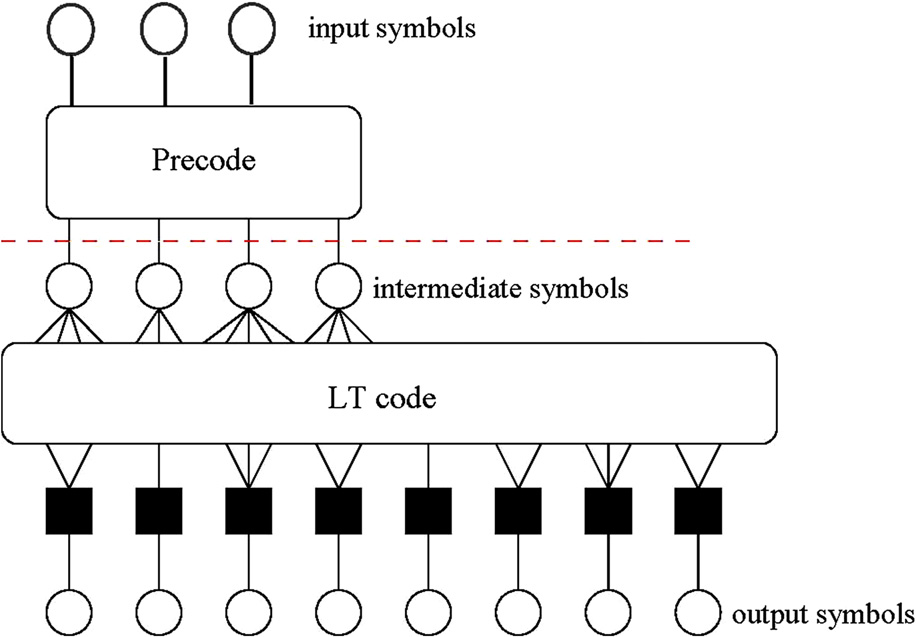}
				\caption{Tanner graph representation of Raptor codes \cite{CIT-060}.} \label{fig:raptor}
			\end{figure}

Raptor codes are a class of fountain codes that achieve linear-time encoding and decoding \cite{shokrollahi2006raptor}. They can be regarded as an improvement of their LT relatives. The key idea of Raptor codes $(k,\mathcal{C},\Omega(x))$ is to concatenate an inner LT code with degree distribution $\Omega(x)$ as in \eqref{eq:LT_code_degree} with a high rate outer code or a precode $\mathcal{C}(n,k)$. Raptor codes can be represented using the Tanner graph as shown in Fig. \ref{fig:raptor}. The outer code can be any erasure correction code, e.g., LDPC codes, which can recover up to a fraction $\delta$ of erasures among the intermediate symbols, that is, the outer codeword symbols. Meanwhile, the LT inner code is responsible for recovering the remaining $(1-\delta)$-fraction of the intermediate symbols. It has been proved that for any $\epsilon>0$, there exists a class of universal Raptor codes such that any subset of symbols of size $k(1+\epsilon)$ is sufficient to recover the original $k$ symbols with high probability. Moreover, the average number of symbol operations per generated encoded symbol and to decode the source block are $O(\log(1/\epsilon))$ and $O(k\cdot \log(1/\epsilon))$, respectively. 

To design Raptor codes with overhead $k(1+\epsilon)$ for achieving good asymptotic performance, the degree distribution $\Omega(x)$ is designed in such a way that
\begin{equation}
\sup\{ x\in[0,1) | 1-x-e^{-(1+\epsilon)\Omega'(x)}>0\},
\end{equation}
is maximized. To take into account the finite blocklength effect, the degree distribution satisfies the following
\begin{equation}
1-x-e^{-(1+\epsilon)\Omega'(x)} \geq \gamma \sqrt{\frac{1-x}{k}},
\end{equation}
for $x\in[0,1-\delta]$ such that the decoding process will continue with high probability until it has recovered all but a $\delta$-fraction of the intermediate symbols, and $\gamma$ is a positive design parameter. 

In addition to the erasure channel, the design of raptor codes has also been carried out for the BSC \cite{1624639}, the BI-AWGN channels \cite{5336849,8030096}, and the fading channels \cite{1576565}, where density evolution and EXIT charts (see Section \ref{sec:ldpc}-\ref{sec:ldpc_tools}) are the main design tools. The benefits provided by Raptor codes were also exploited in a number of communication scenarios, such as wireless relay channels \cite{4350223}, multiple access channels \cite{7725919}, and Gaussian broadcast channels \cite{8691426}. Raptor codes have been adopted in a number of different standards, such as the 3GPP Multimedia Broadcast Multicast Service and the Internet Engineering Task Force (IETF) Request for Comments (RFC) 6330, for file delivery and streaming \cite[Ch. 3.1]{CIT-060}. 

\subsubsection{Spinal Codes}

Spinal codes are a family of rateless codes proposed in \cite{10114520705622070568} in 2011. They have been proved to achieve the capacity of the BSC and the AWGN channel \cite{balakrishnan2012derandomizing}. The rateless feature allows Spinal codes to be naturally adapted to time-varying channel conditions. Compared to other rateless codes, e.g., Raptor codes, Spinal codes have demonstrated advantages in error performance under different channel conditions and message sizes \cite{10114523776772377684}. Different from conventional algebraic coding, Spinal codes employ hash functions and random number generator (RNG) functions.
\\

First, an $n$-bit message $\boldsymbol{m}$ is divided into $n/k$ $k$-bit segments $\boldsymbol{m} = [\boldsymbol{m}_1,\ldots,\boldsymbol{m}_{n/k}]$. Next, the encoder applies a hash function $h(\cdot)$ to sequentially map the message segment to a $v$-bit hash state as
\begin{equation}
\boldsymbol{s}_i = h(\boldsymbol{s}_{i-1},\boldsymbol{m}_i),\; i = 1,\ldots, n/k, 
\end{equation}
where $\boldsymbol{s}_0 = \boldsymbol{0}$ is the initial hash state known by both the encoder and decoder. Then, the $v$-bit has state is input to the RNG function as a seed to generate a pseudo-random sequence of length $c$
\begin{equation}
\text{RNG}: \boldsymbol{s}_i \rightarrow \boldsymbol{x}_{i,j}\in\{0,1\}^c, \; j = 1,2,3,\ldots.
\end{equation}
The Spinal encoder then maps the $c$-bit sequence to a channel input set $\Psi$ to fit the channel characteristics:
\begin{equation}
f: \boldsymbol{x}_{i,j}  \rightarrow \Psi,
\end{equation}
where $f$ is a constellation mapping function.\\

Due to the introduction of a hash function, the decoding of Spinal codes usually has a higher complexity than other classes of rateless codes. By leveraging the tree structure of Spinal codes, a tree pruning decoding algorithm called bubble decoding was proposed for Spinal codes \cite{10114523776772377684}. It has a decoding complexity of $O(nB2^k(k+\log_2 B +v))$, where $B$ denotes the pruning width. To reduce the decoding complexity, a forward stack decoding algorithm was proposed in \cite{7113798}, which divides the decoding tree into several layers and searches the decoding paths in each single layer without going back to the previous layer. However, the complexity reduction is limited. Yet, the development of low-complexity and high-performance decoding algorithms for Spinal codes remains to be a challenging problem.

\subsection{Lattice Codes}\label{sec:lattice}
Lattice codes, constructed from lattices, serve as the Euclidean counterpart of binary linear codes \cite{zamir2014lattice}. 
They possess numerous favorable properties and elegant mathematical structures \cite{conway1999sphere}. The motivation for employing
lattices in channel coding is due to the fact that many elegant properties of the lattices can be carried over to solve practical engineering problems. To preserve lattice symmetry and save complexity, lattice decoding which finds the closest lattice point while ignoring the decision boundary of the code \cite{1019833}, is often used rather than the maximum-likelihood decoding. Remarkably, it was first proved in \cite{1337105} that there exists a sequence of lattice codes that can achieve the AWGN channel capacity under lattice decoding. 

In addition to information-theoretic analysis, another line of research is dedicated to the construction of practical lattice codes with capacity-approaching performance. In general, there are two main approaches. The first involves constructing lattice codes directly in the Euclidean space, e.g., low-density lattice codes (LDLC) \cite{4475389} and convolutional lattice codes (CLC) \cite{5961819}. Another approach is to adapt modern capacity-approaching channel codes to construct lattices, i.e., construct lattice codes from LDPC codes \cite{1705007,Boutros16}, IRA codes\cite{8066336}, and polar codes \cite{8492454}. Their construction methods are based on \cite{conway1999sphere}: 1) Construction $A$; constructing lattices based on a linear code; 2) Construction $D$: constructing lattices based on the generator matrices of a series of nested linear codes; and 3) Construction $D'$: constructing lattices based on the parity check matrices of a series of nested linear codes. These methods allow one to construct lattice codes not only with good error performance inherited from capacity-achieving linear codes, but also having relatively lower construction complexity compared with LDLCs and CLCs. The concepts of lattices have also been used to design constellations with large shaping gains and achieving the full diversity of the Rayleigh fading channel \cite{485720,Oggier:2004:ANT:1166377.1166378} and to design full-rate full-diversity space-time coding for MIMO channels \cite{CIT-016}. 

In addition to point-to-point channels, lattice codes have been shown to outperform Gaussian random codes in multiuser communications and interference management. It is worth noting that one of the key properties of lattice codes is that a linear combination of multiple lattice codes is still a lattice code. Leveraging this property, \cite{6034734} proposed compute-and-forward (C\&F) relaying strategy based on nested lattice codes by harnessing structural interference to achieve significantly higher rates than conventional amplify-and-forward and decode-and-forward relaying strategies. Building on the C\&F framework, lattice network coding has been proposed and studied in \cite{6516165,6582523,Tunali15}. Multiple access based on lattice partition was proposed and investigated in \cite{8291591,8517129,8731926,9535131}, which enables low-complexity treating interference as noise decoding for interference management as opposed to complicated successive interference cancellation decoding. In all of the aforementioned works, higher-dimensional lattices are required to attain the optimal performance, whose implementation complexity scales with dimension. 

\subsection{Sparse Regression Codes (SPARCs)}
SPARCs or sparse superposition codes were first introduced in \cite{6142076,6657788}, which have been proved to achieve capacity over the AWGN channel under maximum-likelihood decoding and adaptive successive decoding, respectively. SPARCs step back from the coding/modulation divide and instead use a structured codebook to construct low-complexity capacity-achieving schemes tailored to the AWGN channel. Later, it has been proved that SPARCs also achieve capacity under approximate message passing (AMP) decoding \cite{9441022}.

\begin{figure}[ht]
				\centering
				\includegraphics[width=\linewidth]{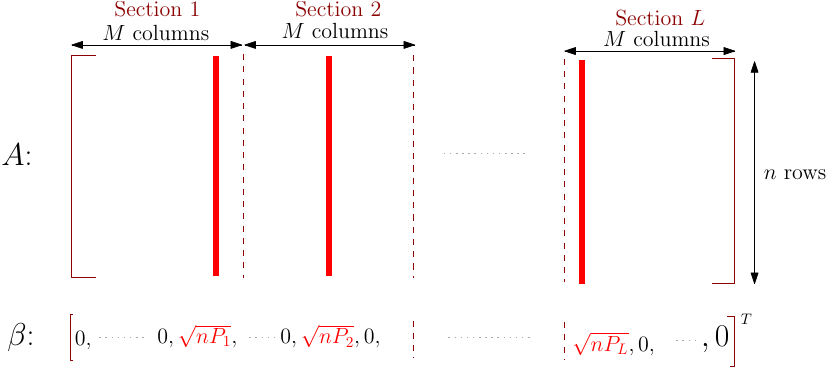}
				\caption{A Gaussian sparse regression codebook of block length $n$ \cite{CIT-092}.} \label{fig:SPARC_1}
			\end{figure}

A SPARC is defined in terms of a design matrix $\boldsymbol{A}$ of dimension $n \times ML$, whose entries are chosen i.i.d. over $\mathcal{N}(0,1/n)$, where $n$ is the blocklength, $M$ and $L$ are integers associated with $n$ and the code rate $R$. An example of a Gaussian sparse regression codebook of blocklength $n$ is shown in Fig. \ref{fig:SPARC_1}. Matrix $\boldsymbol{A}$ consists of $L$ sections with $M$ columns each. 

The SPARC codeword is generated by
\begin{equation}
\boldsymbol{x}=\boldsymbol{A}\boldsymbol{\beta},
\end{equation}
where $\boldsymbol{\beta}=[\beta_1,\ldots,\beta_{ML}]$ is the message vector. It satisfies the following property: there is \emph{exactly one} $\beta_j\neq 0$ for $j=1,\ldots,M$, one $\beta_j\neq 0$ for $j=M+1,\ldots,2M$, and so forth. The non-zero value of $\beta$ in section $l\in\{1,\ldots,L\}$ is set to $\sqrt{nP_l}$, where $P_l>0$ and satisfies $\sum^L_{l}P_l=P$ and $P$ is the average power per input symbol. Since each of the $L$ sections contains $M$ columns, the codebook size is $M^L$. To obtain a rate $R$ code, it is required that
\begin{equation}\label{sparc_rate}
M^L=2^{nR}\Rightarrow R=\frac{L\log M}{n}.
\end{equation}
There are several choices for the pair $(M,L)$ that satisfy \eqref{sparc_rate}. In most cases, $M=L^a$ for some constant $a>0$ \cite{CIT-092}. In this case, the code rate becomes $R = \frac{aL\log L}{n}$. Note that SPARCs are non-linear codes. The choice of $P_l$, known as power allocation, has critical impacts on performance. For example, SPARCs with exponentially decaying power allocation $P_l \varpropto 2^{-2C/L}$ have been proved to be capacity achieving under AMP decoding, where $C$ denotes channel capacity \cite{9441022}. Further optimization of power allocation is required to achieve good finite blocklength error performance.

Several works have improved the original SPARCs. First, notice that matrix $\boldsymbol{A}$ is an i.i.d. Gaussian matrix that may not be suitable for practical implementation. Therefore, the SPARC can be defined via a Bernoulli design matrix with entries that are chosen uniformly at random from the set $\{-1,1\}$ \cite{6776455,7541483}. The capacity-achieving property is still preserved under the optimal ML decoding. In \cite{9433595}, the SPARCs were extended to \emph{modulated SPARCs}, where the information was encoded in both the locations and values of the non-zero entries of $\boldsymbol{\beta}$. To be specific, each non-zero entry of $\boldsymbol{\beta}$ takes values from a $K$-ary constellation. In this case, the code rate of \eqref{sparc_rate} becomes
\begin{equation}
R = \frac{L\log (KM)}{n}.
\end{equation}
Adding modulations introduces an extra degree of freedom in the design of SPARCs, which can be used to reduce the decoding complexity without sacrificing finite-length error performance \cite{9433595}. In addition to power allocation, another way of achieving capacity is by applying spatial coupling \cite{9441022}. Other extensions, such as concatenated SPARCs with different coding schemes, have been shown to be promising in multiuser channels \cite{9432925}.

\section{Other Relevant Topics}
Channel coding is one of the cascaded stages of the transmission process. As the main contribution of coding is to the reliability of the wireless link, along with influences on other KPIs, we need to consider other stages in a communication system.

\subsection{Signaling and Shaping}
Modulation is the process of converting a digital signal into an analog signal that can be transmitted over a communication channel. The choice of modulation scheme can have a significant impact on the reliability and data rate of a communication system. For example, higher order modulation schemes, such as 16-QAM and 64-QAM, can achieve higher data rates compared to low order modulation schemes, such as BPSK and QPSK, but they are more susceptible to noise and interference. 
QAM is widely used in cellular systems, e.g., 2G to 5G systems, because it is relatively simple to modulate and demodulate. However, its constellation points are arranged in equally spaced positions, and the distribution is far from Gaussian. Particularly, in the high SNR regime, QAM has a gap of 1.53 dB to the unrestricted Shannon limit (i.e., not restricted to any signal constellation), which is known as the shaping loss. Amplitude phase shift keying (APSK) is more robust to the non-linearity of the power amplifier (PA) and phase noise than QAM. This makes it a good choice for broadcasting networks and satellite communications. APSK could also be used in 6G systems.

	\begin{figure}[ht]
				\centering
				\includegraphics[width=0.8\linewidth]{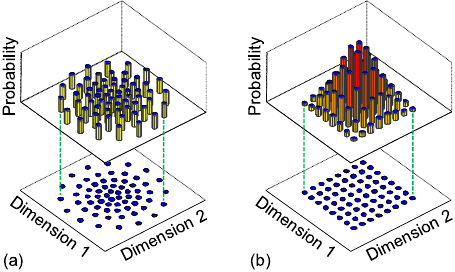}
				\caption{ (a) Geometric and (b) probabilistic constellation shaping \cite{8640810}.} \label{fig:GS_PAS}
			\end{figure}

To close the gap to the unrestricted Shannon limit and to increase the spectral efficiency, signal shaping can be applied. There are two main classes of signal shaping, namely geometric shaping \cite{4557486} and probabilistic shaping \cite{CIT_111}. Illustrations of both shaping methods are shown in Fig \ref{fig:GS_PAS}. In geometric shaping,
the constellation points are arranged in a way to mimic the capacity-achieving signal distribution. Constellations built from lattices (see Section \ref{sec:other}-\ref{sec:lattice}) are an example of geometric shaping. Probabilistic shaping, on the other hand, assigns different probabilities to different constellation points. Compared to geometric shaping, probabilistic shaping builds up on off-the-shelf constellations, hence incurring no additional complexity in system design and implementation.


The best way to improve both the reliability and spectral efficiency of a coded modulation scheme is to jointly design modulation, shaping, and channel coding schemes. This is because the performance of each component depends on the other components. 
However, jointly designing the modulation, shaping, and channel coding schemes is not a trivial job, in particular for a cellular system, where we need to adapt the coding and modulation parameters with the channel condition.

\subsection{The Choice of Waveform} 

Waveform design can play a significant role in improving the KPIs of cellular networks. Some of the ways that waveform design can improve KPIs include: 1) Increased data rates: Waveform design can be used to increase the data rates of cellular networks by increasing the spectral efficiency of the waveforms. This can be done by using techniques such as orthogonal frequency-division multiplexing (OFDM) and its variants. 2) Sensing resolutions: Waveform design can be used to improve the sensing performance of cellular networks by exploiting its time-frequency localization. This can be done by using techniques such as shorter waveforms, lower overhead, and the associated simplified detection algorithms.
Improved coverage: Waveform design can be used to improve the coverage of cellular networks by making the waveform more robust to interference and fading. This can be done by using techniques such as multiple-input multiple-output (MIMO) and beamforming. 3) Increased energy efficiency: Waveform design can be used to increase the energy efficiency of cellular networks by reducing the power consumption of generating the waveforms. This can be done by using techniques such as adaptive power control and waveform selection.

Some of the multi-carrier waveform candidates for downlink in 6G cellular networks include: 1) Orthogonal frequency division multiplexing (OFDM) \cite{nee2000ofdm}: OFDM is a waveform that has been used in cellular networks since 4G. It is also a good candidate for 6G networks as it is efficient in terms of spectral efficiency, power consumption, and easy to integrate with MIMO. However, it has high out-of-band emission (OOBE) and peak-to-average power ratio (PAPR), and it performs poorly at high mobility. 2) OFDM variants such as filter bank multi-carrier (FBMC) modulation \cite{siohan2002analysis}:  FBMC achieves good frequency-domain localization by increasing the pulse duration in the time domain and using carefully designed pulse shaping filters. Among the variants of FBMC, offset quadrature amplitude modulation (OQAM–FBMC), is preferred due to handling interference while allowing dense symbol placement in the time–frequency plane. Compared to OFDM, it has lower OOBE and PAPR at the cost of higher complexity and larger bit error rate (BER). Other OFDM variants also suffer from some disadvantages, such as inter-symbol interference (ISI) due to lack of cyclic prefix (CP), challenging MIMO integration, higher complexity, and therefore larger latency.
3) Orthogonal time frequency space (OTFS) modulation \cite{hadani2017orthogonal} and its improved variant orthogonal delay-Doppler modulation (ODDM) \cite{lin2022orthogonal}: OTFS/ODDM or a general delay-Doppler multicarrier modulation (DDMC) \cite{lin2023multicarrier} are relatively new waveforms that improve robustness in environments with high-frequency dispersion by processing the signal in the delay-Doppler domain where the signals are sparse. Nevertheless, this advantage comes with a relatively high complexity cost. 
Researchers are still working to improve the discussed waveforms and investigate further as candidates for 6G cellular networks. A combination of different waveforms will likely be used in 6G networks to meet the diverse requirements of different applications.


\subsection{Integration with Non-Terrestrial Networks and Laser Links}
In the pursuit of ubiquitous connectivity for 6G networks, integration of non-terrestrial networks (NTNs) \cite{giordani2020non} such as unmanned aerial vehicles (UAVs), satellites, and other high altitude platforms (HAPs) such as balloons with terrestrial networks, along with the utilization of alternative communication technologies such as free space optical (FSO) links \cite{jeon2023free} (a.k.a. laser links) for backhaul/fronthaul becomes imperative. Note that FSO is also an option for point-to-point communications in terrestrial networks; however, due to weather dependency, limited range (as line-of-sight is required), challenges for meeting alignment requirements, etc., they have not been employed. FSO links have already been used for inter-satellite communications and could be an option for linking HAPs to gateways of terrestrial networks. The integration of terrestrial and non-terrestrial networks addresses the limitations of the frequency spectrum available for terrestrial communications, as well as the high data rates envisioned for 6G. The inherent challenges of spectrum congestion and limited bandwidth on earth necessitate the exploration of non-terrestrial alternatives to expand coverage and increase capacity. However, these non-terrestrial technologies may require different design considerations for channel coding and modulation due to factors such as atmospheric conditions, propagation delays, and varying link characteristics due to side effects (namely, absorption, scattering, and turbulence) of the atmospheric channel, which results in fluctuations of the received signal intensity. To model the received intensity distribution according to the turbulence levels, probability density functions such as lognormal, negative exponential, and, in particular, gamma-gamma (G-G) are used. Adaptation of current error correction codes and modulation techniques must account for these factors to ensure reliable and efficient communication over non-terrestrial links. Potential candidates for such adaptations include advanced coding schemes like LDPC codes, polar codes, and turbo codes, which offer robust performance in challenging environments and can be tailored to meet the requirements of non-terrestrial communication channels for both radio frequency (RF) and optical communications. Limited investigation \cite{sandalidis2010coded,chochol2015evaluation} has been carried out on channel coding for FSO where LDPC codes \cite{dixit2020analytical}, rateless codes \cite{abdulhussein2010rateless}, and Reed-Solomon codes \cite{kumar2020impact} have been used. Nevertheless, due to the characteristics of the atmospheric channel, improving channel modeling, channel estimation approaches, equalization methods, and waveform design seem to contribute more than channel coding schemes in link performance. Overall, the integration of non-terrestrial networks and innovative communication technologies presents an opportunity to address the scalability and performance challenges of 5G networks while paving the way for seamless connectivity in diverse environments.

\section{Summary and Conclusion}
We reviewed the specifications of the mobile communication standards from 1G to 5G, in particular the channel coding schemes employed in every generation, the envisioned 6G requirements, and the potential contributions of channel coding to meet these requirements. We then reviewed Turbo, LDPC, and polar coding schemes, their variants, and recent advances. The comparisons were made in terms of error correction performance and the performance of hardware architectures designed for different decoding algorithms.
We also considered other coding schemes, such as fountain codes and lattice codes, and other considerations in the physical layer, such as modulation and waveform.

In our opinion, turbo codes, polar codes, and LDPC codes will remain strong contenders for the next generation of mobile communication standards. The reasons are twofold: 1) These three coding schemes are well established and investigated. Turbo and LDPC codes for code lengths required for data channels are capable of outperforming other codes at reasonable complexity. Similarly, polar codes outperform other codes for the code lengths required for control channels. Based on previous trends, it appears that we need more than the span of one generation of mobile communications, more than a decade, to develop a well-investigated and mature coding scheme for practical applications. 2) Adapting a new coding scheme can be challenging due to the need for standardization and interoperability. The introduction of new coding techniques can require significant modifications to the existing standard, which is costly. Furthermore, the efficient implementation of new codes on mobile devices may present practical challenges. Optimizing the software and hardware for efficient code implementation requires careful consideration of the device's capabilities and constraints.

However, recent advances and promising directions in existing coding schemes, as discussed in Sections \ref{sec:turbo}-\ref{sec:polar}, could lead to some modifications of existing standards. In particular, the spatially coupled and non-binary variants of turbo, LDPC, and polar codes could play a role in some applications.

Furthermore, as discussed in Section \ref{sec:intro} and observed in Section \ref{sec:1g_5g} in particular Table \ref{tab:KPI_tech}, the contribution of coding schemes in improving KPIs is more significant in enhancing reliability, throughput, and coverage. However, these KPIs and others that channel coding has less or no impact on them, can be improved by other components of transmitter/receiver, network architecture, and network management. On the other hand, the decoding process could be the bottleneck of a receiver and further improvement of the decoding algorithm for lower complexity towards terahertz communications is necessary and it is as important as improving the reliability.

\bibliographystyle{IEEEtran}
\bibliography{Ref,refs_ldpc_new_style,Min}




\begin{IEEEbiography}[{\includegraphics[width=1in,height=1in,clip,keepaspectratio]{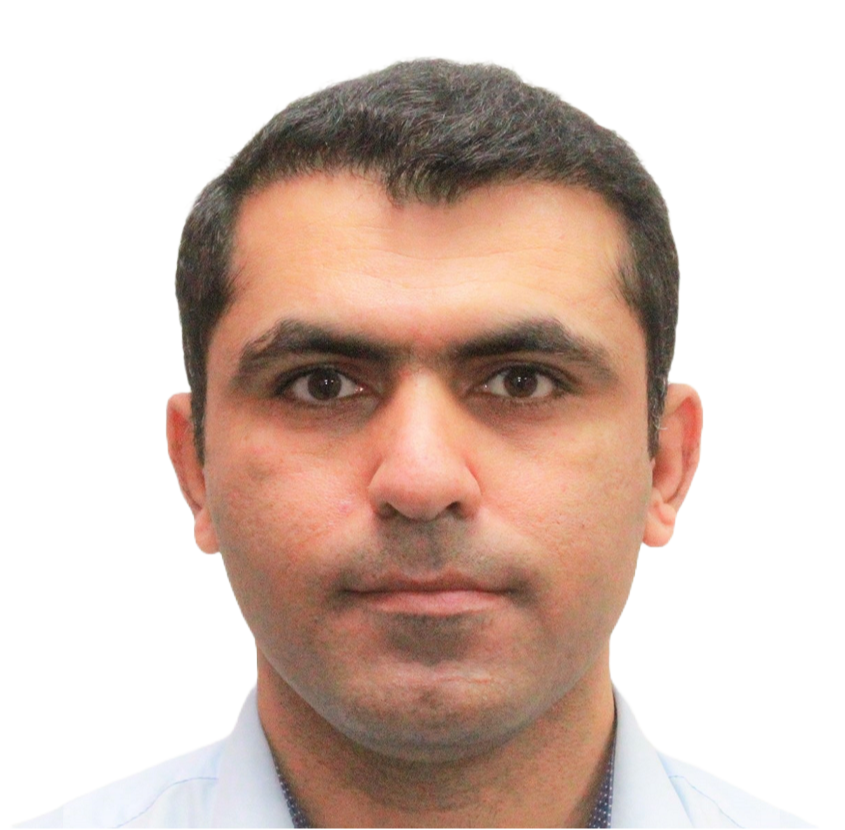}}]%
{Mohammad Rowshan } (S'13-M'22) is currently an Engineering ECA Fellow (Associate Lecturer) at the School of Electrical Engineering and Telecommunications at the University of New South Wales (UNSW) in Sydney, Australia. 
He received his B.Eng. (Hons), M.Sc., and Ph.D. degrees in electrical engineering from the University of Nottingham in 2015 (ranked 1), the Hong Kong University of Science and Technology (HKUST) in 2016, and Monash University in 2021, respectively. 
His research interests include channel coding, signal processing for communication systems, and hardware architecture design.
\end{IEEEbiography}
\begin{IEEEbiography}[{\includegraphics[width=1in,height=1in,clip,keepaspectratio]{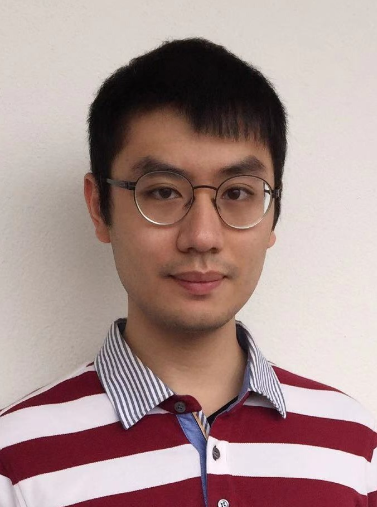}}]%
{Min Qiu } (Member, IEEE) received his Ph.D. degree in Electrical Engineering from the University of New South Wales (UNSW), Sydney, Australia, in 2019. He is a Postdoctoral Research Associate with UNSW. He received the Australian Government Research Training Program Scholarship for the duration of his Ph.D. degree, the Australia Awards-Endeavour Research Fellowship in 2018, and the Chinese Government Award for Outstanding Self-Financed Students Abroad in 2019. He was honored as the Exemplary Reviewer of the IEEE TRANSACTIONS ON COMMUNICATIONS in 2018, 2019, 2021, and 2022, and the IEEE COMMUNICATIONS LETTERS in 2021-2023. His research interests include channel coding, information theory, and their applications for building reliable communication systems.
\end{IEEEbiography}
\begin{IEEEbiography}[{\includegraphics[width=1in,height=1in,clip,keepaspectratio]{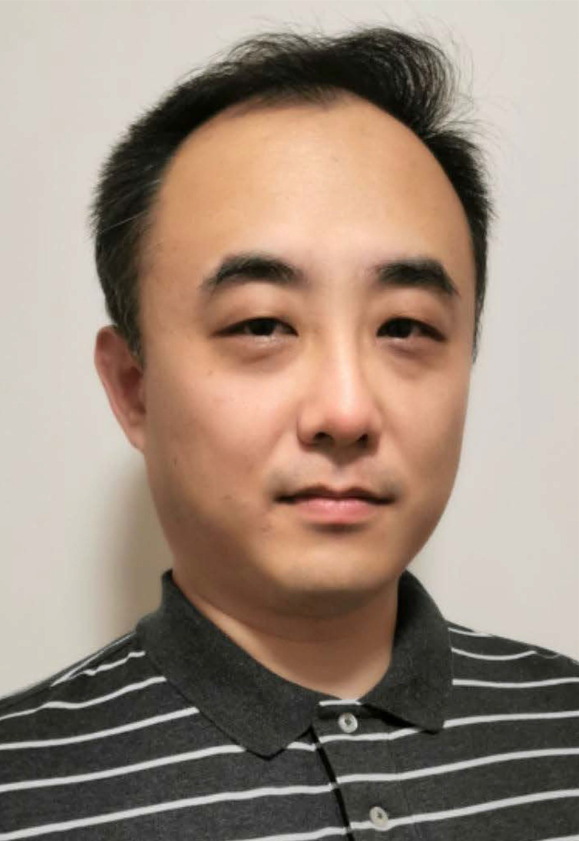}}]%
{Yixuan Xie } (Member, IEEE) earned his Ph.D. in Electrical Engineering from the University of New South Wales (UNSW), Sydney, in 2016. Since then, he has been a pivotal figure, focusing on FPGA implementation of channel codecs and prototyping communication schemes on software-defined radio platforms at the Wireless Communication Laboratory at UNSW Sydney. Currently, he is with the School of Electrical Engineering and Telecommunications at UNSW Sydney. His research focuses on error control coding, iterative detection/decoding methods, digital communication systems and signal processing, and multiple access schemes.
\end{IEEEbiography}
\begin{IEEEbiography}[{\includegraphics[width=1in,height=1in,clip,keepaspectratio]{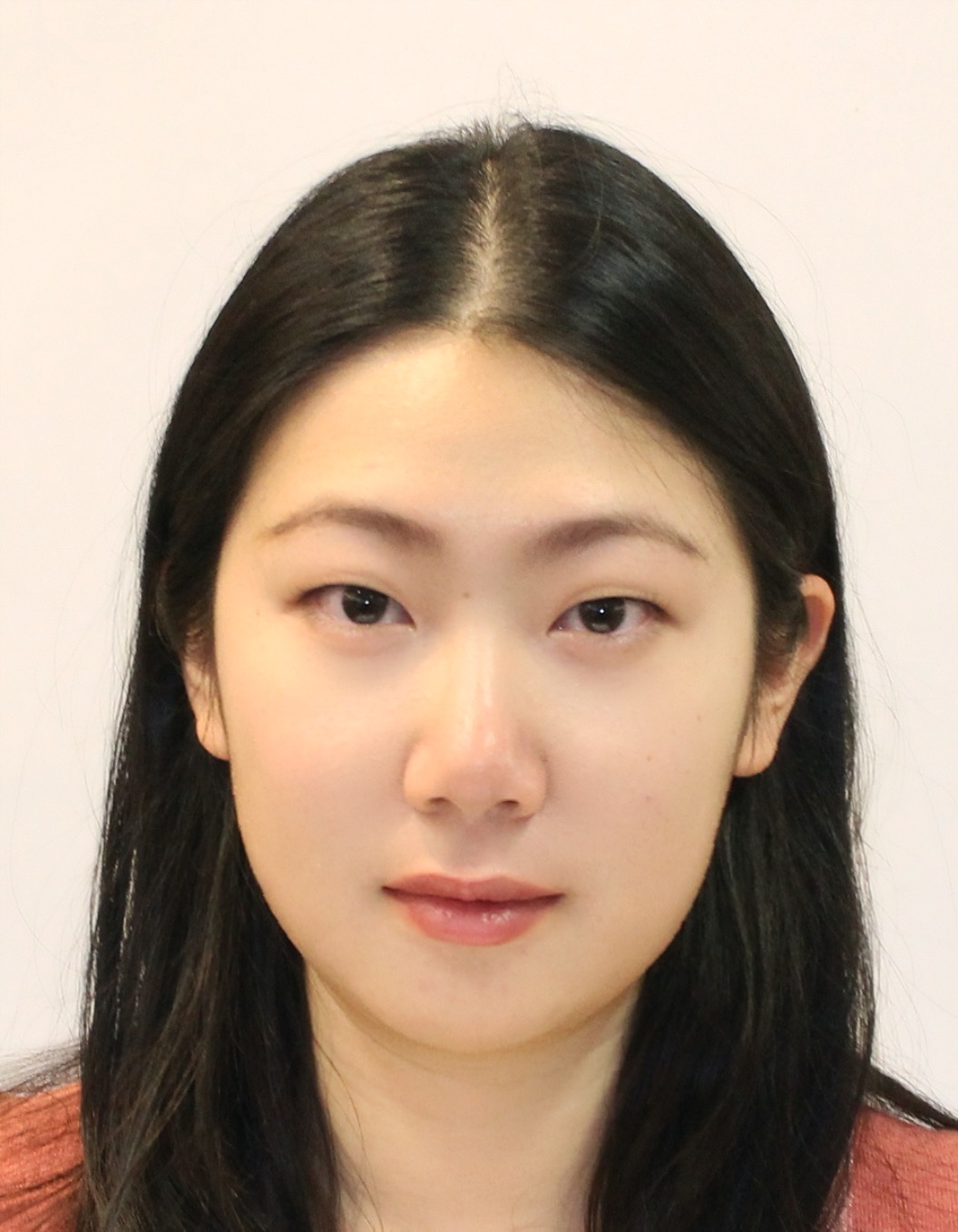}}]%
{Xinyi Gu } (Graduate Student Member, IEEE) received the B.E. and M.Phil. degrees in electrical engineering from the University of New South Wales (UNSW), Australia in 2020 and 2023, where she is currently pursuing her Ph.D. degree with the School of Electrical Engineering and Telecommunications.  Her research interests include channel coding and hardware architecture design for decoders.
\end{IEEEbiography}
\begin{IEEEbiography}[{\includegraphics[width=1in,height=1in,clip,keepaspectratio]{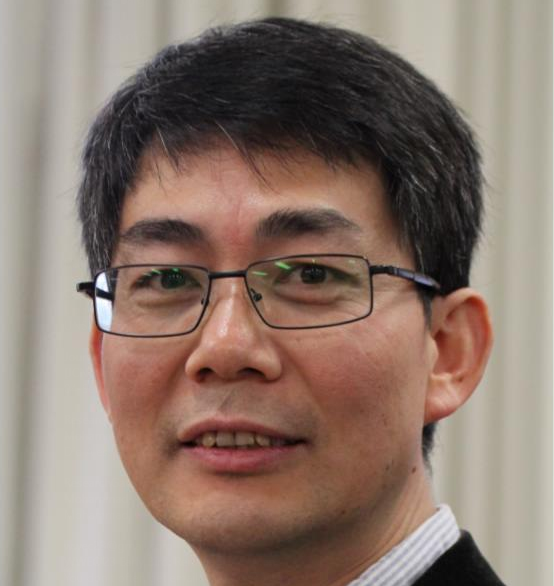}}]%
{Jinhong Yuan } (M'02-SM'11-F'16) received the B.E. and Ph.D. degrees in electronics engineering from the Beijing Institute of Technology, Beijing, China, in 1991 and 1997, respectively. From 1997 to 1999, he was a Research Fellow with the School of Electrical Engineering, University of Sydney, Sydney, Australia. In 2000, he joined the School of Electrical Engineering and Telecommunications, University of New South Wales, Sydney, Australia, where he is currently a Professor and acting Head of School and Head of Telecommunication Group with the School. He has published two books, five book chapters, over 300 papers in telecommunications journals and conference proceedings, and 50 industrial reports. He is a co-inventor of one patent on MIMO systems and four patents on low-density-parity-check codes. He has co-authored four Best Paper Awards and one Best Poster Award, including the Best Paper Award from the IEEE International Conference on Communications, Kansas City, USA, in 2018, the Best Paper Award from IEEE Wireless Communications and Networking Conference, Cancun, Mexico, in 2011, and the Best Paper Award from the IEEE International Symposium on Wireless Communications Systems, Trondheim, Norway, in 2007. He is an IEEE Fellow and currently serving as an Associate Editor for the IEEE Transactions on Wireless Communications and IEEE Transactions on Communications. He served as the IEEE NSW Chapter Chair of Joint Communications/Signal Processions/Ocean Engineering Chapter during 2011-2014 and served as an Associate Editor for the IEEE Transactions on Communications during 2012-2017. His current research interests include error control coding and information theory, communication theory, and wireless and underwater communications.
\end{IEEEbiography}

\end{document}